\documentclass{article}
\usepackage{amsfonts}
\usepackage{geometry}
\usepackage{ragged2e}
\usepackage{fancyhdr}
\usepackage[subfigure]{tocloft}
\usepackage{sectsty}

\partfont{\centering} 
\makeatletter
\newcommand{\l@subfigure}{}
\newcommand{\l@subtable}{}
\makeatother

\renewcommand{\thepart}{\Alph{part}}
\newtheorem{theorem}{Theorem}

\newtheorem{case}[theorem]{Case}

\newenvironment{principle}{\bigskip\noindent\hangindent= 0cm}{\bigskip}
\input{tcilatex}

\begin{document}

\title{Quantum General Relativistic Framework 5: The Physical Foundations of
\ Unconsciousness and Consciousness Matter with Retrocausality}
\author{SURESH K MARAN \\
CEO, Founder, UNITESERVE\\
www.qstaf.com\\
www.sureshmaran.com\\
www.linkedin.com/in/sureshmaran\\
\textit{\copyright\ 2013 - 2026 Suresh K\ Maran. }}
\date{ \textit{All rights reserved. }}
\date{May, 14, 2024 Basic Concepts, \\
Further Revisions by December 2024, and again by March 2026 (details
inside). \\
Sharing and Usage Governed by CC BY-NC-ND \\
Please see change log and license info in foot note 1\\
This document has final versions only.}
\maketitle

\begin{abstract}
This is the fifth update for the quantum gravity framework project. The
research work has been split into two parts: Part A deals with unconscious
matter, while Part B deals with conscious matter. I introduce retrocausality
in both. The separate parts are given in this paper with their own
abstracts. Part A, based on the first five papers, I finalize a three-set of
fundamental postulates for understanding time, quantum reduction, and
consciousness from quantum general relativistic concepts. Then I introduce a
new reduction theory that includes retrocausality and spontaneous reduction,
and further concepts to include conscious and unconscious influence. Then I
discuss in mathematical detail the three postulates in the quantum general
relativistic framework. Toward the end, I discuss sample calculations and
experimental verifications. I summarize various calculations done in the
previous versions of the Quantum General Relativistic Framework. In Part B
of this paper, I propose a set of ideas to explain the phenomena of life and
consciousness in the universe and their relation to the interpretation of
quantum mechanics using my past papers and other important work.
Consciousness is described as a result of specific dynamical configurations
of matter and fields aided by fundamental properties of nature proposed in
this paper. First, I review the various theories of consciousness in current
literature and come up with various hypotheses, definitions, and
propositions to build a physical theory of consciousness. I propose three
sets of postulates on the physical foundations of consciousness from a
quantum general relativistic background, using the proposals in Quantum
General Relativistic Framework 5, part A, to understand time and reduction.
Various new concepts are introduced. The concept of classicum is introduced
for the first time as an analog of quantum, and its relation to conscious
being is discussed. Mathematical models for various ideas are discussed. I
discuss theory and experiments to verify these.
\end{abstract}

\tableofcontents

\pagebreak

\part*{ Part A \hspace{1em}Postulates on Quantum Evolution and Reduction
with Retrocausality for Unconscious Matter (v3)}

\addcontentsline{toc}{part}{Part A ~\thepart\hspace{1em}Postulates on Quantum Evolution and Reduction with Retrocausality for
Unconscious Matter (v3)}

\begin{center}
\begin{minipage}{0.75\textwidth}
 \centering
v1 Date Completed:\ May 4, 2024, Time stamped May 14, 2024,\\
Futher updates v2 (minor) and v3 (concepts added). \\
Final version v3 March 24, 2026.\\
\begin{justify}
Abstract: Please see change log in note 1.
This is the fifth update for the quantum gravity framework project. Based on
the first five papers, I finalize a three-set of fundamental postulates for
understanding time, quantum reduction, and consciousness from quantum
general relativistic concepts, within the canonical quantum gravity
framework for unconscious matter. At the beginning, I review the history of
quantum measurement research and the models and ideas proposed, in the
context of the quantum gravity framework project. Then I\ introduce a new
reduction theory that includes retrocausality and spontaneous reduction, and
further concepts to include conscious and unconscious influence. Then I
discuss in mathematical detail the three postulates in the quantum general
relativistic framework. Toward the end, I discuss sample calculations and
experimental verifications. I summarize various calculations done in the
previous versions of the Quantum General Relativistic Framework. In the
companion paper on the physical foundations of consciousness, complementary
sets of principles are introduced to explain consciousness using the
concepts in this paper. Further advancement of this project requires
theoretical and experimental study which are discussed in this paper. It is
noted that the proper analysis of quantum gravity predicts potential
retrocausal effects. The mathematical formulation of retrocausality is
discussed. Many of the wordings in this update are different from the
previous versions.
\end{justify}
\end{minipage}
\end{center}

\newpage

\section{Introduction\label{1}}

\bigskip Please read 5v3 for all inclusive discussion. Please see change log
in this footnote\footnote{%
Sharing and Usage Governed by CC BY-NC-ND
\par
License Link: https://creativecommons.org/licenses/by-nc-nd/4.0/
\par
Official Link: qstaf.com/qgrf5
\par
Please read version 1.3 only which has all the latest updates. Change Log:
Due to numerous serious engagements I have, I often end up delaying
publication by drifting into other projects. For more information on contact
the author or visit www.qstaf.com/qgrf5-publishing for proof of dates and
explanation.
\par
\begin{itemize}
\item Version 5v1: Date Completed May 4 2024, and Time stamped May 14, 2024:
(5v1, File version GR5.4-Part-A-v29). This version can be skipped for 5v2.
\par
\item Version 5v2: minor updates by 10-December-2024 (5v2, GR5.5-PART A
-v34) and Time Stamped of August 2 2025.
\par
\item Version 5v3: concept added in section 2.3, and some sections revised
on March 24, 2026 (5v3, GR5.5-PART A -v36).
\par
\item All the versions will be available for reference in Academia.edu,
ResearchGate.net and Publications.UniteServe.com and other places listed at
qstaf.com/qgrf5.
\end{itemize}
}

This research is part of the quantum gravity framework project involving the
creation of a conceptual framework to bring together the necessary concepts
at the foundations of science for the unification of general relativity and
quantum mechanics, and help understand life and consciousness. This proposal
is put forward to help as foundations for experiments and further research
that will help to discover the proper scientific concepts. This paper is the
part A of the 5th revision of the quantum gravity framework, \cite{MYP5A},
and the paper 'The Physical Foundations of Life and Consciousness, Framework
1' \cite{MYP5B} is part B of the work.

Beginning in 2013, I started publishing a sequence of papers, discussing the
heuristic set of ideas. As we will see this has evolved into sufficient
ideas for rebuilding the foundations of sciences at the fundamental level.
First I started focusing on solving the problem of time in quantum gravity.
Then it slowly evolved to include measurement problem in quantum mechanics
that involves understanding quantum reduction. The whole project was done in
the context of canonical quantum general relativity to deal with time and
quantum reduction as an integral part of dealing with conceptual problems
with it. Simple applications have been included in various papers.
Eventually, life, consciousness, and its role in structure formation have
been included.

In this update, I will focus on the physical aspects of the foundations of
science, both living and non-living. We will refine and review the basic
ideas to understand physics and life in the universe based on the previous
versions of the quantum gravity framework project. This involves defining
concepts needed to understand time, quantum reduction, decoherence,
foliation, smoothness, consciousness, and life. In `The Quantum General
Relativistic Framework, 5.0', I focus on the physical aspects of matter,
space, and time. In a companion paper, `The Physical Foundations of Life and
Consciousness, Framework 1' \cite{MYP5B}, I discuss the physical concepts
necessary for understanding life and consciousness based on ideas in this
paper. In a separate paper \cite{MYP5C}, I summarize the ideas in `The
Quantum General Relativistic Framework 5.0', and `The Physical Foundations
of Life and Consciousness, Framework 1'.

First, I will bring together various ideas that were discussed in the
previous versions: The Quantum Gravity Framework 0 to 4 \cite{MYP0}, \cite%
{MYP1}, \cite{MYP2}, \cite{MYP3}, and \cite{MYP4}. First I discuss the
overall introduction to various conceptual issues without using the rigorous
mathematical terms needed to include quantum general relativistic
description. In this, I will discuss various important ideas existing to
tackle the issue of time in quantum gravity and quantum measurement and
clarify the various essential elements. Then I will go into mathematical
formalism and rigorous mathematical structures. Then we will discuss the
overall features and conceptual aspects. Then we will discuss the
experimental study of the various ideas proposed.

The physics that we can infer from experiments support the theory of general
and special relativity, the standard model of particle physics, and quantum
field theory. When we put all this together all we get is the
pseudo-Riemannian 4D quantum block universe \cite{Block}. In part A, we
discuss in systematic detail the set of three fundamental postulates that
create the 3+1 formulation of this universe. This universe has all possible
3D semi-classical histories that conscious observers can experience.
Understanding time requires explaining time-parameterized 3D semi-classical
reality as experienced by conscious observers. They not only experience
reality but also react back and modify the physics of space-time and matter
around them. The physical part is done in this paper and the conscious part
is done in the companion paper \cite{MYP5B}.

The general relativity doesn't give a preferred foliation of the 4D
pseudo-Riemannian 4D quantum block universe. We introduce the rest frame
foliation as a probabilistic postulate (introduced first in \cite{MYP2}). In
this foliation, quantum fields are the least changing. I consider this as
the link between general relativity to the Newtonian 3D universe
parametrized by time. Together all these proposals give probabilities for
the 3D semi-classical time parametrized history the conscious observers of
the universe experience. This is different from Everett's many worlds
theory. In this every time a macroscopic superposition of a 3D universe is
created, both universes come into existence along with copies of conscious
observers for each one. But in the proposals, we formulated the conscious
observers just go into one possible history with probabilities given by the
postulates in the two papers. The physics as given by the postulates gives
all possible future histories and their probabilities that the conscious
observers can experience. The conscious observers themselves remain
unchanged or modified or copied, except for the change of information held
by their neural network in which they reside.

In this paper, I don't discuss the smoothness postulate proposed in the
previous versions of the quantum gravity framework project, and I am
assuming that quantum fields of physics are embedded in a smooth manifold.
This postulate which was included as a fourth postulate, was proposed to
describe the smoothness of the physics given the randomness contribution due
to continuous reduction. This fourth postulate to formulate the physics of
quantum gravity will be discussed in future versions of the quantum gravity
framework project.

In the companion paper I\ apply the ideas in this paper to understand
consciousness. The effect of conscious elements in physical reality results
in promoting relational structures. The physics of consciousness acts as a
bridge between Social Sciences and Physical Sciences. I published a paper to
discuss the merger of social sciences and physical sciences \cite{GUK}.
Based on this proposal organizing human knowledge has been discussed \cite%
{OHK}. In general, human knowledge is all about the knowledge of relational
structures in the universe. In the companion paper, I lay the foundations
for understanding relational structures and their connection to the basic
conceptual framework for quantum gravity.

I believe the conceptual framework is ready for application and testing.
Together we have sufficient foundations to take the next major step in the
advancement of sciences. Further advancement requires, nailing down the
mathematical details, working out the proposed physical constants and
precise forms of action terms, experimental verifications, and applications
to solving problems at various regimes of both physical and social sciences.
The quantum gravity framework project is part of the project for rebuilding
the foundations of science, society, and economics \cite{RFS}.

\textit{Unless otherwise specified }$H\ $\textit{corresponds to Hamiltonian
constraint only. Also, I\ am going to assume }$\hbar =1$\textit{,}$~G=1$%
\textit{, and }$c=1$\textit{\ unless otherwise specified. Variables }$q$%
\textit{\ and }$p$\textit{\ relate to free variables and conjugate momenta
of discrete system, and }$\mathcal{\phi }$\textit{\ and }$\mathcal{\pi }$%
\textit{\ stand for fields and their conjugate momenta. }Greek letters are
used as internal field indices for tensor, spinor, fermion, boson, and gauge
fields, while Latin letters are used as indices for non-field space-time
coordinate vectors

\pagebreak

\section{Overview}

\subsection{Physics Introduction}

\subsubsection{Time}

There are two conceptual issues in quantum gravity: 1) The problem of time
and 2) the quantum measurement problem. These two are needed to be properly
addressed to formulate the proper theory of dynamical evolution in quantum
gravity. Let me start with the problem of time:

\begin{equation}
H\left\vert \Psi \right\rangle =0  \label{HamCon}
\end{equation}

This is the fundamental dynamical equation that comes out of canonical
quantum gravity. But in this, there is no time derivative operator, as $t$
the time variable is a coordinate and it is not a physical variable in
general relativity. In general, time is a variable that refers to the
physical state of the clock. The most obvious choice for this variable for a
physical system is the configuration variable dual to its classical
conjugate momentum. This is what I\ call the \textbf{self-time} in \cite%
{MYP0} and \cite{MYP1}. If $x^{i}$ and $p_{i\text{ }}$are the dynamical
variables of an isolated system, then the physical time parameter is $T=x^{i}%
\bar{p}_{i}$, where $\bar{p}_{i}$ is the unit vector along $p_{i}$. In the
case of a quantum system, we need to take the expectation value of its
conjugate momentum $p_{i}=\left\langle \Psi \right\vert \hat{p}%
_{i}\left\vert \Psi \right\rangle $. Since the momentum keeps changing so is
the time direction. This means the physical time parameter can be defined
heuristically by $T=\int dx^{i}\bar{p}_{i}$, where $dx^{i}$ is the change in
the expectation value of the position. This definition looks straightforward
but implementing this rigorously requires foliating the configuration space
of the system.

In the general time evolution postulate that we will propose in this paper,
for a general Hamiltonian constrained system with N degrees of freedom, we
can foliate the phase-space into sequences of N-1 dimensional hyperplanes
such as we do in canonical quantum gravity. Parametrizing the continuous
sequence of the unique hyperplanes by a parameter $t$ can define time. The
details of general evolution for a single-point N-dimensional system were
discussed in \cite{MYP1}. For 4D space-times, we heuristically defined the
generalization of this in \cite{MYP1}, and in further updates. In this
paper, we will do this more rigorously while doing the detailed mathematical
formalism.

\subsubsection{Foliation Dependence and the Rest frame foliation\label{2.1.2}%
}

If we choose the internal foliation and its mapping of hyperplanes to the
time parameter such that the wave function is always at rest, then this is
the rest-frame foliation discussed in \cite{MYP1}.

In dealing with time in 3+1 space-time, we need one of them for each point.
If we have a quantum state $\left\vert \Psi \right\rangle $ of all physical
fields on a 3 manifold, the expectation value of the conjugate momenta of
all the fields at a particular point gives rise to the direction of the
self-momentum to calculate time in self-time formalism as we discussed
before. If there is one time parameter for each point then different rates
of evolution along each of these parameters give rise to different ways to
foliate the space-time manifold.

We need to choose a preferred foliation for the quantum reduction. This is
because the quantum reduction which will be discussed next is a non-linear
process, and depends on the foliation. This is why I\ have defined what's
called the rest frame foliation in quantum gravity framework 2 \cite{MYP2}.

I will discuss now why quantum reduction, depends on the foliation of the
space-time manifold.

\FRAME{ftbpFU}{3.2603in}{2.4241in}{0pt}{\Qcb{Quantum measurement in special
relativity is not Lorentz invariant.}}{\Qlb{FigLrntz}}{%
relativedecoherence.png}{\special{language "Scientific Word";type
"GRAPHIC";maintain-aspect-ratio TRUE;display "USEDEF";valid_file "F";width
3.2603in;height 2.4241in;depth 0pt;original-width 6.5544in;original-height
4.8663in;cropleft "0";croptop "1";cropright "1";cropbottom "0";filename
'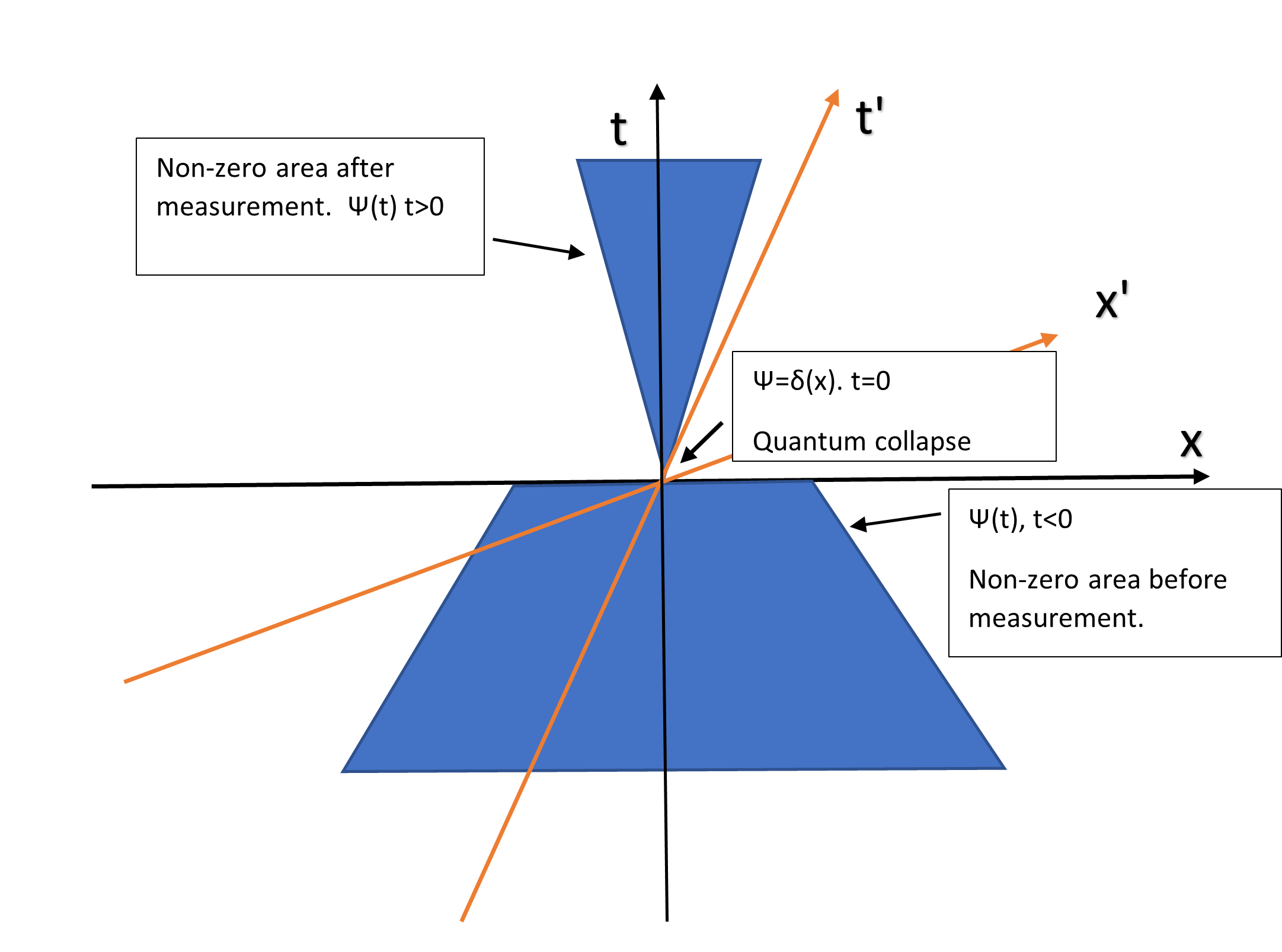';file-properties "XNPEU";}}

The first thing to remember is that when we study a quantum system we study
it with respect to the spatial coordinates of the laboratory and the time in
the lab. This is nothing but the restframe foliation associated with the
laboratory. When quantum particles interact with the laboratory measurement
instruments, the combined system always probabilistically reduces into
semiclassical states. So it appears that the rest-frame foliation is the
foliation in which quantum reduction occurs.

Consider two observers A and B travelling in inertial reference frames with
different velocities. Then the quantum measurement in one of these cannot be
a quantum measurement in a frame-independent manner. Assume the particle is
measured by A. According to A, the wave function gets projected to a certain
point simultaneously in his time. But for B this is not a quantum
measurement in the same spirit as A, as the concept of simultaneity, is
different for B. For B the quantum field looks different. To state this in
an observer-independent manner, we need to say that the quantum reduction is
foliation-dependent and the foliation is dependent on the observer. The
observer is an instrument or human brain, which is made of matter and
fields. This means the foliation depends on the distribution of matter and
fields around the measuring instrument which are at rest with respect to it,
and it leads to the concept of rest frame foliation. The rest frame
foliation can be defined as the foliation in which the fields are least
changing up to a scale factor. In other words, the rest frame (or geodesic
)foliation is in which the pattern of fields is the most preserved. This
kind of foliation is given by the Schwarzschild time in the Schwarzschild
coordinate system for a spherically symmetric scenario and the metric scale
factor in the Robertson-Walker metric in the initial universe. We will make
the various examples of systems discussed in this paper, to be in the
cosmological or static spherical situation and the times and foliation are
defined accordingly.

We will assume the following axiom:

\begin{principle}
Axiom 1: The Time for quantum reduction is determined by the rest frame
foliation
\end{principle}

The rest-frame foliation is also essential for defining time for conscious
observers, which is referred to as the binding problem in neuroscience. The
human brain is extended. Yet we observe the universe in specific time
instants. This requires a 3+1 formulation of the space-time region of the
human brain to define the time instants. Rest frame foliation is the natural
way to define the sequence of these time instants. We will further discuss
this in `The Physical Foundations of Life and Consciousness, Framework 1' 
\cite{MYP5B}.

Later we will discuss mathematically general way to formulate the restframe
foliation. The idea is that any spatial foliation can act as a 3+1 split of
space-time and can serve as the physical basis of time and space. This
foliation is probabilistically chosen such that the higher the measure of
rest frame conditions for the whole space-time in the foliation, the higher
the probability for the foliation to be the physical foliation in which
reduction occurs. This will be mathematically rigorously discussed in the
postulate later in this paper.

\subsubsection{Quantum State Decoherence and Incoherent Macroscopic
Superposition.}

The subject of quantum measurement has been discussed for about a century
since the discovery of quantum mechanics. There have been many attempts to
solve and clarify how to interpret quantum measurement. Let us try to
understand what has been done. Quantum measurement problem involves two
issues: Decoherence and Reduction. Lets first discuss decoherence. Usually,
quantum systems interact with the environment such as background microscopic
radiation, vacuum fields, or a heat bath. If the quantum system is in a
coherent quantum superposition $\dsum\limits_{i}c_{i}\left\vert \Psi
_{i}\right\rangle $,, initially, the coupling with the environment makes the
quantum system decohere, as the coupling results in altering the phase of
wave function randomly, eliminating quantum superposition effects. But the
quantum superposition still exists, only that we cannot see any
superposition patterns as the phases are randomly changing. If the system is
macroscopic, it evolves rapidly into a quantum superposition of many
semiclassical states of different macroscopic configurations.

Usually, macroscopic quantum states decohere rapidly in normal temperatures.
But this doesn't collapse quantum superposition, just that the superposition
is incoherent and we cannot observe any pattern. If the system is made of
many particles the system is evolved in an incoherent superposition of
entangled states. Decoherence is different from reduction, usually, both are
confused with each other. This must be avoided.

Let's consider a gas. Consider the molecules in a gas. They repel each other
when they come close. So the wavefunction of each particle gets localized
after each collision. The wave function of each particle is localized in
between particles and far away from other particles. But this does not
prevent long-range superposition, as this wave function can spread around in
the space between the molecules. So the gas exists as a superposition of
various macroscopic states of different configurations, with random phases.
This is quite easily seen in metals. When atoms come together in metals,
their outer atomic orbitals come together to become bands of different
types. Particular interest is the conduction band in metal and
semiconductors. The electrons occupying these bands are actually in a
non-localized quantum state. There are a huge number of electrons that exist
in incoherent long-range superposition occupying this state as discussed in
standard texts in semiconductors. The incoherence is brought about by
thermal collision among electrons and atoms.

Let me mathematically formulate the incoherent quantum state of a large body
to understand the complexity involved in it. In general, consider a system
of N molecules. This corresponds to simple many-body problems of new
particles. And, for a large number of particles it corresponds to some phase
of matter such as solid, liquid, or gas.

Let $\left\vert \Psi (T)\right\rangle $ be the quantum state of the system,
and $T$ is the time defined in the rest frame foliation. Let for molecules $%
i $, the $r_{i}^{a}$ is the center of mass coordinates of the molecules, $%
l_{i}^{a}$ are the eigenstates of internal degrees of freedom such as
energy, spin, angular momentum, vibrational quantum numbers, etc.\ Also let $%
\left\vert k_{i}s_{i}\right\rangle $ are photons of momentum $k$ and spin
state $s_{i}$ entangled with the system at a given time $T$. Assume that the
particles are trapped in a potential well so that they continue to interact.

Then we have, 
\begin{equation}
\left\vert \Psi (T)\right\rangle =\dsum\limits_{sl}\int_{rk}\Psi
(\{r_{i_{1}}^{a},l_{i_{2}}^{a},k_{i_{3}},s_{i_{4}}\},T)~\dprod\limits_{k}%
\left\vert r_{k}^{a},l_{k}^{a},T\right\rangle ~\dprod\limits_{i}\left\vert
k_{i}s_{i}\right\rangle \Pi dkdrdl,
\end{equation}%
where $\{r_{k}^{a},l_{k}^{a}\}$ denotes all free variables for all
particles. $\left\vert r_{k}^{a},l_{k}^{a},T\right\rangle $ represents the
quantum state of each molecule at a given time $T.$ Let

\begin{eqnarray}
H &=&\dsum\limits_{i}H_{r_{i}}+\dsum\limits_{l}H_{l_{i}}+\dsum_{i\neq
j}H_{I,ij}+H_{EM} \\
&=&H_{r}+H_{l}+H_{I}+H_{\text{em}}
\end{eqnarray}%
denotes Hamiltonian for the whole system.$~H_{\text{em}}$ is electromagnetic
Hamiltonian, $H_{r_{i}}$ is center of mass Hamiltonian for particle $i,$ $%
H_{l_{i}}$ is internal degrees of freedom Hamiltonian for each particle $i,$ 
$H_{I,ij}$ is the interaction Hamiltonian between $i^{th}$ and $j^{th}$
particles interacting through Electromagnetic Fields. $H_{r}=\dsum%
\limits_{i}H_{r_{i}}$, $H_{l}=\dsum\limits_{l}H_{l_{i}}$ and $%
H_{I}=\dsum_{i\neq j}H_{I,ij}$ are sums of the individual Hamiltonians.

The $\left\vert r_{k}^{a},l_{k}^{a},T\right\rangle $ are eigen states of $%
H_{l_{k}}$. Assume

\begin{equation}
\left\vert \Psi (T)\right\rangle =\dprod\limits_{k}\int \Psi
_{0}(r_{k}^{a},l_{k}^{a})\left\vert r_{k}^{a},l_{k}^{a},T\right\rangle \Pi
drdl  \label{QMACSTATE}
\end{equation}%
is the initial state of the system with no photons, which are in the lowest
internal states for $l_{k}^{a}$, and $r_{k}^{a}$ are in Gaussian
semiclassical states with localized wavefunction and random momentums.
Assume that, initially, states are made of non-entangled products of the
quantum state of each particle. Due to the influence of the interaction
terms the particles interact with each other. We assume this interaction is
weak compared to the other two parts of the Hamiltonians. Due to the
influence of $H_{r}$ the molecules freely evolve. But due to continuous
interaction eventually the $\left\vert r_{k}^{a},l_{k}^{a},T\right\rangle $
become entangled with each other, and with the photons. There is a release
and absorption of photons, that entangles with spin, angular momentum, and
vibrational states. If we have a large number of particles, the system
undergoes random interactions with each other, and eventually, the quantum
state is the incoherent superposition of entangled states of matter and
photons.

If we use the Feynman Path integral we have,

\begin{equation}
\Psi (\{r_{i_{1}}^{a},l_{i_{2}}^{a},k_{i_{3}},s_{i_{4}}\},T)=\int_{\text{%
Random Paths}}e^{iS}Dr_{k}^{a}l_{k}^{a}\text{ }\Psi _{0}(r_{k}^{a},l_{k}^{a})
\label{Npartpath}
\end{equation}

This sum has both random collisions between particles and photons. So for
large N, $\Psi (\{r_{i_{1}}^{a},l_{i_{2}}^{a},k_{i_{3}},s_{i_{4}}\},T)$
evolves quite randomly. The quantum state is a highly entangled and
incoherent sum. This sum represents an ensemble of states for a large number
of states described by both classical probabilities due to the deterministic
evolution of the center of mass of the matter, and also quantum
probabilities due to entanglement and superposition. The sum will have
macroscopic states that are macroscopically distinguishable. These are the
Schr\"{o}dinger Cat states. Usually, it is assumed that such quantum states
collapse to avoid macroscopic superpositions. It is not yet established
under what conditions such quantum states will collapse to avoid macroscopic
superposition.

To give an idea let me give some examples. Scientists have
quantum-superposed huge molecules with thousands of atoms \cite{QEXP3}, \cite%
{QEXP}, \cite{QEXP3}. Quantum superposition has been also done at large
distances \cite{QEXP1}. Also, they have superimposed superfluid states such
as in SQUID devices, which are macroscopic quantum states made of quantum
superposition of states of cooper pair condensates \cite{Squid1} \cite%
{Squid2}. A laser is a macroscopic coherent quantum state of photons. Due to
continuous interaction most macroscopic systems are in large-scale
incoherent superposition. Until now there has been no limit that has been
set on how massive, big, or large enough for a quantum-superposed system to
spontaneously collapse. To collapse these states, we need quantum reduction
which we will discuss next.

\subsubsection{Quantum State Reduction}

Decoherence only deals with phases. It is not known what is the theory that
is involved in the collapse of superposed macroscopic quantum states into
the free states of macroscopic observables that are being observed.
Something has to collapse the Schr\"{o}dinger Cat states $\Psi
(\{r_{i_{1}}^{a},l_{i_{2}}^{a},k_{i_{3}},s_{i_{4}}\},T)$. For this, we need
a theory of quantum state reduction. In this paper, we will use quantum
collapse and quantum reduction as synonymous with each other.

Humanity has been dealing with ever-shrinking integrated circuit technology.
The transistor sizes are getting smaller and smaller, and now we are dealing
with the size of 25nms \cite{ICWIKI}, and \cite{ICWIKI2nm} in the designs of
integrated circuits. Various components involved in the design of these
devices are of 5nm scales. The latest smallest transistor design has sizes
of angstroms for various parts \cite{ICTECH}. These sizes are quite close to
scales of the size of atoms and molecules. As I have discussed before
quantum superpositions can exist with molecules made of few atoms to few
thousands of atoms, separated by large distances in atomic scales to
micrometer distances \cite{QEXP3}. Yet these high-tech devices that operate
our mobile phones are quite classical. So somewhere the macroscopic
superpositions are reduced for us to see a classical world. This is where
quantum reduction comes into play, and the exact theory is not known.

Usually in quantum decoherence theory density matrix is used and the
environment is traced over. Let $S$ be a quantum system that is interacting
with the environment such as a heat bath, background radiation, etc. Let $%
\left\vert \text{Env}_{i}\right\rangle $ be the orthogonal states of the
environment which are also eigenstates of macroscopic observables.

If $\left\vert \Psi _{i}\right\rangle $ are orthonormal states of $S$. Let $%
\left\vert \Psi (0)\right\rangle =\dsum\limits_{i}c_{i}\left\vert \Psi
_{i}\right\rangle $ be the initial state of S. If we assume that it gets
entangled with the environment on its interaction, $\left\vert \Psi
(T)\right\rangle =$ $\dsum\limits_{i}c_{i}\left\vert \Psi _{i}\right\rangle
\left\vert \text{Env}_{i}\right\rangle .$ The Von Neumann density matrix is $%
\rho =\left\vert \Psi (T)\right\rangle \left\langle \Psi (T)\right\vert $.
In the standard procedure of decoherence to get the density matrix of $\rho
_{S}$ of $S$, the environment is traced over \cite{Zurek}, \cite{Zurek2}.

\begin{equation}
\rho _{S}=Tr_{\text{Env}}(\rho e^{-\beta H_{\text{Env}}}),
\end{equation}%
where $\beta $ is the standard $\frac{1}{kT_{temp}}$ inverse temperature
term. This tracing is the same as replacing the quantum superposition with
classical superposition. That is replacing $\dsum\limits_{i}c_{i}\left\vert
\Psi _{i}\right\rangle $ with ensemble of collapsed states $\left\vert \Psi
_{i}\right\rangle $ with relative probabilities $|c_{i}|^{2}$. Basically in
decoherence theories the environment is assumed to be a classical ensemble
that is interacting with a quantum system $S.$ The following axiom is
assumed in the decoherence theories.

\begin{principle}
Axiom 2: Environmental Classicality Assumption: The macroscopic states of
the environment are assumed to be orthonormal to each other and continue to
remain classical and any microscopic quantum system entangled with it due to
interaction gets reduced continuously to keep the classicality.
\end{principle}

As these orthonormal states get entangled with the system $S$, due to
quantum interaction, the classical nature of the environment forces the
system to collapse with Max Born probabilities. The tracing operation is the
mechanism that enforces the classicality of the environment and also the
reduction of the states of the quantum system S entangled with it. Usually,
this environmentally induced reduction is assumed in the decoherence models
based on density matrix evolution. I would like to call such models
`Environmentally reduced decoherence Models'.

Tracing the environment leads to the master equation that describes the
evolution of the density matrix of a quantum system. The master equations of
an open quantum system describe the decoherence of a system that is coupled
to a heat bath. The heat bath could be the vacuum of a quantum field such as
the electromagnetic field. The decoherence of quantum systems due to
coupling to the environment was calculated by Joos and Zeh \cite{ZEH1} and
is one of the earliest works of such calculations. The formula gives good
agreement in calculations \cite{DECREV1}. Environmentally induced
decoherence models have been quite successful in applications in
open-quantum systems.

The important thing now is to explain the reduction process of the
environment itself at the fundamental level. According to the Schr\"{o}%
dinger equation, the environment itself becomes the sum of macroscopic
superposition. The wave function $\left\vert \Psi
\{r_{i_{1}}^{a},l_{i_{2}}^{a},k_{i_{3}},s_{i_{4}}\},T\right\rangle $
describes a fully quantum heat bath. The Schr\"{o}dinger evolution only
leads to the incoherent superposition of quantum macroscopic states. Not
classical states. For this, we need collapse models of the quantum state of
evolution.

Initially, there was a lot of philosophical debate on how the collapse
occurred. The many-world theory avoids collapse by saying the universe
itself becomes a superposition of macroscopic universes, and the observer
just goes into one of the universes as dictated by Max-Born probabilities.
Wigner and others associated the process of collapse with conscious
observation \cite{WigNeu}.

During the 1980's there were numerous attempts to describe rigorous collapse
models for quantum evolution, in which both the Schr\"{o}dinger and
reduction processes merged to get quantum stochastic evolution. There are
many models for quantum reduction. There are notable works in these by
Pearle, Ghirardhi, Rimini, Weber, Diosio, Gisin, etc. Some were trying to
modify the Schr\"{o}dinger equation of the state and others trying to modify
the density matrix evolution. Some of the important work is as follows:

\begin{itemize}
\item Gisin \cite{QSDGisin} has derived a stochastic Schr\"{o}dinger
equation using the idea of Ito's lemma. Further, this work has evolved into
the quantum state diffusion theory by Gisin and Percival \cite{QSD1} and 
\cite{QSD2}. We will discuss this later.

\item Ghirardhi, Rimini, and Weber (GRW) model \cite{GRW1} and \cite{GRW2},
where the superposition of atoms is reduced periodically.

\item Diosi \cite{Diosi} and Penrose \cite{PEN1} have been focusing on
gravitationally induced reduction.

\item Penrose and Hameroff's Orchestrated objective reduction (Orch OR) \cite%
{PENHAM1}, \cite{PENHAM2} in which quantum reduction happens when two
superposed states are separated by sufficient gravitational energy of
interaction between them.
\end{itemize}

The interesting part of the Orch OR model is that it combines both the
collapsing of the state and also discusses the involvement of conscious
observation. They propose this collapse gives rise to proto-consciousness.
Understanding the role of quantum collapse in consciousness is highly active
research now. Given the potential presence of macroscopic superpositions,
the potential role of consciousness cannot be dismissed. There has been much
discussion of the relation between consciousness and quantum reduction by
many researchers starting with the discovery of quantum mechanics. The
earliest ones are Wigner and Von Neumann \cite{WigNeu} who were later
followed by researchers such as Stapp, Vitiello, Freeman, Beck, Eccles,
Penrose, Hameroff, etc \cite{QCON}.

Newtonian Physics is quite deterministic, and it suggests a purely
deterministic universe without the possibility of free will. The discovery
of quantum uncertainty has been considered great progress in bringing back
the possibility of free will into consciousness by many of the founding
fathers of physics such as Planck, Bohr, Schr\"{o}dinger, Pauli (and others) 
\cite{QCON}. We will discuss this the later in `The Physical Foundations of
Life and Consciousness, Framework 1' \cite{MYP5B}.

\subsubsection{A General Quantum State Reduction Model}

Consider a macroscopic system whose quantum state $\left\vert \Psi
\right\rangle $ is orthonormal and is an incoherent superposition \ $\dsum
c_{i}\left\vert \Psi _{i}\right\rangle $of many orthonormal quantum states $%
\left\vert \Psi _{i}\right\rangle $, which are macroscopic quantum states. A
general evolution of an N particle system was given in the equation \ref%
{Npartpath}. Assume the macroscopic quantum state is observed. Usually,
physical observables at classical levels are position and momentum, and
their functions. We can include fields, and we know the field values such as
Electric and Magnetic fields are observable.

First, let me assume quantum states evolve in the time defined by the rest
frame foliation discussed before and to be discussed later. In general, let
me assume the Hermitian operator $M$ is the observable that is being
observed, and the system keeps reducing its quantum state randomly in the
Poisson process. We can assume that they collapse at a rate $r$ that is
inversely proportional to the difference in uncertainty squared of some
dynamical variable $M$.

\begin{equation}
r=\frac{\hbar K}{\sigma },
\end{equation}

\begin{eqnarray}
\sigma ^{2} &=&<M-<M>>^{2} \\
&=&\dsum\limits_{i}|c_{i}|^{2}(M_{i}-<M>)^{2},  \label{Quncert}
\end{eqnarray}%
where the expectation values are the quantum expectation values:

\begin{eqnarray}
M &=&\left\langle \Psi _{i}\right\vert \hat{M}\left\vert \Psi
_{i}\right\rangle  \nonumber \\
&<&M>=\dsum\limits_{i}|c_{i}|^{2}M_{i}
\end{eqnarray}%
Here I$\ $have not shown the amplitudes of each quantum state $\left\vert
\Psi _{i}\right\rangle $ explicitly. If $\sigma $ has the units of energy,
then $K$ is a numerical constant, which we can assume is close to $1.$ Later
we will discuss more rigorous mathematical formalism for this.

$M$ could be like the electric field energy of the gauge field or the
integral of scalar curvature, etc. The integral of scalar curvature yields
the suggestion of the Penrose-Hameroff Orch OR model \cite{PENHAM1}, and 
\cite{PENHAM2} as we will discuss later.

We need to deal with the Zeno paradox in this proposal. For an isolated
quantum system, this collapse has to be assumed to happen only when the
quantum uncertainty given in equation \ref{Quncert} reaches significant
values. This could be when the quantum system interacts with a macroscopic
field. This can happen during the process of measurement. If the collapse is
assumed to happen at all times this will result in the Zeno paradox, which
will freeze the evolution of the state. In the case of macroscopic fields,
we need to assume the quantum state evolves into roughly equally possible
quantum states so that it continuously collapses into one of these. This
will help avoid the Zeno paradox.

In Diosi, Penrose, and Hameroff versions, the quantum collapse of the
quantum states is supposed to involve radiation. This has been tested by
experiments recently, and no evidence of radiation was found within the
expected range of Orch Or Models \cite{TEST1}. This doesn't mean collapsed
models are not physically valid. The experimental group suggests that for
radiation to happen macroscopic superposition of the entire human brain
might have to happen before the collapse. This could be positive for the
quantum-conscious models. Also, as I have discussed even though we lose
quantum coherence, we still have a quantum incoherent superposition of
entangled macroscopic states which could be relevant in understanding
consciousness and free will as we will discuss in `The Physical Foundations
of Life and Consciousness, Framework 1' \cite{MYP5B}.

\subsubsection{Density Matrix Formalism\label{2.1.6}}

The major problem with collapse models is that in extended space-time with
infinite degrees of freedom, one cannot sensibly define it properly. For
example, in what region did the collapse happen, and how do you define the
wavefunction? What degrees of freedom or fields to include? These questions
cannot be answered by the naive collapse model described in the previous
subsection. For this, we need to go to the density matrix formalism or the
stochastic Schr\"{o}dinger equation \cite{QSDGisin} formalism, which can
help in the theoretical formulation of quantum reduction in a general
relativistic quantum field theoretic context. We will discuss the
introductory foundations of this now.

We can describe the collapse models using a density matrix or stochastic Schr%
\"{o}dinger equation. Lindlad gave the general form evolution of density
matrix evolution \cite{LND}. Gisin's approach gives the Stochastic Schr\"{o}%
dinger equation, that leads to stochastic evolution \cite{QSDGisin}. These
two approaches are closely related to each other mathematically.

When you have a model for the reduction of macroscopic fields we can deduce
the derivation of the density matrix evolution equation. The heat bath is
the superposition of macroscopic quantum states, whose superposition
continuously undergoes reduction. Because of this, any quantum system that
is coupled to this heat bath gets reduced gradually if we assume a weak
coupling. The density matrix evolution equation can be derived by tracing
out the heat bath degrees of freedom. Due to the weak coupling, the system
also undergoes decoherence as the quantum amplitudes get randomized. But the
system also undergoes reduction due to classicality enforced by the
environment as discussed by the assumption of environmental reduction axiom.
In standard Von Neumann's definition of the density matrix, it represents
the average density matrix for an ensemble of systems of the same type.
Later we will get a better understanding of the density matrix while we
discuss the stochastic Schr\"{o}dinger equation.

The general form of the evolution of density matrix in the Lindblad form 
\cite{LND} is,%
\begin{equation}
\frac{d\rho }{dT}=\frac{i}{\hbar }[H,\rho ]-\frac{1}{2}\dsum\limits_{i}\{%
\rho M^{\dagger i}+M^{i}\rho -2M^{i}\rho M^{\dagger i}\},  \label{MAST}
\end{equation}%
where $T$ is time which I assume to be defined by restframe foliation, $%
M^{i} $ are preferred operators of observation that need not be Hermitian,
which gives the preferred basis for reduction. For example, $M$ could be the
raising and lowering operators of quantum field theory, and then the
preferred states of collapse would be the coherent quantum states.

To understand how this equation reduces the quantum state, assume $H$ is
zero, $M$ is Hermitian, and there is only one $M$.

Let $m_{i}$ and $\left\vert m_{j}\right\rangle $ are eigenvalues and
eigenstates of the system. Then on this basis, the master equation reduces to

\begin{equation}
\frac{d\left\langle m_{i}\right\vert \rho \left\vert m_{j}\right\rangle }{dT}%
=-\frac{1}{2}\dsum\limits_{i}(m_{i}-m_{j})^{2}\left\langle m_{i}\right\vert
\rho \left\vert m_{j}\right\rangle
\end{equation}

We can see the non-diagonal terms become negligible after a duration, while
the diagonal terms remain unchanged. If we assume the environment reduces to
eigenstates that are measured by measurement instruments, which in turn are
entangled with the eigen states of the system $S,$ then the resulting
diagonal density matrix represents an ensemble of quantum systems $S$ that
are in eigenstates $\left\vert m_{i}\right\rangle $. \ Initially if the
quantum state is represented by the pure state $\rho =\left\vert \Psi
\right\rangle \left\langle \Psi \right\vert ,$ then the statistical
probabilities of $\left\vert m_{i}\right\rangle $ are given by the diagonal
value $\left\langle m_{i}\right\vert \rho \left\vert m_{i}\right\rangle
=|\left\langle m_{i}\right\vert \left\vert \Psi \right\rangle |^{2}$, the
Born probabilities.

What we need in this paper is to formulate the density matrix formalism in
the way it is fundamental rather than derived from reducing the classical
environmental degrees of freedom. Later we will do this in general
relativistic mathematical formalism.

\subsubsection{Stochastic Schr\"{o}dinger Equation\label{2.1.7}}

Any quantum system that interacts with macroscopically large fields gets
continuously reduced. This results in the diffusion of the wavefunction of
the system. This is described by the quantum state diffusion equation
formalism \cite{QSD1}, \cite{QSD2}. In this paper, I will refer to the
evolution equation that describes this in \cite{QSD1} and \cite{QSD2} as the
Stochastic Schr\"{o}dinger equation as I assume the stochastic evolution to
be fundamental in nature.

The density matrix formalism equation needs to be considered to be derived
from the Stochastic Schr\"{o}dinger equation rather than the opposite. This
has been discussed in the original papers \cite{QSD1}, \cite{QSD2}, and also
in mine \cite{MYP1}. \ I\ have discussed the stochastic Schr\"{o}dinger
equation in the context of quantum general relativity in my previous
versions of the paper \cite{MYP0} to \cite{MYP4}. Let me discuss this
further in the paper. The stochastic form of the Schr\"{o}dinger equation
was first derived by Gisin \cite{QSDGisin}. Then it was further discussed by
Gisin and Percival in \cite{QSD1}, \cite{QSD2}. The stochastic Schr\"{o}%
dinger equation that is suitable and intuitive for the purpose of this paper:

\begin{equation}
d\left\vert \Psi (T)\right\rangle =iH\left\vert \Psi (T)\right\rangle
dT-\dsum\limits_{i}D_{i}^{\dagger }D_{i}\left\vert \Psi (T)\right\rangle
dT+\dsum_{i}D_{i}dz^{i}\left\vert \Psi (T)\right\rangle ,  \label{SSE}
\end{equation}%
where $D_{i}=M_{i}-<M_{i}>$, $M_{i}$ are the same as in the density matrix
equation to be measured and $T$ is time I assumed to be defined by restframe
foliation. Here $dz^{i}$ are random complex-valued numbers that are assumed
to be uniformly distributed with ensemble average satisfying:

\begin{equation}
M(dz^{i}d\bar{z}^{j})=2\delta ^{ij}dT
\end{equation}

$D_{i}$ contains the wavefunction within it to calculate the quantum
expectation values. So these equations are non-linear and stochastic. These
equations are not continuous, because of the random terms in them. Depending
on the values of $dz^{i}$, the change could be small or big compared to the
initial state. The first term is the Schr\"{o}dinger term, the second term
is the localizing term and the third is the randomizing term. The equation
has both decoherence (first term) and reduction (2 and 3rd term) built into
it, explicitly.

\ The relation between the density matrix equation and the stochastic Schr%
\"{o}dinger equation can be derived as follows. Let $\rho _{\Psi }$ be the
pure density matrix $\left\vert \Psi (T)\right\rangle \left\langle \Psi
(T)\right\vert $,

\begin{equation}
\rho _{\Psi }=\left\vert \Psi (T)\right\rangle \left\langle \Psi
(T)\right\vert  \label{densitymean}
\end{equation}%
Then the change in $\rho _{\Psi }$ is \ 
\begin{equation}
d\rho _{\Psi }=d\left\vert \Psi (T)\right\rangle \left\langle \Psi
(T)\right\vert +\left\vert \Psi (T)\right\rangle d\left\langle \Psi
(T)\right\vert
\end{equation}%
We can calculate the change in $d\rho _{\Psi }$ using equation \ref{SSE}\
and calculate the rate of change:

\begin{equation}
\frac{d\rho _{\Psi }}{dT}=\frac{i}{\hbar }[H,\rho _{\Psi
}]-\dsum\limits_{i}\{\rho _{\Psi }D^{\dagger i}+D^{i}\rho _{\Psi
}\}+\dsum\limits_{ij}D^{i}\rho _{\Psi }D^{\dagger j}dz^{i}d\bar{z}^{j}
\label{RhoPhi}
\end{equation}%
Here I\ have neglected the higher powers of $dT,$ keeping only the
first-order terms in $dT$. Now we define,

\begin{equation}
\rho =E(\rho _{\Psi })  \label{RhoDef}
\end{equation}%
Applying this in equation \ref{RhoPhi}, and substituting $%
D_{i}=M_{i}-<M_{i}>,$ we get the density matrix evolution equation \ref{MAST}%
.

Now the stochastic Schr\"{o}dinger formalism gives clear explicit formalism
for interpreting the density matrix equation. The density matrix of the
system $\rho _{\Psi }$ keeps undergoing stochastic evolution, and in each
step, it gets reduced stochastically as described in the previous section to
the eigen states of $M_{i}$. Basically in terms of $\left\vert \Psi
(T)\right\rangle $ this evolution looks like Brownian motion for the quantum
state. Assume the system is undergoing measurement of operators $M_{i}$ and
the $H$ term is zero or small. If the initial state is not in one of the
eigenvalues of $M_{i}$, then the $D_{i}$ term is large. It evolves to yield
a quantum state which is one of the eigen states of $M_{i}$. The stochastic
Schr\"{o}dinger evolution slowly drifts the state into an eigen state of $%
M_{i}$.

\section{The New Fundamental Quantum Reduction Theory}

\subsection{Bidirectional temporal Evolution and Retrocausality}

\subsubsection{Formulation\label{2.2.1}}

Now assume we have the rest frame foliation and a prescription for quantum
reduction specified. We need to discuss a new issue, which is the presence
of retrocausal evolution. One of the important predictions of quantum
gravity is retrocausality, as we will discuss now.

The important thing to note is that the Hamiltonian constraint equation $%
H\left\vert \Psi \right\rangle =0$ is second order in momenta. Once we have
the time variable $T$ we can formally rewrite the Hamiltonian constraint
equation as

\begin{eqnarray}
-\hbar ^{2}\frac{\partial \left\vert \psi (T)\right\rangle }{\partial T^{2}}
&=&H_{T}^{2}\left\vert \psi (T)\right\rangle  \label{Ham2ord} \\
\left\vert \Psi \right\rangle &=&\int \left\vert \psi (T)\right\rangle dT
\end{eqnarray}%
The $H_{T}^{2}$ stands for the sum of remaining kinetic and potential terms.
The second-order nature of this equation means the initial value formulation
of this equation requires two specifications: 1) The state and its time
derivative for a specific time instance, or 2) specifying the quantum state
at two different times. The first choice is essentially the second choice,
where the difference between two nearby instants is specified as the time
derivative. The second choice of specifying the initial value formulation
leads to retrocausality. The quantum state evolution depends on the future
state. This could answer many phenomena of retrocausality that appear to
happen in many areas such as Wheeler's delayed choice experiments \cite%
{Wheel}, or Libet's free will experiments \cite{Libet}.

Consider we make quantum state measurements of an isolated quantum state
system specified by the second-order Schr\"{o}dinger equation. For a typical
system, consider the non-relativistic quantum system of bulk matter
discussed in Quantum gravity framework 3 \cite{MYP3}. Its evolution is
defined by the form of the equation \ref{Ham2ord}. This is the best example
for discussing physical systems in the most non-relativistic conditions.
Here time is the distance travelled by the center of mass of the group of
particles that are gravitationally bound to each other.

Let$~\left\vert \psi _{1}\right\rangle $ $=$ $\left\vert \psi
(T_{1})\right\rangle $ and $\left\vert \psi _{2}\right\rangle $ $=$ $%
\left\vert \psi (T_{2})\right\rangle $ be quantum measurement at two
different times $T_{1}$ and $T_{2}$. Let me discuss how we can derive the
time evolution of the state $\left\vert \psi (T)\right\rangle $ between the
two different times. Let me take the square root of equation \ref{Ham2ord}.

\begin{equation}
\frac{\hbar }{i}\frac{\partial \left\vert \psi _{\pm }(T)\right\rangle }{%
\partial T}=\pm H_{T}\left\vert \psi _{\pm }(T)\right\rangle
\end{equation}

This resembles the conventional Schr\"{o}dinger equation, but it describes
two different wavefunctions: $\psi _{+}(T)$ evolving forward in time and $%
\psi _{-}(T)$ evolving backward in time. The quantum states described by the
equation \ref{Ham2ord} are the sum of these two different wavefunctions.

\begin{equation}
\left\vert \psi (T)\right\rangle =\left\vert \psi _{+}(T)\right\rangle
+\left\vert \psi _{-}(T)\right\rangle
\end{equation}

We can rewrite this equation using the Hamiltonian Formulation:

\begin{equation}
\left\vert \psi (T)\right\rangle =e^{\frac{i}{\hbar }H_{T}\Delta
T}\left\vert \psi _{+}(T_{1})\right\rangle +e^{-\frac{i}{\hbar }H_{T}\Delta
T}\left\vert \psi _{-}(T_{1})\right\rangle ,  \label{QstatSum}
\end{equation}

where $\Delta T=T-T_{1}$.

We can derive the quantum state $\left\vert \psi (T)\right\rangle $ between
times $T_{1}$ and $T_{2}$ using this:

\begin{eqnarray*}
\left\vert \psi (T_{1})\right\rangle &=&\left\vert \psi
_{+}(T_{1})\right\rangle +\left\vert \psi _{-}(T_{1})\right\rangle \\
\left\vert \psi (T_{2})\right\rangle &=&e^{\frac{i}{\hbar }H_{T}\Delta
T}\left\vert \psi _{+}(T_{1})\right\rangle +e^{-\frac{i}{\hbar }H_{T}\Delta
T}\left\vert \psi _{-}(T_{1})\right\rangle
\end{eqnarray*}

we can invert this equation to get

\begin{eqnarray}
\left\vert \psi _{+}(T_{1})\right\rangle &=&\frac{\left\vert \psi
(T_{2})\right\rangle -e^{\frac{i}{\hbar }H_{T}\Delta T}\left\vert \psi
(T_{1})\right\rangle }{1-e^{\frac{2i}{\hbar }H_{T}\Delta T}}  \label{BIDEREQ}
\\
\left\vert \psi _{-}(T_{1})\right\rangle &=&\frac{\left\vert \psi
(T_{1})\right\rangle -e^{-\frac{i}{\hbar }H_{T}\Delta T}\left\vert \psi
(T_{2})\right\rangle }{1-e^{-\frac{2i}{\hbar }H_{T}\Delta T}}
\end{eqnarray}

From this, we can derive the quantum state between $T_{1}$ and $T_{2}$ using
equation \ref{QstatSum}. To summarize, if we make a sequence of quantum
measurements, then for each pair of two consecutive times we can derive the
quantum state in between them as a function of time using the quantum states
of each of the two consecutive states.

\subsubsection{Radiation Free Collapse\label{2.2.2}}

The recent experiments to test Diosi, Penrose, and Hameroff proposals, have
come up with negative results \cite{TEST1}. They tried to detect the quantum
radiation due to quantum reduction. But no radiation was found. We see that
the retrocausal phenomena associated with quantum gravity discussed in this
paper predict no radiation as there is no discontinuity in wavefunction due
to collapse. At each point of the quantum collapse we see that the total
wavefunction doesn't change, only the positive and negative energy
components change.

To give a simple understanding of the quantum reduction process let the
reduction happen at instants $T_{n}$. Let $T_{n}^{+}$ be the time instant
after quantum reduction at $T_{n}$ and $T_{n}^{-}$ be the time instant
before quantum reduction at $T_{n}$. Let $\left\vert \psi
(T_{n}^{+})\right\rangle $ and $\left\vert \psi (T_{n}^{-})\right\rangle $
be the quantum states before and after reduction at $T_{n}$. Let $\left\vert
\psi _{+}(T_{n}^{+})\right\rangle ~$\ and $\left\vert \psi
_{-}(T_{n}^{+})\right\rangle $ be the positive and negative energy
components at $T_{n}^{+}$. Similarly, let $\left\vert \psi
_{+}(T_{n}^{-})\right\rangle ~$\ and $\left\vert \psi
_{-}(T_{n}^{-})\right\rangle $ be the positive and negative energy
components at $T_{n}^{-}$.

1) Reduction determines the values of $\left\vert \psi
(T_{n}^{+})\right\rangle $ at each time instant $T_{n}$ of reduction. At $%
T_{n\text{ }}$the total wavefunction is continuous.

\begin{equation}
\left\vert \psi (T_{n}^{+})\right\rangle =\left\vert \psi
(T_{n}^{-})\right\rangle
\end{equation}

2) Between each consecutive instants $T_{n}$ wavefunctions evolve by the
equations:

\begin{eqnarray}
\left\vert \psi (T_{n}^{+})\right\rangle &=&\left\vert \psi
(T_{n}^{-})\right\rangle =\left\vert \psi _{+}(T_{n-1}^{-})\right\rangle
+\left\vert \psi _{-}(T_{n-1}^{-})\right\rangle \\
\left\vert \psi _{\pm }(T_{n}^{-})\right\rangle &=&e^{\pm \frac{i}{\hbar }%
H_{T}\Delta T}\left\vert \psi _{\pm }(T_{n-1}^{+})\right\rangle \\
\left\vert \psi (T_{n}^{-})\right\rangle &=&e^{\frac{i}{\hbar }H_{T}\Delta
T}\left\vert \psi _{+}(T_{n-1}^{+})\right\rangle +e^{-\frac{i}{\hbar }%
H_{T}\Delta T}\left\vert \psi _{-}(T_{n-1}^{+})\right\rangle
\end{eqnarray}

3) The value of the quantum state at $\left\vert \psi
(T_{n}^{+})\right\rangle =\left\vert \psi (T_{n}^{-})\right\rangle $ is
determined by the Copenhagen interpretation. If $\left\vert \psi
(T_{n-1}^{+})\right\rangle =\left\vert \psi (T_{n-1}^{-})\right\rangle $ is
the quantum state at $T_{n-1}$, then it evolves to $T_{n}$ as positive
energy quantum state by Schr\"{o}dinger evolution: $e^{\frac{i}{\hbar }%
H_{T}\Delta T}\left\vert \psi (T_{n-1}^{+})\right\rangle .$ There it is
reduced to yield a new quantum state $\left\vert \psi
(T_{n}^{-})\right\rangle =P(e^{\frac{i}{\hbar }H_{T}\Delta T}\left\vert \psi
(T_{n-1}^{+})\right\rangle )$ by Max Born probabilities: $|\left\langle \psi
(T_{n}^{-})\right\vert e^{\frac{i}{\hbar }H_{T}\Delta T}\left\vert \psi
(T_{n-1}^{+})\right\rangle |^{2}$.

When the quantum state is reduced at $T_{n}$, it sends a positive energy
state toward the time of the next projection $T_{n+1}$ and the negative
energy state towards the time of the previous projection $T_{n-1}$. The
first one determines the quantum state at $T_{n+1}$. The negative energy
reflection back in time adjusts the positive and the negative energy states
at $T_{n-1}$, such that the quantum state evolution between $T_{n-1}$ and $%
T_{n}$ is determined by the projections at these instants according to \ref%
{BIDEREQ}.

4) The basis of the state into which $e^{\frac{i}{\hbar }H_{T}\Delta
T}\left\vert \psi (T_{n-1}^{+})\right\rangle $ is projected is determined by
macroscopic fields with which the quantum state of the system interacts. Let 
$\mathbf{P}$ be the projection operator. Let me define $\mathbf{D}$ as the
difference operation due to reduction at $T_{n}:$

\begin{equation}
\mathbf{D}\left( e^{\frac{i}{\hbar }H_{T}\Delta T}\left\vert \psi
(T_{n-1}^{+})\right\rangle \right) =\mathbf{P}(e^{\frac{i}{\hbar }%
H_{T}\Delta T}\left\vert \psi (T_{n-1}^{+})\right\rangle )-\left( e^{\frac{i%
}{\hbar }H_{T}\Delta T}\left\vert \psi (T_{n-1}^{+})\right\rangle \right)
\end{equation}

Then we can rewrite the above equation as follows:

\begin{equation}
\left\vert \psi (T_{n}^{-})\right\rangle =e^{\frac{i}{\hbar }H_{T}\Delta
T}\left\vert \psi (T_{n-1}^{+})\right\rangle +\mathbf{D}\left( e^{\frac{i}{%
\hbar }H_{T}\Delta T}\left\vert \psi (T_{n-1}^{+})\right\rangle \right)
\label{SPROJEQ}
\end{equation}

The $\mathbf{D}$ operator is stochastic and is determined by how the quantum
state interacts with macroscopic fields at $T_{n}$.

These can be used to calculate the evolution of the negative energy
components. Based on the above four conditions, we can calculate the
evolution of wavefunctions. The evolution is radiation-free as there is no
discontinuity in $\left\vert \psi (T_{n})\right\rangle $.

\subsubsection{Relation to the Stochastic Schr\"{o}dinger Equation\label%
{2.2.3}}

Consider that the quantum state undergoes continuous collapse so that $%
\Delta T$, $H_{T}\Delta T$, and $\mathbf{D}\left( e^{\frac{i}{\hbar }%
H_{T}\Delta T}\left\vert \psi (T_{n-1}^{+})\right\rangle \right) $ is small.
Then we can rewrite the equation \ref{SPROJEQ}\ as

\begin{equation}
d\left\vert \psi (T_{n}^{-})\right\rangle =\frac{i}{\hbar }H_{T}\Delta
T\left\vert \psi (T_{n-1}^{+})\right\rangle +\mathbf{D}\left( \left\vert
\psi (T_{n-1}^{+})\right\rangle \right)  \label{SmallSSE1}
\end{equation}

Then comparing this with \ref{SSE} we have

\begin{equation}
\mathbf{D}\left( \left\vert \psi (T_{n-1}^{+})\right\rangle \right)
=-\dsum\limits_{i}D_{i}^{\dagger }D_{i}\left\vert \psi
(T_{n-1}^{+})\right\rangle dT+\dsum_{i}D_{i}dz^{i}\left\vert \psi
(T_{n-1}^{+})\right\rangle
\end{equation}

From the second-order evolution,

\begin{equation}
d\left\vert \psi (T_{n}^{-})\right\rangle =\frac{i}{\hbar }H_{T}\Delta
T\left\vert \psi _{+}(T_{n-1}^{+})\right\rangle -\frac{i}{\hbar }H_{T}\Delta
T\left\vert \psi _{-}(T_{n-1}^{+})\right\rangle  \label{SmallSSE2}
\end{equation}

Then using \ref{SmallSSE1},

\begin{equation}
d\left\vert \psi (T_{n}^{-})\right\rangle =\frac{i}{\hbar }H_{T}\Delta
T\left\vert \psi _{+}(T_{n-1}^{+})\right\rangle +\frac{i}{\hbar }H_{T}\Delta
T\left\vert \psi _{-}(T_{n-1}^{+})\right\rangle +\mathbf{D}\left( \left\vert
\psi (T_{n-1}^{+})\right\rangle \right)  \label{SmallSSE3}
\end{equation}

Comparing the last two equations we have,

\begin{equation}
-2\frac{i}{\hbar }H_{T}\Delta T\left\vert \psi
_{-}(T_{n-1}^{+})\right\rangle =\mathbf{D}\left( \left\vert \psi
(T_{n-1}^{+})\right\rangle \right)
\end{equation}

From this, we can calculate the reflected negative energy state.

\begin{equation}
\left\vert \psi _{-}(T_{n-1}^{+})\right\rangle =(-2\frac{i}{\hbar }%
H_{T}\Delta T)^{-1}\mathbf{D}\left( \left\vert \psi
(T_{n-1}^{+})\right\rangle \right)  \label{NegState}
\end{equation}

The negative energy state component depends on random projections due to
quantum reduction. \ The total wavefunction evolves without any
randomization. This can be considered to mean there is no radiation due to
spikes. The random changes happen only to the positive and negative energy
components of the wavefunction but the total wavefunction evolves smoothly.

\subsection{The New Fundamental Quantum Reduction Theory\label{2.3}}

Neither the density matrix formulation nor the quantum diffusion equation
gives fundamental formulation of quantum reduction. The density matrix
formulation, inspired by environmental decoherence, is not fundamental. It
can be derived by tracing out the environmental degrees of freedom. This is
equivalent to assuming that the environment undergoes spontaneous reduction,
and the interaction also reduces the quantum systems that are entangled with
it. Essentially, the master equation can be derived based on this
assumption. And from this, the quantum diffusion equation can be derived
from this.

But the master equation doesn't capture the original environmental-induced
decoherence. Because the density matrix is assumed to describe an ensemble
of systems in their eigenstates, and with the density distribution given by
its eigenvalues. But this can be describing the physical system if we assume
that the states that are undergoing spontaneous reduction due to
environmental influence always go into orthogonal states. This need not be
the case. For example, coherent states are the best quantum states that
produce quantum physics that is essentially equivalent to classical
phenomena. The coherent states of the quantum states are not orthogonal. So,
the density matrix formulation can only represent a physical system under
restricted conditions, and the reduced environmental states are orthogonal.

The quantum diffusion equation formulation can be considered to be much
closer to the spontaneous environment-induced decoherence because it doesn't
involve averaging. The equation describes a system that undergoes random
evolution. The random variables are assumed to be normally distributed. This
need not be the case, and it can be more general than that.

To fully describe a fundamental quantum reduction, we need to start with
systems that are undergoing spontaneous reduction using the general
reduction model discussed before. The following properties seem to be
closest to describing this.

\begin{enumerate}
\item The reduction is spontaneous.

\item The preferred states of the environment into which the collapse occurs
are the coherent-like states of a group of operators M relating to the
rising and lowering operators of the configuration variables of the field.

\item The evolution of the quantum state is described by bidirectional
temporal evolution between reductions.

\item The reduction happens along the rest frame foliation.

\item A system that starts with a pure state evolves into possible choices
described by mixed states at a rate determined by decoherence constants.

\item The mixed states are converted into a pure state by observation at
rate $\beta _{R}$.
\end{enumerate}

Let me call this reduction the "New Fundamental Quantum Reduction Theory",
or Bidirectional Temporal Spontaneous Quantum Reduction Theory.

The mixed and pure states can be understood using density matrix formalism,
and the decoherence constants are related to the decoherence terms of the
Lindblad formulation. The concept of observer is quite general; it could be
a camera, bacteria, or consciousness, which will be discussed in part B of
the paper, The physical foundations of life and consciousness \cite{MYP5B1.3}%
. The third requirement comes from the need for continuity of the quantum
state. Otherwise, the collapse leads to a sudden variation in the quantum
state, leading to various other issues. This is quite general framework.
Finding the precise formulation requires experimental investigation.

The most important coherent states are those of the gravitational field for
mediating reduction. It's the gravitational field that remains smooth and
classical at the scales of physics we are doing. The gauge fields other than
the gravitational field have quantum properties, and their physical
properties are described by quantum phenomena. For example, for
electromagnetic fields, we have black body radiation, electromagnetic
spectral distribution, etc., and intensities are described by photonic
distribution. Only during highly special conditions do we see its classical
wave properties. Other gauge fields, such as weak and strong fields, are
mostly described by quantum physics. Only the gravitational field remains
classical in most extreme physical phenomena other than singularities. The
quantum fluctuations of the gravitational field require quantum physics
phenomena involving macroscopic matter, which always leads to quantum
reduction. So, the gravitational field is the most important field that
mediates the gravitational field, if there is no other field that is more
classical than it. Penrose is an important proponent of the role of the
gravitational field in quantum reduction.

Among the fields, the longitudinal part of the field can be expected to play
a critical role in the spontaneous quantum reduction. In QED, the
longitudinal field has been considered to be in a quantum coherent state.
Also, the gravitational field itself is mostly longitudinal. In the previous
papers, I have described how the longitudinal fields play an important role
in deriving the rest frame foliation, which I propose to be the foliation in
which the reduction happened. So the coherent state description of the
longitudinal fields of the electromagnetic and gravitational field can
potentially be the preferred state of the spontaneous quantum reduction. In
the case of an electromagnetic field, the coherent state's expectation value
needs to be classically observable to be the preferred state of reduction.

I consider a reduction theory that is based on these six properties to be
the fundamental quantum reduction theory.

The approximate equivalent of these is the density matrix form or quantum
diffusion equation.

We can expect, physically, that the reductive evolution happens as follows:

\begin{enumerate}
\item The initial state of the quantum system is given in the rest frame
foliation at a specific instant.

\item After infinitesimal time, the quantum system evolves through the
positive-going Hamiltonian evolution.

\item Expand the quantum state into coherent states of the fields.

\item The quantum state undergoes spontaneous reduction into one of the
coherent states of the macroscopic quantum fields, according to the Max Born
rule for calculating probability, in which the microscopic quantum systems
interacting with it are also reduced according to the entanglement.

\item Work out the negative and positive going waves according to the
bidirectional temporal evolution form using the initial and reduced state.

\item Repeat this for the next infinitesimal step.
\end{enumerate}

This theory and implementation scheme are just guidelines to have a clear
conceptual framework, but the details need to be worked out. For example,
the following question needs to be answered through experimental and
physical observation.

\begin{enumerate}
\item What are the preferred states of reduction, or equivalently, the
operators whose eigenstates are these states, if they are not the coherent
states and field operators?

\item How often does the collapse happen? For example, is it at a constant
rate, or can it happen within any time interval with a certain probability
(Poisson process)?

\item Do the bidirectional temporal equations have a solution?
\end{enumerate}

\subsection{Classicum, Entanglement Domain, and Reduction features\label%
{claent}}

The way the fundamental reduction theory works depends on the physical
context. Physically we have a sequence of fields and particles that can be
classified as quantum or classical on a grey scale. He is the list from more
quantum to more classical in the physical context.

\begin{enumerate}
\item Photons - highly quantum.

\item Electron -- more quantum

\item Weak Particles, Quarks, and Hadrons -- more classical particles.

\item Higgs hidden as mass.

\item Missing Gap

\item EM fields and Gravity highly classical
\end{enumerate}

The absolutely quantum entities are photons, and those that are absolutely
classical in all known physical context is gravity. EM fields are classical
or quantum depending on the context.

If the gravitational field is assumed to be the most classical, then when
quantum matter interacts with it will be spontaneously reduced. So, the
gravitational field itself is possibly the agent in spontaneous reduction
that induces reduction in matter. For this to happen matter needs to be
massive enough to create a quantum fluctuation in the gravitational field.
For this, lets me define the \textquotedblleft self decohering entanglement
domain\textquotedblright\ (SDED) as a large group of bonded atoms that are
large enough to do this. In SDED, any superposition of one of the particles
entangled with it is reduced.

In the quantum gravity framework formalism, the time is determined by the
rest frame foliation (least variation in general). The rest frame foliation
is determined by the condition in which the longitudinal fields are at rest.
The relative contribution of various fields depends on the constants in
postulate 3, determined by the various constants. If the gravitational field
is the most dominant, then the flow of time is determined by Earth's rest
frame for all terrestrial objects. Instead, if the Electromagnetic field is
the most dominant, then the flow of time is determined by the rest frame of
each object, such as a ball, bat, or human brain. That's each entanglement
domain carries its own time. Reduction phenomena need to be worked out in
the time associated with the rest frame of each object. In this paper, I
propose an experiment to map the rest frame to determine the relative
contribution of various fields. I strongly believe that, consistent with the
assumption of special relativity, the rest frame of any object is the rest
frame foliation of each object in which the reduction happens.

\FRAME{ftbpF}{5.5495in}{3.2474in}{0pt}{}{}{reductions.png}{\special{language
"Scientific Word";type "GRAPHIC";maintain-aspect-ratio TRUE;display
"USEDEF";valid_file "F";width 5.5495in;height 3.2474in;depth
0pt;original-width 6.0001in;original-height 3.4999in;cropleft "0";croptop
"1";cropright "1";cropbottom "0";filename
'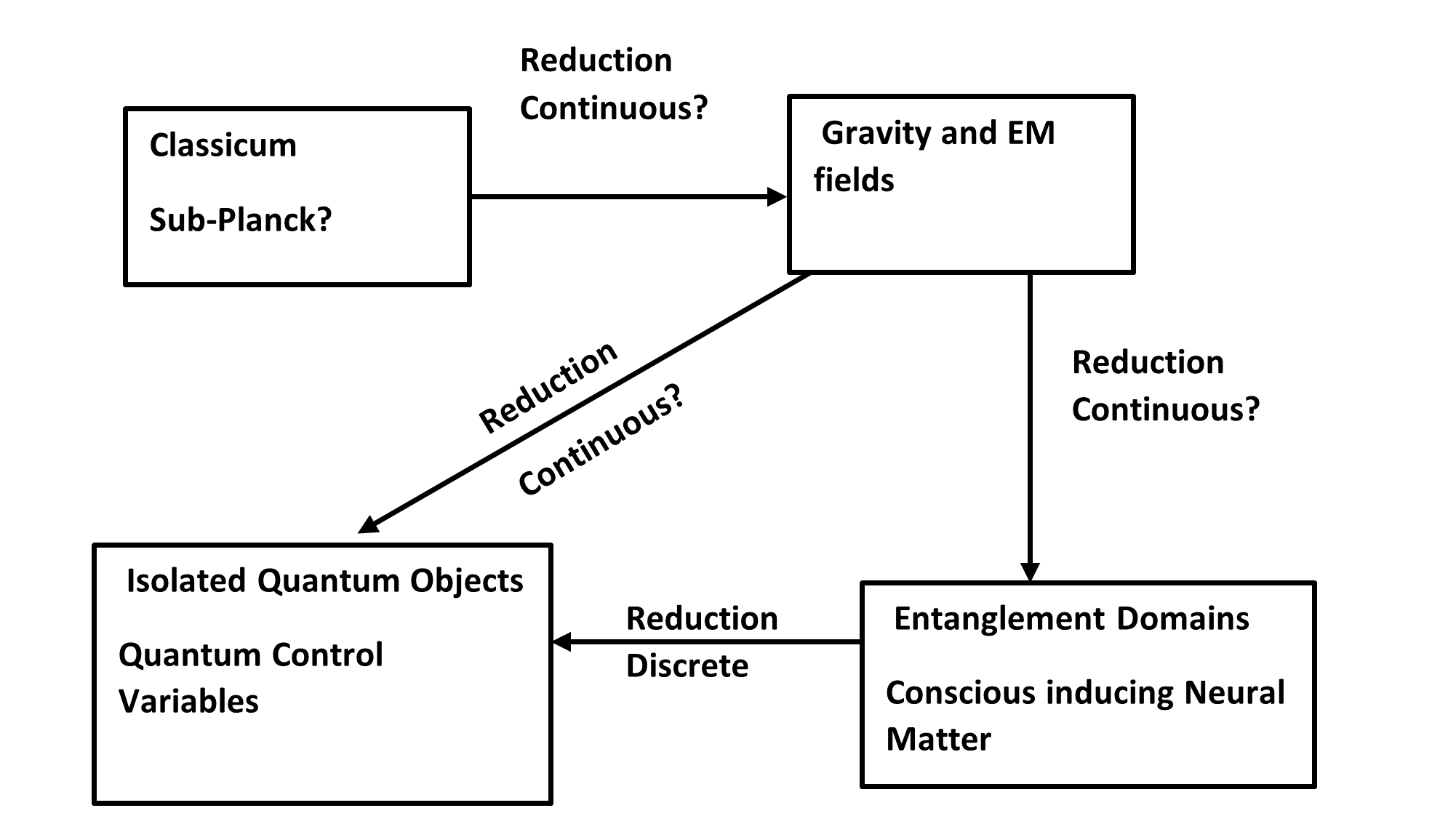';file-properties "XNPEU";}}

If you have a charged particle not bonded to an entangled domain or not
large enough to cause gravitational superposition that will reduce it, then
we can consider the influence of the EM field. Coupling to microscopic
background radiation and other stray EM fields emitted by objects induces
environmental decoherence in the charge particle. There are multiple
research papers in this field.

Gravitational and Classical EM fields can be seen as the ultimate source of
reduction. Now to understand what reduces them, I can coin the term
classicum as a hyper classical entity that observers the these two fields
and induces reduction. We can expect this entity to be sub-Planck, assuming
that gravity is plank scale field that remains classical in most know
context, and anything that can impose a reduction of the superposition of
the gravitational field needs to be sub-Planck.

\FRAME{ftbpF}{5.2901in}{3.5769in}{0pt}{}{}{energyscale.png}{\special%
{language "Scientific Word";type "GRAPHIC";display "USEDEF";valid_file
"F";width 5.2901in;height 3.5769in;depth 0pt;original-width
6.4152in;original-height 3.5302in;cropleft "0";croptop "1";cropright
"1";cropbottom "0";filename '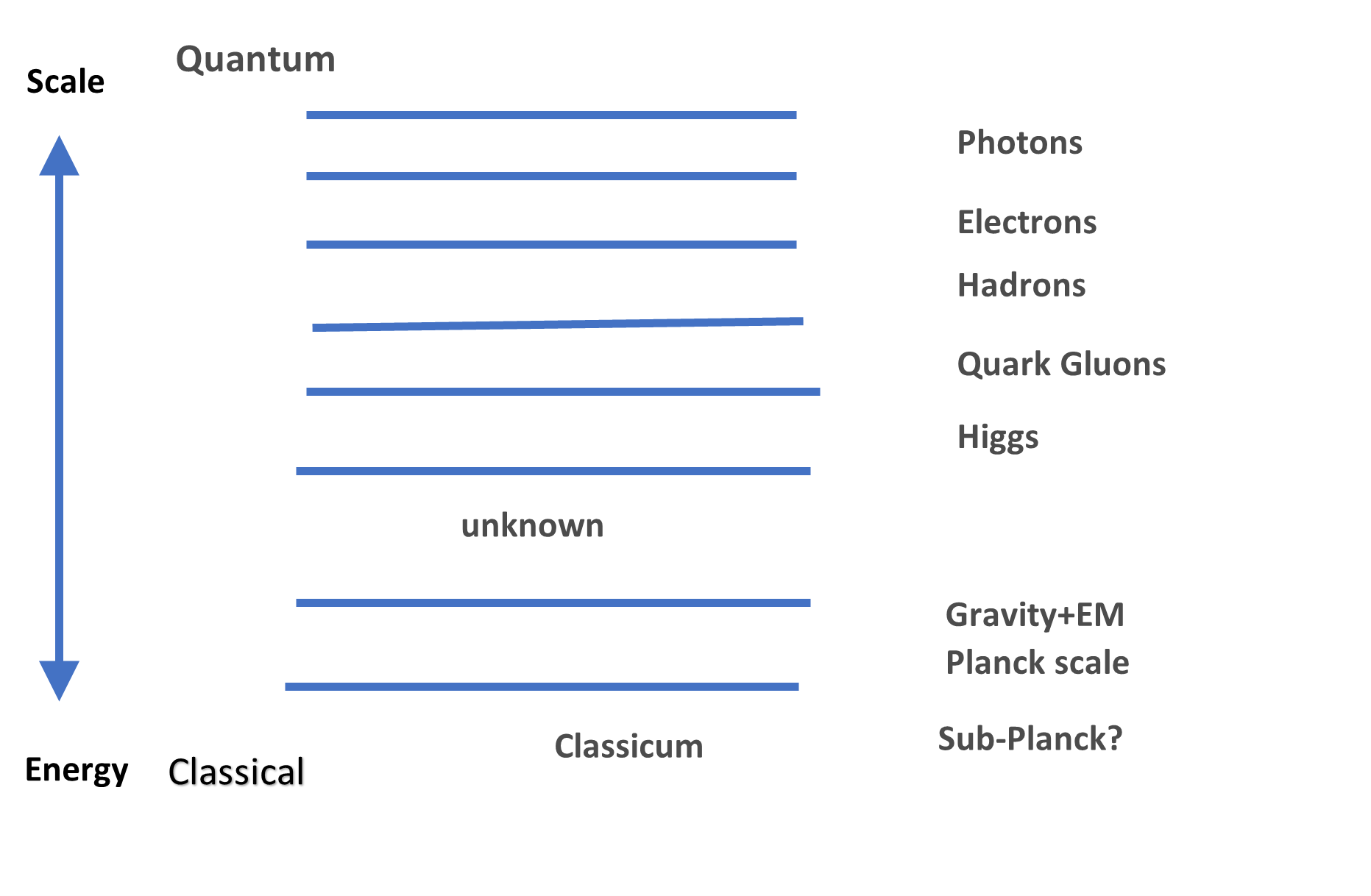';file-properties
"XNPEU";}}

The classicum is considered to undividable extended entity that continuously
captures the information from the quantum world to create the classical
features of nature. They induce a reduction in the gravitational and EM
field; being always classical, we can expect the reduction to be continuous.
This can be modelled by either the density matrix or quantum diffusion
theory. Then, through coupling to SDEDs that are large enough, the latter
are reduced continuously due to numerous spontaneous reductions. The
classicum, as a sub-Planck entity, reduces the longitudinal fields (EM and
gravitational), as they are the massive part of the field. Through that, it
can be expected to reduce the quantum states of the SDEDs that create these
fields. The quantum particles, such as photons or electrons, are reduced by
the coupling to the SDEDs, and this tends to be a discrete reduction. An
example of this is the reduction of electrons when it hits the screen in the
double slit experiment. For this, the discrete model of the bidirectional
temporal evolution in section \ref{2.2.1} may apply, to be tested by
experiment. One has to use the initial and final electron states to
calculate the bidirectional temporal state of the electron.

In case of scattered particles such as cosmic rays, particles in
interstellar and intergalactic space, they may have quantum properties that
build until they become large enough in interaction with other particles to
be reduced by the EM or Gravitational field. The same thing can be said
about gaseous matter. Continuous interaction builds large-scale
superposition among the interacting particles that can be reduced when
strong enough due to the EM or gravitational field. When they interact with
SDEDs that also induce reduction.

Now the quantum particles, specially prepared and isolated enough to
maintain coherence, such as electrons in vacuum or qubits in quantum
computers, maintain superposition for a long time.

The EM and gravitational field, assumed to rapidly reduced by classicum, can
be modelled by quantum diffusion with bidirectional temporal reduction in
section \ref{2.2.3}. The quantum particles under reduction after an extended
time, we can solve for bilateral temporal reduction using equations given in
section \ref{2.2.1} or \ref{2.2.2}.

In the physical foundations of consciousness \cite{MYP5B1.3}, the ideas
regarding these two types of reduction, entanglement domains, and classicum
are discussed in more detail.

\section{Standard Quantum General Relativity}

\subsection{Flat Space-Time}

Let me now discuss the basic setup of Hamiltonian formulation for the
physics of fields and particles in flat space-time using canonical quantum
gravity. I am considering particles and fields separately because both are
necessary to understand macroscopic physics to formulate time evolution and
quantum reduction in quantum gravity. If we separate particles from fields,
the remaining fields have only zero-point fluctuations in the case of
fermions, and zero-point and macroscopic semiclassical fields in the case of
bosonic fields. In this paper, I\ will simply refer to them as continuum
fields unless otherwise specified. In this paper, greek letters are abstract
indices relating to space-time coordinates, while Latin indices $a,b,c,d$
denote spatial coordinates, $i,j,k,l$ denote particles, and $X$ denotes a
set of continuum fields.

Let {$\mathcal{\phi }$}$_{j}^{X}$ denote the set of all continuum fields on
the manifold such as Yang-Mills, bosons, fermions, etc. The $j$ denotes
different fields and $X$ represents indices for these continuum fields. Let $%
q_{i}^{\alpha }$ be particle positions in space-time, where $i$ denotes
different particles, $\alpha $ is space-time coordinates. Let $S$ be the
total action corresponding to the fields and particles on the space-time
manifold $\mathcal{M}$ without including general relativity for now:

\begin{equation}
S=\int \mathcal{L(\phi }_{j}^{X},\partial _{\beta }\mathcal{\mathfrak{\phi }}%
_{j}^{X},\dot{q}_{i}^{\alpha },q_{i}^{\alpha }\mathcal{)}d^{4}x
\end{equation}%
where the $\mathcal{L}\ $is the corresponding Lagrangian density and the $x$
are the space-time coordinates. The particle part of the Lagrangian density
will involve deltas, which on integration, this part reduces into a
summation of Lagrangian of various particles and their interactions.

To get the Hamiltonian formulation, we assume the topology of the space-time
manifold is $R\otimes $ $S,$ where $S$ is a flat $3D$ manifold. The
space-time manifold is foliated by a one-parameter family of spatial
hyperplanes $\Sigma (t)$, parametrized by a (time) parameter $t$. Now, let $%
q_{i}^{a}$ be the (spatial) coordinates of particles on the 3D hyperplanes.
Let $\mathcal{\pi }_{Y}^{k}$ be the conjugate momenta of continuum fields
and $p_{a}^{i}$ are the momenta of particles. For non-dynamical variables in 
$\mathcal{\phi }_{j}^{X},$ the dual conjugate momenta will be zero.

We go to the Hamiltonian formulation using the Legendre transformation. For
this, usually, formally, we rewrite the Lagrangian and the Action as follows:

\begin{eqnarray}
L(\mathcal{\mathfrak{\phi }}_{j}^{X},\partial _{\beta }\mathcal{\mathfrak{%
\phi }}_{k}^{Y},p_{b}^{j},q_{i}^{a}) &=&\int_{\Sigma (t)}\mathcal{L}(%
\mathcal{\mathfrak{\phi }}_{j}^{X},\partial _{\beta }\mathcal{\mathfrak{\phi 
}}_{k}^{Y},p_{b}^{j},q_{i}^{a})d^{3}x \\
L(\mathcal{\mathfrak{\phi }}_{j}^{X},\partial _{\beta }\mathcal{\mathfrak{%
\phi }}_{k}^{Y},p_{b}^{j},q_{i}^{a}) &\mathcal{=}&p_{a}^{j}\dot{q}%
_{j}^{a}+\int \mathcal{\pi }_{X}^{j}~\mathcal{\mathfrak{\dot{\phi}}}%
_{j}^{X}d^{3}x-\int \mathcal{H}(\mathcal{\mathfrak{\phi }}_{j}^{X},\mathcal{%
\pi }_{Y}^{k},q_{b}^{j},p_{i}^{a})d^{3}x \\
H(\mathcal{\mathfrak{\phi }}_{j}^{X},\mathcal{\pi }%
_{Y}^{k},q_{b}^{j},p_{i}^{a}) &=&\int_{\Sigma (t)}\mathcal{H}(\mathcal{%
\mathfrak{\phi }}_{j}^{X},\mathcal{\pi }_{Y}^{k},q_{b}^{j},p_{i}^{a})d^{3}x
\\
S &=&\int Ldt \\
&=&\int p_{a}^{j}dq_{j}^{a}+\int \mathcal{\pi }_{X}^{j}~d\mathcal{\mathfrak{%
\phi }}_{j}^{X}d^{3}x-\int \mathcal{H}(\mathcal{\mathfrak{\phi }}_{j}^{X},%
\mathcal{\pi }_{Y}^{k},q_{b}^{j},p_{i}^{a})d^{3}xdt
\end{eqnarray}%
\ \qquad

Above the $\mathcal{\mathfrak{\phi }}_{j}^{X}$ and $\mathcal{\pi }_{Y}^{k}$
are the collection of the configuration variables and the corresponding
conjugate momentum variables (one set for each point of the manifold) of the
fields and $\mathcal{H}$ is the Hamiltonian density. Usually, $\mathcal{H}$
involves constraint terms $\mathcal{C}_{k}$,

\begin{equation}
\mathcal{H}=\mathcal{H}_{0}(\mathcal{\mathfrak{\phi }}_{j}^{X},\mathcal{\pi }%
_{Y}^{k},q_{b}^{j},p_{i}^{a})+\lambda ^{k}\mathcal{C}_{k}(\mathcal{\mathfrak{%
\phi }}_{j}^{X},\mathcal{\pi }_{Y}^{k},q_{b}^{j},p_{i}^{a})
\end{equation}

Usually, the variational analysis of the action leads to the Hamilton
equations of motion. For any observable $F(\mathfrak{\phi }^{\alpha },%
\mathcal{\pi }^{\alpha },p_{i}^{a},q_{i}^{a}),$ a function of the variables $%
\mathfrak{\phi }^{\alpha },\mathcal{\pi }^{\alpha },p_{i}^{a},$and $%
q_{i}^{a},$

we have,

\begin{eqnarray}
\frac{dF}{dt} &=&\{F,H\}, \\
C_{\alpha } &=&0,
\end{eqnarray}%
where $\{,\}$ is the Poisson bracket. At the quantum level, we can define
the quantum state on $\Sigma (t)$ as follows:

\begin{equation}
\left\vert \Psi \right\rangle =\int \psi (\mathcal{\mathfrak{\phi }}%
_{j}^{X},q_{i}^{a})d\mathcal{\mathfrak{\phi }}dq
\end{equation}

Here $d\mathcal{\mathfrak{\phi }}dq$ formally denotes an integral measure of
the configuration space. Dirac has formulated the way to go from the
classical Hamiltonian formulation to the quantum mechanical formulation \cite%
{DRC}:

\begin{eqnarray}
\lbrack \mathcal{\mathfrak{\phi }}_{j}^{X}(x),\mathcal{\pi }%
_{Y}^{k}(x^{\prime })] &=&i\hbar \delta _{j}^{k}\delta _{Y}^{X}\delta
^{3}(x-x^{\prime }),  \label{3} \\
\lbrack q_{i}^{a},p_{b}^{j}(x^{\prime })] &=&i\hbar \delta _{b}^{a}\delta
_{i}^{j}, \\
\hat{C}_{k}\left\vert \psi \right\rangle &=&0,~\forall k, \\
\lbrack \hat{F},\hat{C}_{k}] &=&0,~\forall k, \\
i\hbar \frac{d\hat{F}}{dt} &=&[\hat{F},\hat{H}], \\
i\hbar \frac{d}{dt}\left\vert \psi \right\rangle &=&\hat{H}\left\vert \psi
\right\rangle
\end{eqnarray}

The $\hat{C}_{k}\left\vert \psi \right\rangle =0$ and $[\hat{F},~\hat{C}%
_{k}]=0,~\forall k$ mean that the physical states and operators are gauge
invariant.

\subsection{Curved Space-Time\label{3.2}}

Now assume we also include gravity in the continuum fields$~\mathcal{%
\mathfrak{\phi }}_{j}^{X}$ and the action. The spatial manifolds are no
longer assumed to be flat and we include metric density in the integrals to
make them invariant. In quantum gravity $H_{0}$ is absent. $H\ $\ is
composed of first-class constraints only, which can interpreted as gauge
transformations on the fields of both space-time and internal coordinates of
the fields. There are three constraints in quantum gravity:\ The Hamiltonian
constraint $H_{N},$ the Diffeomorphism constraint $D_{N^{i}},$ and the Gauge
Constraints $G_{i}.$

Let's discuss the setup to discuss the flow of time in the full physics on a 
$(3+1)D$ space-time based on ADM formalism \cite{ADM}. Consider a space-time
with metric $g_{\alpha \beta }$ and one parameter spacial foliation $%
\mathcal{S}_{t},$ where $\mathcal{S}_{t}$ is the spatial hypersurface for a
given $t.$ This foliation can be specified by function $t(x^{\alpha }),$ $%
x^{\alpha }$ is a point in space-time, with $t=$ constant describing the
surface $\mathcal{S}_{t}$. We can choose $t$ to be the time coordinate.
Consider the vector field, $T^{\gamma }=(\frac{\partial }{\partial _{t}}%
)^{\gamma }=\partial _{t}^{\boldsymbol{\gamma }}$. $T^{\gamma }$ generates a
one-parameter family of space-time diffeomorphism, such that a given initial
surface $\mathcal{S}_{t^{1}}$ is mapped to a different surface $\mathcal{S}%
_{t^{2}}$ of the foliation. If $n^{\boldsymbol{\alpha }}$ are normal to $%
\mathcal{S}$, $\partial _{a}^{\boldsymbol{\alpha }}$ are coordinate bases on 
$\mathcal{S}$, $\partial _{A}^{\alpha }$ are bases for internal degrees of
freedom,

\begin{equation}
T^{\alpha }=Nn^{\boldsymbol{\alpha }}+N^{a}\partial _{a}^{\boldsymbol{\alpha 
}}+\alpha ^{A}\partial _{A}^{\boldsymbol{\alpha }}
\end{equation}%
$N$ is called the lapse and $N^{a}$ is called the shift. The free variables
of the metric in canonical formulation are the spatial metric $h_{ab}$ which
is the projection of $g_{\alpha \beta }$ onto the spatial hypersurfaces.

We require both fields and particles in the theory to understand the flow of
time and decoherence in a quantum system. The Hamiltonian can be written as

\begin{eqnarray}
\mathcal{H} &=&N\mathcal{H}_{N}(\mathcal{\mathfrak{\phi }}_{j}^{X},\mathcal{%
\pi }_{Y}^{k},q_{b}^{j},p_{i}^{a})+N^{a}\mathcal{D}_{a}(\mathcal{\mathfrak{%
\phi }}_{j}^{X},\mathcal{\pi }_{Y}^{k},q_{b}^{j},p_{i}^{a})+\alpha ^{A}%
\mathcal{G}_{A}(\mathcal{\mathfrak{\phi }}_{j}^{X},\mathcal{\pi }%
_{Y}^{k},q_{b}^{j},p_{i}^{a}))  \label{HamAll} \\
H &=&\int \mathcal{H}d^{3}x=H_{N}+D_{N^{a}}+G_{\alpha ^{A}}  \label{HamAll2}
\\
H_{N} &=&\int N\mathcal{H}(\mathcal{\mathfrak{\phi }}_{j}^{X},\mathcal{\pi }%
_{Y}^{k},q_{b}^{j},p_{i}^{a})\sqrt{h}d^{3}x, \\
D_{N^{a}} &=&\int N^{a}\mathcal{D}_{a}(\mathcal{\mathfrak{\phi }}_{j}^{X},%
\mathcal{\pi }_{Y}^{k},q_{b}^{j},p_{i}^{a})\sqrt{h}d^{3}x, \\
G_{\alpha ^{A}} &=&\int \alpha ^{A}\mathcal{G}_{A}(\mathcal{\mathfrak{\phi }}%
_{j}^{X},\mathcal{\pi }_{Y}^{k},q_{b}^{j},p_{i}^{a})\sqrt{h}d^{3}x,
\end{eqnarray}%
where $\alpha ^{A}$ are non-dynamical parameters relating to gauge
transformation on the internal field variables. $\sqrt{h}$ is the 3D spatial
metric density. $\mathcal{\mathfrak{\phi }}_{j}^{X}~$and $\mathcal{\pi }%
_{Y}^{k}$ are assumed to contain all dynamical variables and conjugate
momenta of all fields, including metric, gauge, spinor, and scalar fields. $%
\mathcal{L}$ and $\mathcal{H}$ are assumed not to explicitly depend on time.

Here the time evolution is given by

\begin{equation}
i\hbar \frac{d}{dt}\left\vert \psi \right\rangle =\hat{H}\left\vert \psi
\right\rangle =0,
\end{equation}

The time derivative is zero as the Hamiltonian is purely a sum of
constraints that are supposed to annihilate the quantum state. This is the
Hamiltonian Constraint equation which contains the information about the
time evolution.

\section{The New Quantum General Relativity Framework}

\subsection{Introduction\label{4.1}}

We need to extract evolution from the Hamiltonian Constraint equation, which
is defined as the problem of time in quantum gravity. I will discuss the
various proposals on how to do it next, based on the quantum gravity
framework series of papers 0 to 4 \cite{MYP0}, \cite{MYP1},\cite{MYP2},\cite%
{MYP3}, and \cite{MYP4}.

In the previous version of the quantum gravity framework, I\ discussed a new
conceptual general relativistic formulation of quantum general relativity.
In this paper, I put all the previous discussions together in a clear
framework.

The path integral for quantum general relativity is

\begin{equation}
G(q_{a}^{\alpha },\mathfrak{\phi }_{X}^{\alpha },q_{a}^{\prime \alpha },%
\mathfrak{\phi }_{X}^{\prime \alpha })=\int \exp (ip_{a}^{j}\Delta
q_{j}^{a}+i\int \mathcal{\pi }_{X}^{j}~\Delta \mathcal{\mathfrak{\phi }}%
_{j}^{X}d^{3}x)\delta (\mathcal{H}_{N})\delta (\mathcal{D}_{a})\delta (%
\mathcal{G}_{A})\mu dpd\mathcal{\pi }\mathfrak{,}  \label{PathGen}
\end{equation}%
where I\ assume $\mu $ is a function of phase space variables to impose
orthonormal conditions on the path integral, and $dpd\mathcal{\pi }$ is the
Hilbert measure on the space of momenta. Here $q_{a}^{\alpha },\mathfrak{%
\phi }_{X}^{\alpha }$ are configuration variables in an initial time, $%
q_{a}^{\prime \alpha },\mathfrak{\phi }_{X}^{\prime \alpha }$ are
configuration variables in a future time, and $\Delta q_{j}^{a}$, $\Delta 
\mathcal{\mathfrak{\phi }}_{j}^{X}$ are differences between them.

The path integral in equation \ref{PathGen} is quite general and it doesn't
have a time variable in it. All it does is give quantum amplitudes to go
from configuration values in a hypersurface to configuration values in an
adjacent hypersurface. We can use any configuration variable or a function
of many configuration variables as time. This time parameter can be
considered as the state of a clock. From this, we can calculate the time
evolution of quantum amplitudes for any state defined on the remaining
degrees of freedom with respect to this clock using the path integral. We
can also have multiple clock parameters and can calculate quantum evolution
between different states of them. Let $T(q^{\alpha },\mathfrak{\phi }%
_{X}^{\alpha })$ be the clock function. Then if $\psi (T_{1},q_{j}^{\alpha },%
\mathfrak{\phi }_{X}^{\alpha }),$ is the quantum state at time $%
T_{1}=T(q_{j}^{\alpha },\mathfrak{\phi }_{X}^{\alpha })$ then the quantum
evolution to $\psi (T_{2},q_{j}^{\alpha },\mathfrak{\phi }_{X}^{\alpha })$
is formally given by:

\begin{equation}
\psi (T_{2},q_{j}^{\prime \alpha },\mathcal{\phi }_{X}^{\prime \alpha
})=\int G(q_{j}^{\alpha },\mathcal{\phi }_{X}^{\alpha },q^{\prime \alpha },%
\mathcal{\phi }_{X}^{\prime \alpha })\delta (T_{1}-T(q_{j}^{\alpha },%
\mathcal{\phi }_{X}^{\alpha }))\delta (T_{2}-T(q_{j}^{\prime \alpha },%
\mathcal{\phi }_{X}^{\prime \alpha }))\psi (T_{2},q_{j}^{\alpha },\mathcal{%
\phi }_{X}^{\alpha })dq_{a}^{\alpha }d\mathcal{\phi }_{X}^{\alpha }
\end{equation}

This is quite formal. To make sense of the time in a physical setting we
need a proper mathematical setup for evolution in this context and a way to
physically interpret this quantum evolution. This is what we will do in this
section.

In the new framework for quantum general relativity, we adopt everything
that was discussed in the standard canonical quantization of quantum general
relativity. There are two changes: 1) I add new formulations to understand
time evolution and quantum reduction. 2) I remove the Copenhagen
interpretation. I assume the quantum state is physical and reduction happens
in the way of a density matrix or equivalently stochastic Schr\"{o}dinger
equation.

We assume the universe is filled with fields that behave classically at each
point and is filled with macroscopic objects made of many quantum particles
whose center of mass behaves classically.

Now based on the path integral formulation of quantum general relativity, I
put forward a new framework necessary to understand the dynamics of general
relativity. Here is the overview:\qquad

\begin{itemize}
\item In the new framework for quantum general relativity, we adopt
everything that was discussed in the standard canonical quantization of
quantum general relativity. There are two changes: 1) I add new formulations
to understand time evolution and quantum reduction. 2) I remove the
Copenhagen interpretation. I assume the quantum state is physical and
reduction happens in the way of a density matrix or equivalently stochastic
Schr\"{o}dinger formalism that we discussed before.

\item We assume the universe is filled with fields that behave classically
at each point and is filled with macroscopic objects made of many bound
quantum particles whose center of mass behaves classically.

\item Time evolution happens bidirectional temporally along a foliation both
in continuous deterministic and stochastic evolution. The continuous
deterministic evolution corresponds to Schr\"{o}dinger evolution and the
stochastic evolution corresponds to spontaneous quantum reduction.

\item First, we formulate the continuous deterministic evolution. The
internal space of each of the fields at each spatial point is foliated into
(n-1)+1 form using time constraint, which creates a time parameter in the
foliation space. This will be discussed in the single-point system case in
the next section. Later I will also discuss the case of the multipoint
system of the entire spatial manifold. I will define time evolution for both
positive and negative energies traveling in opposite directions. The center
of masses of the bound systems acts as clocks. This can be used to foliate
the configuration space of position variables of the particles of the bound
systems. This also gives time evolution.

\item Next, we formulate the stochastic part of the evolution using the
density matrix formulation. The density matrix evolution is formulated using
path integral formulation and a distance function is introduced to formulate
reduction. This evolution can be recast into discrete sequences of
projections or continuous reduction that can be reformulated into stochastic
Schr\"{o}dinger form.

\item Both the continuous and stochastic evolution depend on the foliation
of the configuration space of particles and fields that we discussed before.
The time scaling of the quantum evolution of each spatial point results in
different foliations of the physical phenomena in space-time. Here I
formulate the rest frame foliation, which is defined probabilistically using
a function that measures how much the physical fields are least changing up
to a conformal metric scale factor. I propose that the foliation with the
smallest change has the highest probability being the foliation in which
evolution happens.

\item In `The Physical Foundations of Consciousness and Life, Framework 1' 
\cite{MYP5B} I discuss the relation to structure formation and consciousness
in the universe. It describes in detail the dynamical phenomena of life and
consciousness in terms of the three fundamental postulates discussed in this
paper. We introduce an action term that promotes structure formation and the
free-will influence of consciousness on matter.
\end{itemize}

The postulate on the deterministic and continuum limits as the fourth
postulate in the previous versions of the quantum gravity framework project
has been omitted and the discussion regarding it will be included in the
future versions.

In these formulations, we first discuss single-point systems and then we
generalize to full quantum general relativity. The single-point system
refers to a system with one space-time point with an N-dimensional
configuration space internally.

\subsection{Time Evolution}

\subsubsection{A Single Point System}

First lets formulate time evolution in a single point system. Consider the
phase space described by variables of the single-point system discussed in
postulate 2.1 in section 2 of \cite{MYP1}. We have the conventional timeless
single-step propagator for this as

\begin{equation}
G(q^{a},q^{\prime a})=\int \exp (ip_{a}\Delta q^{a})\delta
(H(p_{a},q^{a}))\mu dp  \label{Prop}
\end{equation}%
Here standard definitions apply to the variables. Time is hidden implicitly
in this propagator. In quantum gravity frameworks 0 and 1, \cite{MYP0}, \cite%
{MYP1}, I have introduced a naive time constraint to impose time evolution.

Let me assume the time function is given by 
\begin{equation}
T=v_{a}q^{a}
\end{equation}%
where $v_{a}$ is a constant. This foliates the configuration space into a
sequence of temporal hyperplanes. Let me introduce a general form of the
naive time constraint in the phase space as follows:

\begin{equation}
v_{a}\Delta q^{a}\mp v\Delta T=0.  \label{Tcon}
\end{equation}%
where $v_{a}$ is the unit vector in the metric defined by the kinetic terms.
The minus sign corresponds to positive energy solutions traveling along the $%
+ve$ time direction and the plus sign corresponds to negative energy
solutions traveling towards the $-ve$ time direction. Here $v$ is a
parameter is a function of $t$ that acts as a time scale parameter, like the
lapse. In this paper I\ will set $v=1$, assuming that the time parameter has
been scaled. This will help make the discussions much clearer.

The single-step path integral for the time evolution for positive and
negative energy wavefunctions can be defined as follows: 
\begin{equation}
G_{s\pm }(q^{a},q^{\prime a};\Delta T,v_{a},v)=\int_{p_{a}v^{a}<0}\exp
(ip_{a}\Delta q^{a})\delta (H)\delta (v_{a}\Delta q^{a}\mp v\Delta T)\mu dp,
\label{selprop}
\end{equation}%
where $dp$ is a measure of the momentum space. The $\Delta T$ is assumed to
be infinitesimal. This was first defined in the quantum gravity framework 1 
\cite{MYP1}. The semicolon separates the configuration variables and the
parameters with respect to which the time evolution is defined. The
single-step operator is applied sequentially to get time evolution in a
finite time interval.

\subsubsection{Relative Time Evolution for a Single Point System\label{4.2.2}%
}

If $v_{a}$ is not constant, the foliation of the configuration space is
curved, and we need to define the path integral in a much more general way.
This has been discussed in detail in \cite{MYP1} for a general single-point
system which is the foundation of time evolution in this paper. Let me
discuss this briefly.\ Basically, time evolution can be defined by foliating
the configuration space of the single-point system. Given any smooth
classical path $\eta ~$defined by $Q^{a}(t)$ in the configuration space $%
\mathit{R}^{n}$ of the single point system, one can always define the
quantum evolution with respect to this path using the time-constrained path
integral.

The orthogonality and the raising and lowering of indices are done using the
metric defined by the kinetic term as discussed in the previous papers. If
the metric is defined by the kinetic term is the function of $q^{a}$ then,
we use the value of metric at $q^{a}=Q^{a}(t)$ at time $t$.

\FRAME{ftbpFU}{4.1105in}{2.3341in}{0pt}{\Qcb{Evolution of a single point
system.}}{\Qlb{FigEvol}}{hypersurface_evolution2.jpg}{\special{language
"Scientific Word";type "GRAPHIC";maintain-aspect-ratio TRUE;display
"USEDEF";valid_file "F";width 4.1105in;height 2.3341in;depth
0pt;original-width 7.3172in;original-height 4.1416in;cropleft "0";croptop
"1";cropright "1";cropbottom "0";filename
'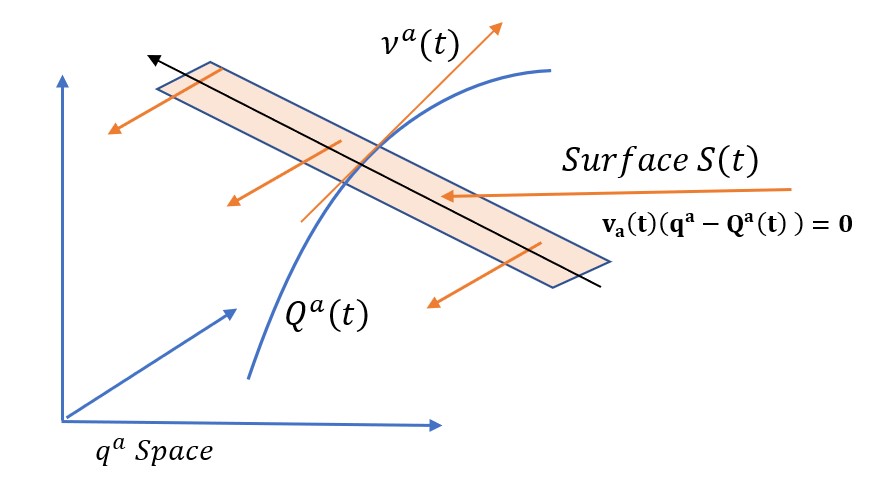';file-properties "XNPEU";}}

Let at each point $Q^{a}(t)$ we define cotangent $v_{a}(t)$ of $\eta $ at
that point. At each point $Q^{a}(t)$ of $\eta ,$ we can define a hyperplane $%
S(t)$ orthogonal to the cotangent $v_{a}(t)$ of $\eta $ at that point. If $%
v_{a}$ is not constant we cannot foliate the internal configuration space
using the sequence of hyperplanes as they intersect with each other. But we
make the following assumption:

\begin{principle}
Assumption A.1: The S(t) do not intersect in the region where the wave
function of the system is non-zero $\psi (x^{b})$.
\end{principle}

\FRAME{ftbpFU}{4.638in}{2.7146in}{0pt}{\Qcb{The foliation doesn't intersect
in the region in which the $\protect\psi (x^{a})$ is non zero.}}{\Qlb{psireg}%
}{psi_region.png}{\special{language "Scientific Word";type
"GRAPHIC";maintain-aspect-ratio TRUE;display "USEDEF";valid_file "F";width
4.638in;height 2.7146in;depth 0pt;original-width 6.0001in;original-height
3.4999in;cropleft "0";croptop "1";cropright "1";cropbottom "0";filename
'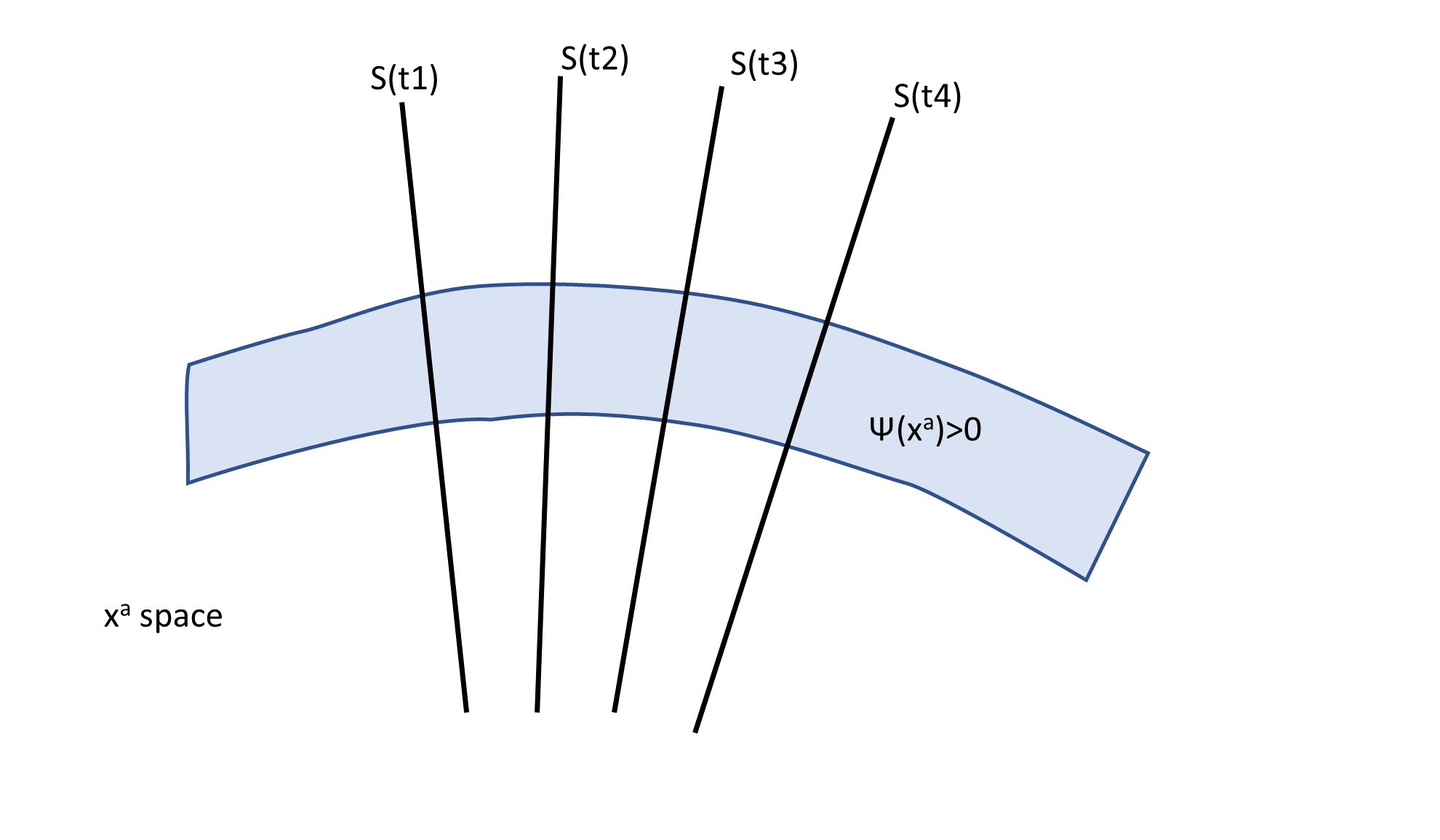';file-properties "XNPEU";}}

If $t$ is the time coordinate, used in doing canonical quantization, then
for each $t$ there will be a configuration space $R^{n}.$We assume that the
cotangent $v_{a}(t)$ defines a foliation $v_{a}(t)q^{a}=T,$ using
hyperplanes for each of the configuration space $R^{n}$ at coordinate time $%
t.$ Different $T$ corresponds to different hyperplanes. This foliation
changes between different $t$ as $\nu _{a}(t)$ varies.

Now we can map $T$ to the coordinate time $t,$ by a function $T(t)$ so that
there is one unique hyperplane $S(t)$ for each configuration space for each $%
t.$ This hyperplane $S(t)$ defines an effective configuration space with $%
T(t)$ as the time parameter associated with it.\ We can define wavefunctions
on these effective configuration spaces to study physics with time defined
by $T(t).$

We can define $S(t)$ by the equation:

\begin{equation}
v_{a}(t)(q^{a}-Q^{a}(t))=0
\end{equation}

Then the hyperplanes $S(t)$ will pass through $Q^{a}(t),$ and the map
between $T$ and $t$ is given by 
\begin{equation}
T(t)=v_{a}(t)Q^{a}(t)  \label{T(t)}
\end{equation}

We can define $q_{_{\bot }}^{a}=$ $q^{a}-Q^{a}(t)$ as free variables on $%
S(t) $. We can define a Hilbert space of functions on each hyperplane $S(t)$
to define quantum states$.$ We can define positive and negative going
quantum evolution of wave functions $\psi _{\pm }(q_{_{\bot }}^{a},t)$ on
the foliation $S(t)$, using the below single-step propagators from $t$ to $%
t+dt,$

\begin{eqnarray}
&&G_{s\pm }(q_{_{\bot }1}^{a},q_{_{\bot }2}^{a};\eta \mathfrak{,}t,\Delta t)
\label{hypEvol} \\
&=&\frac{1}{\left( 2\pi \right) ^{d-1}}\int_{\pm p^{a}v_{a}(t)<0}\exp
(ip_{a}\Delta q_{_{\bot }}^{a})\delta (H)\mu dp \\
&=&\frac{1}{\left( 2\pi \right) ^{d-1}}\int_{\pm p^{a}v_{a}(t)<0}\exp \left[
ip_{_{\bot }a}\Delta q_{_{\bot }}^{a}\mp iH_{\bot }(p_{_{\bot }a},q_{_{\bot
}}^{a})\Delta t\right] \mu dp_{\bot },
\end{eqnarray}%
where $\Delta q_{_{\bot }}^{a}=q_{_{\bot }2}^{a}-q_{_{\bot }1}^{a}$. The $%
dp_{\bot }$ is a measure on projected momentum space. This path integral
depends on $\eta .$

\begin{equation}
\psi _{\pm }(q_{_{\bot }1}^{a},t+\Delta t)=\int_{q_{_{\bot }2}^{a}}G_{s\pm
}(q_{_{\bot }1}^{a},q_{_{\bot }2}^{a};\eta \mathfrak{,}t,\Delta t)\psi _{\pm
}(q_{_{\bot }2}^{a},t)  \label{hypdiff}
\end{equation}%
Here the$\ \delta (H)$ is used to solve for the effective Hamiltonian $%
H_{\bot }(p_{_{\bot }a},q_{_{\bot }}^{a})$ as a function of $p_{_{\bot }a}$
and $q_{_{\bot }}^{a}$. The $\mu =\mu (p_{_{\bot }a},q_{_{\bot }}^{a})$ is a
weight, which is deduced so that $G_{s+}(q_{_{\bot }1}^{a},q_{_{\bot
}2}^{a};\eta \mathfrak{,}t,0)=\delta (q_{_{\bot }1}^{a}-q_{_{\bot }2}^{a}),$
where $\delta $ is the Dirac delta function.

We can define an effective path integral as follows:

\begin{equation}
S_{\pm }(p_{_{\bot }a},q_{_{\bot }}^{a};\eta )=\int p_{_{\bot }a}dq_{_{\bot
}}^{a}\mp H_{\bot }(p_{_{\bot }a},q_{_{\bot }}^{a})dt
\end{equation}%
This action depends on the path $\eta $.

In quantum gravity framework 1 \cite{MYP1} I\ explicitly calculated the
Hamiltonian for a single-point system for a typical Hamiltonian. I\ can
write the differential form of the evolution equation of wavefunction as

\begin{equation}
d_{s}\left\vert \psi _{\pm }\right\rangle =\pm iH_{\bot }dt\left\vert \psi
_{\pm }\right\rangle
\end{equation}

We can choose $\eta $ such that expectation values of $p_{_{\bot }a}$ and $%
q_{_{\bot }}^{a}$ are zero. I define this choice as the self-time evolution,
as its intuitive time direction is given by its momentum \cite{MYP1}. This
makes much more sense in special relativity as the direction of the time
axis in the $3+1D$ space of an observer is its 4-velocity. This also defines
rest frame foliation in the single-point system case.

Relative time evolution is more general than self-time evolution
formulation. The self-time evolution is useful only for simple cases, which
has been discussed in various cases in the previous papers on the quantum
gravity framework project. Relative time evolution is necessary for properly
defining quantum reduction as we will see later.

\subsubsection{Full Space-Time}

Now let us go to the full space-time with particles and fields. The timeless
path integral is given by

\begin{equation}
G(\mathcal{\phi }_{j}^{X},\mathcal{\phi }_{j}^{\prime
X},q_{i}^{a},q_{i}^{\prime a})=\int \exp (iS)dNdN^{i}d\alpha ^{A}\mu d%
\mathcal{\pi }dp,
\end{equation}%
where $S$ is full action with the constraint functions. $d\mathcal{\pi }dp$
is the measure in the momentum space.\ This action depends on $%
N,N^{i},\alpha ^{A}$. They are to be integrated out to get delta terms
imposing constraints in the path-integral.

To define global evolution we need to generalize $\eta $ to include infinite
space-time points with continuum fields and particles living on them. Let $%
\mathbf{\eta }$ collectively denote this generalized $\eta .$ The $\mathbf{%
\eta }$ is a collection of smooth paths in the configuration space of
particle systems and continuum fields, except fermionic fields. We ignore
the fermionic fields because they don't have macroscopic expectation values
for fields. For each point $x,$ $\mathfrak{\Phi }_{j}^{X}(t,x)$ be a set of
classical paths corresponding to continuum fields.

Let $Q_{i}^{a}(t)$ be the center of mass corresponding to a collection of
particles that make bulk matter such as planets, stars, football, physical
clocks, or insects, such that it behaves classically. How small one can go
such that $Q_{i}^{a}(t)$ is classical depends on establishing reduction
constants in the quantum reduction postulate that will be discussed later. $%
Q_{i}^{a}(t)$ are calculated using the weighted averaging of the coordinates
of the particles. This is good for flat space-time and for weak classical
gravitational fields. For objects of small mass in meter scales or smaller
than that this is suffice. In the case of highly curved space-times, when
dealing with bounds systems like planets to stars to black holes, if they
have symmetry that allows for a central point of symmetry then we can set $%
Q_{i}^{a}(t)$ it to equal to that. In other cases, this is harder, but an
appropriate choice of $Q_{i}^{a}(t)$ can be found.

I am going to call these bound systems simple systems as particle systems.
These clocks have to be an isolated piece of bulk matter in a region in
which all bound molecules are included. I am using particle systems because
this captures macroscopic components of fermions, whose field expectation
values are zero at any point. Our discussion here will be based on $%
Q_{i}^{a}(t)$ being the center of mass. But it need not be, it could be
other functions of particle coordinates, in which case we need to adapt the
theory we will be discussing appropriately.

In quantum gravity framework 3 \cite{MYP3} I\ have formulated quantum
dynamics for space-time with one single bulk object and also quantum
dynamics for a continuum field. Both of these cases were treated separately,
and the effective Hamiltonian was derived. In this paper, I\ will combine
them together.

Now $\mathbf{\eta }$ corresponds to the collection of information $%
\{Q_{l}^{a}(t),~\mathfrak{\Phi }_{j}^{X}(x,t)\}$ with tangents given by the
derivatives of $Q_{l}^{a}(t)$ and $\mathfrak{\Phi }_{j}^{X}(x,t)$ upto a
scaling. 
\begin{equation}
\mathbf{\eta =}\{Q_{l}^{a}(t),\mathfrak{\Phi }_{j}^{X}(x,t)\}  \label{FOL-4D}
\end{equation}

Let $\nu _{a}^{j}(t)$ and\ $\kappa _{X,j}(x,t)$ be the tangents to paths
corresponding to paths $Q_{l}^{a}(t)$ and $\mathfrak{\Phi }_{j}^{X}(x,t)$ at
point $t$ on $\mathbf{\eta }.$We can define hyperplanes $\boldsymbol{S}(t)$
in the internal space of fields and particles as

\begin{eqnarray}
\nu _{a,j}(t)(q_{cm,j}^{a}-Q_{j}^{a}(t)) &=&0.  \label{Fol-EQ} \\
\kappa _{X,j}(x,t)(\mathcal{\phi }_{j}^{X}-\mathfrak{\Phi }_{j}^{X}(t,x))
&=&0.  \label{FOL-EQ2}
\end{eqnarray}

where $q_{cm,j}^{a}$ is the center of mass of $j^{th}$ particle system. The
orthogonality and the raising and lowering of indices are defined by the
classical spatial metric value that is included as part of $\mathfrak{\Phi }%
_{j}^{X}(t,x)$. The hyperplane $\boldsymbol{S}(t)$ here is defined as the
topological product of all the hyperplanes for each particle system defined
by equation \ref{Fol-EQ} and those for the configuration space of fields at
each point $x$ defined by equation \ref{FOL-EQ2}.

We can generalize $T(t)$ defined in the single point system by equation \ref%
{T(t)} for the general 4D case. But there will be many $T(t)$ as there are
many curves $Q_{j}^{a}(t)$. The $T_{j}=\nu _{a,j}(t)q_{cm,j}^{a}$ can be
considered as free variable characterizing time for the $j$ system, and its
relation to $t$ can be calculated using \ref{Fol-EQ} as follows:

\begin{equation}
T_{j}(t)=\nu _{a,j}(t)q_{cm,j}^{a}=\nu _{a,j}(t)Q_{j}^{a}(t)  \label{TimJ}
\end{equation}

In this paper, we will only use $t$ parametrizing the foliation of 4D
space-time as a time parameter, and discuss effective Hamiltonians
associated with them to describe dynamics. But we can always set $T_{j}(t)$
as the effective time if it will simplify the analysis.

Let $\mathcal{\phi }_{\perp x}^{X}=\mathcal{\phi }_{j}^{X}-\mathfrak{\Phi }%
_{j}^{X}(t,x)$ are the free variables of the field, with $\mathcal{\phi }%
_{j}^{X}$ being on $\boldsymbol{S}(t)$. Assume $q_{i}^{a}$ belongs to $%
\boldsymbol{S}(t).$ There are two cases:

\begin{itemize}
\item If the particle doesn't belong to any particle system then $q_{\perp
i}^{a}=q_{i}^{a}$\emph{. }

\item If particle $i$ belongs to $j^{th}$ particle system then $q_{\perp
i}^{a}=q_{i}^{a}-Q_{j}^{a}(t)$ are free fields on the hyperplanes $%
\boldsymbol{S}(t).$
\end{itemize}

In the second case, all these $q_{\perp i}^{a}$ are not independent because
of the constraint equations \ref{Fol-EQ}. So one of the particle position
needs to be expressed as a function of others, and the independent variables
need to be properly identified along with the respective conjugate momenta.
One of the free variable that can be expressed as a linear combination of
the $q_{i}^{a}$ of the $j^{th}$ system will be the parallel component to $%
\nu _{a,j}(t),$ which is defined in equation \ref{TimJ}. This is $T(t)$
which can act as a local time parameter.

\begin{principle}
Fundamental postulate I.1\textbf{: Given a set of the classical foliation of
configuration space of particles and continuum fields by }$\mathbf{\eta =}%
\{Q_{l}^{a}(t),\mathfrak{\Phi }_{j}^{X}(x,t)\},$\textbf{\ time-oriented
single-step propagator describing the combined systems in the relative-time
formalism can be defined as follows: }%
\begin{eqnarray}
&&G_{\pm }(\mathcal{\phi }_{\perp x}^{X},\mathcal{\phi }_{\perp x}^{\prime
X},q_{\perp aj},q_{\perp aj}^{\prime };\mathbf{\eta },\Delta t,t)
\label{GTevol} \\
&=&\int_{\forall i,j,\text{ }\pm \mathcal{\pi }_{X}^{j}\kappa
_{j}^{X}(t)<0,~\pm p_{a}^{i}\nu _{i}^{a}(t)<0,}\exp (i\dint\limits_{x}%
\mathcal{\pi }_{,X}^{j}\Delta _{\perp }\mathcal{\phi }%
_{j}^{X}d^{3}x+ip_{a}^{i}\Delta _{\perp }q_{i}^{a}) \\
&&\delta (\mathcal{H})\delta (\mathcal{D}_{a})\delta (\mathcal{G}_{A})\mu d%
\mathcal{\pi }dp\},
\end{eqnarray}%
$\mu $\textbf{\ is determined to make }%
\begin{equation}
G_{\pm }(\mathcal{\phi }_{\perp x}^{X},\mathcal{\phi }_{\perp x}^{\prime
X},q_{\perp aj},q_{\perp aj}^{\prime };\mathbf{\eta },0,t)=\delta (\mathcal{%
\phi }_{\perp x}^{\alpha }-\mathcal{\phi }_{\perp x}^{\prime \alpha })\delta
(q_{ai\perp }-q_{ai\perp }^{\prime }),
\end{equation}%
\textbf{when }$\Delta t=0;\perp $ refers to free degrees of freedom left
over after the application of all the constraints;
\end{principle}

This time evolution depends on $\mathbf{\eta }$. We will discuss how it is
determined in the later postulates. The internal variables $N,N^{i},\alpha
^{A}$are integrated to get the constraints in equation \ref{GTevol}.

We can define an effective action for the path integral as follows:

\begin{equation}
S_{\bot ,\pm }(\gamma _{\bot },\mathbf{\eta })=\int_{x}\mathcal{\pi }_{\bot
,X}^{i}\Delta \mathcal{\phi }_{\bot ,i}^{X}d^{3}x+p_{\bot ,a}^{j}\Delta
q_{\bot ,j}^{a}\mp H_{s}(p_{_{\bot }a}^{j},q_{j_{\bot }}^{a},\mathcal{\pi }%
_{\bot ,X}^{i},\mathcal{\phi }_{\bot ,i}^{X})dt,  \label{ACT3D}
\end{equation}%
where $\mathcal{\phi }_{\bot ,i}^{X}=\mathcal{\phi }_{i}^{X}-\mathfrak{\Phi }%
_{i}^{X}(x,t)$, $q_{\bot ,j}^{a}=q_{,j}^{a}-Q_{i}^{a}(t)$ at each time
instant $t$ and $\mathcal{\pi }_{\bot ,x,\alpha }$, $p_{\bot ,aj}$ are
conjugate momenta corresponding to it. This path integral depends on $%
\mathbf{\eta .}$

Using the path integral we can define the time evolution of the quantum
state on $\mathbf{S}(\mathbf{\eta )}$

\begin{equation}
\psi _{\pm }(\mathcal{\phi }_{j^{\prime }}^{\prime X^{\prime }},q_{i^{\prime
}}^{\prime a\prime },t+\Delta t)=\int_{\mathcal{\phi }_{,x}^{\alpha }}G_{\pm
}(\mathcal{\phi }_{\perp j}^{X},\mathcal{\phi }_{\perp j^{\prime }}^{\prime
X^{\prime }},q_{\perp i}^{a},q_{\perp i^{\prime }}^{\prime a\prime };\mathbf{%
\eta },\Delta t,t)\psi _{\pm }(\mathcal{\phi }_{\perp j}^{X},q_{\perp
i}^{a},t)d\mathcal{\phi }dq
\end{equation}

This defines a single-step evolution in the small time interval $\Delta t$.
To get the evolution over a finite interval we need to divide the interval
into small time intervals and apply the single-step evolution repeatedly.
Then the number of time divisions is increased and the time intervals are
sent to zero to get continuous evolution.

\subsection{Spontaneous Fundamental Quantum Reduction}

\subsubsection{A Single Point System: Density Matrix Formulation}

Consider again a single-point quantum system. We can consider the density
matrix of a simple quantum system as an element of the space of outer
product of the Hilbert Space $\boldsymbol{H}$ and its Hermitian Conjugate
space $\boldsymbol{H}^{\dagger }$ of the system. In the following
definitions, variables with $tilde$ represent operators that only act on the
variables that belong to the Hermitian conjugate space. Let me explain this
in a simple example in one dimension. If $q$ and $\tilde{q}$ are elements of 
$\boldsymbol{H}$ and $\boldsymbol{H}^{\dagger }$. Then we have $\rho (q,%
\tilde{q})$ as a quantum state in $\boldsymbol{H}\otimes \boldsymbol{H}%
^{\dagger }$. Then the Hamiltonian evolution of the density matrix is

\begin{equation}
\frac{d\rho (q,\tilde{q})}{dt}=i\left[ H\rho (q,\tilde{q})-\rho (q,\tilde{q})%
\tilde{H}\right]
\end{equation}%
Above $\tilde{H}$ only acts on $\tilde{q}$.\ The path integral form of this
is formally,

\begin{equation}
<\rho (q_{1},\tilde{q}_{1},t_{1})|\rho (q_{2},\tilde{q}_{2},t_{2})>=\int_{%
\gamma ,\tilde{\gamma}}\exp (iS(\gamma )-i\tilde{S}(\tilde{\gamma}))D\gamma D%
\tilde{\gamma},
\end{equation}%
where $\gamma $ and $\tilde{\gamma}$ are paths from $q_{1}$to $\tilde{q}_{2}$
and $q_{1}$to $\tilde{q}_{2}$ respectively, $S$'s are actions for paths $%
\gamma $ and $\tilde{\gamma},$ $D\gamma $ and $D\tilde{\gamma}$ are measures
corresponding to paths in $\boldsymbol{H}$ and $\boldsymbol{H}^{\dagger }$
space, and, $1$ and $2$ represent initial and final states. This form of
evolution of density matrix was first introduced by Feynman and Vernon \cite%
{FeynVern}. There they discussed the evolution of an open quantum system in
interaction with measurement devices and with the environment. In quantum
gravity framework 4, I proposed a simple reduction formulation to be used as
the fundamental reduction mechanism in quantum gravity systems \cite{MYP4}.
Let me discuss this now.\qquad The $d(\gamma ,\tilde{\gamma})$ can be
considered as a distance function as in the definition of a metric space.
The $d(\gamma ,\tilde{\gamma})$ term helps to implement the reduction. We
can also call $d(\gamma ,\tilde{\gamma})$ as the reduction function. To be
physically sensible we can assume the reduction function contributes only
when $\gamma $ and $\tilde{\gamma}$ are separated by macroscopic distance,
so that only macroscopic reduction happens. So, for isolated simple atomic
quantum systems, this doesn't have an effect. But when they start
interacting with measurement apparatus and the environment they undergo
continuous reduction.

At the fundamental level, the environmental degrees that are to be traced
when deriving density matrix evolution either in the Master equation
derivation \cite{ZEH2} or the Feynman-Vernon \cite{FeynVern} analysis are
replaced by the evolution of the density matrix of the environment according
to the above proposal.

If $d(\gamma ,\tilde{\gamma})=\int d(M,\tilde{M})dt,$ where $M$ are
eigenvalues of an operator $\hat{M}$, whose eigenstates acts as the basis
into which the fundamental reduction happens. The differential form of this
equation heuristically is,

\begin{equation}
d\rho (M,\tilde{M})=i<M[H\rho -\rho H]\tilde{M}>dt-c_{d}d(M,\tilde{M})\rho
(M,\tilde{M})
\end{equation}%
The presence of $d(M,\tilde{M})$ removes the cross terms over time and only
preserves the diagonal terms. As I did before, we assume $d(\gamma ,\tilde{%
\gamma})$ is significantly non-zero only when the paths are separated by
large values, that is the separation can be considered as classical, in
which case the effect of quantum reduction enforces classicality.

\subsubsection{Quantum General Relativistic Density Matrix Formalism}

We can generalize the single point system form of the Lagrangian form of
reduction to the quantum general relativistic case. The density matrix is a
function of configuration variables and its conjugate:

\begin{equation}
\rho =\rho (\mathcal{\phi }_{_{\bot ,j}}^{X},q_{\bot i}^{a};\mathcal{\tilde{%
\phi}}_{_{\bot ,j}}^{X},\tilde{q}_{\bot i}^{a},t)
\end{equation}

For the purpose of quantum relativistic description, I will be defining
density matrix as the ensemble average of the pure density matrix as
discussed in the stochastic Schr\"{o}dinger equation in the quantum state
diffusion formulation discussed before in equation \ref{RhoDef}:

\begin{equation}
\rho =M(\left\vert \Psi \right\rangle \left\langle \Psi \right\vert )
\label{DensityMean}
\end{equation}%
This case was discussed in section \ref{2.1.7}. This helped to relate the
matrix evolution to the stochastic Schr\"{o}dinger equation. Now we can
define the reductive evolution of the system as follows:

\begin{principle}
Fundamental postulate I.\textbf{2: Quantum Reduction of the unconscious
matter is guided by the New Fundamental Reduction Theory:\ The Bidirectional
Temporal Spontaneous Reduction theory. The quantum state goes undergoes
spontaneous reduction} in the relative time formulation based on $\eta $ in
the bidirectional temporal evolution to maintain continuity, with the
decoherence reduction constants of $c_{d}$ depending on each field, and at
observation rate\textbf{\ }$\beta _{R}.$
\end{principle}

This postulate is a quite general framework based on the new reduction
theory in section \ref{claent}. In section \ref{claent}, I discussed that
the new reduction theory can be modelled using the density matrix or quantum
diffusion in case of gravitational or EM fields, and discrete evolution in
the case of isolated quantum objects that decohere slowly. In the case of a
discrete case, the analysis is quite complex, and it will be published in
future updates. In this paper, we will only focus on continuous decoherence
based on the density matrix or Lindblad evolution.

The new fundamental theory was described in section \ref{2.3}. It requires
the definition of the rest frame foliation, which will be discussed in the
next postulate. The new fundamental reduction theory required two constants.
The $c_{d}$ describes the rate at which a pure state is converted into a
mixed state and it depends on the field. The $\beta _{R}$ describes the rate
at which the mixed state is converted into a pure state due to observation
by conscious or unconscious matter. Please note that I have used unconscious
matter, because the interpretation is different for conscious and
unconscious matter. In conscious matter, the reduction is considered to be
performed by the consciousness. More understanding regarding the role of
consciousness and reduction is discussed in The Physical foundations of
consciousness and Life \cite{MYP5B1.3}, which is considered as part B of
this paper. The relation between $\beta _{R}$ and $c_{d}$ needs to be
investigated and will be updated in the next update.

. In the relative time evolution and density matrix formulation based on $%
\eta ,$\ a spontaneous fundamental quantum reductive evolution of the $\rho $%
\ is given by the following path integral,\ \ 
\[
G_{r}(\mathcal{\phi }_{_{\bot ,j}1}^{X},\mathcal{\phi }_{_{\bot ,j}2}^{X},%
\mathcal{\tilde{\phi}}_{_{\bot ,j}1}^{X},\mathcal{\tilde{\phi}}_{_{\bot
,j}2}^{X},q_{\bot ,1i}^{a},q_{\bot ,2i}^{a},\tilde{q}_{\bot ,1i}^{a},\tilde{q%
}_{\bot ,2i}^{a};\mathbf{\eta },\Delta t,t)= 
\]%
\begin{eqnarray}
&&\frac{1}{(2\pi )^{BD}}\int_{\mathcal{\phi }_{\bot ,1}^{X},\mathcal{\phi }%
_{\bot ,1}^{X},q_{\bot ,2i}^{a},\tilde{q}_{\bot ,2i}^{a}}^{\mathcal{\phi }%
_{\bot ,2}^{X},\mathcal{\phi }_{\bot ,2}^{X},q_{\bot ,1i}^{a},\tilde{q}%
_{\bot ,1i}^{a}}\exp \left[ iS_{\bot ,+}(\gamma _{\bot },\mathbf{\eta }%
)-iS_{\bot ,+}(\tilde{\gamma}_{\bot },\mathbf{\eta })-c_{d}d(\gamma _{\perp
},\tilde{\gamma}_{\perp })\right]  \label{DecEvol} \\
&&\mu D\gamma _{\perp }D\tilde{\gamma}_{\perp },  \nonumber
\end{eqnarray}%
where $d(\gamma _{\perp },\tilde{\gamma}_{\perp })$\ is the fundamental
reduction functional and $c_{d}$\ is the reduction constant. I\ am assuming $%
\eta $\ is the same for both $\gamma $\ and $\tilde{\gamma}$\ space. The $%
\mu D\gamma _{\perp }D\tilde{\gamma}_{\perp }$\ is determined to make 
\begin{eqnarray}
&&G_{s+}(\mathcal{\phi }_{_{\bot ,j}1}^{X},\mathcal{\phi }_{_{\bot ,j}2}^{X},%
\mathcal{\tilde{\phi}}_{_{\bot ,j}1}^{X},\mathcal{\tilde{\phi}}_{_{\bot
,j}2}^{X};q_{\bot ,1i}^{a},q_{\bot ,2i}^{a},\tilde{q}_{\bot ,1i}^{a},\tilde{q%
}_{\bot ,2i}^{a};\mathbf{\eta },0,t) \\
&=&\delta (\mathcal{\phi }_{\perp j1}^{X}-\mathcal{\phi }_{_{\perp
j}2}^{X})\delta (\mathcal{\tilde{\phi}}_{\perp j1}^{X}-\mathcal{\tilde{\phi}}%
_{_{\perp j}2}^{X})\delta (q_{\bot ,1i}^{a}-q_{\bot ,2i}^{a})\delta (\tilde{q%
}_{\bot ,1i}^{a}-\tilde{q}_{\bot ,2i}^{a}),  \nonumber
\end{eqnarray}%
\textbf{when }$\Delta t=0;\perp $ refers to free degrees of freedom left
over after the application of diffeomorphism and gauge constraints; $\gamma
_{\bot }$ is the path defined in the phase space $\{p_{_{\bot }\alpha
}^{j},q_{j_{\bot }}^{\alpha },\mathcal{\pi }_{\bot ,X}^{i},\mathcal{\phi }%
_{\bot ,i}^{X}\}$. We can also discuss this using quantum diffusion
formulation, which may be more closer to the new fundamental reduction
formulation.\textbf{\ }

We used this path integral and for specific simple choices of $\mathbf{\eta }
$ as given in the previous version of the quantum gravity framework to give
the self-time evolution descriptions as given in the introduction section %
\ref{1}.

Let me discuss how this works out. First of all please note that this path
integral function depends only on the positive energy action. We can define
the density matrix formulation in simplified circumstances using this path
integral formulation as given in section \ref{2.1.6}. From this, the
stochastic Schr\"{o}dinger evolution can be derived as discussed in section %
\ref{2.1.7}. The relation of the evolution governed by this path integral to
bidirectional temporal evolution is given in section \ref{2.2.3}. Assume you
have a simple quantum system $S$ of a few particles that interact with
macroscopic fields $\mathcal{\phi }_{_{j}}^{X}$ at specific and narrow time
durations. Every time $S$ interacts with the macroscopic fields it gets
reduced. The strength of reduction depends on the coupling. Then by
combining the postulates $I.1$ and $I.2$ we can give a bidirectional
temporal quantum evolution of the system as given in the formulation in
section \ref{2.2.1}, and calculate the positive energy components, and the
negative energy components reflected back.

This time evolution depends on $\mathbf{\eta =}\{Q_{i}^{a}(t),\mathfrak{\Phi 
}_{j}^{X}(x,t)\}$. We will discuss how to determine them later in other
postulates. We might have to restrict the fields in the definition of $%
d(\gamma ,\tilde{\gamma})$ to free fields unconstrained by time constraints.
Usually, the variables that we use for measuring time are always far away
from the region of fields and matter that we study. So the field variables
of the study are not constrained by time constraints. But if the matter or
fields that we study are interacting significantly with the time variables
acting as a clock then we need to restrict $d(\gamma ,\tilde{\gamma})$ to $%
\mathcal{\phi }_{\perp j}^{X}$ and $\mathcal{\tilde{\phi}}_{\perp j1}^{X}$
only, that we use $d_{g}(\gamma _{\perp },\tilde{\gamma}_{\perp })$.

In quantum gravity framework 4, section 4, I\ proposed a definition for the $%
d(\gamma _{\perp },\tilde{\gamma}_{\perp })$ for gravitational decoherence:

\begin{equation}
d_{g}(\gamma _{\perp },\tilde{\gamma}_{\perp })=c_{dg}\int (h_{\perp ab}-%
\tilde{h}_{\perp ab})(h_{\perp }^{ab}-\tilde{h}_{\perp }^{ab})(g\tilde{g})^{%
\frac{1}{4}}d^{3}x
\end{equation}

Here the $d_{g}(\gamma _{\perp },\tilde{\gamma}_{\perp })$ is defined as the
difference between metrics $\gamma _{\perp }$ and $\tilde{\gamma}_{\perp }$
on the hyperplanes $S(t)$. In $d_{g}(\gamma _{\perp },\tilde{\gamma}_{\perp
})$ the origin of $\perp $ space cancels out, and so it wouldn't matter in
this definition.

This coordinate measure is defined to be invariant under the combined
coordinate transformation in configuration space $\gamma _{\perp }$ and its
dual $\tilde{\gamma}_{\perp }$. $c_{dg}$ is small enough that $d_{g}(\gamma
_{\perp },\tilde{\gamma}_{\perp })$ is non-zero only when the difference
between $h_{ab}$ and $\tilde{h}_{ab}$ can be deemed classical. Determining
this requires experimental measurement of $c_{dg},$ to be discussed later.
I\ am using the spatial metric $h_{ab}$ instead of $g_{ab}$ as we need only
the spatial components.

An alternative definition for $d_{g}(\gamma _{\perp },\tilde{\gamma}_{\perp
}),$%
\begin{equation}
d_{g}(\gamma _{\perp },\tilde{\gamma}_{\perp })=c_{dg}\int (h_{ab}-\tilde{h}%
_{ab})(h^{ab}-\tilde{h}^{ab})(\frac{g+\tilde{g}}{2})^{\frac{1}{2}}d^{3}x
\end{equation}%
where the product of metric in the measure is replaced by the sum of metric. 
$g=\det (g_{ab}).$Both $(g\tilde{g})^{\frac{1}{4}}dtd^{3}x$ and $(\frac{g+%
\tilde{g}}{2})^{\frac{1}{2}}dtd^{3}x$ are invariant under simultaneous same
space-time diffeomorphisms in both $\gamma _{\perp }$ and $\tilde{\gamma}%
_{\perp }$.

If we also include other fields such as gauge fields, bosonic, and fermionic
fields we have

\begin{equation}
d(\gamma _{\perp },\tilde{\gamma}_{\perp })=d_{g}(\gamma _{\perp },\tilde{%
\gamma}_{\perp })+d_{f}(\gamma _{\perp },\tilde{\gamma}_{\perp
})+d_{b}(\gamma _{\perp },\tilde{\gamma}_{\perp })
\end{equation}

\begin{eqnarray}
d_{f}(\gamma _{\perp },\tilde{\gamma}_{\perp }) &=&c_{df}\int
\dsum\limits_{Ai}(\psi _{Ai}-\tilde{\psi}_{Ai})\overline{(\psi ^{Ai}-\tilde{%
\psi}^{Ai})}(\frac{g+\tilde{g}}{2})^{\frac{1}{2}}d^{3}x \\
d_{b}(\gamma _{\perp },\tilde{\gamma}_{\perp }) &=&c_{db}\int \dsum_{i}(\phi
_{i}-\tilde{\phi}_{i})\overline{(\phi ^{i}-\tilde{\phi}^{i})}(\frac{g+\tilde{%
g}}{2})^{\frac{1}{2}}d^{3}x \\
d_{m}(\gamma _{\perp },\tilde{\gamma}_{\perp }) &=&\dsum\limits_{i}\frac{1}{2%
}q_{a\bot }^{i}q_{i\bot }^{a},
\end{eqnarray}%
where we have to include the respective traces of the spinor and gauge
fields to make them invariant under these transformations.

$c_{dg},~c_{df}$ and $c_{db}$ are small enough to make the decoherence
effects show only for classical separations between paths. But this affects
all the microscopic quantum systems entangled with it leading to the
reduction of these systems, as discussed in the previous section. This leads
to the density matrix evolution in Lindblad form \cite{LND} or the
Feynman-Vernon \cite{FeynVern} type evolution in the path integral form.

\subsection{Rest Frame Foliation}

Now we need to include quantum reduction globally. First, we need to
determine the foliation as quantum reduction depends on it. There are
various ways to foliate the manifold depending on $\mathbf{\eta =}%
\{Q_{i}^{a}(t),\mathfrak{\Phi }_{j}^{X}(x,t)\}$. They determine both the
foliation of space-time and also foliation of internal fields as discussed
in the example of single-point systems in section \ref{4.2.2}.

The foliation cannot be deterministically defined as there is no global time
parameter in general relativity. Instead, it has to be determine it
probabilistically~through $\mathbf{\eta =}\{Q_{i}^{a}(t),\mathfrak{\Phi }%
_{j}^{X}(x,t)\}$. To understand this please note that $Q_{i}^{a}(t)$, $%
\mathfrak{\Phi }_{j}^{X}(x,t)$ determine the values of time constraints on $%
q_{i}^{a}$ and $\phi _{j}^{X}$ that determine time evolution through
equations \ref{Fol-EQ} and \ref{FOL-EQ2}. So $\mathbf{\eta }$ are quantum
variables themselves that are functions of $q_{i}^{a}$ and $\phi _{j}^{X}$.
They need to be determined by quantum amplitudes. Changing $\mathbf{\eta }$
simply reparametrizes time, transforms the free part of the configuration
variables and so fundamental postulate I.1 doesn't determine quantum
amplitudes of $\mathbf{\eta }.$

Fundamental postulate I.2 modifies the difference between free configuration
variables and the corresponding dual variables as determined by equation \ref%
{DecEvol}. Changing $\mathbf{\eta }$, doesn't affect this difference. It
changes the foliation and so the reduction amplitudes. But it doesn't
determine the quantum amplitudes of $\mathbf{\eta }$ We need a separate
postulate for this. The main act of changing $\mathbf{\eta }$ is altering
the foliation of the 3+1 space-time manifold. The foliation that is most
favored and intuitive for reduction is in which the fields and particles are
most at rest, and so the quantum amplitudes of $\mathbf{\eta }$ must favor
this naturally. Based on this we can postulate the following postulate:

\begin{principle}
\textbf{\ }Fundamental postulate I.\textbf{3 The quantum evolution and
spontaneous fundamental decoherence process depend on }$\mathbf{\eta }$ as
defined in equation \ref{DecEvol}. This process happens \textbf{along a
spatial foliation such that the }$C^{1}$\textbf{smooth functions }$\mathbf{%
\eta =}\{Q_{i}^{a}(t),\mathfrak{\Phi }_{j}^{X}(x,t)\}$\textbf{\ take smooth
values, with relative quantum amplitude given by }$\exp (-c_{f}\Upsilon
_{f}) $\textbf{, where }$c_{f}$\textbf{\ is a fundamental constant, where }

\begin{equation}
\Upsilon _{f}(p_{\perp a}^{i},\mathcal{\pi }_{\perp j}^{X},\mathbf{\eta })=%
\left[ \int \mathcal{\pi }_{jX\perp }\mathcal{\pi }_{\perp }^{jX}\sqrt{h}\mu
(dx^{3})dt+\dsum\limits_{i}\frac{1}{2}p^{i}p_{i},\right]  \label{GLFL}
\end{equation}
\end{principle}

In quantum reduction described in fundamental postulate I.2, we minimize the
difference in the configuration variables $q_{\bot ,i}^{a}$ and $\mathcal{%
\phi }_{\bot ,j}^{X},$ on $\boldsymbol{S}(t)$. In fundamental postulate I.3
we minimize the conjugate momenta $p_{\perp a}^{i},\pi _{\perp j}^{X}$ on $%
\boldsymbol{S}(t).$

The quantum evolution of the state is modified by adding $ic_{f}\Upsilon
_{f} $ to the action. The free variable on which $\Upsilon _{f}$ depends is
the free momentum dual to the free configuration space unconstrained by the
time constraint. $\mu (dx^{3})$ is the appropriate gauge-invariant measure
on the spatial hypersurface coordinates. The smaller the value of\ $\Upsilon
_{s}$ higher the probability for $\mathbf{\eta }$. The quantum evolution of
the density matrix is modified by adding $ic_{f}\Upsilon _{f}+ic_{f}\tilde{%
\Upsilon}_{f}.$ $\tilde{\Upsilon}_{f}$ is a function of free momenta in
orthogonal space.

The fields on which $\Upsilon _{f}$ is defined can be the free dynamical
variables, or it can be also the quantum expectation values of the dynamical
variables after imposing all other proposals and postulates. The right
choice can be only determined experimentally.

The density matrix is defined on the Hilbert and conjugate Hilbert space on $%
q_{\bot ,i}^{a}$ and $\mathcal{\phi }_{\bot ,j}^{X}$. Its evolution depends
on $\mathbf{\eta .}$ Given a pure density matrix to start with, we can
evolve using the first three postulates. The postulates I.1 and I.2 don't
change the trace of the density matrix. But the third postulate will change
the trace. The trace of the density matrix after including the third
postulate gives you the probability associated with $\mathbf{\eta }$.
Essentially, the probability is higher, the lower the expectation values of $%
p_{\perp a}^{i},\pi _{\perp j}^{X}$.

As before we need not include the fermionic fields as the macroscopic effect
of it is taken care of by the momentum terms of the particles. As before the
terms of fields that are to be included are that of bosonic fields which can
have large values. The fermion field density does not lead to large values.
They have the macroscopic effect only as particles momenta.

The fields and particles momenta have two parts. The orthogonal and parallel
components to $\{\kappa _{X}^{j},\nu _{a}^{j}\}$. Minimizing the orthogonal
component makes fields and position variations parallel to hyperplanes of
internal foliation to be minimal.\ Minimizing the parallel component makes
field values and particle position variation parallel to hypersurfaces of
the space-time foliation to be minimal.

In the case of the Schwarzschild space-time, this foliation will be defined
by the time like a killing vector. In the case of the Friedman universe,
this will be the foliation defined by the scale parameter. The orthogonal
field values are almost zero in both of these cases. In the case of the
Schwarzschild, the conjugate momenta of metric is zero.

Let me clarify the need for this postulate. We need a preferred foliation
for decohering, depending on based on the distribution of matter and fields.
This was discussed in section \ref{2.1.2}. It is quite clear from the
discussion on relativistic quantum observation, that quantum measurement
must depend on the preferred foliation depending on matter and fields. Also,
we need that there is no global time parameter that is defined in general
relativity. Trying to introduce a time field to define time evolution wasn't
successful after my numerous attempts.

Fundamental postulate I.3 builds a bridge between Newtonian physics and
Einsteinian physics. In Newtonian physics, there is an absolute time
parameter that foliates the space-time into the sequence of fixed flat
hypersurfaces in a 3+1 space, with each hypersurface representing a time
instant. In Einsteinian physics, that is in General Relativity, there is no
fixed time parameter, and we could choose any spatial foliation to study
dynamics. But quantum reduction depends on the foliation. So, we need a
preferred foliation like in Newtonian physics, but yet the spirit of general
relativity must be respected. The intermediate choice is to do it
probabilistically and dynamically depending on the distribution and flow of
matter and fields. This is what fundamental postulate I.3 achieves.

Making $\Upsilon _{f}$ imaginary will disturb physics by affecting quantum
evolution. So the best way to define a foliation is through classical
probabilities and not quantum probabilities. That is why the exponent in
fundamental postulate I.3 is kept real. The foliation may need not have any
properties such as being smooth, connected, or universal. The foliation
could heavily depend on the particular region. In the intersection between
regions, they could intersect in an indeterminate way. This could lead to
complex time flow in this intersection.

Studying the various possible choices of foliation-dependent quantum
decoherence theoretically and investigating experimentally will help in
understanding quantum reduction. This will be discussed later.

\subsection{Full Theory: The Action}

Now we can write the action for full Quantum Gravity Theory in the density
matrix formulation for unconscious matter as follows:

\begin{equation}
\rho (\mathcal{\phi }_{\bot j}^{\prime X},\mathcal{\tilde{\phi}}_{\bot
j}^{X},q_{\bot i}^{\prime a},\tilde{q}_{\bot i}^{\prime a};\mathbf{\eta ,}%
t+\Delta t)=\int \exp (iS_{\bot }(\gamma _{\bot },\tilde{\gamma}_{\bot },%
\boldsymbol{\eta },t))\rho (\mathcal{\phi }_{\bot j}^{X},\mathcal{\tilde{\phi%
}}_{\bot j}^{X},q_{\bot i}^{a},\tilde{q}_{\bot i}^{a};\mathbf{\eta ,}%
t+\Delta t)\mu \tilde{\mu}d\gamma _{\bot }d\tilde{\gamma}_{\bot }
\end{equation}

\begin{eqnarray}
S_{\bot }(\gamma _{\bot },\tilde{\gamma}_{\bot },\boldsymbol{\eta },t)
&=&\int \{[i\mathcal{L}_{\bot }g-i\mathcal{\tilde{L}}_{\bot }\tilde{g} \\
&&+ic_{f}\mathcal{\Upsilon }_{f}(\mathcal{\phi }_{\bot j}^{X},q_{\bot
i}^{a})g+ic_{f}\mathcal{\Upsilon }_{f}(\mathcal{\tilde{\phi}}_{\bot j}^{X},%
\tilde{q}_{\bot i}^{a})\tilde{g}+i\frac{1}{2}c_{\sigma }\mathcal{\sigma }(%
\mathcal{\phi }_{\bot j}^{X})g+i\frac{1}{2}c_{\sigma }\mathcal{\sigma }(%
\mathcal{\tilde{\phi}}_{\bot j}^{X})\tilde{g}]  \nonumber \\
&&-ic_{r}\mathcal{R}(x)g-ic_{r}\mathcal{\tilde{R}}(x)\tilde{g}\}d^{4}x+\int
ic_{d}d(\gamma _{\bot },\tilde{\gamma}_{\bot })g\tilde{g}dt  \nonumber
\end{eqnarray}

\begin{itemize}
\item $\mathcal{L}-\mathcal{\tilde{L}}$ is the Standard Lagrangian for
particle physics that leads to the unitary Hamiltonian evolution of the
density matrix.

\item $\mathbf{\eta =}\{Q_{i}^{a}(t),\mathfrak{\Phi }_{j}^{X}(x,t)\}$ is the
classical path, which is also the observables with respect to which path
integral is calculated.

\item $\gamma _{\bot }$ and $\tilde{\gamma}_{\bot }$ are phase space
orthogonal to $\mathbf{\eta =}\{Q_{i}^{a}(t),\mathfrak{\Phi }_{j}^{X}(x,t)\}$
and its conjugate. They are the same for both $\gamma _{\bot }$ and $\tilde{%
\gamma}_{\bot }$ spaces.

\item $g$ and $\tilde{g}$ are integral measures on spacial foliation to make
the action invariant on spatial diffeomorphism and gauge transformation.
This is the square of root of the determinant of metric or some other
similar variation. They need not be the same as all the integrands in the
action.

\item Decoherence term $d(\gamma _{\bot },\tilde{\gamma}_{\bot })$ that acts
as distance operator between fields $\mathcal{\phi }_{\bot j}^{X}$,$q_{\bot
i}^{a}$ and $\mathcal{\tilde{\phi}}_{\bot j}^{X},\tilde{q}_{\bot i}^{a}$ but
only on positive going wavefunctions.

\item $\mathcal{\Upsilon }_{f}(\mathcal{\phi }_{\bot j}^{X},q_{\bot i}^{a})$
and its conjugate are the hypersurface terms that select the hypersurface in
which decoherence is happening.

\item $\mathcal{\sigma }_{x}(\mathcal{\phi })$ is the smoothness term that
enforces the continuum limit.

\item $\mathcal{R}(x)$ is the relational harmonic term that promotes
harmonic relationship structures. This will be discussed in `The Physical
Foundations of Consciousness and Life, Framework 1' \cite{MYP5B}.

\item The various $c^{\prime }s$ are the physical constant terms to be
measured.

\item $\beta _{R}$ is the rate of conversion of mixed state to pure state
due to observation.
\end{itemize}

\section{Features}

In this section, I will discuss some important features of formalism in this
paper based on the three postulates. As I have mentioned in the introduction
the experience of the flow of time comes into being only on the introduction
of consciousness. This has been done in the companion paper on `The Physical
Foundations of Life and Consciousness 1' \cite{MYP5B}. The physics in this
paper and ideas on consciousness in the companion paper are summarized in
part B of this paper \cite{MYP5C}.

\subsection{Time and Classicality}

The $\mathbf{\eta =}\{Q_{i}^{a}(t),\mathfrak{\Phi }_{j}^{X}(x,t)\}$ is
supposed to be classical. It has many classical variables one per point and
one per classical system. If none of them are available then there is no
classical time evolution. Such a system is purely quantum and is described
by a quantum state annihilated by all the constraints. Whether a variable is
purely quantum or classical depends on fundamental postulate I.2. In
formulas, conditions for classicality can be expressed as follows, if there
exists $\delta T$ such that,

\begin{eqnarray}
d(\gamma _{\bot },\gamma _{\bot }) &\approx &|\left\langle H_{\bot
}\right\rangle | \\
|\left\langle H_{\bot }\right\rangle |\delta T &\gg &\hbar \\
\frac{\delta \left\langle X_{i}\right\rangle }{\delta T} &\approx &\text{%
finite} \\
\frac{\delta \left\langle P_{i}\right\rangle }{\delta T} &\approx &\text{%
finite}
\end{eqnarray}%
where $\delta T$ is small time in which the expectation values of the
configuration variables and momentum value of field change by only a small
amount in proportion to $\delta T$. Above $T$ is the effective time
conjugate to $H_{\bot }$ in the rest frame foliation. If $d(\gamma _{\bot
},\gamma _{\bot })>|\left\langle H_{\bot }\right\rangle |~$the decoherence
terms dominate and the quantum uncertainty in the physical variables of
reduce. If $d(\gamma _{\bot },\gamma _{\bot })<|\left\langle H_{\bot
}\right\rangle |~$the decoherence terms dominate and the quantum uncertainty
in the physical variables of increase.

The transition between quantum to classical and vice-versa under different
conditions requires much more detailed study.

The definition of the particle systems in defining $\mathbf{\eta =}%
\{Q_{i}^{a}(t),\mathfrak{\Phi }_{j}^{X}(x,t)\}$ is something arbitrary. We
can consider each particle such as a proton or an electron of bulk matter as
an independent particle system. Then this will be a problem because these
systems are quantum mechanical, they tend to be entangled and we will have
too many variables of $Q_{i}^{a}(t)$ resulting in redundancy. If we consider
particle systems such as molecules, even they may not be big enough because
as we have discussed before even molecules with thousands of atoms have
quantum behavior. We need to use particle systems of much larger mass, but
small enough to treat them as independent systems to treat their center of
mass as classical variables. This depends on the decoherence constant $c_{d}$
and $\beta _{R}$. Measuring this is important to identify the
classical-quantum transition. This will help specify the number, size, and
mass of the particle systems necessary to define the $Q_{i}^{a}(t)$
variables.

\subsection{Imposition of Constraints.}

According to Dirac's principles, the first-class constraints of the system
have to be imposed on the wavefunctions on the spatial hypersurfaces. But
this will annihilate the quantum state and eliminate time evolution.
Instead, we keep the constraints as part of the path integral. The foliation
of the configuration space is done first. This will give the free variables
on the orthogonal space of time parameter direction of the configuration
variables at each point and each particle system. The foliation is combined
with the Hamiltonian constraint to get the effective Hamiltonian for the
orthogonal space as done in the first postulate discussed in the last
section. The reduction function in Postulate 2 and the foliation function in
Postulate 3 depend on the free variables in the orthogonal space as defined.
Eventually, on the free variables, diffeomorphism and gauge constraints are
to be imposed to get gauge and diffeomorphism invariant physics.

\subsection{Quantum Measurement Interpretation}

We discussed this in postulate $I.2$. In the proposal I discussed, there is
no need for any fancy interpretation. In this paper I\ considered the
quantum reduction is spontaneous. The postulate $I.2$ can be implemented
through density matrix formalism or stochastic Schr\"{o}dinger equation as a
physical possibility. The density matrix can be considered as derived from
the wavefunction evolution governed by the stochastic Schr\"{o}dinger
equation using the ensemble average $\rho =M(|\psi ><\psi |)$. The collapse
doesn't happen instantaneously, instead, slowly the state evolves into one
of the eigenstates of the measured observable if the observable interacts
with macroscopic fields in a one-to-one manner. Also, every time the quantum
reduction happens we have the reflection of negative time-traveling waves to
keep the wavefunctions $|\psi >$ continuous at moments of the change. Also,
the quantum state reduces continuously at all points of space-time where
there is sufficient macro superposition. So, the wave function in stochastic
formalism is an objective and observer-independent definition of a quantum
state.

The semi-classical history described by the probabilities is the physical
history experienced by the conscious observers. Every time there is
superposition the conscious observer becomes a minute superposition of
various states. As the state continuously collapses into a specific
semi-classical state the conscious observers also go into that particular
semiclassical state. This was discussed in the introduction and put into the
form of a postulate in I.2 and discussed in detail in `The Physical
Foundations of Life and Consciousness, Framework 1' \cite{MYP5B} of the
paper.

\subsection{Observables}

The observables were discussed in Quantum Gravity Framework 1 \cite{MYP1}.
Naive intuition says that observables are functions of the dynamical
variables defined on the hypersurfaces of the rest frame foliation. From
Standard Dirac quantization the observables are supposed to commute with
first-class constraints, in this case, Hamiltonian $H$, Diffeomorphism $\hat{%
D}_{i}$, and Gauge constraints $\hat{G}_{A}$. In our formalism, we have used
the foliation of configuration space to solve the Hamiltonian constraint to
calculate the parallel component of the momentum in the direction of time
flow as a function of free dynamical variables $q_{_{i,\bot }}^{a}$ and $%
\mathcal{\phi }_{_{j,\bot }}^{X}$. This becomes the effective Hamiltonian.
In the framework, the observables are functions of $q_{_{i,\bot }}^{a}$ and $%
\mathcal{\phi }_{_{j,\bot }}^{X}$ in the rest-frame foliation, annihilated
by the remaining constraints.

\begin{eqnarray}
\hat{O} &=&O(\mathcal{\phi }_{_{j,\bot }}^{X},\mathcal{\pi }_{X_{,\bot
}}^{j},q_{_{i,\bot }}^{a},p_{\bot a}^{i}); \\
\lbrack \hat{O},\hat{G}_{A}] &=&0; \\
\lbrack \hat{O},\hat{D}_{a}] &=&0;
\end{eqnarray}

\subsection{Unification and Organization of Human\ Knowledge}

In \cite{GUK} the merger of physical and social sciences was discussed.
Consciousness is the bridge that links physical and social sciences. On top
of the physics of biological materials, the consciousness postulates act to
create living things that are conscious and intelligent. For a detailed
discussion of the unification and organization of human knowledge, please
refer to\cite{GUK} and \cite{OHK}. This will be further discussed in `The
Physical Foundations of Consciousness and Life, Framework 1' \cite{MYP5B} of
the paper.

\subsection{Relation to ADM\ Formalism}

To understand the relationship of\ $\mathbf{\eta }$ to the foliation on
continuum fields let us do a redefinition of components of $\mathbf{\eta }$.
Given a smooth function $\mathfrak{\Pi }_{X}^{j}(t)$ and $P_{a}^{j}(t)$ of $%
t,$ and an initial value $\mathfrak{\Phi }^{X}(x,0)$ and $Q_{j}^{a}(0)$ we
can define a curve $Q_{i}^{a}(t)$ and $\mathfrak{\Phi }^{X}(x,t)$ as
follows. First, let me calculate the rate of change of $\mathcal{\phi }%
_{j}^{X}$ using the Poisson brackets with the constraints:

\begin{eqnarray}
\bar{\kappa}_{X}^{j} &=&\kappa _{X,N}^{j}+\kappa _{X,i}^{j}+\kappa _{X,A}^{j}
\\
\kappa _{X,N}^{j} &=&\{H_{n},\mathcal{\phi }_{j}^{X}\} \\
\kappa _{X,D}^{j} &=&\{D_{n^{i}},\mathcal{\phi }_{j}^{X}\} \\
\kappa _{X,G}^{j} &=&\{G_{\alpha ^{A}},\mathcal{\phi }_{j}^{X}\}
\end{eqnarray}%
This is the function of momentum and configuration variables. Initially set
the value of $\mathcal{\phi }_{j}^{X}$ to be $\mathfrak{\Phi }_{j}^{X}(x,0)$
and conjugate momentum to be $\mathfrak{\Pi }_{X}^{j}(0)$ after evaluating
the commutators. \ Now we can set

\begin{equation}
\frac{d\mathfrak{\Phi }_{j}^{X}(x,t)}{dt}=\kappa _{X}^{j}
\end{equation}

Using this calculate further values of $\mathfrak{\Phi }_{j}^{X}(x,t),$ in
this equation set $\mathcal{\phi }_{j}^{X}$ to be $\mathfrak{\Phi }%
_{j}^{X}(x,t)$ and set $\mathcal{\pi }_{X}^{j}$ to be $\mathfrak{\Pi }%
_{X}^{j}(t)$ to extrapolate the future values of $\mathfrak{\Phi }%
_{j}^{X}(x,t).\ $We calculate $\kappa _{X}^{j}$ as follows:

Similarly, we separate the rate of change for particles into three parts:

\begin{eqnarray}
\nu _{a}^{j} &=&\nu _{a,N}^{j}+\nu _{a,D}^{j}+\nu _{a,G}^{j} \\
\nu _{a,N}^{j} &=&\{H_{n},q_{cm,j}^{a}\} \\
\nu _{a,D}^{j} &=&\{D_{n^{i}},q_{cm,j}^{a}\} \\
\nu _{a,G}^{j} &=&\{G_{\alpha ^{A}},q_{cm,j}^{a}\}
\end{eqnarray}

Calculate $\dot{Q}_{i}^{a}(t)$ using

\begin{equation}
\frac{dQ_{j}^{a}(t)}{dt}=\nu _{a}^{j}
\end{equation}

and substituting $q_{j}^{a}$ by $Q_{i}^{a}(t),$ and $p_{j}^{a}$ by $%
P_{a}^{j}(t)$ in this equation. Using $Q_{i}^{a}(0)$ as the initial value we
can solve the first-order equation to get the future values of $%
Q_{i}^{a}(0). $

Now $Q_{i}^{a}(t),\mathfrak{\Phi }_{j}^{X}(x,t)$ depends on parameters $%
n,n^{i},\mathfrak{\alpha }^{A},\mathfrak{\Pi }_{X}^{j}(t)$, $P_{a}^{j}(t)$
and the initial value $Q_{i}^{a}(0)$ and $\mathfrak{\Phi }_{j}^{X}(x,0).$
The equation used to derive $n,n^{i},\mathfrak{\alpha }^{A}$ maps the
configuration variables from one hypersurface to the next hypersurface after
doing all three transformations:\ gauge, diffeomorphism, and Hamiltonian,
assuming arbitrary smooth momentum values $\mathfrak{\Pi }_{X}^{j}(t)$ and $%
P_{a}^{i}(t)$. Now the relative time evolution is given in equation \ref%
{GTevol} and it now depends on $n,n^{i},\mathfrak{\alpha }^{A}$. The $%
n,n^{i},\mathfrak{\alpha }^{A}$ are related to $N,N^{i},\alpha ^{A}$ in the
same spirit. But the former is external and fixed, and determined
probabilistically, while the latter is the internal variables to be
integrated out in the path integral.

If we choose $n,n^{i},\mathfrak{\alpha }^{A},\mathfrak{\Pi }_{X}^{j}(t)$ and 
$P_{a}^{j}(t)$ and initial values $Q_{i}^{a}(0)$ and $\mathfrak{\Phi }%
_{j}^{X}(x,0)$, such that the quantum expectation values of $\mathcal{\phi }%
_{\bot j}^{X}$ and $q_{\bot i}^{a}$ are zero using the four postulate of
dynamics, then $Q_{i}^{a}(t),\mathfrak{\Phi }_{j}^{X}(x,t),$ $\mathfrak{\Pi }%
_{X}^{j}(t)$ and $P_{a}^{j}(t)$ captures a classical solution to Einstein's
Equations, both configuration and momenta values.

The real functions $Q_{i}^{a}(t),\mathfrak{\Phi }_{j}^{X}(x,t)$ as functions
of $n,n^{i},\mathfrak{\alpha }^{A}$ determine the relative evolution of
fields at various points, and so they determine the foliation. By choosing
the right values for $n,n^{i},\mathfrak{\alpha }^{A}$ we can make the
foliation such that the momenta in fundamental postulate I.3 is minimal or
zero. Minimizing the momenta by appropriate foliation by selecting the right
values of $n,n^{i},\mathfrak{\alpha }^{A}$ in $\Upsilon _{f}$ makes the
variation of configuration fields along the direction of evolution to a
minimum. This will capture the rest frame foliation discussed in the third
postulate. Then the $n,n^{i},\mathfrak{\alpha }^{A}$are the parameters that
are associated with the classical evolution corresponding to $Q_{i}^{a}(t),%
\mathfrak{\Phi }_{j}^{X}(x,t)$. Now we have captured the standard classical
evolution of both dynamical variables and the parameters.

\section{Applications.}

To understand how postulate 1.1 for relative time evolution works, let me
give a simple example. Consider a single-point system of dimension $d.$ Let
the Hamiltonian constraint be

\begin{eqnarray}
H(p_{a},q^{a}) &=&\frac{<p,p>}{2}+V(q^{a})=s^{ab}p_{a}p_{b}+V(q^{a}) \\
&=&\frac{p_{a}p^{a}}{2}+V(q^{a}).
\end{eqnarray}%
The metric is defined by coefficients of second-order momentum terms $s^{ab}$%
. Let me set $s^{ab}=\delta ^{ab}.$

Let me choose $\eta $ by $Q^{a}(t)=t\delta _{0}^{a}$. This means, $q^{0}$ is
the time variable. We can calculate the tangent:

\begin{equation}
v^{a}(t)=\frac{dQ^{a}(t)}{dt}=\delta _{0}^{a}
\end{equation}

We can define orthogonal unit vectors $e_{i}^{a}$ such that $i~$varies from $%
0$ to $d-1$, and $e_{0}^{a}$ $=\bar{v}^{a}=\delta _{0}^{a}$. Let $E$ $%
=-e_{0}^{a}p_{a}$ be the momentum conjugate to $T$ and $P_{I}$ = $%
e_{I}^{a}p_{a}$ be the momentum conjugate to $Q^{I}.$ Then $H$ becomes a
function of $E,P_{I},T$ and $Q_{I}$:

\begin{equation}
H=\frac{1}{2}(E^{2}+\dsum\limits_{I}P_{I}P_{I})+V(T,Q^{I}),
\end{equation}%
Since $v_{a}$ is constant, the propagator can be written as follows:

\begin{eqnarray}
&&G_{\pm }(Q^{I},T,Q^{\prime I},T^{\prime };\eta ,\Delta T)  \label{Cprop} \\
&=&\frac{1}{(2\pi )^{d-1}}\int \exp (i\sum_{I}P_{I}dQ^{I}\mp iEdT)\delta
(\Delta T-|v|\Delta t)\delta (H)\mu dP_{I}dE,
\end{eqnarray}%
where the $\Delta T=T^{\prime }-T$. By integrating over $\delta (H)$, we can
get $E$ as a function of $P_{I},t$ and $Q_{I}$.

\begin{equation}
E_{\pm }=\pm \sqrt{-V(T,Q^{I})-\sum_{I}P_{I}^{2}}
\end{equation}

We assume that $-V(T,Q^{I})-\sum_{I}P_{I}^{2}>0,$ so that $E$ is not
imaginary. The effective Hamiltonian is given by $E$ as follows:

\begin{equation}
H_{\pm }=\pm \sqrt{-V(T,Q^{I})-\sum_{I}P_{I}^{2}}\text{,}
\end{equation}%
in which $H_{\pm }$ is a function of $Q^{I}$ and $P_{I}$. Now,

\begin{eqnarray}
&&G_{\pm }(Q^{I},T,Q^{\prime I},T^{\prime };\eta ,\Delta T) \\
&=&\frac{1}{(2\pi )^{d-2}}\int \exp (i\sum_{I}P_{I}dQ^{I}-iH_{\pm }dT)\mu
dP_{I},
\end{eqnarray}%
defined upto a constant.

The time evolution of the positive and negative energy parts is given by

\begin{equation}
\psi _{_{\pm ,}t+\Delta t}(T^{\prime },Q^{\prime I})=\int G_{\pm
}(Q^{I},T,Q^{\prime I},T^{\prime };v_{a},\Delta t)\psi _{\pm
,t}(T,Q^{I})dQ^{I}\Delta T
\end{equation}

The $v_{a}$ has to be chosen according to the probability given by postulate
1.3. This example is simplistic as the foliation is static. In the next
section, we discuss a dynamic foliation.

Throughout the series of five papers in the project, I have worked out
calculations for various simple applications to help understand time
evolution in quantum general relativity. The other postulates relating to
reduction need to be studied under various constraints. Let me list the
simple applications of time constraint formalism studied in the quantum
gravity framework series 1 to 4 below. Many of these applications are
treated in quite a heuristic fashion.

\begin{itemize}
\item In Quantum Gravity Framework 1[BKVI1]: In this paper, the time
evolution using time constraint formalism was discussed. I studied three
cases in the appendix:
\end{itemize}

\qquad Appendix A: I discussed the case of time evolution in a single
particle system, which I further study in the next section.

\qquad\ Appendix B: Canonical Quantum Gravity. I discuss time evolution
with, 1) B.1 Using super metric 2) B.2 Using normal metric.

\qquad Appendix C: Gravity-Matter Evolution: In this, I discuss time
evolution assuming matter doesn't influence gravity only weakly.

\begin{itemize}
\item In Quantum Gravity Framework 2: In this the general time evolution
discussed in this paper was introduced.
\end{itemize}

\qquad\ Section 2.1: Self Evolution in a Single Point Universe with $n$
dimensional configuration space. I carefully studied general time evolution
for a single-point system used in this paper.

I apply the general time evolution formalism to

\qquad Section 3.1: Cosmological Reduced Model

\qquad Section 3.2: Cosmology with Fluctuations

\qquad Section 3.3: Case of Quantum General Relativity in Newtonian Setup.

In the appendix I discuss simple derivations relating to the general time
formalism, quantum diffusion equations, bidirectional temporal time
evolution, and a definition of time based on the momentum values.

\begin{itemize}
\item In Quantum Gravity Framework 3, I have discussed the derivation of
Non-Relativistic Hamiltonian using time constraint formalism discussed in
QGF1.
\end{itemize}

\qquad\ Section 3.1: Conventional Quantum Evolution in (3+1)D starting from
a relativistic energy-momentum constraint.

\qquad\ Section 3.2: In this the case of time evolution of a Bulk Matter
made of a collection of gravitationally bound particles in an otherwise
empty universe was discussed. The conventional Newtonian formalism was
derived using time constraint formalism in quantum gravity.

\qquad Section 3.3: Field Theory Dynamics. In this gravity coupled to the
scalar field is discussed. Assuming gravity is the dominant field, starting
from timeless quantum gravity, the conventional time-parameterized
Hamiltonian was derived.

\begin{itemize}
\item In Quantum Gravity Framework 4. In this time evolution using density
matrix formalism was introduced and various small discussions in section \ref%
{4.1} were presented. It includes the standard Big Bang cosmology and
spherically symmetric space. It was noted that Lagrangian formalism is
better for resolution of singularities.
\end{itemize}

The various examples discussed in these four papers need to be thoroughly
analyzed further using the formalism discussed in this paper.

\subsection{Relative Time Evolution with Vector Constraints}

Let's discuss a system with a simplified diffeomorphism term. Instead of
full diffeomorphism on spatial manifold, we will restrict to translation
invariance of the whole system. Consider a system of $L$ particles with the
following constraints:

\begin{eqnarray}
H_{N}(p_{a}^{i},q_{i}^{a}) &=&\dsum\limits_{i,a}\frac{1}{2m}%
p_{ai}^{2}+V(q_{i}^{a})=0  \label{HamCB} \\
D_{a} &=&\dsum\limits_{i}p_{a}^{i} \\
H &=&NH_{N}+N^{a}D_{a} \\
&=&N\dsum\limits_{i}\left( \frac{1}{2m}p_{i}^{2}+V(q_{i}^{a})\right)
+N^{a}\dsum\limits_{i}p_{a}^{i}
\end{eqnarray}

Assume the system of particles is bound together and the center of the
system acts as the classical configuration path with respect to which we
derive relative time evolution. Let $v^{a}$ be the tangent to $Q^{a}(t)$
defined in the first postulate.

\begin{equation}
v^{a}=\frac{dQ^{a}(t)}{dt}
\end{equation}%
Let me assume the vector indices are raised and lowered by the flat metric $%
\delta _{ab}$. The normalized $v^{a}$ is

\begin{eqnarray}
\bar{v}_{a} &=&\frac{v_{a}}{|v|} \\
|v| &=&\sqrt{v_{a}v^{a}}
\end{eqnarray}

The center of mass of the system is

\begin{equation}
q_{cm}^{a}=\frac{\dsum\limits_{i}q_{i}^{a}}{L}
\end{equation}

The foliation $S(t)$ is given by

\begin{equation}
v_{a}(q_{cm}^{a}-Q^{a}(t))=0  \label{PartCon}
\end{equation}

We will use 
\begin{equation}
T=\bar{v}_{a}q_{cm}^{a}
\end{equation}%
as the time parameter of the system. Then we have the constraint given below
that relates $T$ to points of the curve $\eta .$

\begin{equation}
T=\bar{v}_{a}Q^{a}(t)
\end{equation}

We need to rewrite action in terms of free variables unconstrained by the
above equation. First, let us write the momentum integral part of the action
using $q_{cm}^{a}$ as a free variable of the system as follows:

\begin{equation}
\sum_{i=1}^{L}p_{a}^{i}dq_{i}^{a}=\sum_{i=1}^{L-1}\tilde{p}_{a}^{i}d\tilde{q}%
_{i}^{a}+p_{a}^{cm}dq_{cm}^{a},
\end{equation}%
where for $i<L,\qquad \tilde{p}_{a}^{i}=p_{a}^{i}-p_{a}^{L}$ and $%
p_{a}^{cm}=\sum_{i=1}^{N}p_{a}^{i}$. For details regarding the calculations,
I refer to quantum gravity framework 3 \cite{MYP3}. The new variables are,

\begin{eqnarray}
\text{for}~i &<&L\text{, }q_{i\bot }^{a}=\tilde{q}_{i}^{a}-Q^{a}(t),\text{ }%
p_{a\perp }^{i}=p_{a}^{i}-p_{a}^{L}, \\
\text{for }i &=&L,q_{L\bot }^{a}=q_{\bot cm}^{a}-Q^{a}(t),~~p_{\perp
,a}^{L}=p_{\perp ~a}^{cm},
\end{eqnarray}%
where $q_{\bot cm}^{a}$ and $p_{\bot cm}^{a}$ projection of $q_{cm}^{a}~$and
\ $p_{a}^{cm}$ orthonormal to $\bar{v}_{a}.$ Let $q_{\Vert cm}$ and $%
p_{\Vert cm}$ projection of $q_{cm}^{a}~$and \ $p_{a}^{cm}$ parallel to $%
\bar{v}_{a}$. Then we have,

\begin{equation}
\sum_{i=1}^{L}p_{a}^{i}dq_{i}^{a}=\left( \sum_{i=1}^{L-1}\text{ }p_{a\perp
}^{i}dq_{i\bot }^{a}+p_{\perp ,a}^{L}dq_{L\bot }^{a}\right) +\left(
\sum_{i=1}^{L-1}\text{ }p_{a\perp }^{i}v^{a}dt+p_{\perp
,a}^{L}v^{a}dt\right) +p_{\Vert }^{cm}vdt
\end{equation}

The left side terms have three parts. The first part is the free part. The
second is a translation part that translates the free variables parallel to $%
v^{a}$. The third part is the Hamiltonian part where $p_{\Vert }^{cm}$ is to
be solved using the Hamiltonian constraint.

We can rewrite the Hamiltonian constraint as

\begin{eqnarray}
H_{N} &=&\dsum\limits_{i,a}\frac{1}{2m}p_{ai}^{2}+V(q_{i}^{a}) \\
&=&\dsum\limits_{i=1}^{L-1}\frac{1}{2m}p_{\perp ,ai}{}^{2}\frac{1}{2mL}%
p_{\perp cm}{}^{2}+\frac{1}{2mL}p_{||cm}{}^{2}+\tilde{V}(q_{_{\perp }i}^{a})
\end{eqnarray}

Now we can solve for $p_{||cm}$ using Hamiltonian constraint $H_{N}=0$, for
the effective Hamiltonian.

\begin{eqnarray}
H_{eff}(p_{a\perp }^{i},q_{\perp i}^{a}) &=&p_{||cm}=\sqrt{\left( \tilde{V}%
(q_{_{\perp }i}^{a})-\dsum\limits_{i=1}^{L-1}\frac{1}{2m}p_{\perp ,ai}{}^{2}%
\frac{1}{2mL}p_{\perp cm}{}^{2}\right) /2mL} \\
\sum_{ia}p_{ai}dq_{i}^{a} &=&\sum_{i=1}^{L}p_{a\perp }^{i}dq_{\perp
i}^{a}+\left( \sum_{i=1}^{L-1}\text{ }p_{a\perp }^{i}v^{a}dt+p_{\perp
,a}^{L}v^{a}dt\right) \pm H_{eff}(p_{a\perp }^{i},q_{\perp i}^{a})dt
\end{eqnarray}

Now the new form of the action after solving for Hamiltonian and the time
constraint:

\begin{equation}
S_{\pm }\left( \tilde{p}_{a}^{i},q_{i}^{a};\eta ,\mathit{\ }dt\right) ==\int
\sum_{i=1}^{L}p_{a\perp }^{i}dq_{\perp i}^{a}+\left( \sum_{i=1}^{L-1}\text{ }%
p_{a\perp }^{i}v^{a}dt+p_{\perp ,a}^{L}v^{a}dt\right) \pm H_{eff}(p_{a\perp
}^{i},q_{\perp i}^{a})dt-N^{a}D_{a}dt
\end{equation}%
\qquad \qquad

The vector constraint needs to be imposed directly on the quantum state $%
\psi (q_{ai})$,

\begin{equation}
D_{a}=\sum_{i}p_{ai}=p_{cm}=0.
\end{equation}%
Consider the quantum state expressed as Fourier transform in the original
coordinates:

\begin{equation}
\psi (q_{ai})=\int \exp (ik_{ai}q_{i}^{a})\psi (k_{ai})\mu (dk),
\end{equation}%
Then imposing the constraint on this results in

\begin{equation}
\psi (q_{ai})=\int \exp (ik_{ai}q_{i}^{a})\psi (k_{ai})\delta
(\dsum\limits_{i}k_{ai})\mu (dk).
\end{equation}%
We need to make the coordinate transformation to $p_{a\perp }^{i}$. This
exercise is only a sample calculation with vector constraint and relative
time evolution.

The $Q_{i}^{a}(t)$ is determined probabilistically by foliation terms for
the quantum reduction purpose as discussed in the third postulate. The
smaller the norm of the $p_{\perp ai}$, the higher the probability of $%
Q_{i}^{a}(t)$, as discussed in the section on determining foliation.

\QTP{Body Math}
\begin{equation}
\mathcal{\Upsilon }_{s}=\mathcal{\Upsilon }(\gamma ,T^{\gamma
})=\dsum\limits_{i}p_{\perp ai}p_{\perp i}^{a}
\end{equation}

\subsection{Reduction in Non-Relativistic Quantum Mechanics}

Let me discuss the case of single-particle dynamics to understand
spontaneous fundamental reduction. Assuming this particle is studied in a
stationary lab on Earth, we can apply the method discussed in the quantum
gravity framework 3, section, 3.2 for the bulk matter. The particle can be
considered as an earth-bound particle. The time is given by the center of
mass movement of the earth. The particle coordinates are free variables, as
the momentum of the center of mass carries the bulk of the time direction in
the derivation. I will assume that it evolves by the conventional
non-relativistic Hamiltonian in the lab, as shown in the previous paper,
section \ref{3.2}

We can use equation \ref{DecEvol} to calculate the impact of reduction on
the quantum state. Let $q$ be the location of the single particle and $p$ be
its momentum. Let me assume the reduction is done by the gravitational
reduction term only, by assuming the net charge due to other gauge
interactions is zero. Then we have

\begin{equation}
\dot{\rho}(q,q^{\prime })=-i\left\langle q\right\vert [H,\rho ]\left\vert
q^{\prime }\right\rangle -3c_{h}\rho \int |\phi _{g}(y-q)-\phi
_{g}(y-q^{\prime })|^{2}d^{3}y,  \label{moddenss}
\end{equation}%
where $\rho =\rho (q,q^{\prime },\tau )$ is the density matrix pertaining to
the Hamiltonian evolution and $\tau $ is the time evolution parameter from
the time constraint. From quantum gravity framework 4, section 4, 
\begin{equation}
V=V(q,q^{\prime })=\frac{4c_{h}}{c^{2}\hbar }\int |\phi _{g}(y-q)-\phi
_{g}(y-q^{\prime })|^{n}d^{3}y
\end{equation}%
is the reduction term, where I\ have used $n$ instead of $2$ as power.\ I\
have included standard $\hbar ^{-1}$ to be part action. $c_{h}$ is the
reduction constant. Here $\phi (x-y)$ is the gravitational potential at
point $y$ due to a particle of mass $m$ at $y$. $V$ is a function of $%
r=|x-x^{\prime }|$. I assume the $\phi $ defined by the following equation:%
\begin{eqnarray}
\phi \left( r\right) &=&\frac{Gm}{r_{m}c^{2}}\eta (r,r_{m}), \\
\eta \left( r,r_{m}\right) &=&\left\{ 
\begin{array}{lll}
\left( \frac{r_{m}}{r}\right) & \text{if} & r\geq r_{m} \\ 
\left( \frac{r}{r_{m}}\right) & \text{if} & r<r_{m}%
\end{array}%
\right. ,
\end{eqnarray}%
where the particle is assumed to be a spherical shell of radius $r_{m}$ with
the mass uniformly distributed inside.\ In all cases, we are going to assume
that $r_{m}$ is much bigger than the Schwarzschild radius $r_{s}=\frac{Gm}{%
c^{2}}$.

\begin{eqnarray}
V(x,x^{\prime }) &=&V(r)=\frac{6\pi c_{h}Gmr_{m}^{2}}{\hbar c^{2}}I(r,n)
\label{Vdec} \\
I(r,n) &=&\int_{-1}^{+1}\left\{ \int_{0}^{\infty }\left\vert \eta
_{m}(q,1)-\eta _{m}(\sqrt{u^{2}+r^{2}-2ruz},1)\right\vert
^{n}u^{2}du\right\} dz
\end{eqnarray}%
The $I(r,n)$ is unitless. The constant factor in $V(r)$ is

\begin{equation}
d_{m}=\frac{6\pi c_{h}Gmr_{m}^{2}}{\hbar c^{2}}=8\pi \frac{c_{h}}{\hbar }%
r_{s}r_{m}^{2}
\end{equation}%
$c_{h}$ needs to be calculated from experimental data. A sample plot $I(r,n)$
is given below:

\FRAME{ftbpFU}{5.2676in}{3.9565in}{0pt}{\Qcb{Plot of decoherence function
with respect to $n$ and $r$}}{\Qlb{DecohFun}}{decoherenceplot.png}{\special%
{language "Scientific Word";type "GRAPHIC";maintain-aspect-ratio
TRUE;display "USEDEF";valid_file "F";width 5.2676in;height 3.9565in;depth
0pt;original-width 8.0004in;original-height 6.0001in;cropleft "0";croptop
"1";cropright "1";cropbottom "0";filename
'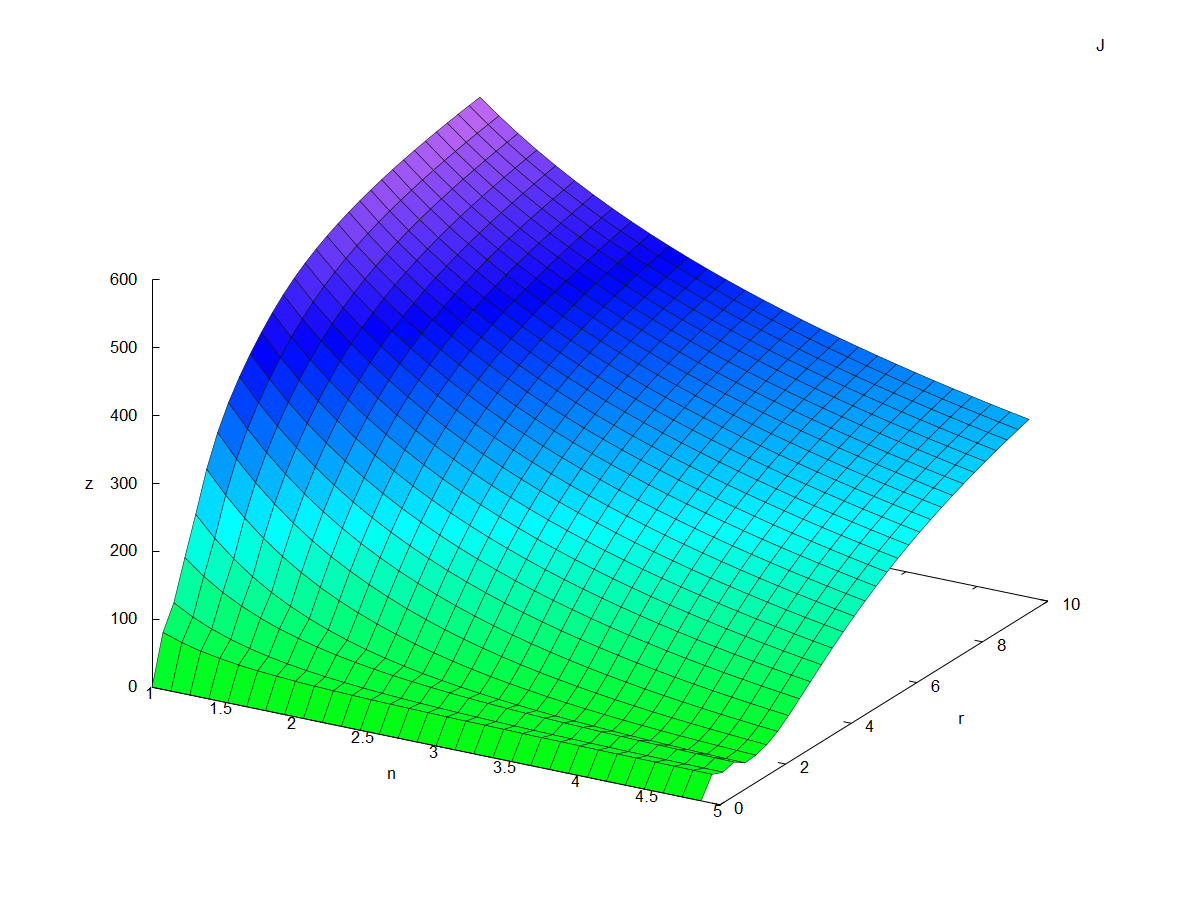';file-properties "XNPEU";}}

We can see that $V(r)$ is overall an increasing function of $r.$

We can solve the density matrix evolution equation for the case of a free
particle:

\begin{eqnarray}
H &=&\frac{p^{2}}{2m} \\
\rho (q,q^{\prime }) &=&\rho _{H}(q,q^{\prime })e^{-d_{h}V(x,x^{\prime })},
\end{eqnarray}%
where $\rho _{H}$ is the solution to 
\begin{equation}
\dot{\rho}_{H}(q,q^{\prime })=i\left\langle q\right\vert [H,\rho
_{H}]\left\vert q^{\prime }\right\rangle .
\end{equation}%
if we define

\begin{equation}
\rho _{H}=|\Psi ><\Psi |,
\end{equation}%
then we have

\begin{eqnarray}
i\frac{d}{d\tau }|\Psi &>&=H|\Psi > \\
|\Psi &>&=\exp (-iH\tau )|\Psi _{0}>.
\end{eqnarray}%
The evolution depends on the initial state $|\Psi _{0}>.$ Explicitly,

\begin{eqnarray}
&<&q|\Psi >=\psi (q,t) \\
D_{t}(q^{\prime }-q) &=&\exp (-d_{h}I(|q^{\prime }-q|,n))
\end{eqnarray}

\begin{equation}
\rho _{t}(q,q^{\prime })=\bar{\psi}(q^{\prime },t)\psi (q,t)D_{t}(q^{\prime
}-q).
\end{equation}

Let us discuss some heuristic estimates. $I$ function is finite and
numerical. For a proton assuming $r_{m}$ is its classical radius we have 
\begin{equation}
d_{h}=\frac{8\ast 3.142\ast 6.673\ast 10^{-11}\ast 1.67\times 10^{-27}\ast
(0.8751\ast 10^{-15})^{2}}{6.6260755\times 10^{-34}\ast (3\ast 10^{8})^{2}}%
\ast c_{h}
\end{equation}

\begin{equation}
d_{h}=3.\,597\,1\times 10^{-50}uc_{h}
\end{equation}

Here $u$ is the unit factor. Consider a spherical object of one Planck mass$%
=2.17671\times 10^{-8}\unit{kg}$. $\ $Consider that density to be $1gm/cc$
which is the same as water. Then its radius is $2.\,\allowbreak 947\,4\times
10^{-4}$ about 300$\mu m.$ Then we have

\begin{eqnarray}
d_{h} &=&\frac{8\ast 3.142\ast 6.673\ast 10^{-11}\ast 2.17671\times
10^{-8}\ast (2.\,\allowbreak 947\,4\times 10^{-4})^{2}}{6.6260755\times
10^{-34}\ast (3\ast 10^{8})^{2}}\ast c_{h} \\
&=&5.\,\allowbreak 318\,6\times 10^{-8}uc_{h}
\end{eqnarray}

For a typical virus we have $m=10^{-20}kg$, $r_{m}=50nm=5\ast 10^{-8}m$

\begin{eqnarray}
d_{h} &=&\frac{8\ast 3.142\ast 6.673\ast 10^{-11}\ast 10^{-20}\ast (5\ast
10^{-8})^{2}}{6.6260755\times 10^{-34}\ast (3\ast 10^{8})^{2}}\ast c_{h} \\
&=&7.\,\allowbreak 031\,7\times 10^{-28}uc_{h}
\end{eqnarray}

We want the virus to be a classical object. This means\ $d_{h}$ to be large.
Assume decoherence time is smaller than the period of visible light. This
means \ $d_{h}$ is an order of $500THz.$

\begin{equation}
7.031\,7\times 10^{-28}uc_{h}>500\ast 10^{12}s^{-1}
\end{equation}

From this \ $c_{h}>7.110\,7\times 10^{41}s^{-1}u^{-1}$.

From this for proton 
\begin{equation}
d_{h}>2.787\,4\times 10^{-10}s^{-1}
\end{equation}

For the Planck mass object

\begin{eqnarray}
d_{h} &>&5.318\,6\times 10^{-8}\ast 7.110\,7\times 10^{41}s^{-1} \\
&=&3.781\,9\times 10^{34}s^{-1}
\end{eqnarray}

This estimate seems reasonable enough to keep proton quantum while the
Planck mass object is classical.

\section{Experimental Verification}

Experimental verification of the theory proposed in this paper is quite
challenging. The quantum evolution of a system is directly related to
quantum reduction. So, verifying the theory requires measuring quantum
reduction and measuring the various terms. Decoherence comes from many
factors. Some of them are the following:

\begin{itemize}
\item Gravitational reduction

\item Collision with gaseous particles, and interaction with solids and
liquids.

\item Background electromagnetic fields such as cosmic microwave and
gravitational radiation background, radiation from matter around, celestial
bodies, etc.

\item Gauge fields such as the electromagnetic and gluon fields.

\item Fermionic fields

\item Bosonic Fields

\item Interaction with vacuum fluctuations/ground states of the gravity,
matter, gauge, and bosonic fields.
\end{itemize}

Giving an adequate theory of reduction requires taking into account all
these things. There is a lot of study regarding quantum reduction in these
topics in the literature. The effects of vacuum fluctuations and
gravitational reduction have been studied before. To understand the
spontaneous fundamental reduction, we need to separate the effect of these
factors. Any quantum system interacts with these external factors, and when
they undergo reduction, it also undergoes reduction. So, measuring the
spontaneous quantum reduction of a simple system by self-decoherence without
the influence of external systems is hard to detect. So, we need to study
the cumulative effect of all types of reduction both external and internal,
and indirectly deduce the nature of spontaneous fundamental reduction.

\subsection{Double Slit Reduction Experiment}

Assuming that we could test the spontaneous fundamental reduction effects of
a particle evolution, let me discuss a double-slit experiment to study it.
Consider that quantum particles are emitted from a slit at $S$. It undergoes
expansion as a spherical wave from left to right as shown in the figure.
There are two slits $A$ and $B$ at a distance $L$\ from $S$. The $A$ and $B$
are separated by distance $D$. The waves from slits $A$\ and $B$\ evolve
toward the right in the figure and eventually, create an interference
pattern in the screen.

\FRAME{ftbpFU}{3.7161in}{2.6636in}{0pt}{\Qcb{Double Slit Experiment to test
decoherence.}}{\Qlb{DSDec}}{single-slit-experiment.jpg}{\special{language
"Scientific Word";type "GRAPHIC";maintain-aspect-ratio TRUE;display
"USEDEF";valid_file "F";width 3.7161in;height 2.6636in;depth
0pt;original-width 5.8496in;original-height 4.1831in;cropleft "0";croptop
"1";cropright "1";cropbottom "0";filename
'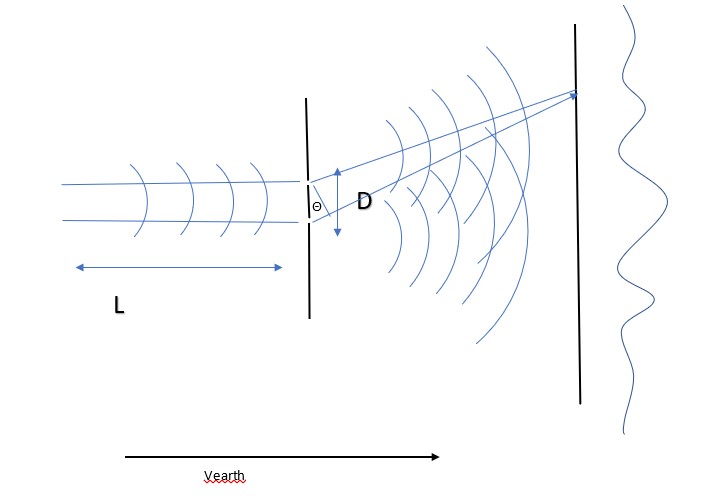';file-properties "XNPEU";}}

Due to the distance $D$ between the slits, the quantum waves lose coherence
due to the gravitational reduction term. Let me assume that the measurement
at the screen happens weakly and so the reflections of negative energy
states are small.

Now we can use the equation for the density matrix to calculate reduction.
The density matrix immediately after the screen near the slits is given by

\begin{equation}
\hat{\rho}=\rho _{AA}\left\vert \psi _{A}\right\rangle \left\langle \bar{\psi%
}_{A}\right\vert +\rho _{BB}\left\vert \psi _{B}\right\rangle \left\langle 
\bar{\psi}_{B}\right\vert +\rho _{AB}\left\vert \psi _{A}\right\rangle
\left\langle \bar{\psi}_{B}\right\vert +\rho _{BA}\left\vert \psi
_{B}\right\rangle \left\langle \bar{\psi}_{A}\right\vert ,
\end{equation}

\begin{equation}
\rho =\left\{ 
\begin{array}{cc}
\frac{1}{2}+\frac{b}{2} & \frac{d}{2} \\ 
\frac{d}{2} & \frac{1}{2}-\frac{b}{2}%
\end{array}%
\right\} ,
\end{equation}%
where $b$ is small and has been introduced to break the degeneracy in the
density matrix. $d$ is due to reduction. If gravitational reduction is only
included, then we have from equation \ref{Vdec}

\begin{eqnarray}
d &=&\exp (-\frac{6\pi c_{h}Gmr_{m}^{2}}{\hbar c^{2}}I(D,n)T) \\
T &=&\frac{L}{v}
\end{eqnarray}

Assuming $b$ is close to zero and $d$ is close to 1 we have the eigenstates
and eigenvalues of the density matrix as follows:

\begin{equation}
A:\left\{ 
\begin{array}{c}
1 \\ 
1%
\end{array}%
\right\} \leftrightarrow \frac{1}{2}+\frac{1}{2}d
\end{equation}

\begin{equation}
B:\left\{ 
\begin{array}{c}
-1 \\ 
1%
\end{array}%
\right\} \leftrightarrow \frac{1}{2}-\frac{1}{2}d,
\end{equation}

I\ have completely neglected $b$ in this after calculations. Due to the
reduction effect, we have two interference patterns superimposed, with
opposite maxima and minima. When $d=1$ the interference pattern is only due
to A. But as d increases the probability of B\ increases, superposing on A's
pattern, washing it away. By measuring the difference we can calculate the $%
d $ and eventually $n$ and $c_{d}.$ Interpreting the data requires fitting
various factors into the theory as discussed before.

\subsection{Double-Slit Retrocausality Experiment}

Let me discuss an experiment to study the retrocausal aspect of quantum
gravity evolution in our framework. Assume there are sensors at the two
slits of the double slit, that detect the movement of particles, say
protons. Also, assume that these sensors are quite weak that the amplitude
of the wavefunction disturbance due to it is quite small.

Let the sensors generate current $I_{L}$ if the proton passes through the
left slit and current $I_{R}$ if the sensors pass through the right slit. If
the particle doesn't pass through the slit then it be will be in the off
state $|$off$>$Let $|P_{L}>$ be the quantum state when the proton passes
through the left slit and $|P_{R}>$ be the quantum state when the proton
passes through the right slit. We can write the Hamiltonian of the
interaction as follows:

\begin{equation}
H_{\text{int}}=\lambda |I_{L}>|P_{L}><P_{L}|<\text{off}|+\lambda
|I_{R}>|P_{R}><P_{R}|<\text{off}|+h.c,
\end{equation}%
where the $\lambda $ is the coupling constant between the protons and the
sensor. The total Hamiltonian is

\begin{equation}
H=H_{P}+H_{\text{int}}+H_{\text{Env}}.
\end{equation}%
Here $H_{P}$ is the Hamiltonian of the proton and $H_{\text{Env}}$ is the
Hamiltonian of the environment.

Let the initial state of the proton be

\begin{equation}
|\psi _{0}>=a|P_{L}>+b|P_{R}>
\end{equation}%
right before the wave packet enters the slits.

We can rewrite this wavefunction as

\begin{equation}
|\psi (t)>=e^{iHt}|\psi _{0}+>+e^{-iHt}|\psi _{0}->,
\end{equation}%
where we resolved this into the positive and the negative energy states. We
will assume that the measurement at the screen is strong so that there is
the presence of a large negative part in the wavefunction due to reduction.

On the screen, we can detect the location of the proton after it is sensed.
Then the final state be 
\begin{equation}
|\psi _{T}>=\exp (-\frac{(x-x_{0})^{2}}{L})
\end{equation}

I\ assume that this state can be considered as the wave function of the
electron after it interacts with the screen. We can resolve this into

\begin{equation}
|\psi _{T}>=e^{iHT}|\psi _{0}+>+e^{-iHT}|\psi _{0}->
\end{equation}

From this and $|\psi _{0}>$, we can calculate $|\psi _{0}+>$ and $|\psi
_{0}->.$From this, we can calculate the wavefunction when it just exits the
two slits

\begin{equation}
|\psi _{\varepsilon }>=e^{iH\varepsilon }|\psi _{0}+>+e^{-iH\varepsilon
}|\psi _{0}->,
\end{equation}%
where $\varepsilon $ is the time taken to transit the slits. Due to the
presence of the sensors, this state gets perturbed. So, the new state is
given in first approximation by

\begin{equation}
|\psi _{\varepsilon }^{\prime }>=|\psi _{\varepsilon }>+i\lambda H_{\text{int%
}}|\psi _{\varepsilon }>
\end{equation}

we can calculate the density matrix corresponding to this wavefunction

\begin{equation}
\rho ^{\prime }=|\psi _{\varepsilon }^{\prime }><\psi _{\varepsilon
}^{\prime }|
\end{equation}

Let me define the current operator by

\begin{equation}
I=I_{L}|I_{L}><I_{L}|+I_{R}|I_{R}><I_{R}|+I_{off}|\text{off}><\text{off}|,
\end{equation}%
where the $I_{L},I_{R}$ and $I_{off}$ are currents relating to the electron
moving through the left slit, right slit and not moving at all,
respectively. The average current is,

\begin{equation}
<I>=tr(I\rho ^{\prime })
\end{equation}

This average current can be measured for an ensemble of protons and averaged
over all the protons. We have assumed that there is a negative energy
wavefunction part present due to the detection/reduction on the screen. We
can calculate the average current without assuming the presence of negative
energy wavefunctions with conventional wavefunctions. But by comparing the
result of this experiment with and without negative energy wavefunctions, we
confirm or deny the presence of retrocausality. Also, if the results of the
experiments confirm retrocausality, this can also help in finding the right
way to calculate the wavefunction as the sum of positive and negative energy
wavefunctions.

\subsection{Studying Rest Frame Evolution}

The restframe evolution has a detectable effect. According to the proposal,
the reduction happens along the direction along which$~\Upsilon _{f}$ is
minimum. The general definition was given in equation \ref{GLFL}. The
gravitational part of $\Upsilon _{f}.$ 
\begin{equation}
\Upsilon _{f}=\int (\frac{\pi _{f}^{ab}\pi _{fab}}{c_{g}})\frac{1}{\sqrt{h}}%
dx^{3}
\end{equation}%
In the case of doing experiments on Earth, $\Upsilon _{f}$ is minimal when
the hypersurfaces are orthonormal to time-like killing fields, assuming we
can neglect the effect of the electromagnetic fields. If the laboratory
frame is on Earth and at rest, then the reduction happens in Earth's proper
time.

Consider the double-slit experiment done in the laboratory. If the
laboratory moves at a certain velocity $v$ along the direction of the
earth's movement, then the laboratory frame undergoes Lorentz
transformation. Assume this direction is the x-axis. Then along the x-axis,
the earth's time is different at different points given by

\begin{equation}
t_{earth}=\gamma (t_{lab}+v\frac{x_{lab}}{c^{2}}),
\end{equation}%
where $t_{earth}$ is the time in earth's reference frame, $\gamma
=(1-v^{2}/c^{2})^{-1/2}$ , $t_{lab}$ and $x_{lab}$ are time and $x$ position
in the laboratory frame respectively. Along the direction of movement earth
time leads and so quantum fields on the right are subjected to more
reduction than the left as shown in the figure. To test this, we can use a
double double-slit experiment.

\FRAME{ftbpFU}{3.4714in}{2.9758in}{0pt}{\Qcb{A Double Double Slit Experiment
to Test Relativistic Decoherence}}{\Qlb{DDSExp}}{%
double-double-slit-experiment.jpg}{\special{language "Scientific Word";type
"GRAPHIC";maintain-aspect-ratio TRUE;display "USEDEF";valid_file "F";width
3.4714in;height 2.9758in;depth 0pt;original-width 5.2252in;original-height
4.4745in;cropleft "0";croptop "1";cropright "1";cropbottom "0";filename
'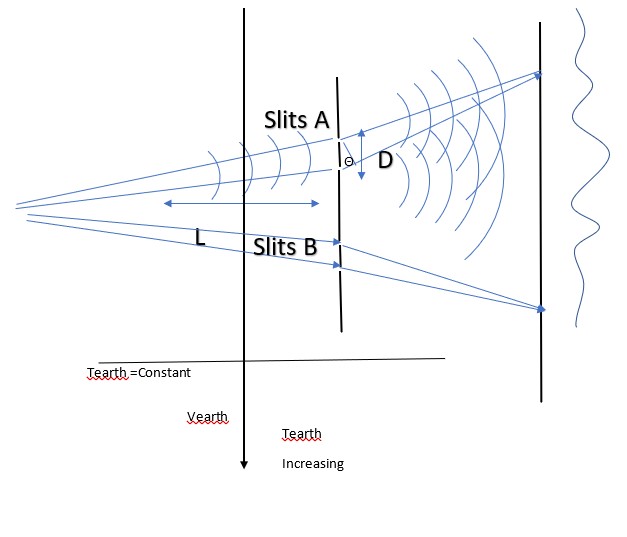';file-properties "XNPEU";}}

The line connecting the slits from A to B is the line of direction of
movement. Then B will be subjected to more reduction than A. Then the
pattern generated by B will be brighter than A. In Earth time the difference
between those slits is

\begin{equation}
\Delta T_{earth}=\gamma (v_{lab}\frac{\Delta x_{lab}}{c^{2}})
\end{equation}

\begin{equation}
\frac{d_{B}}{d_{A}}=\exp (\frac{8\pi c_{h}Gmr_{m}^{2}}{c^{2}}I(D,n)\Delta
T_{earth})
\end{equation}

The difference will be quite small but studying this using an appropriate
relativistic frame will shed light on the frame in which reduction occurs.

\subsection{Experimental Study of Consciousness}

This will be discussed in `The Physical Foundations of Life and
Consciousness, Framework 1' \cite{MYP5B} of the paper.

\section{Conclusion}

In this article, I have discussed three postulates to account for time,
reduction and foliation. I have also discussed sample calculations and
experimental measurements. The major challenge is to do an experimental
study. In smaller scales, the physics is hidden in randomness. As we can see
many of the postulates affect the probabilities. So, verifying the theory
requires extensive data collection from various experimental setups in
various sample applications of the theory. The data initially may look like
noise. But eventually, studying correlations between various parts of the
data can lead to patterns that can help discover the theory behind time and
quantum measurement in quantum general relativity and various features
associated with it. The experimental study needs to be used to fix
mathematical details, calculate the physical constants, and find precise
forms of the action terms, and models. Once this is done the resultant
theory can be applied to study the universe.

\setcounter{section}{0}

\pagebreak

\part*{Part B\hspace{1em}The Physical Foundations of Life and Consciousness,
Framework 1.3}

\addcontentsline{toc}{part}{Part B~\thepart\hspace{1em}The Physical
Foundations of Life and Consciousness, Framework 1.3}

\begin{center}
\begin{minipage}{0.75\textwidth}
    \centering
Date Completed:\ March 30, 2026 \newline
Please change log in footnote 1 \newline
\begin{justify}
Abstract: In this paper, I propose a set of ideas to explain the phenomena of life and
consciousness in the universe and their relation to the interpretation of
quantum mechanics. Consciousness is described as a result of specific
dynamical configurations of matter and fields aided by fundamental
properties of nature proposed in this paper. First, I review the various
theories of consciousness in current literature and come up with various
hypotheses, definitions, and propositions to build a physical theory of
consciousness. I introduce the concept of interaction binding of
consciousness and discuss in detail how the brain gives rise to
consciousness. I propose three sets of postulates on the physical
foundations of consciousness from a quantum general relativistic background
using the Quantum General Relativistic Framework, 5 proposals to understand
time and reduction. A Quantum Reduction Free Will Equivalence Principle is
discussed to link quantum reduction to conscious free will. I discuss the
potential possibility of retrocausal effects on consciousness-influenced
reduction probability. I discuss the concept of conscious being as a more
general concept than conscious identity, as separate from conscious content,
and its nature. I introduce the concept of classicum as an analog of the
quantum, and discuss its relation to consciousness. I discuss mathematical
measures to calculate the complexity of harmonic relational structures. I
put these together in a set of propositions on the nature of consciousness
and its relation to the nature of quantum phenomena. I discuss a
mathematical model of interaction binding, and free will. I discuss theory
and experiments to verify these.
\end{justify}
\end{minipage}
\end{center}

\newpage

\section{Introduction}

Please read v1.3 for all inclusive discussion. Please see the change log in
this footnote \footnote{%
Sharing and Usage Governed by License CC BY-NC-ND
\par
License Link: https://creativecommons.org/licenses/by-nc-nd/4.0/
\par
Official Link: qstaf.com/qgrf5
\par
Change-Log: Due to numerous serious engagements, I have, I often end up
delaying publication by drifting into other projects. Also, I revise with
breaks to improve on concepts. For more information on contact the author or
visit www.qstaf.com/qgrf5-publishing for proof of dates and explanation.
\par
\begin{itemize}
\item Version 1.1: May 4, 2024. Basic concepts were developed on May 4,
2024, and Time-stamped on May 14, 2024
\par
\item Version 1.2v1: December 9, 2024. Extensive Updates added to all
sections.
\par
\item Version 1.2v2: June 20, 2025. Updates are mostly proofreading, adding
references. Time-stamped on August 2, 2025.
\par
\item Version 1.3: March 30, 2026. The concept of conscious identity is
generalized to conscious being. The concept of classicum is introduced and
its relation to conscious being is discussed.
\par
\item All the versions will be available for reference in Academia.edu,
ResearchGate.net and Publications.UniteServe.com and other places listed at
qstaf.com/qgrf5.
\end{itemize}
}

The hypothesis numbers are changed in this version of framework. This
research is part of the quantum gravity framework project involving the
creation of a hypothetical conceptual framework to bring together the
necessary concepts at the foundations of science for the unification of
general relativity and quantum mechanics, and help understand life and
consciousness. This proposal is put forward to help as foundations for
experiments and further research that will help to discover the proper
scientific concepts. This paper is based on the 5th revision of the quantum
gravity framework, \cite{MYP5A}, which is part A, and this paper is part B
of the work.

\subsection{Overview}

First, we will review existing ideas on consciousness, and based on that we
build a full picture of how consciousness from its basic physical concepts
and biological structures arise.

\begin{quote}
\textbf{In this paper, Consciousness is described as a result of specific
dynamical configurations of matter and fields aided by fundamental
properties of nature proposed in this paper.}
\end{quote}

In Quantum General Relativistic Framework 5 (QGRF 5) \cite{MYP5A}, I
postulated various concepts necessary to understand time and quantum
reduction in quantum gravity for non-living matter. This involves defining
concepts needed to understand time, decoherence, and foliations. I proposed
three postulates to explain each of these phenomena. In this paper, we will
focus on the physical aspects of the foundations of the science of structure
of matter, life, and consciousness. We will refine and review the basic
postulates to understand physics, life, and consciousness in the universe in
a single framework.

In this paper first, I discuss an overview of the established theories and
facts about consciousness. In this, I discuss how human consciousness is an
emergent phenomenon built from the elementary consciousness of matter.

\begin{itemize}
\item I discuss the concept of interaction binding, in which interaction
results in elementary conscious experience in the particles undergoing
interaction are bound together.

\item Then I discuss how these convert into wholistic consciousness in the
human brain through binding due to the coherent activity of the neural
network.

\item I introduce the Quantum Reduction Free Will Equivalence Principle,
where quantum reduction is related to conscious free will.

\item The nature of conscious content and conscious being as separate parts
of a conscious entity is discussed. The conscious identity is absorbed into
the conscious being, many more features.

\item I also discuss the presence of possible retrocausality in
consciousness as a direct influence of its presence in the quantum framework
introduced in QGRF 5 \cite{MYP5A} as an effect of the second-order nature of
quantum gravity.

\item I summarize the whole phenomena as the evolution of structures in the
universe, into matter at various scales, consciousness-capable harmonic
relational structures.
\end{itemize}

While discussing these I formulate various hypotheses. These hypotheses are
used later to put forward postulates and propositions to understand
consciousness.

Next, I discuss the theoretical basics to measure the complexity of
relationship structures in physics. Various definitions are introduced. Then
we discuss six sets of propositions.

\begin{enumerate}
\item First, I introduce the concept of quantum entanglement domains which
refer to regions in which the particles are entangled with each other
locally and decohere together.

\item In the second propositions, I discuss the idea of conscious substratum
and its relation to conscious being.

\item In the third propositions, I discuss the role of harmonic relational
structures and their relation to structure formation and life in the
universe.

\item In the fourth propositions, I discuss the physical basis of
consciousness such as time, elementary consciousness, and interaction
binding of the elementary consciousness, and also the presence of
retrocausality.

\item In propositions five, I discuss the physics of large-scale binding in
the human brain, and how it is established through coherence and interaction
binding.

\item The propositions six, I discuss the evolution of life and how
consciousness and survival factors have shaped the formation of the human
mind.
\end{enumerate}

Next, I discuss the mathematical modeling of interaction binding and free
will using a mathematical model. The retrocausal aspects of quantum gravity
discussed in QGRF 5A \cite{MYP5A}, are also introduced in the consciousness
phenomena to explain Libet-like experimental observations.

Then we discuss the experimental verification of the ideas put forward. The
ideas on free will, the inclusion of consciousness-related terms to the
action of quantum gravity, and the ideas on consciousness being are quite
metaphysical in nature. But yet they are necessary to discuss and explore.
By studying the rest of the ideas that are testable in the lab, we can
understand the physical nature of these aspects, or theoretical aspects that
are beyond physics. It's time to investigate this and the theoretical
framework of this paper lays the foundation for this.

The concept of time and measurement is closely connected to consciousness.
The established theories of fundamental physics give only a four-dimensional
formulation of the pseudo-Riemannian quantum general relativistic physics.
The fourth dimension simply differs by a signature, nevertheless, this is
only a four-dimensional block universe description of the universe. Only
consciousness gives the experience of the flow of 3D spatial universes with
one-dimensional time observed by conscious observers. QGRF 5 \cite{MYP5A}
discussed in detail, from the proposed postulates there, how to extract the
3D information from them. In this paper, we will discuss how 4D information
is experienced as a 3D universe with 1D time by the phenomena of
consciousness through the six sets of propositions. We also discuss the
retrocausal effects due to the second-order nature of quantum gravity that
was discussed in QGRF 5 \cite{MYP5A}.

The relation between this paper and the quantum general relativistic
framework project is the following: 1) The rest frame foliation which I
discussed in QGRF 5 \cite{MYP5A} links Einstein's universe to Newton's
Universe is essential for understanding time flow, 2) Retrocausality that
was discussed in quantum gravity framework 5 may be part of conscious
phenomena and its origin is in quantum general relativity, 3) Flow of time
as discussed in QGRF 5 \cite{MYP5A} is a quantum general relativistic
process is a purely conscious phenomenon involving conscious observers, 4)
Quantum reduction discussed in QGRF 5 \cite{MYP5A} and its previous versions
is necessary to understand free will action.

The effect of conscious elements on physical reality results in promoting
relational structures. The physics of consciousness acts as a bridge between
Social Sciences and Physical Sciences. A paper was published to discuss the
merger of social sciences and physical sciences \cite{GUK}. Based on this
proposal organizing human knowledge has been discussed \cite{OHK}. In
general human knowledge is all about the knowledge of relational structures
in the universe. In this paper, we lay the foundations for understanding
relational structures and their connection to the basic conceptual framework
proposed for quantum gravity physical principles along with consciousness
included.

This project is part of the project for rebuilding the foundations of
science, society, and economics \cite{RFS}

Note: $\hbar =1$, $G=1$, and $c=1$ unless otherwise specified.

Note: I have placed $\alpha $ in front of the numbers of propositions to
denote that this numbering is temporary. In the next revision, this ordering
and propositions could be changed. In that case, a different Greek letter
might be placed if again it is considered temporary.

\subsection{Life and Structures}

I have discussed how life evolves in the grand unification of knowledge \cite%
{GUK} and in scientific relationism \cite{MYP4}. I will summarize this now.

The most simple indicators of life are chemical structures and physical
structures that are built by them. We have the following sequence of
complexity in nature relating to life: fundamental particles, atoms,
molecules, organic molecules, amino acids, viruses, bacteria, protozoa,
multicellular colonies, fish, frogs, reptiles, mammals, and humans. In this
sequence from first to last, we have the complexity and quantity of
relational structures of matter increasing. The most important sign of life
is building structures. Life has a balance of two things: Order and Chaos 
\cite{GUK}. Gravity tries to increase order. Heat increases chaos. Wherever
there is a balance of these two things we have in life. Due to increased
gravity things solidify and get compressed. Due to heat complex structures
are broken down to make them and slit into pieces of less simple structures.
The surface of the earth is a place where we have an appropriate balance of
matter and gravity is present to create complex structures of life. When
this balance is met complex structures with symmetry properties evolve,
which I refer to as harmonic relational structures. The dynamic properties
of this consciousness are built on this. Throughout this paper, I will
discuss how structures evolve and how consciousness is built on them using
verbal description and mathematics.

\FRAME{ftbpFU}{4.8836in}{3.5656in}{0pt}{\Qcb{The relationship between life
and death to entropy and order.}}{\Qlb{LIF}}{life_death.jpg}{\special%
{language "Scientific Word";type "GRAPHIC";maintain-aspect-ratio
TRUE;display "USEDEF";valid_file "F";width 4.8836in;height 3.5656in;depth
0pt;original-width 5.4673in;original-height 3.9833in;cropleft "0";croptop
"1";cropright "1";cropbottom "0";filename
'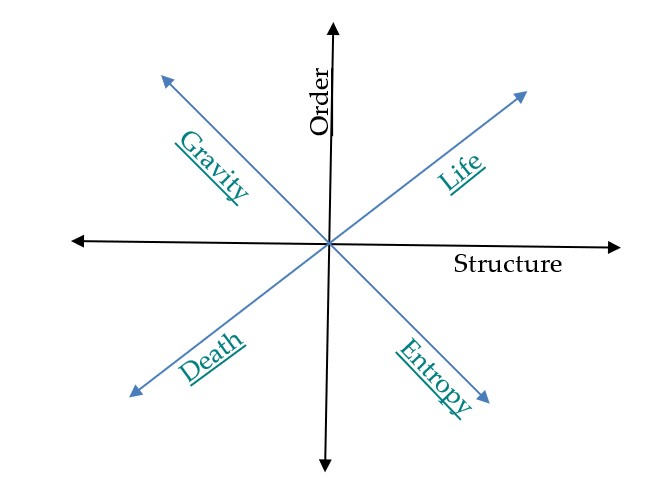';file-properties "XNPEU";}}

\subsection{Consciousness is a Property of Matter}

The Greatest of all mysteries of nature is consciousness. It is an important
aspect of life, particularly for humans. The hard problem of consciousness,
as David Chalmers puts it, trying to find the physical basis of
consciousness, in trying to explain qualia, is one of the most challenging
problems of science. There is extensive debate and proposals have been made
to explain consciousness from physics both classical and quantum. For an
overview and references, I would refer to the Wikipedia article \cite{WQM}
or Stanford philosophy articles \cite{STAN}.

Many theories are out there, attributing consciousness to various phenomena
and aspects. Let me list them briefly as follows:

\begin{itemize}
\item Electromagnetic field in EM field theory of consciousness by James
Mcfadden, and Susan Pocket \cite{EMT1}, \cite{EMT2}, and \cite{EMT3}

\item Quantum mechanics in Orch-Or-Theory of consciousness by Penrose and
Hameroff \cite{OR}, \cite{PENHAM1}

\item Mutual Information in Integrated Information Theory by Tononi \cite%
{IIT}

\item State of matter by Tegmark \cite{TEG1}

\item Neural Synchronization by Koch, and Crick \cite{KOCH}

\item Global workspace theory of consciousness by Bernard Baars and others 
\cite{GWTC}

\item Homeostatic Aspects by Damasio and others \cite{DAM}, \cite{DAM2}

\item Dendritic interactions in Pyramidal neuron firings in Dendritic
Integration Theory \cite{DIT}

\item Recurrent processing in the brain by Lamme \cite{RP1}, \cite{RP2}

\item Tempero-spatial theory of consciousness by Northoff G \cite{TTC1} and 
\cite{TTC2}
\end{itemize}

All these are true to a certain degree, and they are different aspects of
consciousness. A good summary review and strong commonness between these
theories have been discussed in \cite{INT1}. The challenge is to find the
systematic and structured theory behind the phenomena of consciousness, and
until now not much progress has been made in this regard. The research
knowledge accumulated relates to the correlation between phenomena in the
brain and consciousness aspects. This is referred to as the neural correlate
of consciousness by Crick and Koch.

Neuroscientists have tried to find a specific area or organ for
consciousness in the human brain. However, all the research has indicated
that there isn't anything specific like that. People thought consciousness
is associated with the prefrontal Cortex. However, the removal of the entire
prefrontal cortex made no difference for patients undergoing this surgery 
\cite{Solms}, \cite{KOCH2}, in their day-to-day social life. Even the
absence of the entire human cortex seems to not prevent conscious
experience. It only prevents higher functioning but not the experience of
feelings \cite{Solms}. For consciousness, the most critical part seems to be
the reticular formation which is like the primary founding layer of
consciousness. It is responsible for the most basic aspects like feelings,
arousal, controlling sleep cycles, etc. \cite{Solms}.

All these indicate that consciousness is a general property of matter, that
has a special structure like a biological neural network, in which
interactions happen in certain ways. When you touch your skin, you feel.
Somewhere in the human brain chemical reactions happen, due to the stimulus,
leading to the subjective feeling of this touch. The key thing to note here
is the chemical reaction or interaction between molecules somewhere in the
brain leads to the conscious sensation of touch. We can see the same about
other sensory modalities. The integrated collection of the sensations from
all the sensory organs, and the internal memory of the human brain, is
consciousness. This implies simply that interactions in biological neural
networks and their integration give rise to consciousness. So I assume the
following hypothesis:

\begin{principle}
Hypothesis 1: Physical Interactions in matter give rise to consciousness.
\end{principle}

I refer to quantum gravity framework 4 \cite{MYP4} for more discussion of
this where it was introduced. The key idea presented there was, that each
elementary particle is capable of an elementary conscious experience, and it
happens when they interact with each other. We discuss this process in
detail in this paper. The quality and richness of this experience depend on
various aspects as captured by the various theories of consciousness that I
listed before. Our task here is the elementary mathematical and physical
description of these phenomena.

In the human brain, interactions happen at multiple levels: between the
parts of the brain, between the emotional, the cognitive, and the memory
areas, the neural assemblies, between the neurons of different parts of the
brain, between the parts of the dendrites and the axons of the neurons,
within their branches, between the neuron and the surrounding matter, at the
synapses, within the microtubules, between the microtubules and the rest of
the neuron, between the parts of the neuron and the electric brain waves,
between the molecules making up the neurons, between the neurons and the
sensory organs such as the retina, the skin, the cochlea, etc. So, we have a
complex structure of interactions at various scales, that hierarchically
overlaps with each other. These interactions have the potential to create
complex human experiences, and we will discuss how this happens using
theoretical models and various elementary concepts.

The importance of direct physical interaction between the biological
structures is given in the split-brain experiments \cite{split}. Cutting the
brain at the corpus collasm, to stop the communication between the left and
the right brain hemispheres, makes them behave like two different minds. The
information processing happens separately in the two different hemispheres.
At the same time, the person behaves like a single integrated personality as
observed by others, referred to as social ordinaries by Joseph Bogen the
pioneer of split-brain surgery. Roger Sperry himself explains that even
without the corpus collasm, the left, and the right brain can communicate
through other parts of the brain. For a recent review please see \cite%
{Split1}. This merging of the two separate consciousness at a deeper level
to create a unified person may be due to physical interactions that happen
in the lower regions of the brain. We will discuss this further later.

\begin{principle}
Observation 1: Physical interactions strong enough to trigger memory
formation results in consciousness.
\end{principle}

Now, this is a practical assumption regarding consciousness. People can
sleepwalk, respond during sleep, etc. But usually consciousness involves
memory formation. If we are conscious, we are aware of the immediate past.
This means the chemical processes that trigger the memory formation at a
minimum, the working memory is either necessary for consciousness, or is
activated during consciousness. The brain mechanism that turns off and on
consciousness facilitates the turning on and off of memory formation. This
observation will be integrated into various hypotheses in this paper,
particularly the threshold of consciousness.

There is extensive research regarding relation between memory and
consciousness, which I will not go through in this paper. Unless otherwise
stated, we will assume that memory means declarative memory and possibly
more advanced memory as indicated by the context.

\section{Basic elements of consciousness}

\subsection{Simplified Model of the Human Brain \label{simmod}}

To theorize the phenomena of consciousness, let me introduce a simplified
model of the human brain. We have sensory nerves that receive information
from the external world and send it to the human brain and motor nerves that
send information to the muscles, to control muscles to interact with the
external world. The information received is processed by the associative
neural network in the human brain, in which the information is encoded into
a few sets of critical variables. The values of these variables are decoded
into the associative network. The associative network sends back motor
output. This creates a closed feedback and feedforward network. This has
been summarized in figure \ref{SimpModel1}.

\FRAME{ftbpFU}{6.4861in}{2.0513in}{0pt}{\Qcb{Simplified Model of the Human
Brain.}}{\Qlb{SimpModel1}}{simplifiedbrainmodel.png}{\special{language
"Scientific Word";type "GRAPHIC";maintain-aspect-ratio TRUE;display
"USEDEF";valid_file "F";width 6.4861in;height 2.0513in;depth
0pt;original-width 9.1238in;original-height 2.8573in;cropleft "0";croptop
"1";cropright "1";cropbottom "0";filename
'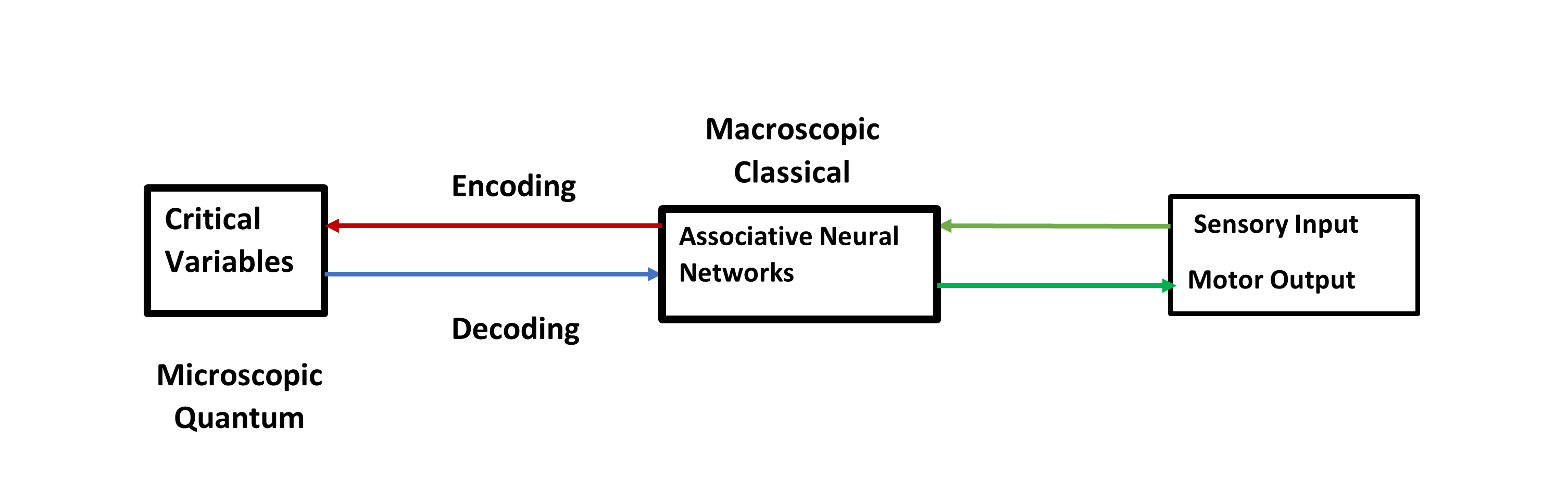';file-properties "XNPEU";}}

We will discuss further details on this model in this paper. Next section I
will discuss the basic hypothesis and definitions relating to this model.
Later I will discuss more detailed models that contain this simplified model
and in terms of which I will discuss the basic hypotheses and assumptions.

\subsection{Basic Hypotheses and Assumptions\label{bashyp}}

Let me put forward the first assumption of the model, which is relevant for

\begin{principle}
Hypothesis 2.1: The control variables are quantum in nature.
\end{principle}

In the neural networks in the human brain, the synaptic connection is
controlled by quantum events relating to the release of neurotransmitters.
So when the encoder network reduces sensory information to a few variables,
its behavior is controlled by quantum phenomena. The decoder's network sends
feedback to the primary areas that control the large-scale neural network
activity which is classical. So we have a feedback connection between
quantum and classical activity. Declaring this, let me describe the
interaction binding of consciousness of a connected physical system using
five sets of hypotheses.

\begin{principle}
Hypothesis 2.2a: Interaction Binding of Consciousness: When a complex set of
molecules interact with each other simultaneously, they become bound by the
interaction, and the conscious experience of each molecule, due to this
interaction, gets locally bound into one single joint experience of the
combined quantum system of molecules.
\end{principle}

\begin{principle}
Hypothesis 2.2b: Integration of Information: The extent of conscious content
of a connected physical entity depends on the long-range binding of
conscious information in all the interactions in the system through
coherence among the quantum interactions between the particles that make up
the system.
\end{principle}

\begin{principle}
Hypothesis 2.2c: Conversion of Information: The various aspects of the
interaction such as energy transfer, details of the interaction, the
structure of the molecules, interrelations among interaction in local
structures, or the global structure, rate, period, frequency, type of forces
involved, etc., are experienced as qualia by the combined quantum system
through the transfer of neural information to perceptual information.
\end{principle}

\begin{principle}
Hypothesis 2.2d: Behavior of the System: The conscious entity interacts with
the medium of its existence through computations driven by Schr\"{o}dinger
evolution, free will, and volition achieved through the quantum reduction of
control quantum variables.
\end{principle}

\begin{principle}
Hypothesis 2.2e: Consciousness threshold (partial): For an integrated
conscious experience to exist the conscious content needs to have feedback
interaction between the quantum control variables exercising free will and
the proper coherent macroscopic processes of the active neural network to
create the qualia.
\end{principle}

These are my basic foundational hypotheses for the elements of
consciousness. Further, I will introduce more assumptions and hypotheses,
and then put them into a set of postulates and propositions. (Hypothesis
2.2.e is a partial definition of consciousness threshold only because it
doesn't take into account other important conditions, such as memory
formation later. We will discuss this and fix this later.)

This is the materialistic description of human consciousness and the idea
that consciousness is an advanced state of chemical interaction in the human
brain existed for a long time. The important thing is to find out the proper
quantitative physical theory behind this phenomena, through theoretical and
experimental study. The theoretical framework to achieve this is the
objective of this paper.

In the first part, I discussed that the entanglement could be coherent or
incoherent, but prevalently incoherent. In the second part, the coherence
could be classical or quantum, prevalently classical.

The first part of the hypothesis describes the local binding of
consciousness, while the second part describes the long-range binding. The
third part of the hypotheses on qualia could depend on the combination of
the first and second, and we will discuss more about it in the propositions
later in this paper. The fourth part of the hypotheses relates the two basic
principles of quantum mechanics (deterministic evolution and reduction) to
the dynamics of conscious behavior involving both deterministic and free
will behavior. Hypothesis 2.2.e tells you whether the system is in a
conscious state or some other non-conscious state such as sleep.

The second part of the hypothesis is important because unless there is
coherent information there wouldn't be any consciousness. In the human
brain, it is well established long-range synchronization of neural firing is
a neural correlate of consciousness (NCC). Here the coherence is classical
in nature, to differentiate from quantum coherence which seems to be highly
unlikely in the human brain due to classical noise, for example \cite{TEG3}.
However, it is possible that coherence may exist at the quantum mechanical
level, as argued by some researchers.

The fifth part, which we call as conscious threshold is critically important
for understanding consciousness, which we will discuss later. If a person is
in NREM sleep or has fainted, they don't experience consciousness. This
implies the conscious experience is in a binary state: off or on. It is
critically important to understand what differentiates the two
quantitatively and why, to get a proper theory of consciousness. All of the
theories on consciousness listed before do not do that. They are mostly
continuous theories. This is necessary to specify the consciousness
threshold and also understand whether the lower animals like fish, cats, and
monkeys, have conscious experience. I will propose some measures to
calculate them later in the section on mathematical models of harmonic
relational structure.

\subsection{Quantum Physics of Interaction Binding}

\subsubsection{Components}

The nature of consciousness was discussed in the quantum gravity framework 4 
\cite{MYP4}. We will further go into details of this. Let me first discuss
the physics of interaction binding discussed in hypothesis 2.2a, using
concepts from quantum mechanics. Any physical interaction between particles
is made of three parts:

\begin{principle}
Hypothesis 3.1a: Time Evolution: The evolution of systems through
interaction-free Hamiltonians, that bring the particles close to each other.
\end{principle}

This is just the natural evolution of particles due to their momentum. In
quantum mechanical treatment, the sum of the free Hamiltonian of particles
is specified by $H_{0}$. Evolution under $H_{0}$ brings particles closer to
each other. In the case of the human brain, we have the ions and organic
molecules floating in water in between cellular layers. The molecules and
ions travel toward each other to collide.

\begin{principle}
Hypothesis 3.1b: Energy Exchange and Entanglement: When the molecules of a
complex life system interact, they exchange energy. Quantum mechanically
they get entangled due to the interaction part of the Hamiltonian. So, the
state now evolves into a superposition of various combinations of states.
\end{principle}

Apart from $H_{0}$, we have interaction of the particles due to the electric
fields of the ions. Due to the electric fields, they collide and either
become neutral particles or get deflected in different directions. During
the collision, they get entangled. This process goes on for a certain
duration and the system turns into an incoherent superposition at the atomic
level and entanglement of the entire system. This aspect of physical systems
for discussed in QGRF 5 \cite{MYP5A}. These interactions are also
synchronized and coherent at the classical level if you are dealing with the
human brain and the brains of other animals, which we will discuss in more
detail later. To complete the interaction, we need the next step.

\begin{principle}
Hypothesis 3.1c: Quantum Reduction: The various molecules of the nervous
system keep interacting producing various macroscopic superimposed states
with random phases. When the reduction happens a random choice of
macroscopic state is produced.
\end{principle}

The last three hypothesis are well tested by experiments and they are part
of the state of the art quantum technologies.

This is the best place to include free will which can modify the outcome of
reduction, which we will discuss further later. Many scientists attribute
consciousness to quantum reduction. Here I am only attributing free will to
reduction, not the whole conscious experience, unlike the various other
proposals that suggest quantum reduction plays a critical role in
consciousness, for example, Penrose and Hameroff \cite{PENHAM1}. In this
context let me propose the following hypothesis which seems to provide much
complete understanding of conscious experience:

\begin{principle}
Hypothesis 3.1d: The whole interaction involving all three stages is
required for consciousness.
\end{principle}

In the case of neuron electrical activity, the three processes occur
simultaneously. Neural Matter undergoes continuous evolution, interaction
(energy exchange and entanglement), and reduction. That means it undergoes,
continuous conscious experience according to hypotheses 3.1.a to 3.2.d.
Incoherent entanglement was discussed in the QGRF \cite{MYP5A}. When the
particles of the neurons interact with glial matter and with each other they
become entangled with each other, and these entangled states are in
superposition. This superposition is not coherent due to high-temperature
thermal interactions. This is what I refer to as incoherent entanglement.
This quantum superposition cannot be observed by us in a lab, but still, it
can be considered as one of the best possible reasons how elementary
consciousness associated with each particle, atom, and molecule is bound
together at the local level. Various authors, particularly Penrose and
Hameroff \cite{PEN2}, have attempted to use quantum superposition is how the
elementary consciousness of each particle is bound together. While this may
be true, it need not be a coherent superposition. I believe the incoherent
entanglement itself is sufficient to bind particle consciousness, as this is
the prevalent nature of quantum entanglement in the human brain due to the
thermal environment.

\begin{principle}
Hypothesis 3.2a: Quantum entanglement binds consciousness.
\end{principle}

This has been proposed in quantum gravity framework 4 \cite{MYP4},
Scientific Relationism \cite{BKVl3}, and also various authors, such as Koch
et al \cite{KOCH3}, and others. Quantum entanglement merges the individual
components into a bigger whole. For example, entangled particles behave as a
single entity. For example, this was the conclusion of the quantum
entanglement group in University of Baltimore County after extensive
research and conferences on quantum entanglement. When it comes to the human
brain the entangled part of the human-brain body system behaves as a single
conscious entity in the framework of our hypotheses.

\begin{principle}
Hypothesis 3.2b: The entanglement in the human brain is mostly incoherent,
this is enough for binding consciousness.
\end{principle}

In the human brain, we have interactions in various ways:

\begin{enumerate}
\item Synaptic Junctions controlled by chemical release

\item Interaction between neurons due to the potential difference

\item Interaction between the microtubules

\item Interaction between and among glial cells and neuron

\item Interaction between axon potential and the microtubules.

\item Interaction among the microtubular components

\item Interaction among Intracellular organelles

\item Molecular interactions among the extracellular fluid

\item Molecular interactions between particles in the extra/intracellular
fluid and the neuron walls and organelles.

\item Interaction between the brain's local potential field and the
particles of the extra/intracellular fluid, neuron walls, and the organelles.

\item Interactions in the dendritic and axonal branches between various
inputs.
\end{enumerate}

As we see there are extensive interactions happening in the brain. Most of
these interactions are thermal in nature which on interaction results in
incoherent binding of the whole brain/body. But when neurons start firing
brain becomes different from the body in two ways: 1) Interaction between
the various cells of the brain is more intense, than the body 2) We have
coherent information present among these interactions. This intensity and
coherence are restricted to the brain only. So, from hypotheses 2.2 and 3.1
this coherent information supplies the binding and information for conscious
perception respectively. We will further discuss the coherence later.

\begin{principle}
Hypothesis 3.3: Interaction in Neocortex is highly intense, complex, and
coherent, that mimics the external environment, compared to the body. The
brain is more conscious than the rest of the body.
\end{principle}

In these interactions, we need to figure out which part is classical and
which is quantum mechanical. These interactions when not quantum reduced
continue to build (incoherent) macroscopic superposition in the brain. When
this superposition increases to critical limits it gets continuously reduced
after a certain point. I suspect the order of quantum to classical
interaction is as follows:

\begin{enumerate}
\item Collisions among the free ions and molecules of extra and
intracellular regions in the brain.

\item Collision between the free ions and molecules, the cell wall of the
neurons, and the intercellular skeletal structures.

\item Collision among the free ions and molecules and the intracellular
support structure of the neuron, namely the microtubules.

\item Interactions in the synaptic joint where the release of
neurotransmitters is controlled by the quantum processes. Then we have
interactions in the currents in the dendrites.

\item Interactions among the microtubular parts.

\item Interactions among the local potential fields of the neurons with the
free ions and molecules.

\item Interaction between the neurons due to local field potentials.
\end{enumerate}

In this list, I can grade the elements from freeness to bondedness. The
freer they are more quantum interactions they can have resulting in extended
quantum wavefunction and entanglement among them. The more bonded they are
narrower the quantum wavefunction. The bonded particles are highly classical
because due to the bonds between them, they are subjected to strong quantum
reduction as per the third rule of QGRF \cite{MYP5A}. The free ones have
strong quantum properties involving incoherent entanglement. They are highly
entangled due to collisions. When they interact with matter and result in
macroscopic superposition, then only they get quantum reduced.

The free ions and molecules are bonded by incoherent entanglement. The cell
walls of the neuron are restricted due to strong bonds throughout the
membrane. But they do react to the nerve impulse and are subjected to
varying electric fields. The microtubules have an intermediate degree of
freedom and they can have extensive electromechanical properties. They tend
to vibrate at various frequencies depending on the potential difference
between the dendritic synaptic junctions and the soma. Microtubules are
under extensive focus of study because of their possible relevance to
consciousness provoked by Penrose and Hameroff's proposal of their role in
consciousness \cite{PENHAM1}, some examples of recent work are \cite{MICRO1}%
, \cite{MICRO2}, \cite{MICRO3}, \cite{MICRO4}, and \cite{MICRO5}.

The neocortical pyramidal neurons have extensive projections of axonal
branches from the other nearby neurons, to the fine dendritic structures.
The dendritic structures have thousands of branches. So the dendritic
microtubular part of the neuron has an extensive complex pattern of
electrical patterns, which in turn trigger vibration in the microtubular
structure. These vibrations are subjected to complex interactions. The ionic
particles are incoherent and entangled so doesn't have much pattern to be
related to qualia. The neuronal walls have minimal movement compared to the
microtubules. The dendritic microtubular structure is one that is best
capable of stable and unique patterns. So I believe the dendritic
microtubular structure is linked to sensory perception.

Neocortex sensory areas are arranged in cortical columns that process
sensory information. These act like fundamental units of conscious
perception. The interconnections between the dendrites and axons in the
neurons of neural assemblies are repeated stable patterns within the sensory
area of the same modality. We can propose the following.

\begin{principle}
Hypothesis 3.4: The neocortical dendritic microtubular structure of
pyramidal neurons of cortical columns of sensory areas has unique
structural, electrical, chemical reaction and mechanical patterns to be
related to different types of qualia.
\end{principle}

Now the hard problem of consciousness is to find an experimental
relationship between the dendritic microtubular structure and the different
types of sensory perceptions. Some obvious patterns are well known such as
how sensory information is processed (for example, retinotopy) and we will
be discussing this later in the propositions. We can assume further that
these dendritic microtubular structural, electrical, and mechanical
properties differ in different sensory areas. These dendritic structures
involving chemical reaction in the synapse also have the mechanism to
establish memory formation necessary for consciousness, which is extensively
researched.

\subsection{Binding of Information\label{bininf}}

Now we will discuss in detail how interaction binding is realized in a
living body-brain mechanism to create the whole person. I will build on
common elements from the various theories of consciousness, and the ideas
proposed in my previous papers \cite{MYP4}, \cite{GUK}, and \cite{BKVl3}.

\subsubsection{Coherence and Synchronization\label{cohsyn}}

Let me review the role of synchronization and coherence in the human brain,
and its relation to consciousness. The role of synchronization and coherence
in animal brains are products of biological evolution. Animal brains have
evolved complex systems for creating dynamic synchronization and modulation
of neural firing. Many features help in synchronization. The interneurons
help in the synchronization of the firing patterns of individual neurons 
\cite{SYNC1}. The connection between neurons helps in the synchronization of
firing patterns. The cortex of the mammalian brain due to the
interconnections between the neurons, is like an electric rubber band and is
tuned naturally towards the gamma band resonance oscillations \cite{GAMMA}.
So the information input to the cortex results in synchronized oscillations.
These oscillations are controlled by neurons emanating from the lower brain
layers that control sleep and arousal, particularly the Basal forebrain and
the Brain Stem.

One of the most important parts of the animal brain is the thalamus, through
which all sensory information, except olfactory, is rerouted through its
various nuclei, to the cortex. It has a looping connection extensively with
the cortical regions, called the cortico-thalamic-cortical loops, which
helps in synchronization, association, communication, modulation, and
coherence of neural firing data. The thalamic reticular nucleus that
envelopes the thalamus, seems to play an important common role to all the
nuclei in this process, possibly playing an important role in integrating
the sensory information through all the modalities in the animal brain.
Initially, the thalamus was considered to be the most important area for
consciousness given its central role. But observations on people with
strokes in the thalamus, and also studies in animals indicate that this is
not true. The lesions in the brainstem are highly correlated with loss of
consciousness \cite{STROKE1}, \cite{STROKE2}, \cite{STROKE3}. The brainstem
has been well studied and understood as the critical center for the arousal
of consciousness \cite{BSTEM0}, \cite{BSTEM1}. The role of the thalamus is
to act as a gateway that cuts off sensory information flow to the cortex
during sleep and prevents the back flow of sensory information from the
thalamus to the sensory organs.

The brain waves in the human brain that are created by neural firing
patterns throughout the cerebral cortex come under various bands. Each band
corresponds to different aspects of the information processing of the brain.
The gamma band corresponding to the 40Hz seems to be related to conscious
perception \cite{Eckhorn}, \cite{KOCH}. This has been extensively observed
and studied in numerous research papers. Overall the level of consciousness
is controlled by the presence of gamma-band oscillations in the cerebral
cortex, which in turn is controlled by the lower brain layers around the
brainstem and basal forebrain. The latter influences sleep-wake cycles and
circadian rhythm-related activities.

Phase Synchronization is considered a Neural correlate of Consciousness \cite%
{KOCH}, \cite{KOCH1}. The coupling between neurons results in phase
synchronization \cite{ROSEN},\cite{PARL}. The neurons that process
information that is being under the attention of the mind, fire in
synchronization. The neurons that process information about the background
of attended objects fire out of synchronization with these neurons \cite%
{KOCH}, \cite{KOCH1}. The information that is being attended may be
distributed all over the brain. The best way to understand this is that
coherence binds the information of different neuronal assemblies together to
make them conscious whole through the interaction binding that I have
discussed before. This coherence/synchronization may exist at various levels
of the human brain: regional, neuronal, and internal neuron structures like
microtubules, also further at its molecular level. All these are under
research \cite{Anirban}.

The synchronization of different parts of the brain is necessary for object
recognition. For example, the v1 visual field neurons that detect lines tend
to fire together. In v2, neurons processing signals from the same object
fire synchronized. This group firing pattern helps separate the different
objects viewed in the visual field and process each of them as a single
unit. So synchronization can be considered as a necessary element of object
recognition which also helps in the conscious binding of objects \cite%
{Cohbinding}, \cite{Cohbinding2}. As I have proposed before physical
interaction between parts is necessary for conscious unity. Thus,
synchronization itself is not sufficient for consciousness, as parts of a
system can be synchronized without direct physical interaction.

We need to discuss the general concept of coherence. We can consider that a
process in a system is coherent if exhibits dynamic behavior that puts
various dynamic elements in temporal and physical order. This includes 1)
phase synchronization, 2) phase locking 3) long-range synchronization
between various parts, 4) interrelations between temporal behavior in
various frequency spectrums such as the brain waves in delta, theta, and
gamma bands, etc, 4) periodic dynamic interrelations between firing pattern
of local features, such as neuronal assemblies, etc. We explain the
relevance of coherence based on the following hypothesis:

\begin{principle}
Hypothesis 4.1: The coherence in neural firing helps in making interactions
in the human brain simultaneous, bringing together a brain-wide interaction
binding of local consciousness to create the wholistic consciousness.
\end{principle}

Again, coherence itself is not a sufficient condition for conscious
experience, as we know that the coherence in the human brain is much higher
while a person sleeps \cite{CohSleep}. Also, too much coherence causes
health problems, such as epilepsy \cite{EPIL}. As we will discuss in the
hypothesis on the consciousness threshold, the critical thing is the
presence of a specific combination of coherent information involving
electrical and chemical interactions between neurons that can elicit
conscious experience.

It is well known that during NREM sleep the coherence in the human brain
wave activity increases significantly, with brain waves dominated by delta
and theta bands (the lowest frequency bands), but the NCC of conscious
activity - the gamma band is small.

\begin{principle}
Hypothesis 4.2: The high-frequency gamma-band brain activity can be
considered as a sign of local processing in the cortical columns, necessary
for interaction binding in neurons, which in turn is necessary for conscious
sensory perception.
\end{principle}

For conscious experience to happen the coherence activity happens under the
flow of sensory information to the various neocortical layers of the human
brain. This firing activity is also receiving input from the lower layers of
the brain such as reticular formation, and also from memories stored in the
synaptic connections throughout the brain. How this sensory information is
bound together by various firing patterns in the brain into various qualia
requires a detailed experimental study. We already know some aspects of
firing patterns in visual and auditory areas, and how they map into the
relevant qualia, which we will put into propositions later. But the full
picture is still missing. Later we will discuss various theoretical aspects
of the propositions to help with this.

During sleep sensory processing is turned off, at the local, which we will
further discus later. The interaction doesn't carry anything that is of
conscious relevance, and so the person doesn't experience anything. In the
case of dreaming the sensory information is generated from memory, and it
triggers specific coherent firing leading to conscious experience.

\subsubsection{The Brain-Body mechanism}

Human beings or other animals are organisms made of a body-brain system. The
brain is divided into three important parts: cortex, limbic system, and
brain stem. The brain stem connects to the body through the spinal cord.
Together these three create the mind and behavior of these organisms.

An animal has two features that alternate in its behavior: Homeostasis and
Critical phenomena. Both of these features create elements of consciousness.
These are like two ends of the spectrum of behavior that an animal's mind
and body together can engage. Homeostasis keeps the systems of a body in a
certain equilibrium. Every time some parameter like temperature or Sodium
content goes from the supposed to equilibrium condition it is brought back
by the body's internal mechanism. Homeostasis is an evolutionary product
that has evolved to keep the animal alive.

Another feature that is quite the opposite of Homeostasis is the critical
phenomena in the human body and brain. The human brain works in various
modes of interaction. The default network mode, attention mode, etc. These
modes are like different phases of matter. Critical transitions can occur
from one state to another depending on the internal and external state.

The brain stem is the part that is involved in creating the homeostatic
condition of the body and is the most important region for the arousal of
consciousness. It links the body to the consciousness arousal mechanism of
the human brain. Together it creates the proto-consciousness. as described
by Damasio \cite{DAM}, \cite{DAM2}, and Parvez \cite{DAM3}. Also, Damasio
and others earlier developed the idea that emotions and feelings as mental
representations of changes in the physiological body states, and those
affect the body's homeostasis. Emotional clues are processed by the limbic
areas and are affected by the nuclei in the brain stem through interaction
with bodily states. On top of this, we have the higher sensory processing in
cortical layers working in various cortical states in concert with the lower
brain layers that function to create the human mind that operates on
critical points flipping between various modes of operation.

Proto-consciousness controlled by the brainstem creates the most basic level
operation of the body-brain mechanism, all the way from wakefulness to
sleep, normal or panic mode of operation. The higher consciousness works
jointly with the neocortex and the limbic system to create the full psyche
of a human person. The limbic system connects the higher cognitive
processing in the neocortex and the lower processing of bodily states at the
brain stem through emotional states. The areas in the limbic system such as
the entorhinal cortex, parahippocampal gyrus, perirhinal cortex,
hippocampus, etc. work together to create long-term memory by consolidating
bodily, emotional, and cognitive information.

The binding among the neural activities in the cortex, limbic system, and
brainstem, and their interaction with the body create the whole human
mind-body complex. Our role in this paper is to figure out the physical
basis of these interactions.

\subsubsection{Binding in Neurons}

Let me discuss how the interaction in the brain works. For this, we need to
understand the cortex of the human brain. The cortex is full of various
types of neurons. The largest of them are the pyramidal neurons which are
considered to play an important role in consciousness, as shown in figure %
\ref{CERNER}. The cortex is about a few millimeters in size and heavily
folded and has about six layers, with different constituents. Figure \ref%
{CORINTBIND} shows the six layers and the illustrative presence of pyramidal
neurons along these layers. The pyramidal neuron dendrites have hundreds to
thousands of branches. The dendrites coming out of the apex are called the
apical dendrites and those coming out of the soma as basal dendrites. The
apical dendrites themselves have many types of branches: Proximal, Distal,
and Tuft Dendrites. Proximal and distal are closer to soma. Tuft dendrites
branch from the edge of the apex. They form the topmost layer where these
dendrites make connections with each other. Layers two and three are usually
combined as a single layer in discussions and referred to as a 2/3 layer.
They have axons from other neuronal types which make lateral connections
between various cortical areas. The fourth layer is the input layer which
receives inputs from various sensory organs routed through the specific
nuclei of the thalamus in the primary sensory areas. The fifth and sixth
layers are output layers. The fifth layer connects the various cortical
regions laterally, and the sixth layer sends axonal output to the thalamus.
The non-specific nucleus sends the output to the top layers and receives
input from the lower layers of the cortex. This is a typical set of
connections in the cortex, and need not be strictly like this. The top
layers are involved in higher-order processing, and neural information is
feedbacked from one area to the next. In the bottom layers, the neural
spiking data is feedforward from one area to another in the opposite
direction. This is shown in figure \ref{CORLNGRNG}.

\FRAME{ftbpFU}{5.6688in}{3.3754in}{0pt}{\Qcb{A typical pyramidal neuron in
the cerebral cortex}}{\Qlb{CERNER}}{pyramidal_neuron_word.png}{\special%
{language "Scientific Word";type "GRAPHIC";maintain-aspect-ratio
TRUE;display "USEDEF";valid_file "F";width 5.6688in;height 3.3754in;depth
0pt;original-width 6.5509in;original-height 3.8847in;cropleft "0";croptop
"1";cropright "1";cropbottom "0";filename
'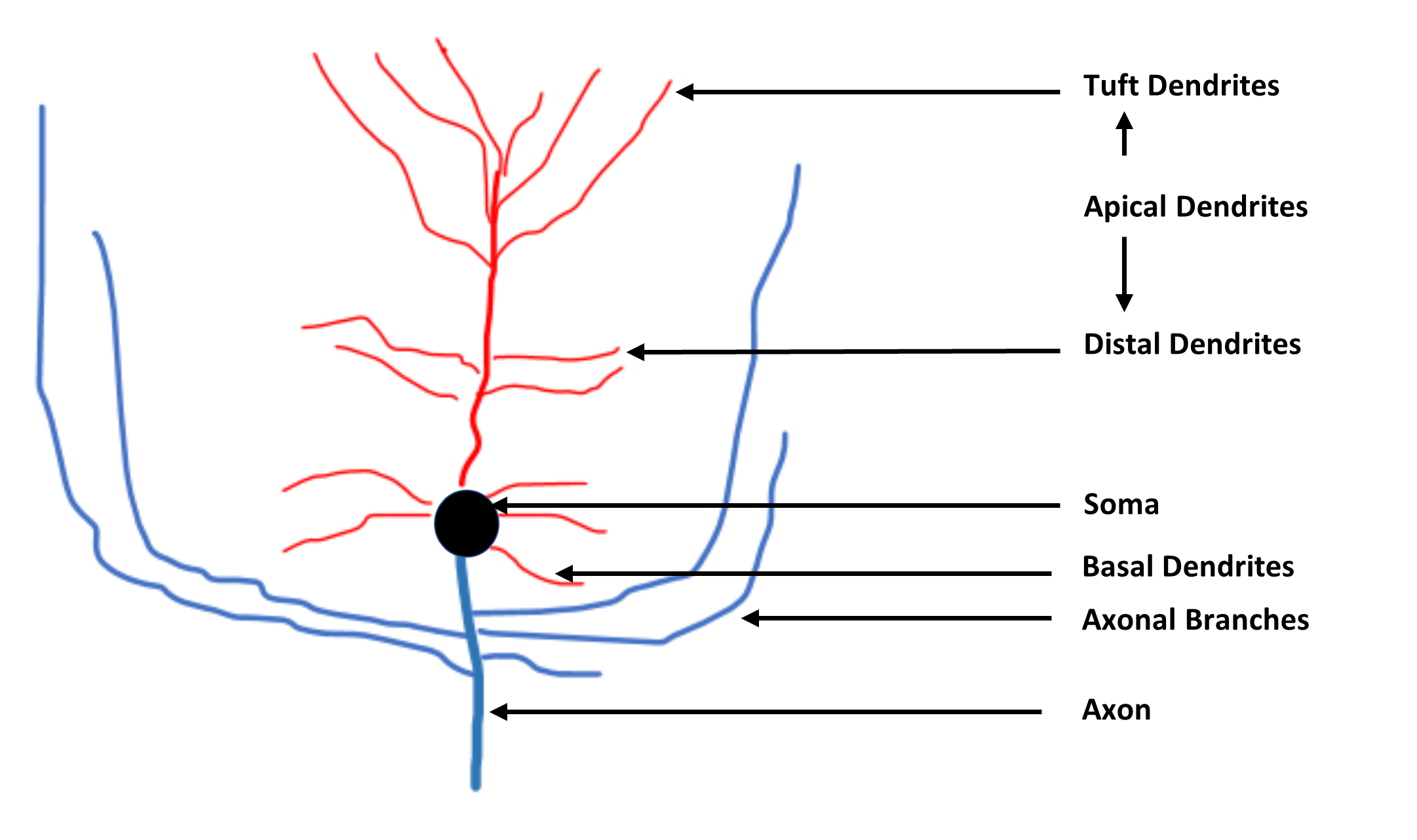';file-properties "XNPEU";}}

The cortical layers are full of local field potentials (LFP) due to neuron
spiking. The cortical layers due to internal connections within them
generate coherent electric oscillations in LFP. This is modulated by sensory
input. The top layers are considered as the context layer which creates the
long-range coherence in the LFP. The fourth layer is of short-range
correlation due to direct sensory input.

In the interaction binding the fourth layer creates the local binding. The
top layer creates the long-range binding. The spike due to the top layers
travel downwards the tuft dendrites. The spikes in the input layer travel
upward due to the back propagation from the basal/lower apical dendrites
towards the tuft dendrites. In the dendritic integration theory (DIT) of
consciousness \cite{DIT} the interaction between these two spikes is
considered central for consciousness, as when a person is unconscious due to
anesthetics, the flow of top-to-bottom current is absent. The DIT is
consistent with our interaction binding.

This integration between feedback and feedforward can be related to various
forms of integration and also memory formation. It is considered a top-down
and bottom-up integration of data between more abstract content from higher
associative regions, which integrate all the sensory information into coded
information, in the feedback network and sensory content in the feedforward
network \cite{TDBU}. Integration is also thought of as the merger of the
state of consciousness and the content of the consciousness \cite{DIT4}, 
\cite{TTC3}. The merger of spatial information in the cortex and temporal
information in the cortex is discussed in the spatiotemporal theory of
consciousness by Northoff \cite{TTC1}, \cite{TTC2}. Spatial information is
contained in the high-frequency gamma-related firing and the temporal
information is in the low-frequency Theta and alpha firing \cite{FFFB1}, 
\cite{FFFB2}. The combination of these creates a full picture of conscious
experience as discussed in the theory.

When a person is awake and conscious, the interactions in the upper layer
and lower layer bind the sensory information and the higher-order processed
information in the top layer. This is necessary, but more things are
happening in the cortex. When unconscious we have a slow spike, no
amplification, and high coherence in the LFP, which we can assume to violate
the consciousness threshold (to be discussed later) necessary to achieve
consciousness in the cortex system, so there is no binding. While awake,
when the interaction between the top and bottom layers is happening, they
are fully active and they are interacting with each other, we see fast
spikes in the LFP, and the Fourier spectrum of this signal is quite rich in
details \cite{OBSCON1}, \cite{DIT} in all bands of brain waves. Also, the
signals in the sensory layer are amplified. The neuron spiking has a
long-range correlation in the top layer, which binds the context and
higher-order data. The neural firing in the lower layers binds the local
details directly from the sensory data relayed through the thalamus. Also,
various sensory data of different orders of processing interact with each
other, which are bound together as shown in figure \ref{CORLNGRNG}. The
various levels of information relating to the sensory data from concrete to
abstract meaning are bound together.

We refer to relevant research work for more details.

\FRAME{ftbpFU}{5.9283in}{3.9427in}{0pt}{\Qcb{The six-layer architecture of
cerebral cortex made.}}{\Qlb{CORINTBIND}}{interactionbindingword.png}{%
\special{language "Scientific Word";type "GRAPHIC";maintain-aspect-ratio
TRUE;display "USEDEF";valid_file "F";width 5.9283in;height 3.9427in;depth
0pt;original-width 7.1278in;original-height 4.7297in;cropleft "0";croptop
"1";cropright "1";cropbottom "0";filename
'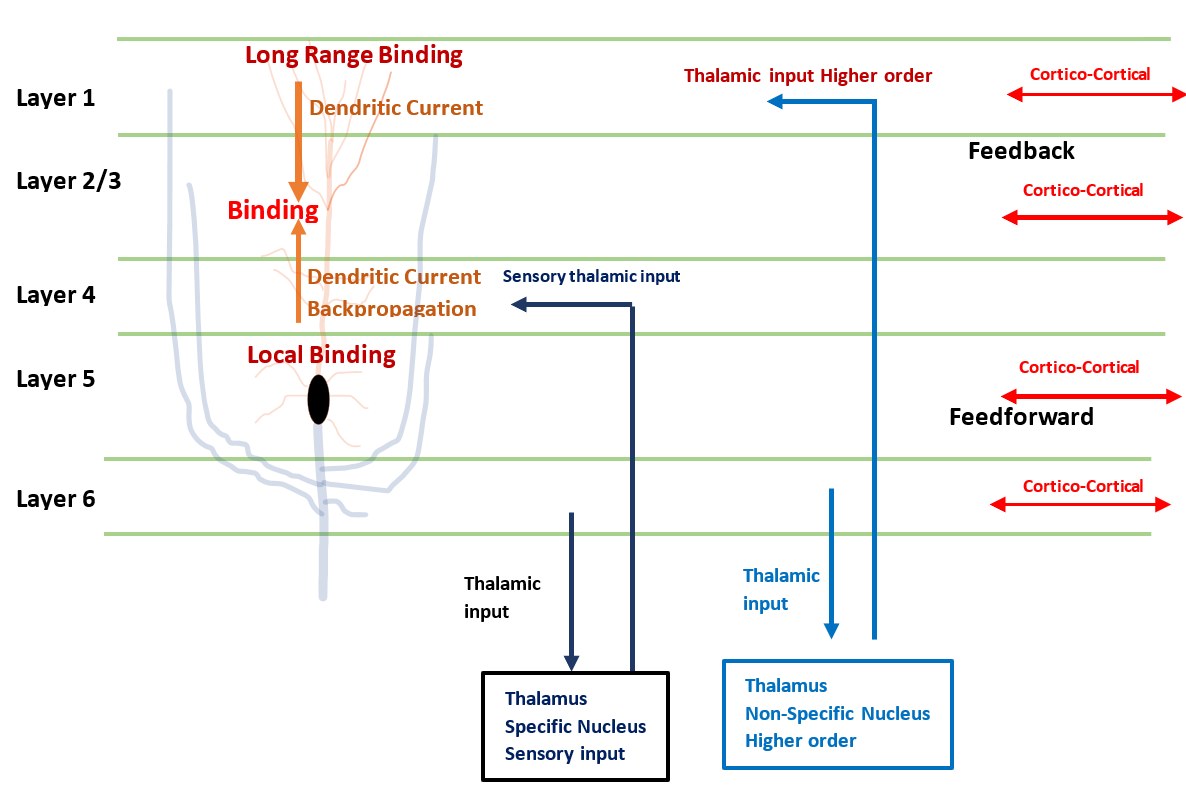';file-properties "XNPEU";}}

\FRAME{ftbpFU}{6.0805in}{3.4722in}{0pt}{\Qcb{Feedback and Feedford mechanisms%
}}{\Qlb{CORLNGRNG}}{corticalmultiorderintegrationword.png}{\special{language
"Scientific Word";type "GRAPHIC";maintain-aspect-ratio TRUE;display
"USEDEF";valid_file "F";width 6.0805in;height 3.4722in;depth
0pt;original-width 7.0794in;original-height 4.024in;cropleft "0";croptop
"1";cropright "1";cropbottom "0";filename
'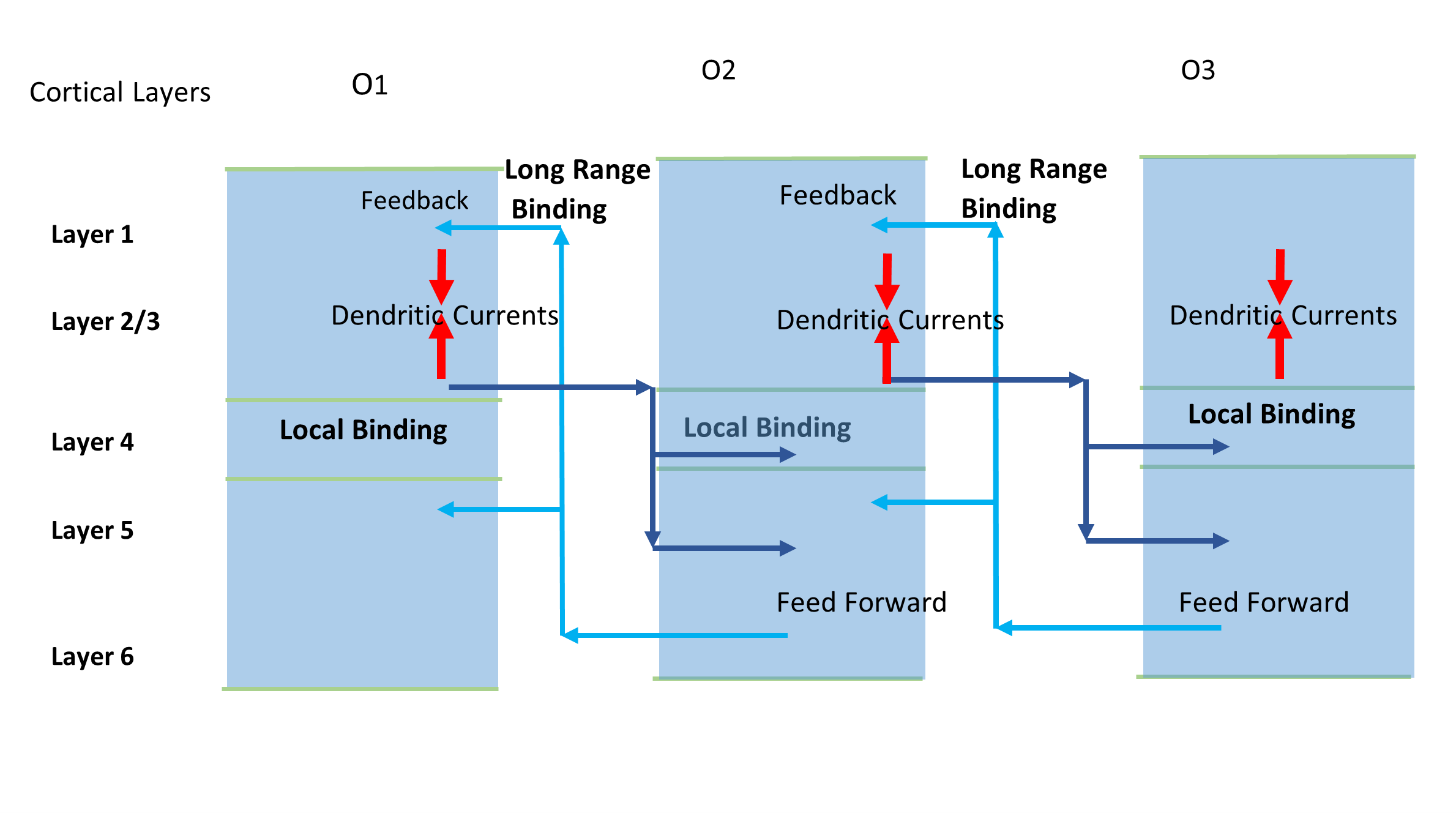';file-properties "XNPEU";}}

One important theory is the proto-consciousness described by Antonio Damasio
and others, which involves the brainstem and its role in homeostasis \cite%
{DAM}, \cite{DAM2}, and \cite{DAM3} which I mentioned before. The brainstem
is the most critical component of the consciousness arousal mechanism. It
combines the bodily and mental states and regulates the overall
mental-physical state of the body. The Brainstem has various nuclei that
project neurons relating to various neurotransmitters to various parts of
the brain (for example \cite{BSTEM}). It works along with the basal
forebrain, hypothalamus, and suprachiasmatic nuclei (SCN) to control the
circadian rhythm. Through them, it modulates the entire cortical firing
activity. Through this, it can make the brain sleep or arouse to wakefulness.

Damage to the brainstem can put the person in a coma. There is a strong link
between brainstem nuclei activity and the oscillations of the cortex. The
brain stem contains the default ascending arousal network \cite{BSTEM1} that
controls the level of arousal of the higher brain layers which has been
mapped in detail recently. Also, the feeling of emotions is considered as a
response to bodily physiological changes as indicated by research proposal
by Damasio \cite{DAM}, \cite{DAM2}, \cite{DAM3}, and others.

We can say that the interaction binding of chemical interactions at the
neural and cellular level in the body, brainstem, limbic areas, and
neocortex together creates the human cognitive/emotional processes and
memory formation. The physics of how these bindings happen is what we have
been discussing and further discuss in the rest of these papers.

\subsubsection{Predictive Coding and Consciousness\label{precod}}

The human brain has evolved to map the external world into mental
information that captures it. To map the world from the environment to the
cortex, interaction must happen. The interaction must happen precisely to
match the physical characteristics of the objects in the environment, such
as shape, color, movement, etc. So evolution when it promotes precision in
mapping the environment promotes the elements of consciousness. There is the
evolution of predictive coding through the generation of feedback from the
higher associative layers to the primary layers, which is indicated by
various research \cite{PCOD1}, \cite{PCOD2}, \cite{PCOD3}, \cite{PCOD4}. In
the animal brain, the sensory information goes from various sense organs
through thalamic relay nuclei and it undergoes layers of deep neural
networks and, finally layers of associative layers that combine the
modalities. This bidirectional processing helps in predictive coding that
helps in the identification of sensory objects.

In the human brain, there are extensive feedback circuits in opposite
directions, such as in the feedforward and feedback loop in the cortical
layers discussed before. The feedforward circuits carry raw information and
get coded, and the feedback loop carries decoded information rising from
long-term memory areas such as the hippocampus and limbic layers. In this
process, we have memory formation being active, which was proposed to be
linked to conscious awareness in observation 1, earlier in this paper, and
the conscious threshold conditions.

The feedback circuits help amplify the information of relevance \cite{FEED1}%
, and \cite{FEED2}. The feedback can be assumed to originate from the
central associative locations with few neurons whose firing captures the
sensory content in an encoded form. The firing of these neurons directly can
bring coherence in the firing of neurons that are involved in the visual
processing of the encoded object. This parallel coherent interaction
happening all the way from the highest associative layers to the primary
layers, binds all the information in the interaction bonding. I summarize
this in the following hypothesis:

\begin{principle}
Hypothesis 4.3: The interaction binding required for sensory consciousness
matches the interactions necessary to map the external world in the neural
system of animals through various neural processing involving memory and
predictive coding.
\end{principle}

\begin{eqnarray}
\text{Sensory Consciousness} &\equiv &\text{Interaction Binding} \\
&\equiv &\text{Memory Formation} \\
&\equiv &\text{Predictive Coding}
\end{eqnarray}

I believe evolution is driven by the relationship between the four. This
shapes every detail of linkage between neurons interaction through the
dendrites. The pyramidal neurons in the cortex have thousands of axonal
branches and dendritic branches that connect to other neurons. The dendritic
branches have their own computing capabilities. All these connection
complexities driven by the relation between sensory consciousness,
interaction binding, and predictive coding, have evolved to map the external
world.

\begin{principle}
Hypothesis 4.4: The combined presence of consciousness, memory formation and
predictive coding is due to evolutionary mechanism.
\end{principle}

\begin{principle}
Hypothesis 4.5: The interaction binding necessary to do these is due to
anthropic effect.
\end{principle}

I strongly believe that the relation between the four might have mostly an
effect of the evolution to support survival. Because memory formation
without consciousness doesn't have a big picture and also the emotional
reactions encoded in it. This is necessary to support survival. Also
predictive coding is integral connected to memory formation, and as they
both need each other. The interaction binding necessary to simultaneously
engage these three may have been related to anthropic effect, as this is
deeply fundamental. The engaging of this interaction binding may have been
due to discovered by life during evolution.

\subsection{Behavior}

\subsubsection{Conscious and Computational Interplay}

In the human brain, the coherence is brought about by feedback loops between
various parts. The coherence happens in time, and space, between the brain
and environment both in receiving sensory information and interacting with
the environment. Consciousness uses coherence for object recognition,
attention, and binding. This happens in parallel with computations in the
neural network that help with coherence, and processing of the sensory
information and the reactions of the various parts of the brain. So, we can
state the following:

\begin{principle}
Hypothesis 5.1: Consciousness and computations are coupled together in the
human brain to create behavior.
\end{principle}

Let's discuss this hypothesis in more detail. The coherent set of neuronal
assemblies that process conscious information keeps shifting all over the
brain. This shifting of neuronal coherent assemblies can be considered as
free will action. The role of consciousness is not in just perceiving the
information but also in the ability to shift the coherence around the brain
such as in triggering actions by influencing motor areas and in shifting
attention. An account of the dynamic interaction between the neural network,
consciousness, and synchronization is discussed by Stuart Hameroff in his
model of conscious pilot theory \cite{CONHAM}. Hameroff suggests that the
interaction between adjacent neurons synchronizes the neural firing (gamma
synchrony) and the shifting of the synchrony as triggered by activation of
various brain layers relates to conscious control or more specifically as he
mentions, " the conscious pilot". We will go into further detail about this
later from the point of interaction binding.

One of the important features of a conscious mind is the capability of
attention. There is a continuous stream of information coming into the human
mind both internally and externally. Human consciousness has to choose to
attend to a particular focus on this stream. It has been well documented
that the neural firing of the neurons that process the data of objects it
attends are well phase synchronized. The objects that consciousness doesn't
attend to have neural patterns out of phase with those that are in
attention. This phase coherence helps concentrate the interaction binding of
the neural information relating to the attending object. The human
consciousness has to continuously steer the phase coherence as objects it
attends to changes. The higher associative areas that process the meaning
and abstract information of objects play a role in this.

We can model high-level wakeful activity in the human brain using a system
model using the following elements as shown in the above picture. This is a
more detailed feedback model of what was described before in section \ref%
{simmod}.

\begin{itemize}
\item Brain Stem: It controls the level of arousal of the cortical system.
It can put the cortex to sleep or make it awake. The brainstem controls the
cortex through various neural projections from its various nuclei.

\item The Associative areas combine the primary visual data into the coded
form and send feedback processing to the primary sensory areas.

\item The primary sensor areas capture information from the sensory organs
and send it through forward processing to the higher associative areas.

\item The higher associative areas link with the limbic system and connect
to memory.

\item The limbic system sends back feedback to the higher associative areas.

\item Consciousness arises out of the integration through the coherence of
the neural information processing.

\item Consciousness directs the higher associative areas through volitional
feedback through critical parameters that control the flow of higher models
of data processed.
\end{itemize}

In this model consciousness and computational system together control the
behavior of the organism through recurrent processing. Computational
processing does everything to match the sensory data to models stored in
memory, through predictive coding which we will discuss further later.
Hypothetically, consciousness receives perceptual data from the processes of
integration through coherence. Then it sends back neural control influences
by changing critical areas in the human brain to create behavior, mental or
physical. This alters the sensory data generated internally or externally
that is processed compositionally by the brain. This results in recurrent
processing between conscious and computational systems.

\subsubsection{Free Will}

Now we want to understand the fundamental physical concepts behind the human
behavioral phenomena. One of them is the Free will. Free will is a
philosophically charged term and may be subjective. Usually, people have a
certain logical reason to do something based on the past. In this case,
there appears to be no free will. Also if they make the decision without
reason, then it is a purely random act. This also appears to be of no free
will. So it is difficult meaningfully define free will.

We really don't know whether consciousness has free will capability. The
conscious process could be purely controlled by computation processes, and
our sense of the existence of free will could be an illusion created by our
mind. The explanation for various human behavior is discussed in the theory
of human behavior \cite{BKVI1}, \cite{BKVI2}. As discussed there most
behavior seems computational. Understanding whether free will exists depends
on separating the unexplained parts of the model in the book, and accounting
for them.

I am going to assume the following about human free will without going into
a deep philosophical discussion about it:

\begin{principle}
Hypothesis 5.2: The human free will actions are the actions that we carry
out with our own effort with or without any a priori influence.
\end{principle}

We psychologically know what this is. Examples of these free will actions
are such as: moving our body, making choices, paying attention, etc. The
`without a priori part is critical' because, if every action we do is a
consequence of something that happened before then there is no free will.

Also, we can assume the following:

\begin{principle}
Hypothesis 5.3: The human consciousness has free will capability.
\end{principle}

Free will actions have survival relevance. This comes from cognitive and
affective influence. Unlike a computer which needs someone to program it,
conscious entities are driven by reasons, feelings, and emotions that direct
the behavior of the organism to attain survival needs. This capability is
directly programmed into the human brain in its computational processing
through a billion years of evolution. The presence of affective influence
directs the organism to do goal-oriented activities. The conscious free will
guided by the neural computational capabilities organizes the human brain to
solve problems of survival. This could be such as fight or flight behavior,
hunting, human thinking activity, etc.

\subsubsection{Dynamics and the Brain-Body Mechanism}

Now we will discuss the behavior of the conscious human brain from the
dynamical aspects of the fundamental physical principles using quantum
theory.

Now assuming any object both living and non-living as quantum objects, they
undergo two types of evolution: Deterministic Schr\"{o}dinger Evolution of
the wavefunction of the object and Random Stochastic evolution that
decoheres to prevent macroscopic superposition.

In the case of macroscopic objects such as a planet, ball, or stone this
evolution is leading to straightforward evolution that can characterized by
classical physics,

\begin{eqnarray}
\frac{dx_{i}}{dt} &=&\{x_{i},H\} \\
\frac{dp_{i}}{dt} &=&\{p_{i},H\}
\end{eqnarray}

Where the brackets are poison brackets.

In the case of a rigid body future macroscopic evolution is determined by
its macroscopic physical configuration variables. The influence of the
statistical thermal interaction among its atoms and molecules is extremely
small on its macroscopic evolution. Also, the system keeps undergoing
incoherent superposition as discussed in the first part. The incoherent
superposition gets large enough to cause a macroscopic difference, it gets
reduced. It undergoes continuous stochastic reduction, that only affects
microscopic local variation.

In the case of living things such as animal species, these processes are
complex. The system is not a complete rigid body. The bones and cartilages
are rigid. But muscles and neural activity are not rigid. There is a direct
coupling between microscopic and macroscopic variables, as I pointed out in
Hypothesis 2.2. The microscopic process has a direct effect on the
macroscopic processes.

The physical configuration variables needed to determine future mental
processes and physical movement are numerous. In the case of physical
movement, it can be approximated by some finite number of variables, say
about 100. But in the case of mental processes, we need to take into account
the state of billions of neurons.

In the case of a conventional classical computer, the design is based on
bits 1 and 0. It is so designed that any analog variation due to the effect
of thermal oscillation or continuous quantum reduction wouldn't have any
effect. Its future is 100 \% determined by its current logical state.

In the case of a classical computer, the number of logical inputs to the
logical gate in its circuits tends to be few, and small variations in the
input current have not much of an effect on its output. The timing of its
firing is controlled by the synchronous clock, so it's a pure logical state
machine undergoing predictable transitions. In the case of neurons, the
situation is different. We have a threshold that when breached will fire in
a predictable pattern. It's asynchronous, as there is no rigid clock that
controls its firing times. The variations in its input, thermal
fluctuations, and reduction probabilities can appreciably directly alter its
firing times. There are extensive inputs to the fine tuft dendritic
branches, and there non-linear effects in them that drastically affect the
inputs. Also, the release of neurotransmitters at the synapses is a
molecular phenomenon, that involves quantum processes \cite{QSYNAP}. So the
neural firing pattern is subjected to extensive computational, thermal, and
quantum effects.

In QGRF 5 \cite{MYP5A} of the paper, I discussed various quantum
superposition experiments, where large-scale coherent superposition of large
organic molecules made of about 10,000 atoms \cite{QEXP2} can be maintained.
The mass threshold to determine the macroscopic superpositions is quite
large on the atomic mass scale. The number of neurotransmitter molecules
released is quite finite, per neural firing from synaptic vesicles much
smaller than 10,000 atoms, within a range of 25 ms, the typical period gamma
firing that is characteristic of awake processing. This means that the
neural inputs from each synapse are subjected to quantum superposition. Beck
and Eccles have worked out a quantum model of synaptic transmission \cite%
{QSYNAP}. But really don't need this here at all to come to the conclusion
of quantum superpositions. Given the current status of experimental masses
that can undergo quantum superposition, it must be considered as a fact that
synaptic processes undergo quantum superposition \cite{QEXP1},\cite{QEXP2},
and \cite{QEXP3}, which we discussed in QGRF 5 \cite{MYP5A}.

Scientists such as Penrose and Hameroff are focused on microtubules very
much with emphasis on coherent quantum superposition. However, given the
role of incoherent superposition, the source of quantum superposition is
everywhere such as synaptic vessels, dendritic branches, ionic gates,
microtubules, volume-released neuromodulator molecules, etc. One of the
reasons the microtubules and coherent superposition are alluded to is
because of the infinite power of potential microtubule quantum computing.
But this is not the only way for this such power. The dendritic computing in
the thousands of dendritic branches and their connections with other
dendrites can result in such powerful computing capability also.

So the brain as a whole is always under quantum superposition of various
possible future states, and it gets quantum reduced only if the neural
currents lead to macroscopic bodily configurations. The number of molecules
needed to determine whether this superposition is macroscopic is certainly
much larger than 10,000 atoms. In the case of the brain, the superposition
is incoherent superposition. As I have discussed in the QGRF 5 \cite{MYP5A},
the molecules such as the neurotransmitters interact with the thermal
background and slowly lose the quantum superposition it has initially. In
the decoherence models, they assume that the thermal background removes this
superposition. This is incorrect. During the thermal interaction, the
background molecules and the neurotransmitters as a combined quantum system
evolve into a superposition of combined states with random phases. The
random phases come from the thermal nature of the background molecules. The
incoherent superposition builds and is removed only when it has macroscopic
effects on the brain-body mechanism, from the point of view of the second
proposal in the QGRF 5 \cite{MYP5A}. How large enough in terms of mass
depends on the reduction constant introduced in the postulate I.2 of the
QGRF 5 \cite{MYP5A}. This has to be discovered experimentally as discussed
in the QGRF 5 \cite{MYP5A}.

\subsubsection{The Quantum Reduction Free will Equivalence Principle}

Currently, it is well established that quantum physics governs the molecular
world, and the human brain functions on the complexity of molecular
chemistry. Now consider your brain is undergoing this quantum mechanical
superposition. You are now in a superposed state. In the course of
interaction between neurotransmitters and thermal background say in a time
period of 0.25s (the period of gamma synchrony), an estimated time interval
it needs to act, the human brain-body system evolves as a superposition of
quantum histories of various potential future macroscopic states. The
probability weights are not purely random due to interactions with thermal
fluctuations. However given that neural circuitries are evolved to promote
mental processes and physical movements that are of survival relevance, the
quantum probability weights for the states that can lead to survival
relevance are high. Here I am talking about the brain-body system because
the central nervous system constantly interacts with the body through the
brainstem. As I discussed before emotions are considered as reflections of
physiological mechanisms (Damasio and others). Various locations in the body
relevant to determining brain states such as emotions, also undergo joint
quantum superposition in the process, along with the neurons in the brain.

Consider your own brain-body mechanism. From an outsider's point of view,
based on the postulates in QGRF 5 \cite{MYP5A}, the brain-body quantum
incoherent superposition states will result in the spontaneous stochastic
quantum reduction of the brain quantum superposition, leading to the
particular macroscopic future history of brain-body processes. From an
insider's point of view, it is you who are doing the reduction. The
brain-body system is you and who is choosing your future state. Otherwise,
the quantum reduction is your free will choice. This is what I call the
Quantum Reduction Free will equivalence principle. I can summarize this as
follows in two parts:

\begin{principle}
Hypothesis 5.4a: Quantum Behavior Equivalence Principle: The quantum
dynamics of the brain-body system involving quantum evolution and reduction
is equivalent to the behavior of the person that it characterizes, both its
mental and physical state, both calculated and free will actions.
\end{principle}

\begin{principle}
Hypothesis 5.4b: Quantum Reduction Free will Equivalence Principle: The
quantum reduction the brain-body system undergoes is the free will choice
made by the person.
\end{principle}

In this, I remove the dichotomy between mind and body. The mind is part of
the brain, it has both conscious and unconscious processes, and the
processes that the mind does is one to one related to the processes the
mind-body system does. Since the probability weights are determined by the
brain-body survival mechanism, the probability weights are high for actions
of survival relevance, if they are created by the brain neural computing
during the period. We need to note that in the case of classical computers,
they also have incoherent quantum superposition. But when it comes to
quantum macroscopic effects, there is only the history that is determined by
the logical operations designed by the engineers that will have the ultimate
effect.

\subsubsection{Consciousness Threshold, Wakefulness, and Sleep\label{conthr}}

One of the important problems to solve in formulating the theory of
consciousness is accounting for the phenomena of sleep and wakefulness, and
also unconsciousness brought about by anesthesia. It appears consciousness
can be turned off and on. We need to explain these phenomena using the
interaction binding model.

\begin{principle}
Definition A.1, The minimal quantitative and necessary qualitative activity
in the brain that is necessary for conscious awareness is the conscious
threshold.
\end{principle}

We need to understand the consciousness threshold using the interaction bind
model. In the neural network of the human brain this threshold could depend
on many things:

\begin{principle}
Hypothesis 5.5a: The conscious threshold is a function of coherence, mutual
information, and total information in the firing activity of the neural
network.
\end{principle}

This is not complete and we will discuss this further later to make a more
comprehensive list of threshold conditions.

As I have hypothesized the local binding is done by quantum entanglement
during interaction. But to perceive something, we need the long-range
binding through coherence. Without coherence in the interaction, there is no
conscious content. When there is no brain activity that is necessary to keep
a person alive, local interactions are thermal in nature, largely containing
thermal fluctuations. This would not create coherent content. If there is
too much long-range coherence, then the local interaction gets weakened. For
example, when neurons fire simultaneously, they interact weakly as there is
not much potential difference. In that case, we could assume that thermal
fluctuations dominate. We have discussed the physical and neural aspects of
this in hypotheses 4.1 and 4.2, respectively.

To summarize, both the complete lack of coherence and too much coherence
result in the lack of coherent content that leads to a lack of conscious
content. But for consciousness to occur we need a balance of these two. This
is our conscious threshold in a qualitative form. We will work out this
condition later quantitatively using a mathematical form.

\begin{principle}
Hypothesis 5.5b: A proper balance of global coherence and local interactions
is required for consciousness.
\end{principle}

\begin{principle}
Hypothesis 5.5c: Coupling between microscopic control variables and the
macroscopic sensory regions is necessary for consciousness.
\end{principle}

Here I have split Hypothesis 2.2e into three parts: Hypotheses 5.5a to 5.5c.
To explain the relation between hypotheses 5.5b and 5.5c we have the
following hypothesis:

\begin{principle}
Hypothesis 5.5d: Coupling between microscopic control variables and the
macroscopic sensory regions is required to maintain the necessary coherence
for consciousness.
\end{principle}

\begin{principle}
Hypothesis 5.5e: Ability to form memory of the processes neural information
is necessary for consciousness.
\end{principle}

The brain may be evolved so that hypothesis 5.5e may be able to satisfy all
the 5.5a to 5.5b.

Let me discuss how these hypotheses work.

If you want to have free will, the best bet is, that it has to be based upon
quantum reduction, which gives free choice to choose between various
possibilities. The magnitude of the wave function of the brain gives a
measure of the influence of various brain processes relating to 1) Thinking
2) Emotional 3) Memory, etc. We can call this the deterministic bias. They
directly cannot influence action. Because if it happens then there is no
free will (hypothesis 5.3 is contradicted).

\begin{principle}
Definition A.2: Deterministic bias is the influence of the deterministic
processes due to memory in the synaptic connections in the brain on the
actions of a conscious entity.
\end{principle}

The link between deterministic bias to action has to go through free will.
Free will has the ability to choose anything, but usually, the choice is
probabilistic, with influence from deterministic bias. The probability of
reduction of the microscopic part of the quantum state of the conscious
brain is dictated by the modulus squared of the wave function determined by
deterministic bias. That is, the link between the wavefunction to action
goes through reduction. This result of the reduction is not chosen a priori.
This suggests a clear link between reduction and free will.

The most important process of free will is the attention. The human brain is
influenced by various sensory inputs and information from memory. There are
so many things to pay attention to. As I have discussed before the neurons
that process neural data from an object that it pay attention to fires phase
coherently. The free will action by consciousness needs to change the object
of attention consciously. This means free will has to alter the focus of
coherence of neural firing to the input neural firing stream of different
objects. This implies that quantum reduction chooses the objects for which
the processing neurons have to be coherent.

At each instant, the higher associative areas which have the micro variables
that have the encoded information can be considered to be in a superposition
of various future possibilities. This follows from hypothesis 2.1, that the
control variables are quantum in nature. When they get amplified by the
feedback circuits to the primary sensory areas, or toward the motor areas,
continuous free will reduction happens, which triggers certain variables or
the sensory regions to become coherent in processing a specific object. The
reduction happens continuously, which means equivalently the free will
action happens continuously. This is the link between coherence and the
coupling between micro and macro control variables explaining hypothesis
5.5d. This also explains hypothesis 5.5c as coherence is essential for
consciousness as stated in hypothesis 4.1. Also, the proper balance between
coherence and local interaction is required for linking local information
into a meaningful whole as we have discussed in the section on integration
of information as stated in hypothesis 5.5b. Hypothesis 5.5a is just a
general statement of hypothesis 5.5b and hypothesis 5.5d. This also has been
discussed in the section on coherence before, and we will further discuss
this later in a more comprehensive list of conscious thresholds conditions.

\subsubsection{Rest Frame Foliation, Time, and Reduction}

In the context of quantum gravity, there is no absolute time variable. I
have proposed the rest frame foliation as the foliation in which quantum
states evolve and reduction happens, which was postulated in QGRF 5 \cite%
{MYP5A} and was introduced first in quantum gravity framework 1 \cite{MYP1}.
This assumption is quite essential to consciousness, which integrates
sensory information throughout the brain into one simple experience at
specific time instants. This requires a physical foliation to identify time
instants. The rest frame foliation is the most obvious choice. This has been
discussed in quantum gravity framework 4 \cite{MYP4}.

I have indicated that rest frame foliation determines the flow in the
interaction bonding. Let me elaborate more on this. The rest frame foliation
or most probable least variation foliation in general, was introduced in the
quantum gravity framework 2 \cite{MYP2}, as the preferred foliation for
continuous quantum reduction of quantum states. As I have pointed out, in
QGRF 5 \cite{MYP5A}, there is no covariant quantum collapse of
wavefunctions. The way the measurement is formulated in Copenhagen
interpretation, the quantum collapse is not covariant.

In the QGRF 5 \cite{MYP5A}, I discussed in detail in postulate I.2 how the
density matrix is continuously reduced along the rest frame foliation. The
rest frame foliation in a region where there is matter is the foliation in
which the matter is at rest. We can justify this from the point of view of
consciousness, the free will quantum equivalence principle.

First of all due to interaction binding various particles of a big molecule
interacting through electrostatic forces undergo the joint experience. This
was proposed in hypotheses 2.2 and 3.1 on interaction binding. Now we need
to understand the time and space of experience. The interaction happens
through the exchange of photons that correspond to the transverse part of
the electromagnetic field. This interaction converts into qualia. Assuming
that the interactions are triggered by an external source that releases
timed triggers, this qualia will also follow the same timing. These joint
molecules experience a set of sensations that are distributed in space and
time. The spatial distribution of the atoms can be related to the spatial
distributions of sensations. The most natural time difference between these
sensations will be measured in the rest frame of these particles. Since
these particles are continuously interacting they need to be bonded to each
other, and so they all are traveling relatively in the same velocity and
direction. The relative velocity has to be quite non-relativistic and can be
considered as part of the interaction information so converts into qualia.

If we consider a large structure like our brain, the interactions happening
throughout the brain need to be bound together into one continuous temporal
stream 3d conscious information. The natural foliation to experience this is
the rest frame foliation of the particles of the brain. This can be
approximated by the center of mass inertial reference frame of the brain.
But if we want to be more precise then we have to use the condition of
minimal variation of the fields in the brain. I have also proposed that this
foliation is determined stochastically, the higher the probability of a
foliation, the less the variation of the fields as proposed in postulate I.2
in quantum gravity framework 2 to 5.

\subsubsection{Longitudinal vs. Transverse Gauge Fields}

The electric field and gauge fields can be resolved into longitudinal and
transverse fields. In the case of any electric field, it is the transverse
field that is responsible for photons. The longitudinal field is the static
electric field of a charged particle. It has substantial energy in this
field, along with the contributions of the Higgs coupling. There is energy
from other gauge fields that make up this also. The rest mass energy of the
longitudinal fields contributes to rest mass. Any macroscopic superposition
of longitudinal fields undergoes continuous reduction. The details of the
transverse field interaction create the interaction binding, and qualia, and
are quantum in nature. If this leads to macroscopic superposition, that is
the superposition of the longitudinal fields, it undergoes reduction. The
longitudinal fields are like the classical observer and the reference frame
in which these fields are at rest presumably creates a space-time arena for
the consciousness created by the interaction binding.

\begin{principle}
Hypothesis 5.6a: The interactions between particles through transverse gauge
quanta create the conscious binding.
\end{principle}

\begin{principle}
Hypothesis 5.6b: The least variation (rest frame)foliation of the
longitudinal fields is the foliation that defines time: That is the
foliation in which 1) the Reduction happens, and 2) the Conscious binding of
interactions of various spaciously separated particles are bound.
\end{principle}

\subsubsection{Retrocausality and Incoherent Entanglement}

Now as I have discussed in QGRF 5 \cite{MYP5A} there is bidirectional
evolution and retrocausal aspects in quantum gravity due to the second-order
nature of the Hamiltonian constraint. Between two consecutive quantum
reductions, the state of the nervous system is determined by both future and
past states at the quantum reductions. Recently Penrose \cite{PEN2} has made
suggestions regarding the possibility of retrocausal influences. Now based
on the analysis that I have discussed in QGRF 5 \cite{MYP5A} regarding
retrocausality, this seems to be a theoretical possibility.

Libet experiments \cite{LIBET} indicate that an unconscious neural process
of about 0.5 seconds in the future needs to happen to experience a skin
stimulus at present, experienced about 20ms after the stimuli. This means
possibly quantum reduction happens 0.5s into the future to complete the
interaction. At this time no experiments have set limits on reduction times
or masses involved. While the coherence of the wave function of matter is
lost rapidly due to interaction between atoms and molecules in any form of
matter, the incoherent superposition can continue for a long time as I have
discussed in QGRF 5 \cite{MYP5A}. For a human brain, it could be 0.5s,
during which the quantum state of the human brain, superposes into multiple
incoherent quantum histories, to be reduced finally. This reduction results
in being felt at the present moment when the retrocausal wave of reflection
reaches the previous reduction, which seems to be a plausible
interpretation. The experiment suggested in QGRF 5 \cite{MYP5A} can help
explore the physical possibility of retrocausality induced by quantum
gravity. Later we will discuss the retrocausal influence in the quantum
mechanical model of free will action.

\subsection{Evolution, Consciousness, and Harmonic Relational Structures}

Let us focus on why living things such as humans need consciousness at all.
Consciousness requires the presence of complex systems in humans, which are
also present in primates and mammalian groups. So we can conclude that human
consciousness evolved from other lower creatures through a gradual
upgradation over the course of millions of years. It is clear that human
consciousness depends much on processes in the cortical layers, which are
heavily evolved in humans. If we go back in evolution all the way to
reptiles, the cortex is quite small with the brain being dominated by the
reptilian layers such as the brain stem and cerebellum. These layers are
mechanical in nature. The lower layers such as the cerebellum don't
correlate with consciousness, as people born without cerebellum, or
cerebellar stroke, can have full consciousness. But the role of organs such
as the cerebellum is the unconscious processing such as during movement. The
dominance of the cerebellum, and only a small layer of cortex in reptiles
indicate that their behaviors are largely mechanical with little conscious
perception if existed. But between reptiles to humans, the neocortex
develops, and the physical aspects of the brain correlated with
consciousness increase in complexity. Why is this? What is the importance of
consciousness for evolution?

The explanation possibly lies in the interaction binding theory proposed in
this paper. I discussed before that consciousness evolved in parallel with
the predictive coding to map the external world into a mental picture in the
living brain. The predictive coding may explain the necessity of sensory
consciousness. But there are many issues involved here. Why the evolution of
coherence, higher consciousness, and the evolution of consciousness at all?
The neural correlate of consciousness is quite elaborate and involves the
integration of information from areas throughout the brain, as argued by the
neuronal global workspace theory, and the integration of processing of the
neural data through phase coupling between neural firing at different
frequencies, for example, \cite{PhLock1}, and further references in the
paper. Creatures such as fish, lizards, and snakes have survived for
billions of years, with just primitive computational capabilities and no
higher consciousness. Why does the evolution of conscious beings occur in
parallel tracks along them, competing with them?

The answer to this question could be that we may be living in a region of
the universe in which the physical principles are made to prefer the
evolution of human consciousness, from the point of view of anthropic
principles. Otherwise, we wouldn't exist in this universe. So our question
would be is the evolution of human consciousness built into known physics or
new physics is needed? Certainly, there is a lack of physics that will link
neural firing to conscious experience. We mostly have some general ideas. We
can look for essential features of life and consciousness that are uniquely
different from non-living things.

We can notice two features of consciousness and life.

1) Life is built on complex structures of organic molecules that have
extensive symmetry properties. And consciousness is built on more complex
biological structures, that have interesting complex and rhythmic dynamics.

2) Life and consciousness tend to build complex structures that have
extensive symmetry properties, particularly in humans.

The most general feature of these complex structures is the `harmonic
relational structure' property.

We can define harmonic relational structures as follows:

\begin{principle}
Definition A.3 Harmonic relational structures are those that are made of
relationship structures of simple elements that have symmetry properties in
time and space.
\end{principle}

\begin{principle}
Definition A.4 Simple harmonic relational structures are entities made of
the complex relationship between molecules that have symmetry properties in
time and space.
\end{principle}

\begin{principle}
Definition A.5 Complex harmonic relational structures are those that are
made of putting together in time and space in simple relationship structures
in a proportional fashion through nesting, associating, multiplying, or
other means.
\end{principle}

The molecules that makeup living things are extensive relational structures
that have symmetry properties. They built structures by combining together
with other structures creating nesting structure and associative structures
that has symmetry properties. The neural networks themselves are interesting
relationship structures that have harmonic dynamical properties that are
neurally correlated with consciousness. Consciousness beings such as humans
are attracted to music and art. Which by themselves are harmonic relational
structures in both time and space. In other words, the essence of life and
consciousness is the harmonic relational structures.

Consider that reptiles and fishes have survived for a long time without any
attraction to music and art, or complex harmonic activity based on the
cerebrum. Yet the evolution of living things such as humans that do and have
these seems to be something more than the necessary Darwinian biological
evolution. That is survival is the most important characteristic of
Darwinian evolution. For humans evolution, it may be a part that brought
them along and they are involved in activities that may not be related to
evolution, such as destroying the environment and building highly risky
technology and weapons. They spend a lot of time listening to music, and
comedy, and talking about useless stuff, that has nothing got to do with
survival. This tells us that survival itself is not the main aim of this
species. Instead building harmonic relational structures seems to be the
main activity.

Let me put this in the form of a hypothesis:

\begin{principle}
Hypothesis 5.7: The evolution of conscious beings is not motivated by
survival as the ultimate aim, but by building harmonic relational structures
as a factor that promotes this species.
\end{principle}

The important question is whether this connection between life and harmonic
relation structures is already provided for by the existing established
physics or if we need to modify the physics to add this connection. We need
to investigate this experimentally. Later when I discuss mathematical
measures of harmonic relational structures, using that I add a new term to
the action term to make this connection as a proposal.

\section{The Postulates of Consciousness}

Now based on discussion until now and QGRF 5 \cite{MYP5A} we will come up
with the postulates to understand consciousness.

\subsection{The Zeroth Postulate of Physics\label{zerpos}}

In quantum gravity framework 4, I brought consciousness and structure
formation into the framework. This is essential to have complete physics of
both living and non-living matter. Also, I have proposed that quantum
evolution promotes relational structures as a fundamental process of
space-time. Given the large number of complex molecules and rare series of
accidents needed to create biological structures involved in creating life.
Humans have a fundamental inclination to build structures, and so it may be
possible that this originates from the innate nature of matter itself. The
ultimate existence of the universe and its laws has been learned through our
human consciousness. Human consciousness is a fascinating phenomenon, and it
requires many very precise conditions, for it to rise. To account for this,
I propose the following:

\begin{principle}
The Zeroth Postulate of Physics: The action and the dynamics of the universe
are chosen to promote harmonic relational structures that will lead to Human
consciousness.
\end{principle}

This directly follows from hypothesis 5.7. The idea of the harmonic
relational structure was introduced in my book \cite{BKVl3}. Harmonic
relational structures are relational structures of matter and ideas in the
human mind that have various symmetry properties in the relational,
temporal, and spatial domains. The universe is filled with it. Most objects
such as stars, planets, or galaxies have circular symmetry. In the human
brain, harmonic properties exist in the coherent firing of neurons, the
brain waves. Organic matter has extensive symmetry properties. Complex
structures are made of organic molecules that make the human body and brain
leading to consciousness. The principle is an anthropic principle \cite{ANTH}
that summarizes the innate nature of the universe and its properties that
were fine-tuned to the rise of consciousness.

This postulate is essential to understanding the dynamic and consciousness
postulates that have been described in quantum general relativity framework
5 and all the detailed definitions, hypotheses, postulates, and propositions
that are essential to understanding consciousness. There are too numerous
conceptual ideas in this volume it is impossible to logically deduce them
from a few sets of rules. That is why the anthropic principle to describe
the precise nature of these ideas is necessary for understanding concepts.
In the next section, we will describe the list of ideas that are needed to
understand consciousness and later put them all into a hypothetical set of
postulates.

\subsection{Quantum General Relativistic Dynamics and Behavior}

In quantum gravity framework 5, I discussed three postulates to formulate
quantum general relativistic dynamics. There are three complementary
postulates necessary to understand the dynamics of the universe based on the
three postulates to understand the rise of human consciousness. I will
discuss them in this section.

\subsubsection{Interaction and Consciousness}

The universe is made of matter at various levels of organization and
complexity. An isolated piece of matter is made of fundamental particles
that are quantum mechanically entangled and bound together by gauge forces.
They keep interacting with each other in a specific manner to the effect of
heat, internal organization, and external influence.

\begin{principle}
Postulate II.1 The coherent information in the interactions of the entangled
quantum structure is experienced as conscious information (qualia) by the
structure.
\end{principle}

This highly loaded postulate has been discussed in detail in the last
section. In the human brain, the matter is organized in a specific manner
such that interactions lead to highly coherent information, which in turn is
experienced as conscious information. This conscious information is what we
use to study the universe. Ultimately as humans, our observations about the
universe need to be understood in terms of this conscious experience.

\subsubsection{Behavior}

The human brain is an integrated information processing system that is
designed to map the external environment into its surroundings as discussed
in the previous section. The human brain and body mechanism together have
dynamical properties that originate from conscious control. Now we can link
conscious control to the dynamics of matter that it is made of. We will
rewrite the postulate I.2 of the quantum gravity framework 5 as follows.

\begin{principle}
Postulate II.2 Quantum Behavior Equivalence Principle: The quantum dynamics
of the brain-body system involving quantum evolution and reduction as
described by postulate I.2 is equivalent to the behavior of the person that
it characterizes, both its mental and physical state, both calculated and
free will actions. The reduction for conscious matter is interpreted to be
performed by consciousness at a rate $\beta _{R}$.
\end{principle}

This postulate has two parts. One is the deterministic evolution. It happens
unconsciously based on Postulate II.1. This is just a deterministic process
used to think and act. The other one is the stochastic evolution. The free
will choices are made by conscious entities all the time. The human brain
when looked at in detail has incoherent quantum superposition of entangled
quantum states of the matter that makes it. Every instant it undergoes
reduction as given in postulate II.2. This is equated with free will actions
of the conscious entity. This will be discussed in more detail in the next
section.

\subsubsection{Time of Experience}

Time in general relativity is given in the block universe sense. While the
universe itself is the 3+1 form in the quantum general relativity framework
built from postulates I.1 to I.3, the flow of time comes from conscious
experience.

\begin{principle}
Postulate II.3 The rest frame or the least variation foliation is the
foliation that defines the time of conscious experience.
\end{principle}

The human brain is made of multiple parts that entangle together into one
single conscious entity. The temporal flow of processes in each fundamental
particle must be integrated to be one single-time variable of conscious
subjective experience. The best foliation in which this integration can
happen is the least variation foliation that was defined in postulate I.3,
discussed in quantum gravity framework 2 to 5. As discussed there the least
variation foliation is determined probabilistically. The lower the variation
of fields in a foliation as measured by an integral variation measure, the
higher the probability for this foliation to occur. As I have discussed in
QGRF 5 \cite{MYP5A} in section 2.3, depending on whether the gravitational
field or electric field dominates the calculation of rest frame foliation,
the frame in which the reduction happens is the rest frame of earth, or the
rest frame each object like ball, bat, or the brain of the living animal
(entanglement to discussed later).

As retrocausal influence is built into the QGRF 5 \cite{MYP5A}, we also can
suspect retrocausal influence in the conscious processes. Retrocausal
influence has been suggested by Libet's experiments, and it appears their
existence may be needed to explain the free will existence as discussed
before.

\subsection{Free will}

The quantum reduction is assumed to be performed by the consciousness in
case of superposition associated with the consciousness states of a
consciousness capable neural matter, consistent with bidirectional-temporal
reduction.

The quantum evolution and conscious reduction can produce free will action.
Consciousness relies on the neural net to give all possible actions. The
quantum evolution of the neural systems guides the consciousness of all
various possible choices for self-induced quantum reduction. But
consciousness requires guidance for it to happen. The square of the relative
quantum amplitude in the superposition of the various states is a good way
to suggest the intention to act in that state, leading to action. The action
along this suggestion is consistent with the Max Born rule. This means that
quantum laws can be consistent with free-will capable consciousness.

\subsection{Retrocausality\label{retro}}

The experiments on free will by Libet and others suggest that the readiness
potential for doing something builds even before conscious perception of
intention. This obviously must be the case because intention to act can only
be sensed after the building of a neural firing pattern that will make you
sense it. This means any intention to do anything by a person has to be
already decided by the neural network. This means the possibility that there
can never be true free will. The only way freewill can happen in this
process is if the quantum processes that start the neural firing of
readiness potential can be triggered by conscious intention, retrocausally,
then free will is possible. This relates well to the spontaneous
bidirectional-temporal form of quantum reduction applied to conscious
matter. Consider the superposition associated with the superposition of the
quantum control variables of the conscious matter. 1) If this is reduced by
conscious influence, 2) produces retrocausal effect that can influence the
past, then free will is possible.

\FRAME{ftbpF}{5.7268in}{3.3512in}{0pt}{}{\Qlb{reductions}}{reductions.png}{%
\special{language "Scientific Word";type "GRAPHIC";maintain-aspect-ratio
TRUE;display "USEDEF";valid_file "F";width 5.7268in;height 3.3512in;depth
0pt;original-width 6.0001in;original-height 3.4999in;cropleft "0";croptop
"1";cropright "1";cropbottom "0";filename
'GR5Images/reductions.png';file-properties "XNPEU";}}

For retrocausal to happen, we can use the bidirectional reduction discussed
in the quantum gravity framework 5 \cite{MYP5B1.3} discussed in section 3.2
and in its implementation in section 3.3. When quantum control variables are
maintained quantum enough, then the bidirectional reduction can pass on
information backward in time. Further theoretical study needs to be done.
Penrose and Hameroff are proponents of retrocausal influence, and they focus
on microtubules. Here, we can use the bidirectional temporal evolution in
the quantum gravity framework 5 to study retrocausal effects and
experimentally verify them.

\FRAME{ftbpF}{5.4829in}{2.9066in}{0pt}{}{}{intent.png}{\special{language
"Scientific Word";type "GRAPHIC";maintain-aspect-ratio TRUE;display
"USEDEF";valid_file "F";width 5.4829in;height 2.9066in;depth
0pt;original-width 6.5025in;original-height 3.4333in;cropleft "0";croptop
"1";cropright "1";cropbottom "0";filename
'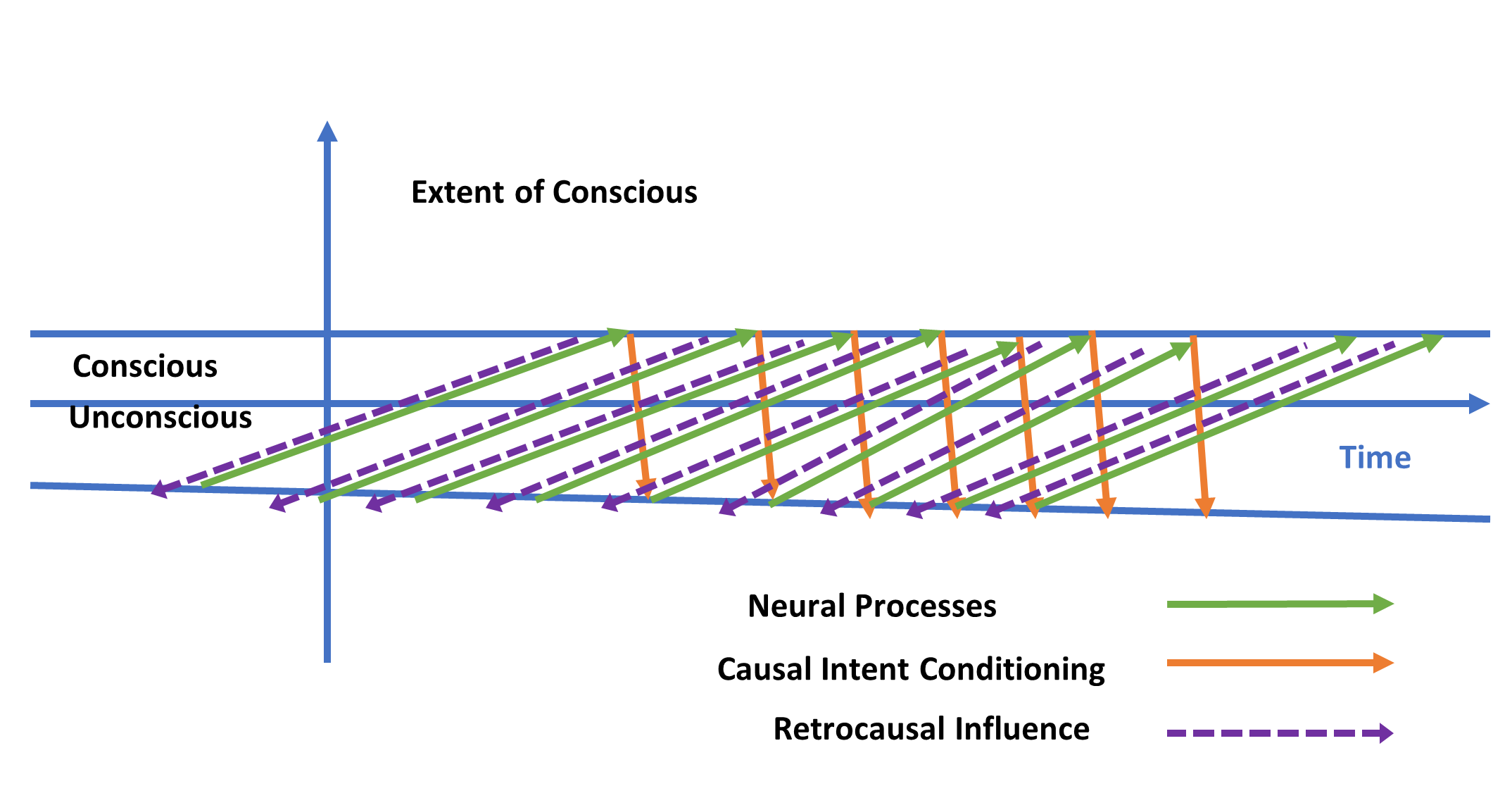';file-properties "XNPEU";}}

The spontaneous bidirectional-temporal evolution may be necessary for
conscious evolution. Whether this is sufficient depends on whether the
theory allows the backward influence necessary for quantum reduction, which
requires experimental and theoretical study. But clearly, the
consciousness-induced bidirectional-temporal reduction postulate can
conceptually help consciousness to have free will.

The diagram describes how consciousness influences the future and past
states. Consider the study of free will using the time of intent and
readiness potential by Libet. It has been consistently reported that the
readiness potential onsets several 100 milliseconds Tdel before the intent
becomes conscious, which may depend on the person. During this time, the
unconscious processing helps build the necessary neural firing pattern
required for the consciousness of the intent. This can be related to the
hypothesis setting the necessary processes for consciousness, involving
predictive coding, memory, feedback loops, and coherence necessary for
consciousness, discussed in section \ref{precod}.

\FRAME{ftbpF}{5.6074in}{2.9724in}{0pt}{}{}{bidirectionalintent.png}{\special%
{language "Scientific Word";type "GRAPHIC";maintain-aspect-ratio
TRUE;display "USEDEF";valid_file "F";width 5.6074in;height 2.9724in;depth
0pt;original-width 6.5025in;original-height 3.4333in;cropleft "0";croptop
"1";cropright "1";cropbottom "0";filename
'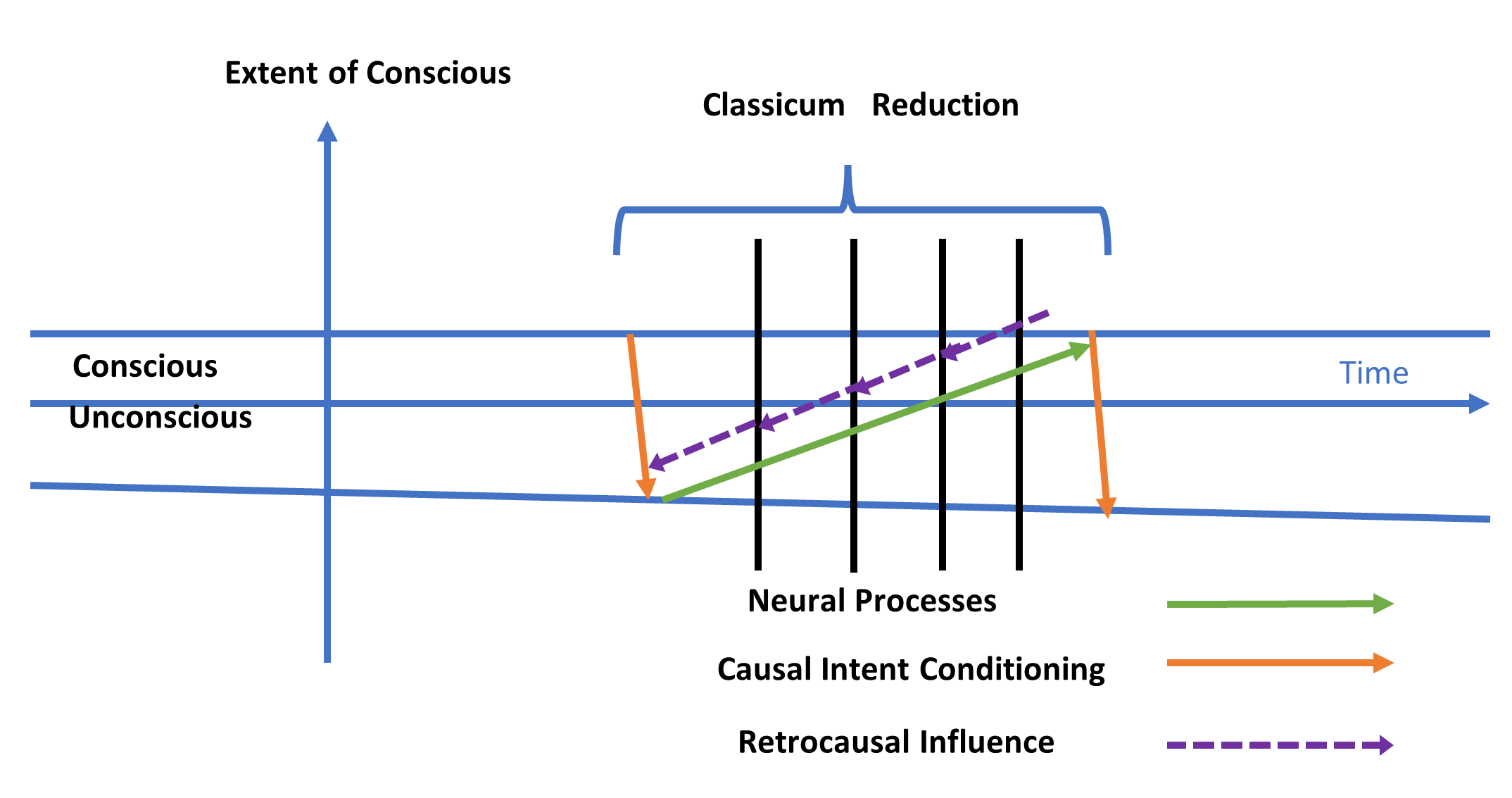';file-properties "XNPEU";}}

During the conscious awareness of the intent, two possible actions can
happen.

\begin{enumerate}
\item Retrocausal influence that influences the onset of the readiness
potential.

\item Procausal influence of awareness of the observation of action.
\end{enumerate}

There is a sequence of activities that happen by human brain. 1)awareness of
the task, 2)onset of readiness potential, 3) awareness of intent,
4)awareness of action, 5)awareness of action consequence, 6) again awareness
of the task\ldots . This process goes in the loop as shown in the diagram %
\ref{IC}.

\FRAME{ftbpFU}{6.0597in}{3.5466in}{0pt}{\Qcb{Intent Cycle:\ The diagram
shows a model of how conscious awareness interacts with the world.}}{\Qlb{IC}%
}{intentcycle.png}{\special{language "Scientific Word";type
"GRAPHIC";maintain-aspect-ratio TRUE;display "USEDEF";valid_file "F";width
6.0597in;height 3.5466in;depth 0pt;original-width 6.0001in;original-height
3.4999in;cropleft "0";croptop "1";cropright "1";cropbottom "0";filename
'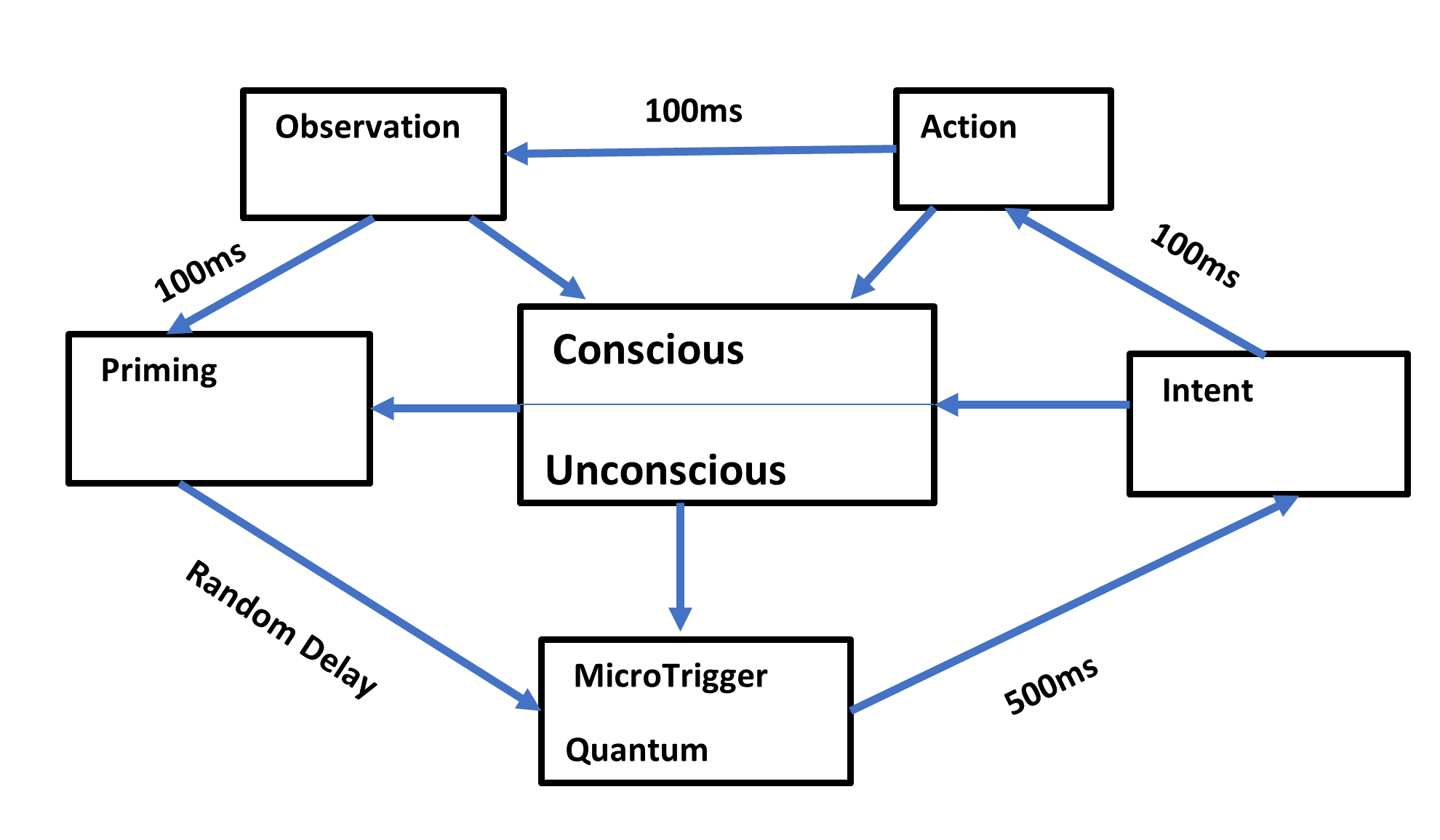';file-properties "XNPEU";}}

This conscious awareness of the experiment forces the quantum reduction of
all superpositions to constraint to carry out the activities that are
necessarily that is focused on that moment: that only focuses on making the
choice and carrying out the various action possibilities of the activity
that one has in mind, for example, the experiment. The intent at that moment
doesn't decide action choice, but constrains the brain activities to carry
out what one is focusing on. For example, writing activity-\TEXTsymbol{>}
decide on the details of writing, relaxing activity -\TEXTsymbol{>} just
focus on random thoughts, singing activity -\TEXTsymbol{>} choose a song.
The awareness makes you focus on the activity, and the brain processes carry
out the task by making a choice and acting.

The retrocausal influence due to bidirectional-temporal reduction can allow
intent to influence the quantum reduction at the onset of the readiness
potential. If we assume that during Tdel, the consciousness acts on the
entire reduction events through bidirectional-temporal evolution to carry
out the intent, then this is possible.

To summarize, inducing bidirectional-temporal evolution can make
consciousness have free will. The awareness of the task attributes conscious
free will forward in time, and the intent produces free will through
retrocausal influence. Both can create born rule guided reduction that leads
to free will.

\section{Consciousness:\ The Content and the Being}

\subsection{Historic Introduction}

One of the important questions that remained after publishing \cite{MYP4}
is\ the question of the observer. In the clubhouse discussion in 2023 on
quantum measurement \cite{Club}, the discussion moved on to consciousness,
then to conscious being. What exactly links you to your body? Various
answers were given, for example, neural structure. In my paper \cite%
{MYP5B1.1}, I talked about conscious content being different from the
conscious observer. While the conscious content keeps changing, turns on and
off, during sleep, during sedation, temporary coma, the conscious observer
remains the same. This has been the subject of religion and philosophers,
right from the time people started questioning about `what am I'. There is
Descartes, who says `I think therefore I am', and John Locke, who talks
about various forms of personal identity relating to the observer. Hinduism
clearly assumes the existence of the soul, which reincarnates into a
different person after death. Christianity assumes the soul goes into hell
or heaven, depending on one's life. The religion assumes the soul has
memory, which we know is part of the body, the brain. So strictly speaking,
the observer, or the soul, is the `I' part that exists without memory or
emotions generated by the brain. The Descartes argument focuses more on the
thinking part, even though he is alluding to your capability to think.

The theories of consciousness listed in the introduction before don't talk
about the `I' as different from the phenomena they theorize. IIT assumes
continuity of the `I', due causal link to past conscious states. All these
theories, and usually materialistic theories of consciousness, implicitly
assume the brain is the seat of the phenomena we are observing, and which
also contains the `I' glued to it, giving continuity.

In the Physical Foundations of Life Conscious 1.1 \cite{MYP5B1.1}, I
introduced the concept of conscious identity, conscious being, and conscious
content. In Physical Foundations of Life and Conscious 1.2 \cite{MYP5B1.2},
I expanded on conscious identity and conscious substratum. In this
identity-conscious identity is absorbed into conscious being. In this paper,
I discuss what exactly the physical characteristics of the `I', the
so-called `conscious being', that exists as part of the human body, are
different from conscious content, as mentioned in my previous version of
this paper.

\subsection{The Conscious Being}

Our awareness begins somewhere in infancy when we become smart enough to
record our life experiences and remember them. While we may have conscious
experiences before we didn't have memory of them. Our life begins when our
brain becomes complex enough to record experiences and convert them into
episodic memory, through which we can recollect or explain to others. But
all these are experienced from the point view of a particular person, the
`I'.

Let me discuss the `I'. Every time we wake up after sleep, we always wake up
in the same body. There is a connection between our mind and the body, that
exists whether we are conscious or not.

I will be referring to the `I' in us as the conscious being, to be different
from the conscious content which is the information experienced by the
conscious being. It is constant throughout the life of a person. The same
person is attached to the body, regardless of the conscious experience
interrupted by sleep, coma, or fainting. The free will needs a conscious
being to be meaningful.

From our hypothesis 2.2e, if a piece of matter satisfies the consciousness
threshold, we have the existence and the experience of `I'. To understand
the relation between conscious content and conscious being we need to know
the quantitative criteria of the consciousness threshold defined in
definition A.1. When the threshold of consciousness is met, then the
conscious content and the conscious being are coupled, if not they remain
decoupled.

There are three important parts to the conscious being:

1. Identity: The conscious being unique identities the person who
experiences the conscious content, and gives it continuity.

2. Observer: The conscious being is the observer who observes the
information in the conscious content.

3. Free will: The conscious being is the one who acts on free will based on
the information in the conscious content.

These three are very important parts of the conscious being that can be
separated from the conscious content.

Let's try to understand the physics behind the consciousness being. The
conscious being doesn't depend on particles, energy exchange, and the
structure of the physical neural structure of the human brain. Because the
fundamental particles are symmetrized or anti-symmetrized with each other,
they cannot be specifically identified. Next energy exchange patterns are
always changing. The chemical relational structures are always changing. So,
the being cannot depend on these three. But we can come to some general
conclusions if we assume that the conscious being is a physical part of a
conscious entity as follows:

\begin{enumerate}
\item The conscious being is localized in the conscious structures.

\item It interacts with matter and gauge fields and has energy states whose
frequency equivalents are in the range that includes 40 Hz.

\item It has properties associated with usual fields in standard quantum
field theory to interact with matter and also other properties that explain
the existence of perceptions.

\item Each conscious being is distinguishable from other conscious being and
there is (possibly) one being per conscious brain.

\item The physical extension of the conscious being, if it exists, has a
physical extension whose width is the same as the size of the human brain,
and possibly the maximum extent of the human body.
\end{enumerate}

\subsection{Conscious Content Model\label{concon}}

There is some general information we can say about how multimodal sensory
information is integrated. They are based on chemical reaction that generate
feelings and emotions. The chemical processes that helps in memory formation
and so in consciousness as noted in observation 1, should be related to
reactions involving various neurotransmitters and other molecules, that
creates the various conscious emotions and feeling.

\begin{itemize}
\item Emotional information. This relates to neurotransmitter release. For
example, when the neurotransmitter dopamine is released by specific areas of
the brain, it results in a feeling of happiness. This neurotransmitter acts
on certain regions of the brain. The chemical property of dopamine and its
receptors, and its effect on neural firing creates the feeling of sensation
of happiness. This feeling is felt at various places in the body, sometimes
on the body, head, face, chest, etc. So, the coherent information in the
tactile area binds with the coherent information in the emotional areas to
integrate this sensation. We can also say the same thing about taste and
smell.

\item Taste and the feeling of tactile sensation are located where sensation
is triggered. This could mean that the entire sensory neurons and their
interactions with various triggers at their sensory tips may be involved in
the integration and provision of sensory information.

\item Smell: The smell is not located externally but in the nose. The smell
is the reaction of our sensory receptors in our nose to various chemicals in
gaseous form. These molecules of gas don't have origin information unlike
sound or vision, so our brain cannot decipher its origin. So, the feeling is
felt only at our nose.

\item Vision and Tactile sensation are sensed on a screen of a continuum
manifold created by our brain. Sound is also associated with this continuum
manifold to point out its origin. The creation of this continuum manifold
along with 3D information including depth is a great evolutionary adaptation
that helps in our survival. The integration of all the sensations from both
the left and right hemispheres to make one single manifold happens with the
help of the corpus-collasm that connects to these hemispheres.
\end{itemize}

The various sensory information integrates together through the multimodal
integration areas through modulation of their firing through active binding
as discussed in proposition $\alpha $III.6a. There are in general several
important pieces of information,

\begin{itemize}
\item $y$ which denotes the sensory continuum,

\item $m$ which denotes the sensory modality,

\item $s$ denotes the strength of the sensation,

\item $f$, which denotes information like color, depth, taste, etc,
\end{itemize}

The whole sensory experience (Qualia)can be represented by $%
C=\{y,s_{i}(y),f_{i}(y)\}$. This $C=\{y,s_{i}(y),c_{i}(y)\}$ can be referred
to as the $C$ field.

In general $C$ is union of triplets $\cup _{M}(\Omega ,S,F)_{M}$
characterized by four parts:\ The $\Omega $ manifold, $S~$strength, $F$
flavor, which depends on $M$ sensary modality . \ These three depends on the
sensory and motor systems for each modality. The $y$ is point in $\Omega ,~f$
is an information in $F$, $m$ is the sensory modality $M$, and $s$ is a
value in $S.$

\subsection{Quantum Physical Properties\label{quaphy}}

Let me split a conscious entity into two bare parts: the bare conscious
being and the bare conscious content. The first one is an element that has
the potential to observe, has identity, and free will. The latter is the
electrochemical pattern. When they encounter each other, they become the
conscious being that exercises the three functions: free will, observation
and identity, and the conscious content is generated.

Let me discuss the properties of bare conscious being and bare conscious
content, and their interaction.

$%
\begin{tabular}{ll}
\textbf{Conscious Being (classicum)} & \textbf{Conscious Content (quanta)}
\\ 
Wholistic and extended & Dividable and entangled \\ 
Conscious substratum & Harmonic relational Structure \\ 
Bound to body or infinite in open space & Brain Matter or other dead
physical state \\ 
Unique & Made of identical particles \\ 
Classical-\TEXTsymbol{>} No Superposition & Quantum-\TEXTsymbol{>} Undergoes
Superposition \\ 
Probabilistic Reduction Influenced by possible free will & Deterministic \\ 
Retrocausal influence and extended in time? & Causal evolution and localized
in time. \\ 
Non-linear & Linear \\ 
Non-covariant due to Conscious Reduction (3D+1) & Quantum General Covariant
4D(Bare) \\ 
one world on reduction & Many world evolution (Bare) \\ 
Reacts & Passive \\ 
Turn on and off between bareness and experience & Neural firing always there
\\ 
Seeks Harmonic Relationship Structures & Is Harmonic Relationship Structure
\\ 
One\ per Harmonic Relationship Structure & Multiple Parts \\ 
$B=$ $\{z_{i,x},Oi,x,C_{x}\}$ & $\left\vert \Psi (T)\right\rangle $ \\ 
$C_{x}$ map, unknown & Known Standard Model Fields%
\end{tabular}%
$

Bare conscious content is contained in the atoms and molecules that evolve
through the laws of quantum evolution. This is made of a very large number
of quantum particles put together by quantum entanglement and interaction.
In contrast, the bare conscious being is holistic, that is, undividable and
extended. This is because there is only one identity contained in it, and
having multiple parts implies otherwise. It is an extended object that
engulfs the brain areas that produce the conscious content. It is a
non-local object, but unlike the atoms and molecules, quantum entanglement
is not required to bind it, as it undividable.

The matter that produces the bare conscious content is made of atoms and
molecules, so it undergoes superposition. But the bare conscious being
cannot have this superposition; we all would experience multiple realities.
That is, the bare conscious being is purely a classical object. This will
help resolve the quantum measurement problem. That is when the superposition
in bare conscious content encounters the bare conscious being; it gets
reduced to one reality. This is consistent with the Copenhagen
interpretation.

The bare conscious content experiences the bare conscious content on
coupling according to a fundamental conversion function that most probably
depends on the anthropic principle and evolution. The bare conscious content
is purely an electrical and chemical signal due to the neural firing in the
human brain. When it encounters the bare conscious being, it is converted
into conscious experience, after undergoing reduction. Only after this, the
neural firing pattern becomes colorful, emotional, and spatial information.

To completely characterize the being, we need more fields, that helps
convert the classical field values into perceptual content, and also
characterize its description. Let me call this field as the $C_{x}$ fields,
yet to be discovered as par to unified physics, which we further discuss
later.

When the bare conscious observer and bare conscious encounter each other,
the bareness is lost. They both become the conscious observer and conscious
being. During sleep, coma, or sedation, this connection is lost. The
conscious being becomes the bare conscious being. That is, the human brain
can couple and decouple the two. This will be discussed using the concept of
conscious threshold later.

The conscious being reacts back to the consciousness content. When the
reduction happens, the conscious observer exercises its free will. The
reduction depends on the wavefunction probabilistically. In the case of
macroscopic objects, for which the conscious content generator is highly
peaked around various classical values, the reduction appears continuous and
deterministic. But at critical points of decision making, the branching
happens, into alternative possible classical histories. This reduction is
randomized by thermal and quantum probability. The later if there is genuine
free will, has to be influenced by the reaction of the conscious being to
the conscious information.

\begin{principle}
Hypothesis 6.1: In terms of the postulate I.2 of quantum gravity framework 5
and postulate II.2 in this paper, the conscious content undergoes reduction
to create mixed states at rate given by the decoherence constants $c_{d}$,
and the conscious being converts the mixed state into pure states at rate $%
\beta _{R}$.
\end{principle}

This proposal will be discussed further in the future updates.

The bare conscious content is purely determined by the evolution of the
wavefunction of Matter by the Hamiltonian, and it is generally Covariant.
But as conscious beings, we only observe a 3D universe. This is consistent
with the assumption that conscious observation by the being reduces the
general covariant quantum evolution into a non-covariant 3D state, evolving
according to the sense of time of the conscious observer. When the bare
conscious content evolves into many worlds universe of Everett, and when it
is observed by the being, it gets projected into 3D information for one
universe in which the conscious being is present.

The bare conscious being lives in conscious capable systems. This is what I
consider the harmonic relationship structures, such as the human brain. The
bare conscious being gets trapped in it, and possibly only when the
structure is destroyed does it come to an end, or it also gets destroyed.
There is only one bare conscious being as part of each physical living thing.

For the possibility of free will, we can infer two important things about
the classicum and quantum. Free may require retrocausal influence during
collapse. This process happens between the onset of readiness potential and
consciousness of intent. This suggests that the classicum potentially could
act on these processes during the entire time interval and create evolution
of the wave function to maintain quantum state continuity during reduction,
and link the beginning of the readiness potential and current intent
retrocausally, and link task awareness to future intent procausally. The
quantum evolution of the quantum systems evolves forward, and the evolution
is determined by the quantum state at each instant, which is localized in
time.

Classicum may involve retrocausal influence and extended in time. Classicum
is possibly extended in time between the initiation of readiness potential
and the consciousness of intent, performing retrocausal reductions. \
Irreversible and cumulative sequence of state information obtained by
non-linear action on a quantum state at specific instants.

Quantum is involves causal evolution and localized in time. Quantum is
defined at each instant extended in space; it evolves causally forward in
time, also retrocausally in the negative direction, if necessary, from the
instant of reduction. It involves continuous and linear evolution of the
quantum state through a Hamiltonian operator.

Classicum is an operational entity that possesses the properties of
identity, observation, and reaction with respect to quantum information,
which is conscious when coupled to an appropriate dynamic physical field
extended in space and time. When unconscious, it does spontaneous reduction,
and when it is conscious, it does conscious reduction.

In the energy scale quantum phenomena occurs in low energies, while the
classical features appears in the high energy scale.

\FRAME{ftbpF}{6.5008in}{3.4471in}{0pt}{}{}{energyscale.png}{\special%
{language "Scientific Word";type "GRAPHIC";maintain-aspect-ratio
TRUE;display "USEDEF";valid_file "F";width 6.5008in;height 3.4471in;depth
0pt;original-width 6.4999in;original-height 3.4333in;cropleft "0";croptop
"1";cropright "1";cropbottom "0";filename
'GR5Images/energyscale.png';file-properties "XNPEU";}}

Classicum is possibly purely a sub-Planck object as it reduces the most
classical of all fields, the gravitational field, as pointed out in the
quantum gravity framework 5, \cite{MYP5B1.3}, section 3.3. This sub-Planck
entity is the conscious being, which couples to the longitudinal part of the
gravitational and EM fields to experience the conscious content.

Let me define C-link $L$ as a connection between the conscious being and
conscious content. The C-link could be a map that maps information in the
conscious content to the conscious experience of the conscious being. Or it
could be a field that couples the field variations of the conscious content
to the experience by the conscious being. To discover the nature of it
requires experimental and theoretical investigation.

$L$ is a map between the neural data $S$ to Conscious field $C:$%
\[
L:S->C 
\]

C-Link could be mostly made of the longitudinal part of the gravitational
and electromagnetic fields (and possibly other unknown fields), because of
their potential relation to quantum reduction. In Part A of this paper, I
have explained that the gravitational field may be best suited to act as
mediator of the quantum reduction because of its persistent classical nature
in all known environments except may be singularities. Also, the
longitudinal fields of both gravitational and electromagnetic fields are
indivisible as they continue throughout the neural network as fields, unlike
the transverse quanta, which are divisible as discrete units. Also, the
gravitational field is non-linear when it comes to quantization. This means
the quantum state of the gravitational field exists as a single continuous
entity, for example, the spin network formulation in loop quantum gravity.
The stable neural matter can act as a scaffold in which the longitudinal
quantum fields of gravitational and electromagnetic field embeds itself. The
varying fields due to interaction are mediated by transverse photons, and
particles act as the conscious content measured by the longitudinal quantum
field. So, it clearly appears that the classicum hosting the conscious being
is intimately connected to the longitudinal fields and the conscious content
to the transverse fields.

In the section \ref{retro}, based on the new reduction framework in \cite%
{MYP5B1.3}, I discussed that both discrete and continous model can be used
to understand reduction. Please see figure \ref{reductions}. Each quantum
control variables may undergo discrete reduction, helping retrocausal
influence. There, I also introduced the quantum entanglement domain, which
will be further discussed next. Let's now assume that the parts of the
neural network in mediating conscious creates an self decohering
entanglement domain (SDED) that undergoes continuous reduction as a whole
due to numerous reductions. Then, we can use the quantum diffusion theory to
characterize the being and the content as a toy model. 
\begin{equation}
d\left\vert \Psi (T)\right\rangle =iH\left\vert \Psi (T)\right\rangle
dT-\dsum\limits_{i,x}O_{i,x}^{\dagger }O_{i,x}\left\vert \Psi
(T)\right\rangle dT+\dsum_{i,x}O_{i,x}dz_{x}^{i}\left\vert \Psi
(T)\right\rangle  \label{QDE}
\end{equation}%
where $D_{i,x}=O_{i,x}-<O_{i,x}>$, $O_{i,x}$ are observables, discussed in 
\cite{MYP5A}$.$ The quantum state $\left\vert \Psi (T)\right\rangle $ is the
bare conscious content. It is described by quantum fields that are peaked
about classical values. That is, it is semi-classical quantum fields. The
bare conscious observer is made of the Gaussian random fields $z$, and the
classical expectation value of the fields, and any other necessary fields.
The observable operators $O_{i,x}$ are to determined. The summation is done
over all points $x$ of each instant of the rest frame foliation, and over
all the observables (to be determined). The summation is integration with
proper measure over the spatial manifolds defined by the rest frame
foliation.

The $O$ can have two different parts: Physical Reduction, Conscious
Reduction.

The physical reduction part of \ $O$ comes from the field terms in Quantum
Gravity Framework 5 postulate $3$. This reduction rate is determined by the
decoherence constants $c_{d}$ in that are part of $O$ that are not shown.

The second possible contribution for $O$ is the map that projects from space
of sensory data $S_{Q,i,x}$ as expressed in terms of physical fields in the
neural network to corresponding brain state $S$ made of \ conscious state $C$
made of both qualia $Q$ and actions of motor neurons $A$. This occurs at the
rate of $\beta _{R},$ the rate the mixed states are converted into pure
states for observation. Mathematical expression of this projected is as
follows:

\[
O_{C,x}=|S_{C,i,x}><S_{C,i,x}| 
\]%
where $C$ are internal variables over range of consciousness state, replaces 
$i$ in \ref{QDE}. In this the $C$ is part of the eigenstates of C-link $L$
that couples, sensory data and Qualia. This choice of $O$ projects $%
\left\vert \Psi (T)\right\rangle $ into one of the Qualia states. This is
consistent with the Copenhagen interpretation. The wavefunction of the
system you observe, collapses into something compatible with what you
perceive.

Free will experiments suggest that retrocausal influence is required for
existence of the free will. This means the conscious projections has to
occur using the bidirectional evolution described in part A. In the rest
frame foliation at each instant of conscious reduction, we have 3 surface.
Between the 3 surfaces with have bidirectional evolution sandwiched using
the theory developed in the part A in section 2.2 in v2 and v3 of \cite%
{MYP5A} or section 1.3 in v1 of \cite{MYP5A}.

Now including the two possible contribution to characterize the conscious
being we need the $B$ Complex, $B=$ $\{z_{i,x},Oi,C\}_{\iota }$. To during
observation $Oi,x$ is converted to it expectation values. The $z_{i,x}$
stand for free will, $Oi,x$ stand for observation, $C_{x}$ stand for
Conscious information. All this three is unique to each conscious being,
with unique identity $\iota $. $\iota $ determines the spatial domain in
which the fields are to restricted for a unique identity.

During conscious observation, the classical expectation values are exercised
as the conscious content according to the conversion formula. The $z_{i,x}$
fields should have been influenced by the reactions of the conscious
observer. Also, $O_{i,x}$ is influenced by what is being observed.

\begin{principle}
Hypothesis 6.2: The $O_{i,x}$ are determined by C field.
\end{principle}

This is consistent with the process of quantum measurement, because the
observer chooses what they measure. In consciousness, the being chooses what
it measures. The precise dependence of $O_{i,x}$ on $C$ is to be researched.
We can say it depends on what the being is experiencing at each instant of
the rest frame foliation. The C-link are determined by the conversion map
from neural firing data and the perceptual data experienced by the being.

The two parts of the quantum diffusion equation on the right side equation %
\ref{QDE} influence the evolution of the quantum state in a stochastic
manner. If we assume the evolution of matter is determined by the quantum
diffusion form, then the universe is filled with bare conscious beings,
exercising reduction in matter. This can happen both in a conscious
supporting system, like a biological neural network, or outside it. If it is
outside the bare conscious being, the $z$'s are purely random. If it is
inside a biological neural network, its $z$'s may have been influenced by
the free choice due to conscious reaction.

So, the full picture of the universe is the matter made of quantum state of
fields $\left\vert \Psi (T)\right\rangle $ \ interacting with the $B$ fields.

\begin{principle}
Definition: B.1 We can coin two terms, the classicum and classica, as an
equivalent of the quantum and quanta, for classical fields.
\end{principle}

The quantum fields are made of quantum particles or the quanta, and
similarly, we can consider that the classical part of the universe is made
of many classica of classical fields or the classica. Each classicum have
the properties described for the bare conscious content: 1) holistic, 2)
extended, 3)Non-local, and 4)Concentrates in complex structures. If the
structured is vaporized then the classicum spreads into infinite free space,
until is absorbed into another finite complex solid structure. This means
the open space is filled with many classica.

We will further discuss classica and classicum later. The classicum are
characterized by the C-link~$L$, or in general the $B\ $Complex. Lets
connect the classicum and the conscious being through the following
hypotheses.

\begin{principle}
Hypothesis 6.3: The classicum holds the conscious being and its properties.
\end{principle}

This can be split into multiple parts as in the following hypothesis:

\begin{principle}
Hypothesis 6.3split: a) Classicum reduces the quantum superpositions on
contact b) Classicum is the observer c)\ Classicum holds the identity.
\end{principle}

The bare conscious being is converted to conscious as follows:

\begin{principle}
Hypothesis 6.4: The C-link $L$ couples the coherent energy interaction in
neural region to the conscious content on the classicum (conscious being)
b)\ The region of the classicum coupled to neural regions that satisfies
conscious threshold is conscious. c) The interaction binding of two parts of
neural region extends the conscious region of classicum.
\end{principle}

The hypotheses 6.2 to 6.4 describe some of the properties of the classicum.

The role of interaction in binding conscious content was discussed before.
It involves quantum entanglement and energy exchange. One needs to relate
this binding to the non-locality of the classicum (the bare conscious
being). The above hypothesis does this. The interaction binding had to
extend the extends the conscious regions of classicum (the bare conscious
being), and the C-link couple the coherent energy interaction to classicum.
The $z$ fields helping making reaction on the neural network that generates
the interaction by reduction. In the process, the classicum, which was the
bare conscious being, gets converted into a conscious being. So through the
human brain and body, wherever the interaction binding operation the
classicum is extended.

\FRAME{ftbpFU}{6.4273in}{3.7671in}{0pt}{\Qcb{Coupling between the conscius
being (classicum) and the conscious content.}}{\Qlb{ccoupling}}{%
c-coupling.png}{\special{language "Scientific Word";type
"GRAPHIC";maintain-aspect-ratio TRUE;display "USEDEF";valid_file "F";width
6.4273in;height 3.7671in;depth 0pt;original-width 6.0001in;original-height
3.4999in;cropleft "0";croptop "1";cropright "1";cropbottom "0";filename
'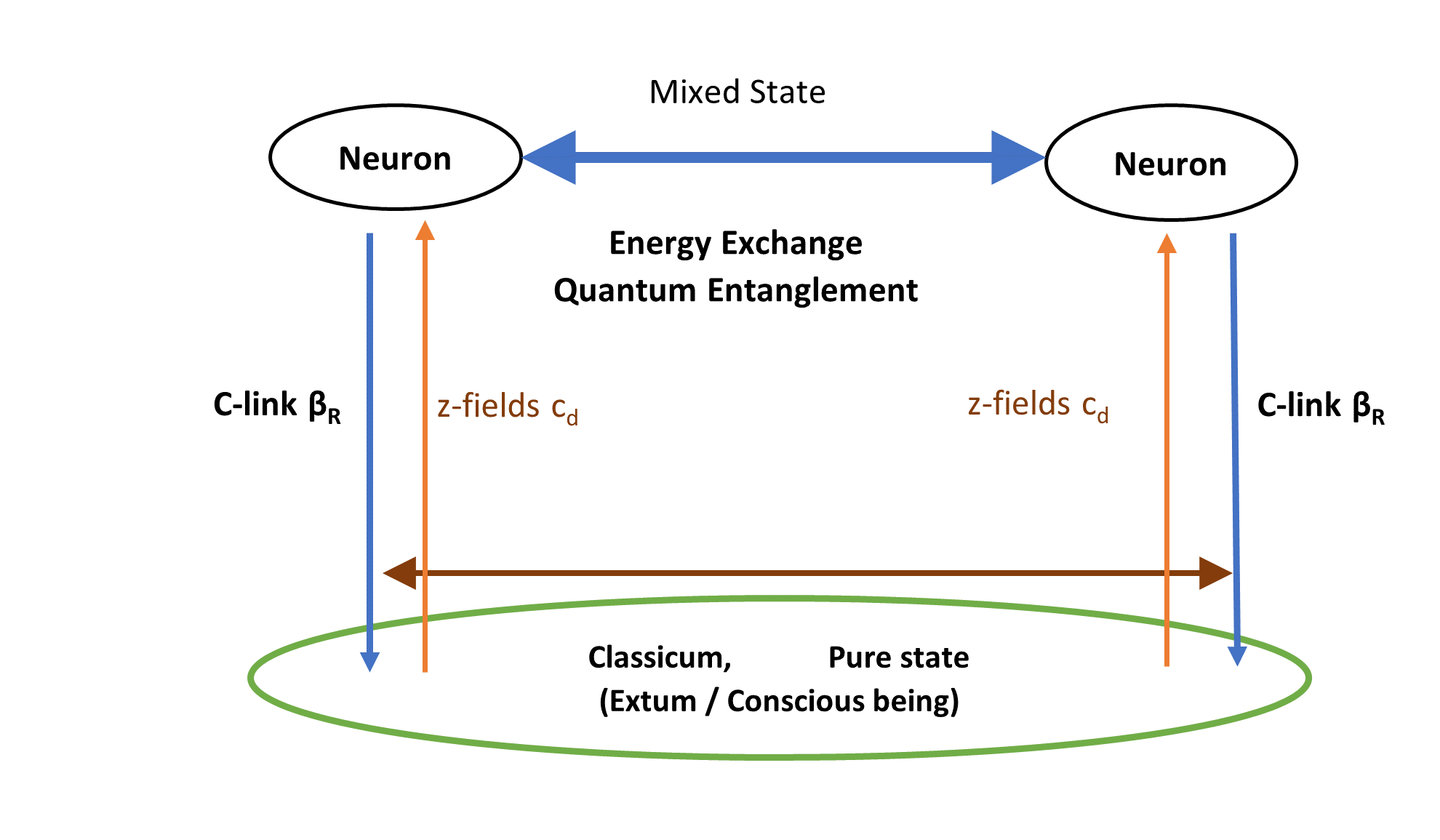';file-properties "XNPEU";}}

\subsection{Birth, Death and the classicum state\label{birdea}}

Now we can contemplate birth and death. I discussed that an classicum
concentrates in an Harmonic relational structures (HRS). Basically, after
death, the classicum associated with the person, expands to include the more
entanglement domain of the surrounding matter with strong HRS characters,
and keeps expanding until it encounters a discontinuity in entanglement. If
it encounters a strong HRS, it merges with it. This could be the emergence
of conscious into a foetus. Also the classicums that keeps expanding into
free space does the spontaneous reduction of the quantum matter, as given in
postulate I.2 of the quantum gravity framework 5. \cite{MYP5A}.

\begin{principle}
Hypothesis 7: a) On birth classicum becomes part of a harmonic relational
structure capable of consciousness to have conscious experience, b) On death
the classicum becomes part of the free space that does the spontaneous
unconscious reduction of quantum fields, c) Classicum exists in the on or
off state depending on whether it is coupled to the conscious content by
C-link or not respectively.
\end{principle}

The classicum is in the off state when it is not part of the conscious,
capable HRS. When it is part of the conscious capable HRS, and the conscious
threshold is met, it is in the ON state, converting the conscious content of
the HRS into conscious experience. So after death, sleep, coma, or sedation,
the classicum goes into the off state.

\subsection{Conscious Substratum}

Let me introduce the concept of conscious substratum.

\begin{principle}
Definition B.2: The conscious substratum is the part of a conscious physical
entity that holds the conscious being (classicum) of the conscious entity
apart from the physical systems that process and creates the conscious
content.
\end{principle}

Here by conscious entity I\ mean the physical entity, such as the human
brain, if that is the one which holds the conscious entity. The conscious
substratum is the physical part of this entity, that holds the conscious
being only, apart from the systems that processes the content, and creates
it. This should be the one that contains the $B_{i,x}$ complex described
before. Once this content is created on combining with the being becomes the
conscious content. Here I\ am making an important assumption:

\begin{principle}
Hypothesis 8.1: The conscious being is contained in the physical embodiment
of the conscious entity.
\end{principle}

I\ am assuming in contrast to the mind body duality argument, the body of
the human being also has the physical aspects that completely characterizes
the conscious being.

Now as the conscious substratum contains the conscious being it also the
three aspects of it discussed before:\ the observer, identity and free will.

The substratum can be possibly considered as the composite quantum state
made of particles and gauge fields, and unknown fields yet to be discovered,
that make up a conscious entity, apart from the quantum information that is
associated with the conscious content.

The part of the conscious entity that links its presence in a particular
body is the body itself. But the body itself including the brain, keeps
evolving and changing at levels right from its conception: embryo genesis,
growing up, formation of neural connections, development of organs, changing
of DNA through epigenetics, mutations, replacement of cells through mitosis,
aging process, etc. In the brain, the firing patterns change continuously
depending on sleep, fight or flight response, awake, coma, vegetative state,
etc. This means the conscious substratum is the body itself apart from these
gradual physical changes, or changes in the neural firing pattern. It is
often thought that the conscious entity in totality (content +substratum) is
the electrical firing pattern in the human brain. But we know the people who
are thought of as brain dead, a state with very small brain electrical
activity, have recovered after a long time. This indicates that the human
body itself plausibly is the substratum that experiences conscious content.
The conscious being is tied to the body and gradual changes in it do not
affect this. This is contrary to what is believed, that a person's soul, the
colloquial term for the conscious substratum we are discussing, is different
from the body, and it leaves the body on death.

Now the question is whether the conscious substratum is restricted to the
brain or extends to the whole body. It is difficult to answer this question.
But from the idea of interaction binding, we infer that the conscious entity
is the set of particles that are entangled with each other through
interaction. This means a person's entire body is involved. This is
consistent with the body-mind as one complete system.

If we subtract the part of the brain activity that produces the conscious
content, the conscious substratum is simply the bundled matter that makes up
the human body, made of fundamental particles. This means any bulk piece of
matter could act as the conscious substrate and when its atoms and molecules
are made to interact with appropriate activity based on the consciousness
threshold, it may experience conscious content. In other words, this line of
inquiry suggests that every object such as a stone, chair, TV, or computer
can become a conscious substrate and the interactions within it can lead to
infinitesimal conscious experiences.

What kind of objects can act as conscious substratum? For this, we consider
the following hypothesis.

\begin{principle}
Hypothesis 8.2: Any object made of quantum mechanically entangled particles,
and massive enough to self-decohere can act as a conscious substratum.
\end{principle}

The massive enough part of the substratum comes with the necessity of the
interaction binding, which requires quantum reduction. The continuous
quantum reduction is the free will action of the object as I have proposed
in this paper. For the conscious substratum to experience consciousness it
has to have the proper chemical bonds in the atoms and molecules inside it
and interactions among them, and must satisfy the consciousness threshold
that we discussed in the interaction binding.

While there is no problem associating the differences in the quantum brain
state with the different conscious content experienced, we can ask many
questions. Why would one conscious substratum at the fundamental level be
associated with different quantum states? Would not these different quantum
states create different conscious beings? So we need to associate the
conscious being with the overall quantum state of a conscious entity, apart
from the different specific configurations that give rise to different
conscious content experienced. The quantum state of an object is a
mathematical construct, that puts together the elementary wavefunctions of
identical particles. So to identify the quantum state of an object with a
conscious substratum we need to assume this quantum state has more features
that we don't know in current physics that can uniquely identify the
conscious being. This is why used the word `host' rather than assume that
the known matter or fields are associated with the conscious substratum.

\subsection{Entanglement Domains}

To discuss conscious substratum we need to describe it in an objective form.
For this, I use the concept of `entanglement domains'.

\begin{principle}
Definition B.3: Entanglement domains are regions in space and time in which
atoms and molecules are strongly entangled through bonding.
\end{principle}

I\ will not define here what it means to be strongly or weakly entangled,
but leave it to experimental investigation to come up with appropriate
measures and thresholds. They can assumed to have the following features:

\begin{enumerate}
\item The particles of entanglement domains are strongly entangled

\item They have boundaries, like the boundaries of magnetic domains in a
metal.

\item They are weakly entangled at the boundaries.

\item The particles of the entanglement domain undergo reduction together.

\item The matter within an entanglement domain can act as a conscious
substratum for any consciousness-capable harmonic relational structure like
the human brain-body system.
\end{enumerate}

The entanglement domains that is large enough to undergo self decoherence we
call as the self decohering entanglement domain (SDED).An appropriate
interaction activity can give rise to conscious experience in it. It can
exist in two states: Off states and On state of conscious experience.
Activities above the consciousness threshold are required to make it go into
a consciously aware state.

As I have discussed in QGRF 5 \cite{MYP5B1.3} in section 2.3, depending on
whether the gravitational field or electric field dominates the calculation
of rest frame foliation, the frame in which the reduction happens is the
rest frame of Earth, or the rest frame of each entanglement domain. I
strongly believe the frame in which reduction happens is that each
entanglement domain, with electric fields being dominant in determining the
rest frame foliation. Consistent with special relativity, the physical time
in which physical phenomena happens including the quantum reduction, is the
rest frame of the entanglement domain.

To associate conscious entities with quantum states, we need to describe the
universe as follows. The state of the universe splits into various pieces of
entanglement domains. Entanglement domains could fill the entire universe.
They might contain the human-body system, a colony of microbes, a mammal, a
stone, a group of molecules in the air or in the vacuum of space that is
entangled together momentarily, etc. The last group keeps undergoing dynamic
configuration change.

We can consider any solid object with an entanglement domain to support a
conscious substratum. But we could also possibly include the portion of
molecules of matter such as a gas or the vacuum of space filled with an
extreme light density of particles as a conscious substratum. As the
molecules of the gas interact they can entangle together and can become
large enough to decohere by itself. The only problem is this set of
molecules keeps changing rapidly, and it violates the slow change condition
of the conscious substratum. So it is difficult to say whether this could
constitute a conscious substratum.

\subsection{Thresholds\label{thr}}

To understand the relation between conscious substratum and entanglement
domain we need what I refer to as the observer threshold.

\begin{principle}
Definition B.4: The observer threshold is the necessary condition for an
entanglement domain to become a conscious substratum.
\end{principle}

Even though I only mention the observation part of the conscious being, I
assume that when this threshold is satisfied, the creation of a conscious
being, with properties of observation, identity, and free will, occurs.
Important here is that the observer need not be the same as the fully
conscious being, like a human mind, for that we need the conscious threshold
discussed in the next section to be satisfied. In a conscious substratum,
the consciousness minimally exists only with is observer threshold is
satisfied. Conscious threshold needs to reach full awareness. Between the
minimal conscious and the full conscious, there are various levels of
activity. In minimal consciousness, sensory awareness may be there, but no
memory. In full consciousness, the memory and sensory awareness are present.

Let me discuss how to characterize this observer threshold. The observer
threshold could be anything such as those defined in the following
subdomains:

\begin{principle}
Hypothesis 9.1a: Observer threshold 1: An object made for a group of
entangled particles with more than a critical mass for self-decoherence
(SDED) constitutes a conscious substratum.
\end{principle}

\begin{principle}
Hypothesis 9.1b: Observer threshold 2: A minimal existence of electrical
activity with an appropriate amount of coherence in the human brain to
maintain the conscious substratum.
\end{principle}

\begin{principle}
Hypothesis 9.1c: Observer threshold 3: An object made for a group of
entangled particles with sufficient capability to form memory constitutes a
conscious substratum.
\end{principle}

These three are sufficient conditions for consciousness as inferred from
observation 1 (earlier in the paper), which requires memory formation.
Usually if Hypothesis 9.1c is satisfied then 9.1a and 9.1b are also
satisfied for living things made of earthly organic matter. But this is not
necessary, because the observer threshold is satisfied, but there it will
not have consciousness. For example, a person in permanent coma, or dead
body that may contain an being without consciousness.

We also add two more.

\begin{principle}
Hypothesis 9.1d: Observer threshold 4: A minimal existence of metabolic
activity in the human brain or the body.
\end{principle}

\begin{principle}
Hypothesis 9.1e: Observer threshold 5: An existence of C-fields and
properties that are not yet accounted for by the standard quantum mechanics.
\end{principle}

For a body that constitutes an entanglement domain to become a conscious
substratum, these conditions may have to be satisfied. When the conscious
threshold is also satisfied we have the body to be conscious. These observer
conditions are necessary, but not automatically satisfied by any
entanglement domain. If the entanglement domain conditions are not
necessary, then when the body dies, the observer continues to exist and it
may combine with another body. At this moment we cannot say anything about
the observer conditions and further research is necessary. But I will
propose a simple assumption:

\begin{principle}
Hypothesis 9.2: Simple Observer Assumption - Every entanglement domain that
has minimal mass for self-decoherence is a conscious substratum.
\end{principle}

This is quite possible because of simplicity. In this picture, a conscious
substratum is associated with an entanglement domain, that can
self-decohere, which could be a stone, a creature, a skeleton, etc. In this
picture, if a person dies the conscious substratum remains with the body.
When the body decomposes and mixes with the surroundings, the conscious
substratum is restricted to the skeleton. When the skeleton decomposes, it
becomes part of the surrounding matter and expands to include the minimal
mass necessary for self-decoherence.

There are other possibilities for the conscious substratum:

\begin{principle}
Hypothesis 9.2alt: a) The conscious substrata are more fundamental than the
entanglement domain, and b) the conscious substratum encompasses an
entanglement domain always.
\end{principle}

This possibility is consistent with hypothesis 9.2. In this, an observer
drifts from body to body, body to object, liberated from an object, and
moves along in an air or vacuum of space until it encounters an object of
sufficient mass for to satisfy the observer threshold or become part of the
object that grows to become a conscious object that encompasses a brain. The
consciousness happens only if the conscious threshold is breached, such as
becoming part of a brain in which this can happen.

\begin{principle}
Hypothesis 9.2alt2: The conscious substratum is linked to a human brain if
the observer threshold is satisfied only.
\end{principle}

This was consistent with hypothesis 9.2, but in this the link between the
conscious substratum and the observer holds only for living things such as
mammals or humans. This is consistent with the idea of spirit in religious
interpretations, and also common belief. In this the observer is established
with the observer threshold is breached in the brain activity, such as in an
infant is being born and the brain becomes active. When the brain dies the
link goes away.

\begin{principle}
Hypothesis 9.2alt3: Each conscious substratum is linked to a human brain of
certain dynamic and static harmonic relational structure that defines the
conscious being.
\end{principle}

In this possibility each conscious substratum is identified by one or more
of various factors such as 1) a specific part of the neural firing pattern
2) a specific pattern in the DNA, and 3) a specific pattern in the neural
connections. In case whenever this pattern emerges in the universe a
person's conscious being comes into existence, and disappears when the
pattern dies, due to the death of the body.

All these alternative possibilities could be debated, researched, and
developed further. One of these could possibly define our situation after
death. These ideas could have implications for how we bury a body.

\subsection{Conscious Thresholds}

Now we discuss the conscious threshold in more detail. Lets combine ideas in
various places before in Hypotheses 5.5, 9.1 and 9.2 into clear list of
thresholds.

\begin{principle}
Hypothesis 10: We can split the conscious threshold into five parts as
listed below:
\end{principle}

1. Coherence threshold: The proper level of coherence that is needed to
experience consciousness.

2. Representation threshold: The proper extent of representation that the
information fed into a neural structure needs to take for it to be perceived
to be a certain qualia.

3:\ Memory threshold:\ The sufficient capability of the brain to form memory
of the conscious content.

4. Integration threshold: The extent of integration of information needed to
create a complete perception.

5. Control threshold: The proper level of superposition of alternative
possibilities of actions and decoherence that determines the future course
of action.

6. Observer threshold: The proper level of conditions that is necessary to
have a being.

Each one of these represents a physically different attribute. The coherence
threshold was discussed in section \ref{cohsyn}. We need a proper level of
coherence. Too much coherence leads to a loss of consciousness, for example,
in sleep. Also, too little coherence leads to less consciousness.

Representation of information in the human brain is specified by the
biological network that controls neural firing data into qualia. Like I
discussed before, this is specified by the zeroth law of physics, which is
related to the anthropic principle. In this universe, we have a specific
space-time configuration of data that gives rise to certain qualia in the
conscious being. This needs to be found experimentally. It could be a
combination of geometric and biochemical structures specific to each qualia.
Finding the representation threshold is basically the hard problem of
consciousness.

The Integration threshold is the extent of neural data in the brain that has
to be integrated to be perceived as conscious data. The integration was
discussed in section \ref{bininf}. The data in the brain structures, all the
way from the reticular formation to limbic structures to higher cortex
structures, coherently integrates to create the perceived information. The
question is how minimal an extent is required to elicit a conscious
experience.

The control threshold differentiates between a purely classical
deterministic system, a random system, and a purely random system. In a
deterministic system, the future is determined by its past state. The best
example is the classical computer (excluding random number generating
devices). A pure random system has no predictability, like a roll of dice. A
quantum system's future course of events is predicted by probability based
on the wave function's squared value. In the human brain, at the synapses,
we have quantum superposition, which controls neurotransmitter release. So,
as a whole, the brain's future state is predicted by the quantum
superposition of the joint system of control variables discussed before. As
I have proposed, this reduction is equivalent to free will action.

The observer threshold was discussed before and is purely a hypothetical
concept for now. It requires experimental proof, and I am not sure how it
can be studied. With the existence of conscious substrata, there is no
spirit/soul in the system to experience reality.

Together, all these describe the conscious threshold necessary to elucidate
a living entity that experience consciousness.

\subsection{Conscious in Unified Physics}

Let us look at how consciousness plays out in the hypothetical unified
physics we have not yet discovered. The activity in the cerebrum that leads
to consciousness is as far as we know, is mostly described by biochemical
reactions. We cannot straightforwardly relate this 1-to-1 to the conscious
experience because of random noise. Is there a physical component of the
matter that 1 to 1 makes to conscious experience? Most matter in the human
brain itself is mostly static and is coupled to the local gravitation fields
that vary very mildly according to the local densities. When the neural
firing happens in a synchronized fashion, they create variations in the
gravitational field. But these variations are chaotic as described by
Poisson equation for gravitational field.

\FRAME{ftbpFU}{5.7873in}{4.4391in}{0pt}{\Qcb{Unified Physics and
Consciousness}}{\Qlb{unicon}}{consciousunifiedfield2.png}{\special{language
"Scientific Word";type "GRAPHIC";maintain-aspect-ratio TRUE;display
"USEDEF";valid_file "F";width 5.7873in;height 4.4391in;depth
0pt;original-width 9.3538in;original-height 7.1693in;cropleft "0";croptop
"1";cropright "1";cropbottom "0";filename
'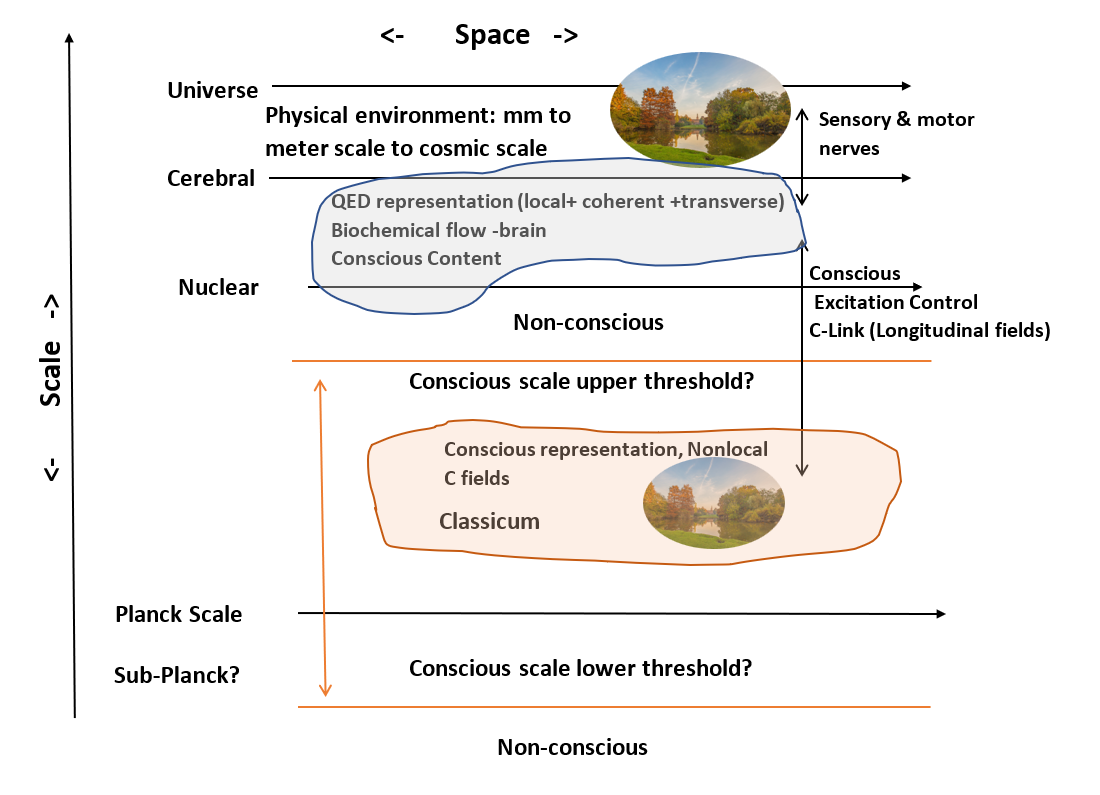';file-properties "XNPEU";}}

We need C-link $L$ which are weakly coupled to matter variations that picks
up the consciousness information 1 to 1 to the conscious entity
hypothetically made of the C fields and Classicum. To understand this, think
of a musical drum knocked by a hammer. The musical instrument vibrates
harmonically. The harmonic vibrations can 1-to-1 map to the conscious
experience in the conscious entity. The C-link $L$ and C-Field C that maps
to conscious experience could possibly have physics between the atomic to
Planck scales to Sub-Planck Scale. The coherent firings of matter certainly
create harmonic mechanical vibrations in matter. These firing could couple
to the C-link to transfer the conscious information to the B complex $B=$ $%
\{z_{i,x},Oi,x,C\}_{\iota }$.

One of the important things to note is that the C-link $L$ and C-field need
not be a new fundamental in nature but emergent. Assume there are more
fundamental fields that show up at higher energy scales than are known (or
even in the lower energy scales, possibly relating to the fact that our
consciousness only have about mm resolution). Together, they can create an
emergent quantum field that maps 1-to-1 to the perceived conscious
information. From the point of a unified field theory of everything, they
are excitations of the unified field due to electro-biochemical neural
activity.

The picture \ref{unicon} explains what to be expected in the unified
physics. At the Planck scale, the quantum fields can be represented by loop
quantum gravity, strings, twistors, or other unknown fields. There is more
to these fields than capturing the conscious experience 1-to-1. Some of the
properties for these physical attributes that capture physical experience is
summarized in the following hypothesis.

\begin{principle}
Hypothesis 11: The consciousness in the unified physics yet to be discovered
is given by coupling of biochemical activity in the brain to physical
features of quantum fields C in the space-time on the classicum that is part
of the region in space-time in which the brain and body lives. Its
characteristics are as follows:
\end{principle}

\begin{enumerate}
\item C field State captures the conscious state 1-to-1: The entire
conscious experience is embedded point by point in the state.

\item C field State is wholistically represented, in that each point has
information that the whole state of the conscious experience.

\item The C field is the same as the observer field or is coupled to the
observer field discussed before.

\item The C field may be an emergent field made of yet to be discovered
fundamental fields.

\item The Extent of the State is restricted to the human body (classicum).

\item The conscious threshold describe the conditions for active coupling
involving C-link $L$ between the electro-bio-chemical coherent activity in
the human brain to the $C$ fields on the conscious substratum, particularly
the memory formation part.
\end{enumerate}

\subsection{Summary}

We can summarize the discussions in this section as a model as shown in
figure (\ref{ConMod}). \ The conscious experience has three parts: The
Conscious superstrata, Conscious Substrata, and the Physical Environment.
The Conscious Superstrata is the brain that is made of physical matter and
fields that creates the physical processes that maps the information about
the surrounding world into specific conscious patterns. The conscious
substrata have the observer fields that uniquely identifies a conscious
entity like human being and the $C$ fields that capture the conscious
experience described in the section \ref{concon}. The physical process in
the brain links the conscious superstrata to the conscious substrata to
create conscious experience. This model is a semantic equivalent of
materialistic discussion given in figure:\ref{SimpModel1}:\ the critical
variables are controlled by conscious substrata, the conscious superstratum
relates to the associative neural network in the brain, and, the sensory
input and motor output are relates the environment.

\FRAME{ftbpFU}{5.5391in}{2.3618in}{0pt}{\Qcb{A Simplified Conscious Model}}{%
\Qlb{ConMod}}{consciousmodel.png}{\special{language "Scientific Word";type
"GRAPHIC";maintain-aspect-ratio TRUE;display "USEDEF";valid_file "F";width
5.5391in;height 2.3618in;depth 0pt;original-width 6.0001in;original-height
2.5425in;cropleft "0";croptop "1";cropright "1";cropbottom "0";filename
'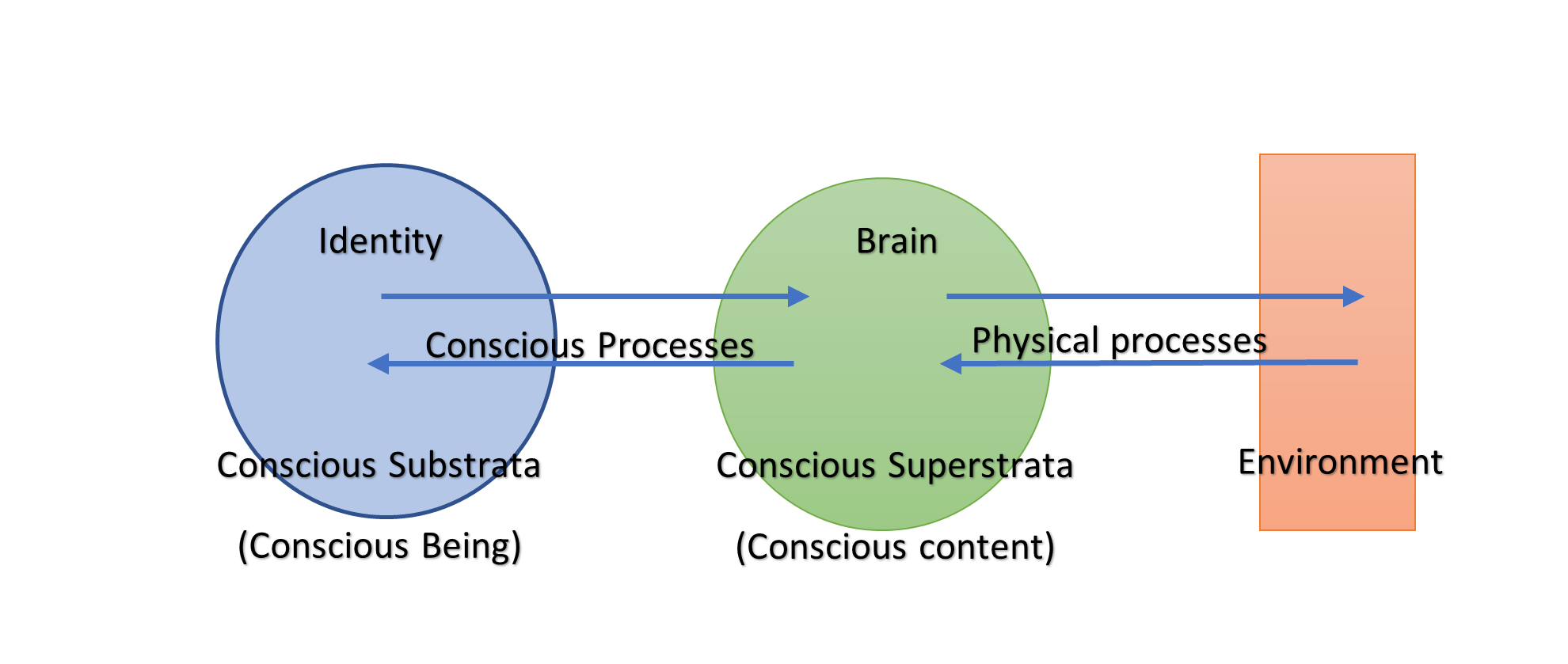';file-properties "XNPEU";}}

From discussions before in section on the conscious substratum, we can the
summarize the properties of observer domains as follows:\ It may be made of
new type of fields that has the following nature:

\begin{enumerate}
\item New state of Matter, other than fermions and bosons.

\item It observes matter.

\item It is coupled to the B fields.

\item Extended in Nature.

\item Exhibits classical superpositions.

\item Tend to localize around harmonic relational structures.

\item Promotes coherence in matter:\ This mean it results creating harmonic
relational structures, leading to life and consciousness.

\item They repel each other when they accumulate coherence.

\item It evolves along block universe from past to future, through present.

\item Undergoes reduction and deterministic evolution as a part of block
universe evolution is equivalent of freewill action.

\item Permeates all observable universe.

\item It is activated to be conscious, when the conscious threshold is
satisfied.
\end{enumerate}

The ideas 5, 6 and 7 define how these fields create observer domain by
localizing within the conscious creature it is part of.

Progress on this subject of identifying the nature of the observer
domain/conscious substratum requires more experimental and theoretical
research on this subject.

\section{Consciousness:Important Features}

\subsection{Levels of Consciousness}

To clearly understand the existence of conscious experience in living things
I can need to identify four levels of behavior:

\begin{itemize}
\item Intelligent: An entity is intelligent if it can respond to
environmental triggers. It is only intelligent and not conscious if it is
not satisfying the observer threshold.

\item Sentient: An entity is minimally intelligent and can also satisfy the
conscious threshold, but has no knowledge of itself but only some sensations.

\item Sapient: An entity is intelligent and can also satisfy the observer
threshold, has knowledge of itself but has minimal sensations.

\item Conscious: An entity is intelligent, satisfies the consciousness
threshold, and has knowledge and sensations.
\end{itemize}

To quantitatively differentiate between the systems that are only
intelligent and not the other three we need the quantitative measures to
calculate the consciousness threshold as specified before. When we are
asleep we are sentient, because we respond to various triggers. Most other
animal life other than humans are certainly sentient but we don't know
whether they are conscious.

Usually in air molecules or metals, during interaction, we have random
collisions, and over the course of time, the molecules superimpose in
incoherent superposition, before collapsing. In this interaction, there is
quite a limited amount of coherent classical or quantum information, and we
can assume that the consciousness threshold is not satisfied. The conscious
experience generated can be canceled out due to thermodynamic randomness,
both classically and quantum-mechanically.

The coherence/synchrony through selection, attention, amplification, and
helps in increasing the interaction binding involving the neural information
that is being consciously attended to. Previously I described various
mechanisms by which coherence/synchronization happens. Pyramidal neurons are
considered as important for conscious perception by many theories listed
before. The pyramidal neurons are the largest set of neurons in the cortical
layers and are placed vertically. They lay between the top layers that have
feedback circuits and the lower layers that are feedforward circuits. All
these suggest the pyramidal cells are designed to amplify the interaction
binding involving the neural firing pattern of the object being attended to,
and is involved in conscious binding.

The way the binding and experience happen may depend on the entity. In
protozoa such as paramecia, the conscious experience if it exists, according
to this hypothesis depends on how the chemical interaction between its
organelles, microtubules, cell membrane, and cilia happens. Particularly the
cilia exhibit coherent behavior due to the synchronized movement to propel
itself forward, which may make it sentient. In multicellular organisms such
as plants there is interaction between the contents of a cell and the
interaction is minimal between cells. According to the interaction binding
hypothesis, they may have a minimal amount of shared sentience such as
feelings of surroundings, density, heat, sunlight, etc. But still, to
understand whether they have conscious experience, they need to meet the
consciousness threshold.

Unlike protozoans, in higher animals, as we know, the consciousness, if it
exists, is related to their nervous system. In this, the complex interaction
between molecules is mediated by neural mechanisms, which create dynamically
complex interacting neural systems, that create complex qualia of conscious
experience. There is a complex interaction between computation and conscious
systems influencing the conscious experience and behavior of the beings.

\subsection{Critical Brain}

The processes in the human brain under high consciousness in interaction
with the internal states and external stimuli constantly change the
large-scale neural processing in the brain. Small variations can bring big
variations in the neural network firing and the areas engaged. Brain in
constant change of mode of operations ( \cite{NETMOD1}, and references in
the article). If it is in default network mode (review article \cite%
{DEFNETMOD}) it contemplates itself or will be at rest. In the attention
state it attends something. There are many modes of operation.

With this rapid switching between many modes of operation, it can choose to
focus on various mental activities like sleeping, attending, thinking,
daydreaming, calculating, etc. This can be considered to be engaged through
volitional activities driven by the Quantum Reduction Free will equivalence
principle. But as we have discussed consciousness can have an effect on the
neural firing pattern due to the behavioral and free will equivalence
principles in hypotheses 5.4.a and 5.4.b. Through this, it has to control
the various activities. This can be done if consciousness targets those
neurons critical in changing the modes of operation and directing various
neuronal processing. Consider the higher associative areas. They send
feedback signals to the primary sensory areas. The number of neurons in the
associative areas that are needed is as small as the amount of information
that is being dealt with in an encoded form. So the effects of firing
patterns of such few neurons will be able to direct the neural processing
required to be performed by the brain in the various modes. Also, the higher
associative area contains the most abstract content in consciousness and so
these areas can trigger the volitional modification of neuronal firing by
modifying the physical processes.

This model is called the critical brain theory and is closely related to the
ideas we discussed here \cite{CBRAIN1}, \cite{CBRAIN2}, \cite{CBRAIN3}. The
critical brain asserts that the brain is always in a critical state
transitioning into various states continuously. This critical change can be
controlled by certain regions of the brain. It seems that consciousness
influencing the physical processes as discussed before can control the flow
of these states.

\begin{principle}
Hypothesis 12: The human brain is always in a critical state of transition.
\end{principle}

\subsection{Consciousness and Harmonic Relational Structures}

The relational structure of the human brain has harmonic features that are
essential for consciousness. In the brains of animals and humans, harmonic
activity is extensively present in brain firing patterns, in various forms.
The most popular knowledge is the brain waves in bands such as alpha, beta,
gamma, etc. This relates to what I call the harmonic relational structures. 
\cite{BKVl3}, \cite{MYP4}, and \cite{GUK}. These are relational structures
that exhibit dynamic or structural harmonic features.

From simple to complex living things the harmonic relational structures
increase in complexity and quantity. We can consider this as an increase in
the level of consciousness from simple to complex. The synchronized firing
pattern of the nervous system is present in most living things with some
form of centralized nervous system. Even in protozoan organisms such as
paramecia, we have synchronized activity in the thousands of cilia. These
may be a source of rudimentary consciousness in them. All living cells have
microtubules which under vibrations interact with structures within the
cell. This could mean the living cells each one has a certain minuscule
amount of consciousness.

The extent of consciousness in a harmonic relational structure depends on
how large the structure and how harmonic the interaction activity is. We
need to differentiate between two defining features of harmonic relational
structures.

Static feature of Harmonic relational structures: These correspond to
chemical structures that are arranged harmonically in the chemical bonding.
Examples of this are proteins, DNA, cells, etc.

Dynamic features of Harmonic relational structures: These features are the
dynamically changing structures made of elements such as electric field,
electric current, etc. The interactions in the neural network in the living
brain are the very example of it. The neurons with dendrites and axons are
placed in a uniform pattern in the human brain. These dendrites have
synaptic connections that are connected in a stochastic pattern created by
memory patterns. But the variation of the electric fields and currents are
synchronized harmonically, to create dynamic harmonic activity.

Living things have various combinations of these two features. Plants have
more static features, while animals have dynamic features in their brains.
These dynamic features have heavy elements of interaction activity and are
responsible for consciousness in human beings and possibly in other animal
species with similar processes. Microbes such as the paramecia have dynamic
features associated with their cilial movement.

Later we will discuss measures of relational complexity of harmonic
relational structures due to these two features to help understand the
evolution of life and consciousness.

\subsection{Summary of Life and Evolution}

Life is a harmonic relational structure as I have discussed. They experience
conscious states when they interact with coherent information. Assuming the
survival instinct is primary in all living organisms, it motivates behavior
related to needs and wants. Now over the course of human evolution, the
conscious and computational capabilities have evolved together tightly in
ways to promote survival needs, increase the wants, and achieve the wants.

One of the important proposals, that is connected to the zeroth principle
discussed before, I would like to make is that the harmonic relational
structures have the innate tendency to promote the creation of more complex
harmonic structures, driving evolution. This capability is important because
it helps explain the presence of thousands of complex organic compounds,
that eventually played a role in creating advanced life forms. This has been
discussed as part of scientific relationism \cite{BKVl3}.

To understand the evolution of life, we can identify several stages of
evolution:

\begin{enumerate}
\item Formation of large-scale structures in the universe: for example,
stars, and galaxies.

\item Creation of planets where complex stable microscopic molecular
structures can exist.

\item Existence of conditions for the creation of harmonic relational
structures, with largely static features such as amino acids, proteins,
organelles, viruses, bacteria, and multicellular, organisms, etc.

\item Evolution of dynamic features of the harmonic relational structures
such as neural firing that leads to complex behaviors, that evolve along
with the static features leading to the evolution of life on Earth-like
planets.

\item Evolution of coherence in the dynamic aspects of the neural firing
pattern of advanced life that leads to consciousness.
\end{enumerate}

This, the evolution of life and consciousness is the story evolution of
harmonic relational structures which was discussed as founding principles in
scientific relationism.

\section{Mathematical Measures of Structures}

\subsection{Setup}

In quantum gravity framework 4, I discussed life, consciousness, and its
role in structure formation in the universe. We introduced the measure of
consciousness in a quantum system motivated by Giulio Tononi's work \cite%
{IIT} but the definitions, basic assumptions, concepts, and interpretations
are different. Tononi's theory gives a measure of consciousness. Here our
purpose is to understand the physics of consciousness. We will develop more
detailed concepts to understand the physical behavior of living complex
systems.

Here we will be focussing on systems such as a neural network or a living
matter such as a part of tissue or simple organisms. In this, we can have
variables such as the neural potential, location of parts, electric
potential within the parts, etc. Using these variables, we can calculate the
relational complexity of the system. We are going to assume these variables
have certain periodic behavior combined with random behavior that is
correlated to other variables due to interconnections, such as in a neural
network, or chemical interactions in general. Life is not just relational
structures, and they all have certain periodic maintenance activities to
revive themselves, like reproduction, digestion, excretion, internal
regulation, etc. This is why I refer to such systems as harmonic
relationship structures (HRS), or life. Ideas regarding this were developed
in \cite{GUK}, \cite{BKVl3}, and \cite{MYP4}, I will describe more detailed
measures of these relational structures in this paper.

\subsection{Basic Definitions\label{basdef}}

Consider a general system defined by a set of variables that can take
specific values.

Here we consider all possible systems such as gas, liquid, solid, plasma,
living things, organic matter, neural networks, electric circuits, digital
networks, etc. We will discuss mathematical quantities to measure the
complexity of relational harmonic structure in these. These variables could
describe the location of particles, potentials at the various points of a
neural network, the boolean state values of a computer system, field values
at the various points of a system, the location of organelles of cells, or
the cells of a living colony.

Consider the joint probability density of the variables of a system: 
\begin{equation}
P=P(\{s_{i}\}),
\end{equation}%
where the left side is the function of all $s_{i}$ at each point. Where $J$
stands for joint probability. This probability can be derived from the
density matrix of the systems that I discussed in Quantum General
Relativistic Framework 5 of the paper and statistical thermodynamical
properties.

The $s_{i}$ stands for different things relating to the static and dynamic
features of the harmonic relationship structures. Relating to the static
features this refers to locations $q_{i}$ are positions of each part.
Relating to the dynamic features they measure power flow $\sigma _{i}$ in
the electrical and chemical interactions, including the quantitative
features of these interactions such as molecules mediating reactions,
frequency, amplitude, etc. In general, for all systems $s_{i}$ is the
combination of both:

\begin{equation}
s_{i}=(q_{i},\sigma _{i})
\end{equation}

We need to give a detailed description of both physical and energy transfer
configurations.

The best classical feature that quantifies the power flow is the \textbf{%
stress-energy tensor}, which is also covariant. In general, in a
relativistic context $s_{i}$ needs to be described by 4D locations and
energy-momentum tensor.

\begin{equation}
s_{i}=(q_{i}^{\alpha },\sigma _{i}^{\beta \gamma }),
\end{equation}

where the Greek letters indicate the indices of 4D space-time.

Giving a proper account of the interaction binding between the consciousness
of each neuronal element requires a detailed study of interaction. This
should include all detailed of interaction both at classical (stress-energy
tensor) and quantum mechanical (quantum states) as discussed in the
definition of the interaction of binding.

In \cite{MYP4} I heuristically used the rate of change of voltage as a
measure of energy transfer in a neural network. But it is too simplistic.
The rate of energy transfer is more important than quantities like voltage
or electric field because they measure the intensity of interactional energy
necessary to exhibit interaction binding. In this paper, we will not go into
details of this as it requires separate treatment. We can simply assume $%
\sigma _{i}$ simply measures the amount of power transferred between each
neuron and its surroundings when it fires for the purpose of this paper.
Each neuron binds with another neuron by chemical interactions with the
intermediate glial matter. This is much stronger than interactions in the
synaptic connections.

Let us further discuss the measurement of relational harmonic structures. In
this, we take a simple case of dynamic harmonic relationship structures,
with $v_{i}$ as the variables describing the state of each part of the
system. The joint probability is a function of all $v_{i}.$

\begin{equation}
P=P(\{v_{i}\})
\end{equation}

From $P$, we can derive the probability distribution for each variable.

\begin{equation}
P_{i}(v_{i})=\dsum\limits_{\{v_{j},~\forall j\neq i\}}P(\{v_{j}\})
\end{equation}

The Shannon information associated with the entire system is

\begin{equation}
I=-\dsum\limits_{\{v_{i}\}}P(\{v_{i}\})\log _{2}P(\{v_{i}\})
\end{equation}

This is total joint information held by the system. The Shannon information
associated with each variable $v_{i}$ is

\begin{equation}
I_{i}=I(P_{i})=-\dsum\limits_{v_{i}}P_{i}(v_{i})\log _{2}P_{i}(v_{i})
\end{equation}

The total self-information carried by all the variables is

\begin{equation}
I_{sT}=\dsum\limits_{i}I_{i}
\end{equation}

The mutual information that is associated with connectivity between the
variables is as follows.

\begin{equation}
I_{mT}=I_{sT}-I_{J}
\end{equation}

The average self-information per variable is 
\begin{equation}
I_{s}=\frac{I_{sT}}{N}
\end{equation}

where $N\ $is the number of variables. The average mutual information per
variable is

\begin{equation}
I_{m}=\frac{I_{mT}}{N}
\end{equation}

The average joint information per variable is:

\begin{equation}
I_{j}=\frac{I_{jT}}{N}
\end{equation}

\subsection{Measuring Harmonic Relationship Structures\label{meahar}}

A given number of variables of a system are in an active relationship if
they are not connected by a pure deterministic relationship and neither are
completely independent of each other. In the first case, they are not many
variables but essentially just one variable. The activity of the related
variables will be based on sporadic activity based on some intelligent
relationship, for example, such as in points of a logic circuit or a neural
network. In the case of conscious brains, this kind of intelligent relations
in which interactions can be turned on or off is critical for processing
dynamically changing sensory data. Using this insight we can define a
measure of the relationship structure using the basic formulas we defined as
follows:

\begin{principle}
Definition C.1: The strength of the relational network in the system is the
geometric mean of the last two.%
\begin{equation}
I_{rT}=\sqrt{I_{mT}\ast I_{jT}}
\end{equation}
\end{principle}

To understand what this means we need to understand the interaction between
two variables. If they are fully independent of each other $I_{mT}=0$. If
they are fully dependent on each other then $I_{jT}$ is a small value. Both
correspond to situations where there is no life in the system. Life exists
whenever there are interconnections between two variables that have the
dependence between these two extremes. For example, neural firing patterns
between various nearby neurons are not fully independent, nor fully
dependent (This is necessary for a healthy conscious brain there must be a
balance between coherence and randomness as I discussed on the subject of
coherence in the introduction). This implies there is some form of
interaction that happens only in certain conditions indicating the presence
of some form of intelligent dependence. $I_{rt}$ measures the existence of
this dependence. $I_{rT}$ is only a simple measure, it doesn't give a full
understanding of harmonic relational structures or life.

\begin{principle}
Definition C.2: The packaging of relational structure is%
\begin{equation}
I_{r}=\frac{I_{rT}}{N}=\sqrt{I_{m}\ast I_{j}},
\end{equation}%
where $I_{rT}$ is a measure of the total structure of the system.
\end{principle}

\begin{principle}
Definition C.3: Another way to get the total packaging of the relational
structure, which I call the intensity of the relational structure:%
\begin{equation}
I_{i}=\frac{2I_{rT}}{I_{s}}=\frac{2}{I_{s}}\sqrt{I_{m}\ast I_{j}}
\end{equation}
\end{principle}

A factor of 2 is used to make the maximum value as $1$.

Let $N$ be the total number of variables. Let $N_{s}$ be the number of
states these variables take. Let me assume the states are equally possible.
Assume there are $N_{c}$ connections.\ Then we can calculate the following:

\begin{eqnarray}
I_{s} &=&N_{s}N \\
I_{m} &=&NcN_{s} \\
I_{j} &=&(N-Nc)N_{s} \\
I_{i} &=&\frac{2\sqrt{Nc(N-Nc)}}{N}
\end{eqnarray}%
Maximum happens when $N_{c}=N/2.$Then the maximum value is $1.$

$I_{i}$ ignores variables that are not dynamic, and remain constant. It only
takes into account the variables that have dynamism. $I_{i}$ and $I_{r}$ are
the same if all variables are equally free and dynamic, that is $I_{s}=N$. 
\FRAME{ftbpFU}{2.9136in}{3.1834in}{0pt}{\Qcb{Sample systems for
calculational of relational structures. The dark ones are connected and take
identical values. White ones are independent.}}{\Qlb{SPLSYS}}{%
consciousmeasure.png}{\special{language "Scientific Word";type
"GRAPHIC";maintain-aspect-ratio TRUE;display "USEDEF";valid_file "F";width
2.9136in;height 3.1834in;depth 0pt;original-width 4.0084in;original-height
4.3837in;cropleft "0";croptop "1";cropright "1";cropbottom "0";filename
'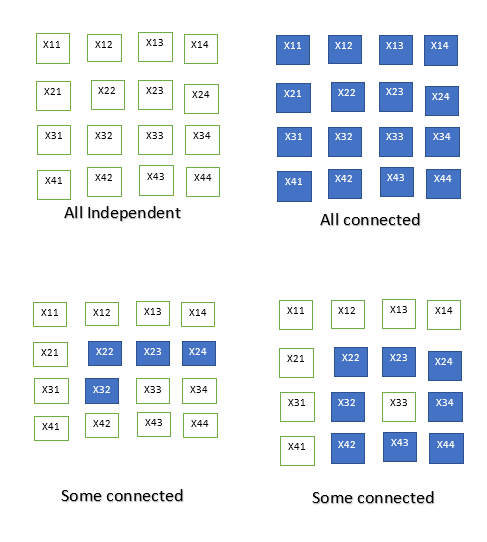';file-properties "XNPEU";}}

To understand what $I_{r}$ means consider the samples in figure \ref{SPLSYS}%
. Consider a set of 16 variables. \ Each variable can take values 0 or 1
with equal probability. Some of them are connected so they take equal
values. The connected ones are shown in blue in figure \ref{SPLSYS}. Then we
can calculate the relational measure of structure in the system. The table
below gives the values of various measures, that will give an idea of what
they mean. The higher the $I_{r}$ more the relational structure in the
system.

$%
\begin{tabular}{llllll}
Case & $I_{sT}$ & $I_{jT}$ & $I_{mT}$ & $I_{rT}=\sqrt{I_{jT}\ast I_{mT}}$ & $%
I_{r}=\frac{\sqrt{I_{jT}\ast I_{mT}}}{N}$ \\ 
A)\ All independent & $16$ & $16$ & $1$ & $4$ & $0.25$ \\ 
B)\ All connected & $16$ & $1$ & $15$ & $3.\,873\,0$ & $0.241\,88$ \\ 
C)\ Some connected & $16$ & $13$ & $3$ & $6.245\,0$ & $0.390\,31$ \\ 
D)\ Some connected & $16$ & $11$ & $7$ & $8.775\,0$ & $\allowbreak 0.496\,08$%
\end{tabular}%
$

\subsection{Measures of Relational Connectivity}

$I_{r}$ doesn't fully capture the relational structure within the system, as
it doesn't look at the parts. Let me give further definitions to fix this.
Let me discuss here the two definitions that were discussed in quantum
gravity framework 4.

\begin{principle}
Definition C.4: If there are entities with their state described by
variables $s_{1}$ and $s_{2}$ then we can calculate the relational
connectivity as follows.

\begin{equation}
C_{12}=\frac{I_{m,12}}{I_{s,12}}
\end{equation}
\end{principle}

Here $I_{m,12}$ and $I_{s,12}$ are the mutual and the self-information
associated with the combined system described only by the variables $s_{1}$
and $s_{2}$. This can be derived from the probability function obtained by
summing over all the variables of the joint probability of the system except 
$s_{1}$ and $s_{2}$. We can define the connectivity within a system as
follows:

\begin{principle}
Definition C.5A: Consider we have a system $S$ with many entities described
by variables $s_{i}$. We divide the system into two regions $S_{1}$ and $%
S_{2}$ by a cross-sectional area $\Sigma .$ Then the total connectivity
between the two parts on the two sides of $\Sigma $ is given by%
\begin{equation}
C(S_{1},S_{2})=\sum_{k_{1}\in S_{1},k_{1}\in S_{2}}C_{k_{1}k_{2}}
\end{equation}
\end{principle}

\begin{principle}
Definition C.5B: We can define the relational connectivity of a system as 
\begin{equation}
C(S)=\dprod\limits_{\forall S_{1}US_{2}=S}C(S_{1},S_{2})
\end{equation}
\end{principle}

\begin{principle}
Definition C.6: A system is relationally connected if $C(S)$ is non-zero. If 
$C(S)=0$ then the system is not relationally connected.
\end{principle}

$C(S)\neq 0,$ it means the system $S$ cannot be divided into two parts $%
S_{1} $and $S_{2}$ such $C(S_{1},S_{2})=0$.

Consider a $S$ system that is fully relationally connected. Let $S_{n}$ be
the set of all n points in the system $S$. Consider one particular set of n
points $X=\{s_{l_{1}}....s_{l_{n}}\}$ and $L=\{l_{1}....l_{n}\}$. For each
two variables in $X,$ we can calculate connectivity. Then we assign a
connectivity to the whole $X$ by multiplying the connectivity of all the
pairs in $X.$

\begin{equation}
C_{X}=\dprod\limits_{s_{1},s_{2}\in X,s_{1}\neq s_{2}}C_{s_{1}s_{2}}
\end{equation}

\begin{principle}
Definition C.7: The extent of connectivity of relational structure is 
\begin{equation}
\mathbf{N}(S)=\frac{\dsum\limits_{S_{n\text{ }}\in S_{\text{ }%
}}\dsum\limits_{X\in S_{n\text{ }}}nC_{X}}{\dsum\limits_{S_{n\text{ }}\in S_{%
\text{ }}}\dsum\limits_{X\in S_{n\text{ }}}C_{X}}
\end{equation}
\end{principle}

The higher the $\mathbf{N}(S)$ more connected the relational harmonic
structure in the system. It is just a weighted average of $n.$

\subsection{Measures of Relational Complexity}

Let me discuss the measure of relational complexity using the variables
discussed in the last subsection. Consider a system that is fully connected.
We can derive the probability distribution $P_{L}(X)$ of the n variables by
summing out the other variables in the full probability distribution. We can
calculate the measure of relational structure associated with these
variables using the $I_{r}$ formula for these variables.

\begin{equation}
I_{r}(L,n)=\frac{2\ast \sqrt{I_{mT}(L,n)\ast I_{jT}(L,n)}}{I_{s}(L,n)}
\end{equation}%
This depends on $n$ and $L$.

\begin{principle}
Definition C.8: For the whole system, we can calculate global relational
complexity $I_{rS}$ as%
\begin{equation}
I_{rS}=\dsum\limits_{n}\dsum\limits_{L\in S_{n}}I_{r}(L,n)
\end{equation}
\end{principle}

This definition is a powerful measure of the relational structure in the
system taking into account all the connections between the various parts of
a system.

There is another more local definition. Consider a set of variables attached
to the points of a discretized space within the region of a continuum
manifold of dimension 3 that contains a physical system. Consider the system
is divided into the smallest independent parts, such that all the variables
inside a part $i$ are fully dependent so that we associate one variable $%
s_{i}$ for each part, from which the values of the other variables of the
part can be calculated. Assume that the field variable of a part interacts
only with the variables of the parts adjacent to it. Let us call this the
elementary relational structure of the system. This could be the neural
network structure of a central nervous system in a living thing, or the
circuit diagram with the voltages in an integrated circuit. Let us call each
independent Quantum General Relativistic Framework 5 along with the variable 
$s_{i}$ as an elementary component (for example, a soma or a dendritic
branch of a neuron, or an input and an output of the components of a digital
circuit.)

Let $s_{y}$ be an elementary component that includes point $y$. Let $S_{y}$
be elementary systems made of a set of elementary components adjacent to the
elementary component $s_{y}$, including $s_{y}$. Let $X_{y}$ be variables
associated with $S_{y}$. We can calculate the probability distribution $%
P(X_{y})$ from the density matrix and statistical probabilities within the
duration $\Delta T$ during which the system has a stable probabilistic
behavior Using this one can calculate the measure of the relational
structure of $S_{y}$ by taking into account the local inter-elementary
connections using the joint probability distribution $P(X_{y})$:

\begin{equation}
\mathbf{R}_{y}=\frac{2\sqrt{I_{mT}(S_{y})\ast I_{jT}(S_{y})}}{I_{s}(S_{y})}
\end{equation}

\begin{principle}
Definition C.9: The spacial density of the relational complexity is%
\begin{equation}
\mathcal{R}(x)=\frac{\mathbf{R}_{y}}{^{\Delta V_{y}\Delta T}}\Delta _{s}(x)
\end{equation}%
where $\Delta V_{y}$ is the volume associated with the elementary system $%
s_{y}$ and $\Delta T$ is the duration for which the probability densities
are calculated. $\Delta _{y}(x)$ is zero if $x$ is outside $s_{y}$ and
constant inside $s_{y}$ such that $\int \Delta (x-y)d^{3}x=1$.
\end{principle}

\begin{principle}
Definition C.10: Let $S$ be a fully connected system. Let $L$ be the number
of links in the system and $N$ be the number of variables in the system. Let 
$V$ be the volume of the system Let $r$ be the average value of the
relational strength. The number density of relational connections in the
system is 
\begin{equation}
\mathcal{N}(S)=\frac{rL}{VN}
\end{equation}
\end{principle}

\begin{principle}
Definition C.11: The net local relational complexity of a system $S.$%
\begin{equation}
\mathbf{R}(S)=\dsum\limits_{\forall s_{y}\in S}\mathbf{R}_{y}=\dsum\limits_{%
\forall s_{y}\in S}\frac{2\sqrt{I_{mT}(S_{y})\ast I_{jT}(S_{y})}}{%
I_{s}(S_{y})}
\end{equation}
\end{principle}

$R(S)$\ seems to be better suited to understanding the relational complexity
of a relational harmonic system because of its local nature. It is a measure
of the life in the system. I have given some ways to calculate the
complexity of the relational structure of systems. This will be used to
discuss various propositions to understand consciousness next.

The most general possible way to understand the relational structure
dependence of a system transition amplitudes is, to add a new term for the
action of the system.

With measures for consciousness developed let me discuss various properties
of consciousness discussed in quantum gravity framework 4 \cite{MYP4}.

\subsection{Consciousness Threshold Measurement}

In section \ref{conthr} I\ introduced the conscious threshold for conscious
experience. Let me make it into a mathematical form. In the consciousness
threshold we had many different thresholds described in section 4.8. Let us
focus on the coherence threshold.

\begin{principle}
Definition C.12: Coherence threshold is considered as the minimum amount of
information input, coherence in the information input, and a proper balance
between them necessary to have the conscious experience.
\end{principle}

Now the three quantities are considered as the following using definitions
in sections \ref{basdef} and \ref{meahar}:

\begin{itemize}
\item Information input:\ Total Information input $I_{jT}$

\item Coherence as mutual information $I_{mT}$

\item The Balance between information input and coherence is given by the
relational complexity $I_{i}=\frac{2\sqrt{I_{mT}\ast I_{jT}}}{I_{sT}}$
\end{itemize}

As I have discussed for $I_{rT}$ = $I_{mT}\ast I_{jT}$ is maximum when there
is a balance between $I_{jT}$ and $I_{mT}$ is equal to $I_{sT}/2.$

Let me assume all the other thresholds for consciousness as described as in
section \ref{thr} are satisfied, except the coherence threshold with depend
on the firing pattern, which in turn depends on the functional state of the
brain.

Now depending on how we define the threshold for $I_{jT},$ $I_{mT},$ and $%
I_{i}$ different systems can be considered conscious or not.

\begin{case}
$I_{jT}>0,$ $I_{mT}>0$ and $I_{i}\symbol{126}1$
\end{case}

In this case, many animals are conscious as long there is a balance between
information input and coherence. This could be primates, dolphins, mammals,
etc.

\begin{case}
$I_{jT}>I_{c},$ $I_{mT}>I_{c}$ and $I_{i}\symbol{126}1$
\end{case}

In this case, we have a minimum information input. Heuristically, while we
are in NREM sleep $I_{jT}$ is too low, \ $I_{mT}$ is high, and $I_{i}$ is
not approximately 1. In this case, we don't have consciousness. Either in
REM\ or wakeful condition, we have consciousness threshold satisfied, so
conscious experience.

Depending on the cutoff, most animals can be made to be not conscious.\ They
just look conscious, but not conscious.

We need an experimental study to properly evaluate the consciousness
thresholds.

\section{Propositions on Human Consciousness}

Until now we have discussed a wide range of existing theories, experimental
facts, definitions, and hypothetical ideas and models to understand the
phenomena of consciousness. In this section, I will be putting them together
in a set of propositions to explain the phenomena of consciousness, and also
summarize what we discussed before regarding each the proposition.

The human brain is a highly evolved conscious structure. It perceives the
world as a sequence of perceptions. The perceptions are nothing but a
combination of sensations, which I would like to refer to as a sensational
complex. This has been introduced in chapter 3 of \cite{BKVI1}. We need to
understand how the sensational complex is built in terms of the concepts of
the quantum general relativistic framework. Let me frame the following
propositions from the founding ideas presented in \cite{MYP4}, \cite{GUK},
and \cite{BKVl3}.

\subsection{The Universe and Entanglement Domains}

The first set of propositions is about the entanglement domains given in
Definition B.2. Let me state them and discuss them.

\begin{principle}
Proposition III1.a The universe is divided into entanglement domains that
encase highly entangled particles within a 3d spatial region.
\end{principle}

The universe is made of objects such as crystals, metals, stones, neutron
stars, planets, living things, etc. They are made of fundamental particles
bonded together by interactions with gauge particles. The interactions
result in entanglement between the fundamental particles that make up these
objects. In the case of solid objects, they have well-defined boundaries. In
objects like liquid or gas, we have particles entangling with each other
continuously they don't have well-defined boundaries. But we can expect
rapid formations of local groups of particles that entangle much more
relative to the area surrounding them. This can be considered as dynamic
rapidly fluctuating entanglement domains. We can refer to this as diffuse
entanglement domains.

The $C$ fields described in section \ref{quaphy} as described is a part of
matter that captures conscious experience is part of the entanglement
domain. We will assume $C$ fields obey physical laws as any other fields.

In the quantum gravity framework 5, I discussed describing matter in
space-time using the density matrix and it undergoes continuous quantum
reduction according to postulate I.2. In the entanglement domains, we have
particles reaching macroscopic superposition occasionally. This rapidly gets
quantum reduced by the postulate I.2. The same thing can be said for diffuse
entanglement domains. When the superposition reaches the macroscopic level
in a certain narrow region of space made of entangled particles, it
undergoes reduction.

When the entanglement domains are under continuous reduction, it reduces
other particles and domains that are entangled with them. For example,
consider a diffuse entanglement domain that is entangled with an
entanglement domain such as a large solid object. The solid object has
higher mass and so it undergoes rapid reduction, driving a much faster
reduction in the diffuse entanglement zone also.

\begin{principle}
Proposition III1.b: A self decohering entanglement domain undergoes time
evolution through continuous quantum reduction together as an entity along
the rest frame (or in general least variation) foliation.
\end{principle}

The continuous quantum reduction also requires a foliation in which this
happens. The obvious choice is the rest frame foliation or (most probable
least variation foliation in general). We also need a foliation to
understand the binding problem of consciousness. Conscious brains need a
foliation to combine all the data in the neural firing into 3+1 form, the
obvious choice is the rest frame foliation.

The reduction can be understood in terms of longitudinal and transverse
fields of gauge fields. The transverse fields such as photons of
electromagnetic field entangles matter (fermionic) particles. The
longitudinal fields contain much of the mass (energy) of the particles. When
the entanglement reaches the macroscopic level, the longitudinal field acts
as the observer and reduces the fields. This has been discussed before in
quantum gravity framework 2 where I introduced rest fame foliation, and also
early in this part.

\begin{principle}
Proposition III1.c The reality is in the form of an evolving block universe,
that is made of a quantum evolving block of entanglement domains evolving
along the rest frame foliation.
\end{principle}

This is the best way to put together general relativity, Newtonian gravity,
quantum mechanics, and the conscious flow of time. I have introduced this
idea in quantum gravity framework 2, based on previous existing ideas of the
block universe models of the universe \cite{Block}. This is shown in picture
1 in QGRF 5 \cite{MYP5A}. We can interpret `the past' as classical in nature
as it has been already reduced. The present is the 3-dimensional represented
by the most probable least variation foliation. The future is purely quantum
and it can be evolved using postulate I.1 in QGRF 5 \cite{MYP5A}.

\subsection{Conscious Substratum and Being}

Here I state three propositions on conscious substratum directly as a
summary of our discussion before. Please know the conscious substratum
contains the classicum plus other matter and fields.

\begin{principle}
Proposition III2.c: A conscious substratum is a fundamental entity
associated with a conscious being, that contains the $B$ fields taking in
account the free will and observation, and $C$ fields that captures
conscious information 1 to 1. \ 
\end{principle}

The $B$ fields and $C$ fields where discussed in section \ref{quaphy}. This
is the basic definition of consciousness substratum and it can be considered
more fundamental than the quantum entanglement domain. Each conscious
substratum has an classicum contained in it, which we discussed before.
Before we discussed various possibilities for defining it, and what happens
between birth and death.

\begin{principle}
Proposition III2.b: A conscious substratum cannot be created or destroyed.
\end{principle}

This proposition is a direct result of the conscious being attached to the
same human body irrelevant of any gradual changes in the structure of the
body, or the extent of electrical activity. This also means the classicum
cannot be destroyed. I have extrapolated this to also include time before
birth and after death.

\begin{principle}
Proposition III2.c: Each conscious substratum encompasses an entanglement
domain.
\end{principle}

This has been discussed before in the section on the conscious substratum.
However, the relation between entanglement domain, observer, $B$-fields and
the conscious substratum is complex, which was discussed before. But we
assume this due to simplicity. These three propositions are consistent with
the birth and death concepts introduced in the section \ref{birdea}.

\subsection{Life and Relational Structures}

Postulates on life and relational structures directly follow from the
principle of physics. Let me discuss here propositions on the nature of
relational structures and their relation to life.

\begin{principle}
Proposition $\alpha $III.3a The extent of life in a relational structure in
quantum entanglement domains is measured by R(S).
\end{principle}

This is our first proposition. The R(S)measures the total extent of
relational harmonic structure in the entanglement domain. In other words, we
can come to the following corollary:

\begin{principle}
Corollary: $\alpha $III.3a' Life is a relational harmonic structure.
\end{principle}

We can deduce various measures of life such as the following: $\mathcal{R}%
(x),$ $\mathcal{N}(S),$\textbf{\ }$\mathbf{N(S),~}$\textbf{and }$\mathbf{R}%
(S)$ are measures of the life of a relational harmonic structure. $\mathcal{R%
}(x)$ stands for spatial density, the number density of the system $\mathcal{%
N}(S),$\textbf{\ }$\mathbf{N(S)}$ stands for depth, and $\mathbf{R}(S)$
stands for the total life of a system.

These quantitative measures of relational structure are possible candidates
for many other possible definitions available in the literature. Systems
that have low life measures, for example, are solid and gas. In the
macroscopic scale, for a solid, if we assume that the configurational
variables are the locations of the nuclei, it has low joint information, as
the locations of the nuclei are fixed with respect to each other. For a gas,
it has zero mutual information, but high joint information. Even liquids
that have free-moving molecules have limited mutual information. But
biological matters have high values of both mutual and joint information. So
have high values for the life measures.

\begin{principle}
Corollary $\alpha $III.3a\textquotedblright\ $\mathbf{R(S)}$ is an
increasing function of time for a living system and is a decreasing function
of time for a dying system.
\end{principle}

From the initial universe, from a highly thermal state, as time passed by,
most matters such as free gas, liquids, solids, plasma, etc. formed. Under
suitable conditions for the existence of highly complex molecules, they
formed complex relational harmonic structures. On Earth, this led to the
development of life. These are viruses, bacteria, unicellular,
multicellular, plant, and all animal species that evolved. $\mathbf{R(S)}$
keeps increasing as living systems absorb chemicals to create more complex
harmonic structures. This could be a growing tree, a growing microbial
colony, or a growing baby. But when a structure is dying then $\mathbf{R(S)}$
decreases as the chemical structure breaks down, such as dying microbes,
trees, or an animal.

\begin{principle}
Proposition $\alpha $III.3b The action of the quantum general relativistic
physics is precisely the one that leads to the evolution of a relational
harmonic structure that hosts a human consciousness.
\end{principle}

This was discussed in section \ref{zerpos} as part of the anthropic
principle. Human consciousness requires extensive evolution and also
physical processes that support them. Without the anthropic principle, it is
quite difficult to explain this.

But we can also go a step further. The most general possible way to
understand the relational structure dependence of a system transition
amplitudes is, to add a new term for the action of the system:

\begin{equation}
S->S-ic_{r}\int \mathcal{R}_{s}(x)d^{3}xd\tau
\end{equation}

The $c_{r}$ is a new constant that is to be determined experimentally if
this model works.

The $\mathcal{R}_{s}(x)$ term is a new dynamical action term that takes into
account the presence of elements of life in a physical system including
consciousness. It promotes the formation of relational and harmonic
structures in the system. For more information, please refer to \cite{MYP4}, 
\cite{GUK}, and \cite{BKVl3}.

A possible need for this term comes from the fact that life is intricately
connected to structure formation. At any time in human history building
structures is the most significant symptom of human presence. It indicates
life and human consciousness promote structure formation. Can this be simply
explained by the evolution of biological structures by standard physical
principles? I don't think so, as it wouldn't explain art and music. What is
the evolutionary need for art and music? Consciousness is driven by
attraction to these and the influence of consciousness promotes the
formation of harmonic relational structures (such as art and music).

The $\mathcal{R}_{s}(x)$ can be set to be the measure $\mathcal{R}(x)$ of
the relational harmonic structure that was discussed before. This $\mathcal{R%
}(x)$ is highly non-linear in nature by our definition. It is calculated
based on the local behavior of systems using a density matrix over the
course of time. For a big system like a neural network, we need to integrate
this out to get $\mathbf{R(S)}$ to get a measure of full relational
complexity as defined before. However, this is an assumption because it
doesn't fully explain the promotion of the dynamic harmonic activity that
elicits consciousness. The $\mathcal{R}(x)$ probabilities need to take into
account both temporal and spatial joint probability for this. This makes the
quantum field theory defined on a block universe model with field equations
in not instantaneous form in conventional field theory. We need further
research on this.

The observer and C field quanta discussed in section \ref{quaphy} plays a
role in promoting structures, and so understanding the physics of it is
necessary to find the current functional form of the structure-promoting
terms. This action may include observer and $C$ fields that are necessary to
explain issues such as dark matter and dark energy terms interacting with
normal matter and gauge fields to promote dynamic structures that are
consciousness capable.

Let me emphasize that, it is not clear to me whether the last proposition is
fundamental or a direct consequence of other principles such as time
evolution, decoherence, and selective Darwinian evolution. The proposition
must be experimentally tested to answer this question.

\subsection{Physics of Conscious Observers}

The propositions in this section are based on Hypotheses 8.2, 1, 2.2, and
3.1, postulate II.3, etc.

Human and animal brains are connected relational structures that encode
information about the external world, through their interaction with it. The
encoding is integrated into the structure such as in the memory elements in
a computer or through the altering of interconnections in the neural network
of a living thing through neural plasticity (Hebbian Phenomena).

\begin{principle}
Definition D.1: Observers are connected relational structures that record
observations by modifying their own structure as classical information.
\end{principle}

Here I am talking about observers in general such as a camera, Geiger
counter, or a human being. They exist as physical systems with particles and
fields. They are described by a joint density matrix of relational
structures that make them. In case of the human brain the conscious
observers is the entire brain with neural net along with the conscious
substratum.

We can consider this as a general definition of an observer in physics. The
universe is made of observers which are relational structures. It could be a
stone, a planet, a bacteria, a virus, a cat, or a conscious human being. As
the observer moves through the rest frame foliation, the joint density
matrix keeps evolving. The spontaneous quantum reduction discussed in
postulate I.2 in QGRF 5 \cite{MYP5A} keeps the quantum state corresponding
to the joint density matrix without macroscopic superposition. The observer
is continuously forced into one of the macroscopic states probabilistically.
This creates a record of its history in its physical structure. The extent
of information depends on the complexity of the structure. For example, from
the earth's geological properties, we can see the record of its history from
its creation.

\begin{principle}
Definition\ D.2: An energy interaction in a simple and connected relational
structure can be defined as i) an energy state change of its joint quantum
system made of itself and the environment involving quantum self-evolution
of its parts, ii) quantum interaction between its parts leading to the
superposition of its entangled states, and iii) the spontaneous quantum
reduction of this state.
\end{principle}

We can consider this as a general definition of what a full interaction is,
which we discussed in detail before in the introduction. A physically
connected relational structure is made of many quantum systems. These
quantum systems first evolve by their free Hamiltonians to get close to each
other. Then through the interaction part of the Hamiltonian the state of the
system becomes a superposition of multi-particle entangled states. In each
of these entangled states, we have a unique energy exchange between the
parts of the systems. Then this superposition evolves and interacts with the
rest of the relational structure and the environment. This leads to a
macroscopic superposition of entangled states, and this eventually gets
quantum reduced to remove the macroscopic superposition. This is the
complete interaction discussed before.

\begin{principle}
Definition D.3: An animate observer is a quantum entangled harmonic
relational structure in which information regarding the external world is
mapped into energy interactions within it, in a form that will help in its
specific functions such as to meet its survival needs.
\end{principle}

Now combining postulate II.1 and the definition D.3, we can define a
conscious entity.

\begin{principle}
Proposition $\alpha $III.4a A conscious entity is an animate observer, 1)
that satisfies the conscious thresholds, 2) the coherent energy interactions
at the biochemical level in the brain is converted into the C field
information, 3) which in turn is coupled to the observer relating to its
conscious substratum experiences through the C field information as qualia.
\end{principle}

We can combine the last definition and proposition to describe the
experience of a conscious entity. A conscious entity contains a harmonic
relational structure, that is built on a conscious substratum, that receives
sensory information from the external world and engages in harmonic activity
to create a coherent pattern of energy interaction in a continuous fashion
to process this information and satisfy all the consciousness thresholds
defined before.

The coherent information in the energy interaction in a relational harmonic
structure is jointly experienced as a sensation at multiple scales of
interaction. We describe the nature of these two propositions by two more
propositions.

\begin{principle}
Proposition $\alpha $III.4b: The extent of sensation and binding depends on
the magnitude of the average energy exchange at each scale within the
relational structure.
\end{principle}

\begin{principle}
Proposition $\alpha $III.4c: Different complex structure of energy
interaction creates different sensations.
\end{principle}

This proposition was already described in \cite{MYP4}, with some
differences. Any quantum system of reality, like a particle or a molecule,
is capable of elementary sensation and the sensation happens when it
interacts with a different quantum system. The sensation is shared by both
systems. This will be more clearly discussed in the next subsection. The
interaction among the relationally connected structure may contain a wide
variety of information including noise.

The complex structure involves details of energy interaction. The various
aspects of the interaction such as energy transfer, details of the
interaction, the structure of the molecules, or the whole system, rate, type
of forces involved, etc., are experienced as qualia by the whole connected
system.

The interaction happens at many scales: At the neuronal assembly level
representing sensory information mapped from the external world, at the
neuron level firing pattern within neurons and between neurons, at the
molecular level, at the particle level, etc. At each scale, we can describe
the coherent information as the average of interaction information at the
next smaller scale. With the neurons, we have microtubules which also engage
in energy interaction between them and the rest of the neuron, and with
electromagnetic fields from other neurons. Many other forms of interactions
may exist in further smaller scales.

Now we try to link the conscious experience with the 3+1 formulation we
discussed in the QGRF 5 \cite{MYP5A}. The three postulates that we discussed
in the QGRF 5 \cite{MYP5A} reduced the 3+1 formulation into potential
possible classical histories. Every time there is a quantum superposition
that evolves into a macroscopic superposition, classical history splits into
many possible histories. The probability of a certain classical history $%
\eta =\{Q_{i}^{a}(t),\Phi _{j}^{X}\}(x,t)\}~$is given as determined by the
three postulates in QGRF 5 \cite{MYP5A} and postulate II.1c and II.3c in
this part.

\begin{principle}
Corollary $\alpha $II.3a: The time of flow of a conscious entity is given by
the rest frame foliation.
\end{principle}

This directly comes from postulate II.3.

In more detail, this corollary contains the following:

I) A conscious entity experiences the 3+1 quantum block universe of the
fields and matter as a 3D experience proceeding over time. It traverses
through one possible history of the semi-classical harmonic relational
structures with certain probability given by the three postulates in QGRF 5 
\cite{MYP5A}, and postulates II.1c and II.3c in this part.

II)The time of flow of observers is given by the rest frame foliation. The
observer state on each of the hypersurface of the rest frame foliation is an
instant of observation.

This proposal is different from the many worlds universe formulation. In
many world theories of Everett, every possible classical history coexists as
reality. In this, the conscious being gets split into many possible
conscious beings following different experiences. But in our interpretation,
each conscious being follows only one possible classical history of many
possible classical histories as determined by the three postulates of QGR5.

In QGRF 5 \cite{MYP5A} I described the variables with respect to which the
quantum evolution of the general relativistic description of a system was
formulated. Under the rest frame evolution condition, these variables
capture the classical information in the system, and this classical
information is captured by the observer. These variables are
probabilistically described by postulates in QGRF 5 \cite{MYP5A}. The lesser
the variation of continuum fields and particles along the foliation, the
higher the probabilities of and, the better the capturing of classical
information. For the observer, the more the number of classical degrees of
freedom, the more the classical variables assigned to these, better its
description of the classical part of the reality around it. Using the
variation of the classical variables in the rest frame foliation as the
clock, the quantum evolution of the remaining degrees of freedom can be
calculated using the various postulates in QGRF 5 \cite{MYP5A}.

From postulate II.2 we have the following corollary:

\begin{principle}
Corollary $\alpha $II.2a Free will: The behavior of the conscious system is
determined by calculated and free will actions that are identical to quantum
evolution and quantum reduction respectively.
\end{principle}

This was discussed in detail before in the overview. A relational structure
at the fundamental level keeps evolving quantum mechanically leading to a
quantum superposition of macroscopic states. This gets continuously quantum
reduced to one of the states, according to postulate I.2 discussed in QGRF 5 
\cite{MYP5A}. To exercise free will, the sensations, thoughts, and ideas
that are experienced by a relational structure need to be able to choose the
final state. These influential factors are defined as deterministic bias
before and they are represented as the amplitude of the wavefunction as
determined by the physical evolution of the relational structure. The entity
reduces itself to one state as determined by this amplitude. The
mathematical modeling of this will be discussed in more detail in the next
section on modeling consciousness.

To take into account the reduction being influenced by the reactions, we
need to take into account the effect of the $C$ fields discussed in the
section of discussion on the conscious being.

\begin{principle}
Corollary $\alpha $II.2b Retrocausality: Free will actions are influenced by
both future and past reductions.
\end{principle}

From QGRF 5 \cite{MYP5A}, in quantum gravity, quantum states are governed by
bidirectional evolution: they are determined by both past and future
reductions. So, this means the free will actions computed based on quantum
states have influence from both the future and the past reductions. The
human brain has multiple parts overlapping with the environment. The
microscopic variables that critically determine the evolution of the
macroscopic quantum state of the system, evolve quantum mechanically, and
its extrapolation leads to the macroscopic superposition of the quantum
states in the future. The quantum reduction of these states results in
negative energy states traveling towards the past quantum reduction, to
shape the evolution of the quantum states as determined by the future and
the past quantum reductions. If the conscious part of the quantum
information in a macroscopic quantum state, before collapse, also has
negative energy states, this means the observer is also aware of the future.
The free will action takes into account the negative energy reflections from
the future reflections.

When we do the mathematical modeling of free will in the next section, we
will discuss how to implement this retrocausal aspect of free will
mathematically.

\subsection{Sensational Complex}

In the previous section, we discussed what a relationally connected system
is, and how it becomes a conscious entity. We will use this in the
propositions in this section. Also, in this section, we are going to expand
ideas from hypothesis 2.2c.

To have a continuous stream of experience a relational structure of a
conscious entity needs to receive sensory information from the outside world
and engage in harmonic activity to experience it as sensations.

Simultaneous energy interactions at various scales in a physical and
relationally connected system of a conscious entity create an integrated
sensation experienced by the whole entity called the sensational complex.
Now let me discuss the propositions to explain the integration of neural
firing to create the sensational complex.

There are two types of binding between two points of a harmonic relational
structure of a conscious entity:

1. Active Binding: An example of this is two or more neurons that are firing
in synchronization. They interact with each other through direct synaptic
connection, and indirectly through their synaptic connections with other
neurons, through the cortico-cortical loop. They also interact through an
electromagnetic field such as brain waves.

2. Silent Binding: An example of this is two neurons nearby that are bound
together by electrostatic forces, but not actively bound together. Due to
these forces, they are quantum-mechanically entangled. This entanglement
usually involves incoherent superposition as they interact thermally. We
discussed the incoherent superposition in QGRF 5 \cite{MYP5A} and in the
introduction in this part.

\begin{principle}
Proposition $\alpha $III.5 The sensory information is bound as follows: a)
Active binding creates a sensational unit b) The specific organizations of
actively bound parts create sensations of the specific modality (visual,
sound, taste, etc.). c) Silent binding fuses together sensational units to
create the sensational continuum.
\end{principle}

The above proposition has three different sub propositions -- a, b, and c in
it.

The harmonic relational structure of a conscious entity is bound by
electrostatic forces, atomic level, or nuclear forces at the nuclear level.
This creates the sensational continuum or a screen on which the conscious
experience is had. This sensational continuum is bound by the silent
binding. On top of this, we have active binding happening due to the
interaction between the atoms and molecules due to the interaction activity,
which creates a conscious experience, like a painting on the screen.

Relationally connected systems could be the various areas of the human
brain, each with its neural assemblies. When harmonically and relationally
connected systems such as a human brain, simultaneously interact within
their parts and with each other, their integrated sensation merges to become
a complex experience, which I would like to refer to as sensational complex
as defined in the proposition, and first defined before in \cite{MYP4}, \cite%
{GUK}, \cite{BKVl3}, and was further discussed in the introduction. The
sensational complex with its many parts is bound together by energy
interaction according to our propositions. Later we will discuss how the
integration happens in more mathematical and physical states.

\subsection{Modalities and Feelings}

We will further discuss hypotheses 3.3, 3.4, and 4.1 to understand the
modalities and feelings of a conscious entity.

\begin{principle}
Proposition $\alpha $III.6a Each continuum set of sensational units of the
same type creates different modalities with the following properties i) The
higher the sensational unit density, the higher the information content and
neural frequency, the lesser the feeling, and vice versa. ii) Different
frequencies in the harmonic activity result in different modalities of
sensation such as mood, emotions, etc. iii) Modulation of harmonic activity
of two different modalities results in their association.
\end{principle}

This proposition is about various modalities. Let me explain the integration
of sensory data happening in animal brains that are capable of consciousness
using the last three propositions.

The synchronized firing of the various elementary systems of a complex
relational structure results in integrated consciousness. This is the case
of synchronous firing in gamma frequency in the human brain. The sensational
complex of a human being is made of a combination of visual, sound, tactile,
nasal, and visceral sensations. It also contains the sense of time,
identity, etc.

Synchronization and the phase coherence of interaction in a relationally
connected system help in binding and amplification of the impact of related
integrated information. For sensation to integrate into a sensational
complex experienced at a given instant, all the energy interactions involved
in it have to happen simultaneously according to our proposition, (time
determined by rest frame foliation). Phase coherence and synchronization
help in this. As I have discussed before this not only helps in conscious
integration but also object recognition in animal brains. In the
introduction, I discussed in detail various ways the synchronization is done.

The visual cortex has the most neural density, so more sensational unit
density. It's felt as information, and less as feeling such as fear, anger,
or taste. We can order the sensational unit density of various modalities as
follows: Vision, Tactile, Gustatory, Hearing, Nasal, and Emotions. The
feelings increase in this order and the complexity of information decreases
in this order also. The emotional modality is internal and generated by
lower layers of the brain, such as the reticular activating system,
amygdala, etc.

In the cortex, the sensational unit is made of neural columns. They are
present for all sensory modalities. They are connected to each other through
feedback neural systems through active binding. A sensational unit of
modality is experienced through long neural circuits distributed throughout
the brains which are actively bound. These circuits start and end with
cortical neural columns that receive connections from the thalamus relaying
sensory information, and connect with each other through the
cortico-cortical loops. Through feedback and feedforward loops connect to
the higher associative areas that bind sensory information from all
modalities. These circuit units silently bind as a continuum of sensational
complex, and they cross-modulate each other and associate the sensory data
with each other.

Neural frequency depends on modality. Visual Cortex has the highest neural
density. This is followed by sensory, auditory, and other systems. The
neural firing frequency increases with the information content. The visual
system has typical gamma band firing during conscious processing and has the
highest neural firing rate. Emotional processing has neural connections
whose number is low and connects to areas throughout the brain, the neural
firing corresponding to emotions shows up as low-frequency brain waves in
EEG. According to the proposition, the cross-modulation of neural firing
binds the sensations in the various modalities into a single stream of
consciousness.

Consider the visual cortex. It has a feedback loop of actively bound neural
circuits, which merge together to create the vision modality. The same thing
applies to tactile sensation. Now these two modalities can modulate each
other's harmonic activity to merge their continuum. For example, you feel
the location of a sensation of touch in a specific location in the vision
space. The same applies to olfactory, gustatory, and auditory modalities.
These sensations are felt in the visual space created by the visual cortex.
Much of these points in the visual space are felt, not seen, such as
feelings inside of nose or tongue, or in our back.

\begin{principle}
Corollary $\alpha $III.6a The concentration of the processes associated with
consciousness depends on a) the Region of highest relational density b) the
region of highest dynamic relational activity, and c) In the frequency
domain the frequency of maximal coherence/synchronization is the dominant
modality around which other modalities are organized.
\end{principle}

The complex structure and details of energy interaction in a conscious
harmonic relational structure of a conscious entity correspond to different
feelings, emotions, or qualia of a conscious system and are experienced as
sensational complex. This proposition is self-explanatory. The third part
refers to the gamma synchronization. The neural firing patterns relating to
conscious experienced objects are synchronized at this frequency band,
particularly visual, and all other modalities such tactile, auditory, nasal,
and emotions are organized around vision.

The perception in a relational structure is a process in which an evolved
conscious harmonic relational structure through conscious computational
processes inputs information into a form useful for its future behavior of
survival significance.

\begin{principle}
Proposition $\alpha $III.6b Understanding: When the conscious harmonic
relational structure processes the sensory information by unconscious
computational process, the existence of perceptual information in the
structure is felt as feedback of feeling of understanding and confidence.
\end{principle}

This explains what understanding means. If we observe something, if the
brain's computational process has sufficient information to get the useful
information necessary regarding this observation, then we get a feeling of
confidence or familiarity. This is how understanding, comprehension, or
insight about something relates to just feedback feeling.

Qualia are sensational complexes mapped to objects by living entities
(harmonic relational structures) through their memory systems and reproduced
on demand by their neural firing patterns. The objects could be external
objects such as trees or stones. The objects could be another sensational
complex such as a word image or sound. The higher the of a relational
harmonic system more complex the conscious perception is.

From the principle of physics, we have the following corollary.

\begin{principle}
Corollary $\alpha $Z A relational harmonic biological system evolves to
adapt a conscious being that experiences the conscious content as an effect
of the influence of structure formation due to conscious elements of nature.
\end{principle}

This can be considered due to the influence of conscious elements that
promote structures, that help amplify the consciousness. This can be
considered as a direct effect of the propositions $\alpha $III.1 to $\alpha $%
III.3. The structures needed to create conscious structures that can enjoy
art and music, which have no survival relevance, necessarily have to come
from the consciousness elements of nature. More on this is discussed in the
bookset \cite{BKVI1} and \cite{BKVl3}.

From the Zeroth postulate of physics, we can have the following proposition.

\begin{principle}
Proposition $\alpha $III.6c A conscious relational complex system promotes
an increase in harmonic complexity of relational harmonic structures and is
felt as sensations. An increase is felt as positive emotions such as
happiness and a decrease is felt as negative emotions such as sadness.
\end{principle}

This is evolutionary in nature. When there is the evolutionary possibility
of relational structures, if some of the structures create sensational
complexes that motivate activities that promote survival, they increase in
number. This leads to the biological evolution of advanced neural networks
that have complex mental capabilities, and relevant consciousness-related
features.

Here the harmonic complexity can be considered to mean the aesthetic and/or
the rhythmic quality. These sensations could be fundamental as coming from
the first postulate or evolutionary in nature-based standard physics. Those
systems that have evolutionarily linked to happiness and sadness, promote an
increase in them, by promoting itself through various behavior patterns
promoted by evolutionary adoption. This idea was introduced in Quantum
Gravity Framework 4 \cite{MYP4}. More information regarding this is in my
other publications \cite{GUK}, \cite{BKVI1}, and \cite{BKVl3}.

\section{Modeling the Conscious Matter}

The quantum state of conscious entities has to be macroscopic and have
complex physical configurations. The fact that this quantum state
experiences the perceptions means it is simply more than known matter and
gauge field described by conventional quantum field theory, but something
more than that. This may require new physics that extends the concept of
quantum state and explains issues such as dark matter and dark energy quanta
so that they have sufficient degrees of freedom to fully explain the
perceptual experiences. This quantum state has a conventional degree of
freedom associated with features such as the charge, spin, etc, that will be
interacting with an entirely different space that corresponds to the
perceptual world, where the perceptual entities are elements of the
conscious experience of the conscious entities. The new quantum physics as
associated with state links known matter and gauge states to the perceptual
world through interaction terms and also interacts back with them possibly
through free will. We can consider each person's conscious state as an
excitation of this quantum state with unique quantum numbers with the ground
state of it describing the conscious substratum.

Now I can write a general formula for conscious beings heuristically as
follows:

\begin{equation}
|\Phi >=\dsum\limits_{\alpha _{i}}\pm \dprod\limits_{k}\left\vert \Psi
_{k}(\{\alpha _{i}\})\right\rangle  \label{QCON1}
\end{equation}

\begin{equation}
|\Psi _{k}(\alpha _{i})>=\dsum\limits_{\{\beta _{l}\}}\left\vert \psi
_{k}(\{\alpha _{i}\},\{\beta _{l}\}\right\rangle \left\vert i_{k}(\{\beta
_{l}\}),I_{k}\right\rangle  \label{QCON2}
\end{equation}

The $k$ refers to a particular conscious being. The signs in the summation
depend on the permutation of bosons and fermions. The $\left\vert \psi
_{k}(\{\alpha _{i}\},\{\beta _{l}\},I_{k}\right\rangle $ are the quantum
states of matter $\{\alpha _{i}\}$ and any other composite related quantum
number $\{\beta _{l}\}$. $\left\vert i_{k}(\{\beta
_{l}\}),I_{k}\right\rangle $ is the quantum states of the other $C$ fields
described in section \ref{quaphy} representing conscious information 1-to-1,
which entangled to $\left\vert \psi _{k}(\{\alpha _{i}\},\{\beta
_{l}\},I_{k}\right\rangle $ through the summation $\dsum\limits_{\{\beta
_{l}\}}.$ There would be internal quantum numbers $I_{k}=\{\iota _{k}\}$
that uniquely identify the conscious experience in $C$ fields which is same
as or is coupled to the conscious substratum. The $\{\alpha _{i}\}$ keeps
changing. But the internal variables $I_{k}$ do not change. As I\ said the
nature of the fields that define this state needs to be identified by
further research and development.

\section{Mathematical Modeling of Interaction Binding.}

The consciousness experience is built on coherent information in the energy
interaction at various scales. Mapping coherent information to the conscious
experience is the hard problem of consciousness put forward by David
Chalmers. To do that first we start with various scales. At the highest
scale, coherent information is present in the primary sensory areas. For
example, in the visual cortex, an external image is projected into the
primary visual cortex. The neurons there fire together to identify the
patterns in the image to identify edges, lines, and curves. Each line is
represented by coherent firing of nearby neurons, and in higher areas, this
is used for object recognition. Here the coherent information is the pattern
in the image that is being sensed. Due to the coherent firing of nearby
neurons, the energies exchanged sum up to a higher amplitude. The energy
interactions say at the individual neurons, or atoms, there is quite little
coherent information, if we remove the coherent information from the image
level. So, when the pattern of information at the various scales is
integrated the image information at the highest scale is the most dominant
sensational complex. However, there is other information that the lower
scales may add to the image information at the highest scale, such as color,
depth, direction, etc.

The entire brain is processing multi-modal sensational complex. The coherent
information in the various parts of the brain, visual, tactile, emotional,
etc., combines together as a single experience, as these areas interact
within the feedback loops, and there is interaction binding. To get to the
proper interaction between various parts of the brain to understand the
proper experience of the human brain, we need evolution to do the setup.

To mathematically describe the conscious experience, we can start with the
potentials at each point of the human brain. Let's do this without involving
the issue of the conscious being part. The potential keeps changing due to
the movement of ions between parts of neurons and their outside due to ionic
channels. The variation of these voltages is a clear measure of the energy
interaction among particles and electromagnetic fields. The static electric
field of the particle described by the gradient of the coulomb potential is
quantum mechanically entangled with the location of the particle. The
particles in a brain are superposition of its location states and the
electric field is entangled with it.

Let's mathematically describe the entanglement in simplified terms. Let $%
|P_{i},x>$ be the quantum state of the $i^{th}$ particle along with the
electric field. Let $|i,x>$ be the quantum state of the $i^{th}$ particle
being position x. Let $\psi _{i}(x,t)$ be its position wavefunction as bound
to a molecule or a material. $|EM,x>$ be the quantum state of the electric
field, $A_{a}(y)$ be the vector potential at point $y^{a}$, and $%
E^{a}(x^{b}) $ be the coulomb electric field. Then we have the following.

\begin{equation}
|P_{i},t>=\int \psi _{i}(x^{a},t)(|i,x^{a}>|EM,x^{a}>)d^{3}x^{a}
\end{equation}%
\begin{equation}
E^{a}(x^{b})=\frac{q}{4\pi \varepsilon |x|^{3}}x^{a}
\end{equation}

We can roughly describe $|EM,x>$ in the following form:

\begin{equation}
|EM,x>=\int_{A_{a}(z)}\phi _{A}(\{A_{a}(z)\})e^{i\int
(E^{a}(x^{b}-y^{b}).A_{a}(y))d^{3}x}\dprod\limits_{z}(|A_{a}(z)>dA(z))
\end{equation}

where $\phi _{A}(\{A_{a}(z)\})$ is a Gaussian like wave function of $%
A_{a}(z) $ peaked at a classical field value of $A_{a}(z).$ Here q is the
charge of the particle.

The quantum state of the matter in the brain is as follows:\qquad 
\begin{equation}
|\Psi ,t>=\int_{\{x^{a}\}}\Psi
(\{x_{i}^{a}\},t)\dprod\limits_{x^{a}}(|i,x^{a}>|EM,x^{a}>)d^{3}x^{a}
\end{equation}

$\Psi (\{x_{i}^{a}\},t)$ is the combined state of particles of the whole
brain. You can see they are highly entangled, as they interact. In the next
section, we discuss an abstract model to describe free will using this. Now
we will focus on understanding the integration of sensory information.

Due to complex feedback loops between parts of the brain quantum, states of
the particles are entangled together in a way that helps in integration. Let 
$\phi (x)$ be the net electric potential in the brain related to the
electric field by $-\nabla \phi $ $=$ $\vec{E}$ or charge by coulomb
equation $\Delta \phi =-\frac{\rho }{\varepsilon }$. This represents the
electric energy of the electromagnetic field and particles. The total energy
is constant. The rate of change $f(x)=\frac{d\phi (x)}{dt}$ gives an idea of
energy interaction among fields and particles. Then we can now average out $%
f(x)$ using a smearing function $s_{y}(x)$ to eliminate the details about
variation of potential due at the molecular and neuronal level. Let me call
this as $F(x)$.

\begin{eqnarray}
F(x) &=&\int f(y)s_{y}(x)d^{3}x \\
&=&\int \frac{d\phi (y)}{dt}s_{y}(x)d^{3}x
\end{eqnarray}
We need to multiply this by the average number of particles $n(x)$ engaged
in energy transfer to represent a particular feature of sensory data, such
as an edge in the image captured by the primary cortex. Then we have a
measure $E(x)$ of the coherent energy transfer to describe the sensational
units of the continuous stream of the sensational complex:

\begin{equation}
E(x)=en(x)\int \frac{d\phi (y)}{dt}s_{y}(x)d^{3}x
\end{equation}

How our conscious experience maps to coherent information in the brain
depends on evolution. Nature puts together various physical phenomena in
different molecular material that maps coherent information in its
interaction phenomena to certain sensory experiences, to make the whole
experience. This has occurred through evolution over billions of years, with
complex intricacies. So experimental investigation is the best guide to
finding this map.

$E(x)$ contains physical information that converts to conscious experience. $%
E(x)$ represents entangled interaction information distributed throughout
the brain due to feedback loops. It contains coherently entangled
information about the feelings in reticular formation, auditory, tactile,
visual, etc. This mapping needs to be done in detail through experimental
investigation to complete the theory. Adding to $E(x)$, which may help sense
patterns in black and white, there should be further coherent information
regarding color, taste, touch, etc., present in the neuronal data. Let me
call the later $p_{i}(x)$, the coherent neural physical/chemical information
that captures this information. The physical information in our brain that
contains our sensational complex is $\{x,~E(x),~p_{i}(x)\}$.

The physical coherent information in our brain is represented by $%
\{x,E(x),p_{i}(x)\}$. So hard problem of consciousness is to map $%
\{x,E(x),p_{i}(x)\}$ to $\{y,s_{i}(y),c_{i}(y)\}$. \ The latter is the
conscious information described in section \ref{concon}. The $x$ continuum
points are mapped to the $y$ continuum point by passive binding as proposed
in proposition $\alpha $III.3a.. The other information is mapped by active
binding based on structural complexity $p_{i}(x)$ of components to sensory
modality unit $c_{i}(y)$ as proposed in proposition $\alpha $III.3a. We need
to work out the details through experimental guidance to figure out the
precise mapping between physical and conscious experience.

Let me include the conscious being and write a full expression of the
quantum state of the human brain using equations (\ref{QCON1}) and (\ref%
{QCON2}) as follows:

\begin{equation}
|\Phi >=\dsum\limits_{\{x,E(x),p_{i}(x)\}}\int \pm
\dprod\limits_{k}\left\vert \Psi _{k}(\{x,E(x),p_{i}(x)\})\right\rangle
\end{equation}

\begin{equation}
|\Psi _{k}(\alpha _{l})>=\dsum\limits_{\{x,E(x),p_{i}(x)\}}\int \left\vert
\psi _{k}(\{x,E(x),p_{i}(x)\},\{y,s_{i}(y),c_{i}(y)\}\right\rangle
\left\vert i_{k}(\{y,s_{i}(y),c_{i}(y)\}),I_{k}\right\rangle
\end{equation}

\section{Modelling of Consciousness Free Will Action}

The human brain is a full-scale feedback system, which is continuously
changing what it is processing. This is controlled by upper-level processing
that can shift the focus of the system to certain objects and certain
processing. This change happens continuously, and it is always at a tipping
point of branching to various directions of processing with various
different outcomes. The variables of the neural system, to which this
transition is highly sensitive we can call the control system variables that
we discussed in hypothesis 2.1. Microscopic changes to these variables can
change what is being attended to and processed in the neural system at the
macroscopic level. Quantum superposition of these variables will lead to Schr%
\"{o}dinger cat-like states made of macroscopic superpositions of
configurations. Eventually, these superposition needs to be quantum reduced.
The new macroscopic state is related to the old macroscopic state through
quantum mechanics and classical evolution.

Let's give a formula for how consciousness can influence the system in which
it is present, without involving the conscious being related matters. We can
split the conscious system into three parts: 1) Macroscopic configuration of
system variables $M_{i}$ such as the potential associated with each free
neural component, such as soma, axon, or dendritic branch, 2) Control system
variables: Microscopic variables $c_{i}$ that can influence the direction of
processing, 3)Molecular Background: This includes configuration variables $%
u_{i}$ of molecules of the system that make up the system. The molecular
background evolves through incoherent superposition due to extensive
collisions.

Let the total Hamiltonian of the system be

\begin{equation}
H=H_{M}+H_{c}+H_{u}+H_{Mc}+H_{cu}+H_{Mu}
\end{equation}

$M$ stands for macroscopic, $c$ stands for the control system, and $u$
stands for molecular background. The first three $H^{\prime }s$ on the right
side in the above equation stand for the independent Hamiltonians for these
three and the next three stand for $H^{\prime }s$ that represent the
interaction between these three. In the control system we assume that the $C$
fields of 4.4 are factored in.

Let the state of the system be $|M_{i}>$ at time $t_{n}$. The control system
and molecular system states are in a quantum superposition, and it depends
on $M_{i}.~$The state can be written as

\begin{equation}
|\Psi (t_{n})>=|M_{i}>\int_{c_{j},u_{k}}\Psi
(t_{n},c_{j},u_{k};M_{i})|c_{j},t_{n}>|u_{k},t_{n}>\dprod dc_{j}du_{k}
\end{equation}

In this, we don't include the observer part of the wave function discussed
in the introduction for the sake of simplicity. But free will influence of
observer will be included later in the model in the probability calculations.

After some time, the control system and molecular system states get
entangled with the superposition of macroscopic states.

\begin{equation}
|\Psi (t_{n+1})>=\int_{M_{l}^{\prime },c_{j},u_{k}}\Psi
(t_{n+1},M_{l}^{\prime },c_{j},u_{k};M_{i})|M_{l}^{\prime
}>|c_{j},t_{n+1}>|u_{k},t_{n+1}>\dprod dc_{j}du_{k}
\end{equation}

We can assume that the $C$ fields together with Matter reduces the quantum
state, leading to free will action. Due to quantum reduction, the
macroscopic state goes from $|M_{i}>$ to $|M_{l}^{\prime }>$ with
probabilities expressed in standard quantum physics: 
\begin{equation}
P_{s}(M_{l}^{\prime }|M_{i})=\int_{c_{j},u_{k}}|\Psi (t_{n+1},M_{l}^{\prime
},c_{j},u_{k};M_{i})|^{2}\dprod dc_{j}du_{k}
\end{equation}

This probability defines the influence of computational circuits on the free
will action as discussed in the section on free will. I\ referred to this as
the deterministic bias before. This probability could depend on many factors:

\begin{itemize}
\item The energies of macroscopic state: $E_{M}=\left\langle \Psi
(t_{n})\right\vert H_{M}\left\vert \Psi (t_{n})\right\rangle $

\item The conscious affective reaction factor $R(M_{i},M_{l}^{\prime })$
associated with the transition from $M_{i}$ to $M_{l}^{\prime }$. This is
depends on the $C$ fields of the conscious being discussed before in the
section on conscious beings. It also includes the effect of the $B$-fields
discussed before along with $z$ variables that helps in stochastic reduction.

\item The integration factor that describes how integrated conscious
information is in the conscious system is $I(M).$
\end{itemize}

This stochastic evolution of the neural system can be approximated by
continuous Schr\"{o}dinger evolution $M(t).$ The evolution of the state of
the neural system $M(t)$ is influenced by the combination of deterministic
evolution by semiclassical quantum physics and the influence of microscopic
factors as follows (classically):

\begin{equation}
\frac{dM}{dt}=\{H,M\}+f_{c,}
\end{equation}%
where $f_{c}$ is stochastic force taking into account the influence of
consciousness ($C$ fields + conscious substratum helps in reduction).

Alternatively, the quantum state can be described by the stochastic Schr\"{o}%
dinger equation as follows:

\begin{equation}
d|\Psi (t_{n})>=\frac{i}{\hbar }H_{M}|\Psi
(t_{n})>-\dsum\limits_{i}M_{i}^{\dag }M_{i}|\Psi
(t_{n})>+\dsum\limits_{i}M_{i}z^{i}|\Psi (t_{n})>
\end{equation}%
where $M_{i}$ describes the operators that reduce the system to macroscopic
conscious states. The $z^{i}$ are complex variables of the stochastic Schr%
\"{o}dinger equation that have both stochastic and conscious influence on
the system by $C$ fields and conscious substratum.

We can include a simple bidirectional evolution to the model. This means
writing the quantum state as a function of both negative and positive energy
states. If $t_{n}>t>t_{n+1},$

\begin{eqnarray}
|\Psi (t) &>&= \\
e^{iH(t-t_{n})}\int_{M_{l}^{\prime },c_{j},u_{k}}\Psi
_{+}(t_{n},M_{l}^{\prime },c_{j},u_{k};M_{i})|M_{l}^{\prime }
&>&|c_{j},t>|u_{k},t>\dprod dc_{j}du_{k}+  \nonumber \\
+e^{-iH(t-t_{n})}\int_{M_{l}^{\prime },c_{j},u_{k}}\Psi
_{-}(t_{n},M_{l}^{\prime },c_{j},u_{k};M_{i})|M_{l}^{\prime }
&>&|c_{j},t>|u_{k},t>\dprod dc_{j}du_{k}  \nonumber
\end{eqnarray}

This is subject to the condition,

\begin{eqnarray}
|\Psi (t_{n}) &>&=\int_{c_{j},u_{k}}\Psi (t_{n},M_{l,t_{n}}^{\prime
},c_{j},u_{k};M_{t_{n}})|M_{t_{n}}^{\prime }>|c_{j},t>|u_{k},t>\dprod
dc_{j}du_{k} \\
|\Psi (t_{n+1}) &>&=\int_{c_{j},u_{k}}\Psi (t_{n},M_{t_{n+1}}^{\prime
},c_{j},u_{k};M_{t_{n}})|M_{t_{n+1}}^{\prime }>|c_{j},t>|u_{k},t>\dprod
dc_{j}du_{k},
\end{eqnarray}%
where $M_{t_{n}}^{\prime }$ and $M_{t_{n+1}}^{\prime }$ are quantum states
at instances $t_{n}$ and $t_{n+1}$. We can use these to derive the negative
and positive energy components as discussed in Quantum General Relativistic 
section 2.2 in v2 and v3.

The evolution of the quantum state depends on the future and the past
states. So, the free will actions at $t_{n+1},$ start their influence on the
quantum state at $t_{n}$ itself, indicating retrocausal effects. Many of the
details of how much contribution the retrocausal effect has to the past
depend on how strong the reduction is, and how long the coherence can be
maintained between $t_{n}$ and $t_{n+1}$ in determining the gap length $%
t_{n+1}-t_{n}$. The latter depends on the interaction terms. This depends on
the model I described here and how it relates to the various physical
features of the human brain.

\section{Discussion}

\subsection{The Quantum Measurement Problem}

We discussed the reduction of a quantum state using density matrix formalism
in Postulate I.2 in QGRF 5 \cite{MYP5A}. The density matrix formalism
discussed there reduces the quantum state spontaneously and fundamentally,
as it evolves and keeps it semiclassical. So, there is no quantum
measurement problem. The density matrix is a function of the free
configuration variables of continuum fields and particles on the rest frame
hypersurfaces. It evolves to become more mixed as time goes on. This means
the quantum state of the observer follows one of the quantum states of an
ensemble whose ensemble average of the pure density matrix corresponding to
the quantum states is equal to the density matrix governed by postulate I.2
described in QGRF 5 \cite{MYP5A}.

When an observer's brain or his instruments interact with an external
quantum system, it doesn't directly interact with it. The quantum system
interacts with a measuring instrument, which in turn interacts with the
environment. This decoheres and spontaneously reduces the quantum state of
the system. When the observer observes the quantum system, they are actually
observing the measuring instrument, reduced by the environment. Here there
is no measurement problem.

In postulate II.2c I discussed the action of free will on matter. Other than
the external spontaneous reduction imposed on the observer by physical
processes there is free will imposed probabilities as proposed in this
postulate. Whether/How this happens needs further investigation. The
influence of neural correlates of conscious influence is in infancy. A
thorough investigation of neural correlates of consciousness will shed light
on this. We will assume that this influence is through modifying the outcome
of reduction as stated in postulate II.2c, and the modeling discussed in the
last subsection. These postulates can help in experimental verification and
advancement of the theory.

\subsection{Interpretation of Quantum Mechanics}

Now we can discuss this in the context of the interpretation of quantum
mechanics. The classical mechanics, before the arrival of the quantum
revolution, suggested a mechanical world without free will. The world was
like a fixed clockwork mechanism, with determinism as the defining
principle. The arrival of quantum mechanics brought freedom as it is not
fully deterministic. Many of the founders of quantum mechanics and
philosophers were happy about this. I myself, until I learned quantum
mechanics, felt depressed as the universe looked fixed and rigid with no
place for free will action. But quantum mechanics brought me relief.

To link quantum mechanics to human brain functioning, the best version is
Von Neumann's presentation of quantum theory. There are three parts:

R1: The conscious observer chooses a physical variable to measure, which
relates to the Hermitian operator in quantum mechanics.

R2: The universe is governed by a quantum state.

R3: The universe responds by giving probabilistic results for the physical
variable.

Henry Stapp \cite{HSTAP1}, \cite{HSTAP2} has linked free will to these three
principles. According to Henry Stapp R1 contains the element of free will,
where the observer has the complete choice in determining the free variable
to measure. According to Henry Stapp, the following are elements of free
will:

1.Ability to choose the operator to measure in the R1 process

2.Ability to choose a time to measure and do the measurement

This consistent with the hypothesis 6.2, where the $C$ determines the
observables $O_{x,i}$ measured. According to Henry Stapp, the conscious
observer using these two can control the human brain, for example, the
quantum Zeno effect. However, I will discuss my simpler interpretation of
the role of the dynamics of the human brain to physical principles using the
objective quantum collapse proposal. This is in line with Roger Penrose,
Diosi, and also the proposals in the quantum gravity framework discussed by
me in the quantum gravity framework proposals. Based on this we can simply
assume quantum collapse of physical systems happens when the Schr\"{o}dinger
evolution results in macroscopic superposition. This is Postulate I.3 in
QGRF 5 \cite{MYP5A}. This process was used in the discussion in coming to
discuss the brain-body system.

Henry Stapp considers R1 as a separate form of the brain-body mechanism, and
it subjects the brain-body quantum state to reduction to measure a variable
of its choice and in its chosen time. In the Quantum Reduction Free will
equivalence principle that I have proposed, the R1 processes are considered
as made of R2 and R3. The free choice of R1 is just the R2 and R3 processes
of the brain-body mechanism. When the R1 process happens, that is when the
conscious observer decides what to do, both R2 and R3 happen. The R2 is
involved in the computational process of possible states of measurement and
possible times. The R3 finally decides the state and timing.

\begin{equation}
R1\equiv R2+R3
\end{equation}

Here we can formulate a hypothesis:

\begin{principle}
Hypothesis 13: The R1 capability of human consciousness is the direct result
of R2 and R2 processes involving its behavior.
\end{principle}

This clearly expresses the view of the equation. Further study needs to be
done to theoretically model this and experimentally verify this.

\subsection{Unification and Organization of Human Knowledge}

In \cite{GUK} the merger of physical and social sciences was discussed.
Consciousness is the bridge that links physical and social sciences. On top
of the physics of biological materials, the influence of consciousness acts
to create living things that are conscious and intelligent. For a detailed
discussion of the unification and organization of human knowledge, please
refer to \cite{GUK} and \cite{OHK}.

\subsection{Experimental Verification}

In this paper, \cite{MYP4}, \cite{GUK}, and \cite{BKVl3} I have discussed
the nature of consciousness and how it influences physical processes. The
experimental test of consciousness involves studying various aspects as
discussed in the ideas in this paper.

The first aspect to study is its fundamental nature in promoting harmonic
relational structures. In biological systems, one can calculate the
probabilities of the formation of various structures. We need to look for
the spontaneous generation of structures that are not accounted for by
deterministic evolution (postulate I.1) and spontaneous reduction (postulate
I.2). This will help to find out how consciousness influences structure
formation as stated in Postulates II.1.

The second aspect to study is the role of neural firing patterns and their
relation to quantum processes. If these quantum processes are not accounted
for by postulate I.1, I.2, I.3, and I.4 in QGRF 5 \cite{MYP5A}, then we can
measure the influence of consciousness on quantum processes, and any other
missing piece of physical processes.

We can study the strength of relational structures in various biological
structures and their correlation to features of consciousness as discussed
in the ideas in this paper. Other aspects of consciousness need to be
studied similarly.

There are two parts to the study of the physics of consciousness based on
the ideas of this paper: 1) The standard physics QGRF 5 \cite{MYP5A}, and 2)
Non-standard physics.

The standard physics part focuses on the binding problem, where the sensory
information from the external and internal memory of neural networks is
converted into a united conscious experience. There is quite a bit of
research on this, but they all focus on a particular modality of
consciousness. Still, we don't have the complete picture. We need a clear
understanding of how the energy interaction in the human brain converts into
conscious experience. This requires mathematical modeling of energy
interaction, and then experimental study of which part of this energy
interaction converts into conscious information, and all the binding
processes involved in it.

The second part relates to ideas on the role of consciousness influenced
structure formation, free will modification of quantum probabilities, the
issue of the B and $C$ fields and the conscious being. These four may be
quite closely connected to each other. These four are non-standard physics
and point direction into a search for new physics that is needed to explain
consciousness. The understand the existence of and role of these we need to
thorough study of the first part of the experimental study based on standard
physics. Anything that cannot be explained in the first part may be related
to the second part.

We can use artificial neural networks to study the dynamics of the neural
phenomena and its relation to consciousness. In artificial networks, one can
change the energy interaction between the neurons and connection, and so
understand the influence of this on the neural activity. This could help how
much the average energy interaction influences the neural processes, and
whether there is a free will aspect of conscious binding influencing the
neural firing dynamics.

\section{Conclusion}

In this article, I have summarized the nature of consciousness as discussed
in the various articles in the reference. I have discussed various
hypotheses, definitions, postulates, and propositions to account for time,
and consciousness in this paper. In this paper, we discussed how to
understand the aspects of consciousness using the conceptual formulation in
quantum gravity framework 5 \cite{MYP5A}. The various ideas linking
consciousness to time, measurement, and quantum gravity have to be debated
and studied further. Experimental verification and further study of the
ideas have to be pursued. Further discussions will be discussed in future
papers.


\begin{thebibliography}{999}
\bibitem{MYP5B} Suresh K Maran, The Physical Foundations of Life and
Consciousness 1.x. This is the series of below 3 papers \cite{MYP5B1.1}, 
\cite{MYP5B1.2} and \cite{MYP5B1.3} with incremental updates. Please read
version 1.3 for the latest updates. Previous versions are presented for
reference. All the versions will be available for reference in Academia.edu,
ResearchGate.net and Publications.UniteServe.com and other places listed at
qstaf.com/qgrf5.
\bibitem{RFS} Suresh K. Maran, Rebuilding the foundations of society. To be
published.
\end{thebibliography}
\end{document}